\documentclass[a4paper,10pt]{report}
\usepackage{graphicx}
\usepackage{amsfonts}
\usepackage[mathscr]{eucal} 
\usepackage{caption2} 
\usepackage{psfrag}
\usepackage{fancyhdr,graphicx}
\usepackage{latexsym,amssymb,epsfig}
\usepackage{amsmath,amssymb}

\newcommand{\Lcal}{{\cal{L}}}
\newcommand{\fr}{\frac}
\def\slash{\!\!\!\!/ \ }
\def\slashs{\!\!\!/ \ }

\usepackage{axodraw,color}

\begin{document}

\thispagestyle{empty}
\begin{center}
{\Huge{\textsc MESON ELECTROWEAK\\
\vskip 0.2cm
INTERACTIONS IN MULTICOLOR\\
\vskip 0.6cm
QUANTUM CHROMODYNAMICS}}\\
\vspace{2.5cm}
{\Large{Oscar Cat{\`a} Contreras\\
\vspace{0.5cm}
\emph{Institut de F{\'i}sica d'Altes Energies}}}
\end{center}

\vskip 1.3cm
\begin{figure}[!h]
\centering
\includegraphics[width=2.4in]{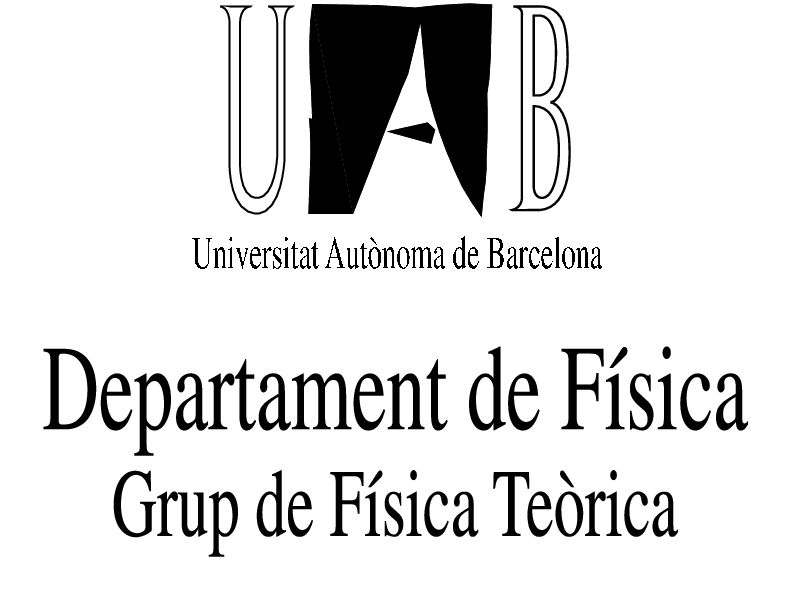}
\end{figure}
\vspace{2.5cm}

\begin{center}
{\Large{
\emph{Universitat Aut{\`o}noma de Barcelona}\\
\vspace{0.5cm}
$\cdot$ July 2005 $\cdot$}}
\end{center}

\newpage
\thispagestyle{empty}
\mbox{ }
\newpage

\pagenumbering{arabic}
\setcounter{page}{1} \pagestyle{fancy} 

\renewcommand{\chaptermark}[1]{\markboth{\chaptername%
\ \thechapter:\,\ #1}{}}
\renewcommand{\sectionmark}[1]{\markright{\thesection\,\ #1}}

\newpage

\addtolength{\headheight}{3pt}    
\fancyhead{}
\fancyhead[RE]{\sl\leftmark}
\fancyhead[RO,LE]{\rm\thepage}
\fancyhead[LO]{\sl\rightmark}
\fancyfoot[C,L,E]{}

\tableofcontents      

\chapter*{Foreword}

This PhD thesis has been the work of 5 years at the IFAE $\&$ F{\'i}sica Te{\`o}rica group of the Universitat Aut{\`o}noma de Barcelona. I would like to thank all my colleagues, which have collaborated to create a warm atmosphere of work and usually also of friendship over this time. Among them I want to mention Xavi Espinal and Xavier Portell, with which I have enjoyed many good times; Rafel Escribano, with whom I have shared the office for the last three years; Jaume, Josevi and Otger, for their help with computers; and all those crazy guys who have joined the group in the last years: Javier Virto, Oriol Romero, Antonio Pic\'on, Javier Redondo, David Diego, Leandro da Rold, ... and the postdocs Carla Biggio, Gabriel Fern\'andez and Ariel M\'egevand.  

Obviously I cannot forget the rest of my friends, who have suffered and understood (I hope) my periods of isolation, especially during the last months. I apologize for not including their names: they know it is not because of forgetfulness.

Special thanks to my PhD advisor, Santi Peris, for the patience, understanding and confidence towards me during these 5 years. Under his tuition I have learned, I am quite sure, more than he believes. At this point I also want to express my gratitude to my former professors Anna Fern\'andez (I really miss her wise advice and extraordinary piano lectures) and Mois\'es Lostao (his enlightening lessons when I was a teenager played a key role in my later decision to study physics).

I would also like to acknowledge the warmth and hospitality I have found abroad in my short stays in Benasque, Frascati, Marseille and Praha. Every time I have been there I felt like being at home. Thank you.

The last two years have brought considerable changes in my personal life. My admiration to my father, brothers and grandmother for how brave they have endured adversity. This PhD thesis is dedicated to the memories of my grandmother, my grandfather and especially to the memory of my mother.  
\newpage
\thispagestyle{empty}
\begin{center}
\end{center}
\newpage
\thispagestyle{empty}
\vspace*{3in}
\hfill
\parbox[c]{5cm}{
\it{To my mother} 
}
\newpage
\thispagestyle{empty}
\begin{center}
\end{center}
\newpage

\chapter{Introduction}

One of the major achievements in theoretical physics in the past 40 years was to find out that not only the electroweak interactions can be properly described through gauge theories, but that the same principle is obeyed by the strong interactions. 
The reason why QCD had for so long eluded theorists' efforts is mainly due to its most peculiar distinguishing feature, {\it{i.e.}}, confinement. Spectroscopy experiments in the 1960's already made it clear that the zoo of new particles discovered at accelerators could not be elementary. The successful Eightfold Way of Gell-Mann and Ne'eman strongly relied on the existence of quarks as constituents, but even for Gell-Mann at that time their physical meaning was more than dubious.
However, in the long run quarks were accepted to be elementary particles, which, endowed with a dynamical $SU(3)_C$ gauge group, build up the so-called QCD Lagrangian, which is purported to yield the proper description of the strong interactions. Confinement is then responsible for the fact that a quark or gluon cannot be detected in isolation, but only colourless combinations of them, {\it{i.e.}}, hadrons. Nowadays we still lack a proof of confinement (we only have some persuasive hints from lattice simulations), but the whole scheme has been thoroughly and successfully tested in several experiments.

The fact that the QCD Lagrangian is expressed in terms of quarks and gluons instead of hadrons makes the determination of physical observables inside the Standard Model a highly non-trivial task. At high energies, asymptotic freedom renders the strong coupling constant small enough to allow the use of perturbation theory methods. However, below 1 GeV, confinement enters the game and binds quarks and gluons together, thus requiring the introduction of non-perturbative techniques. Common strategies are to resort to models (Nambu-Jona-Lasinio, Constituent Quark Model, ...), to rely on numerical simulations (lattice QCD) or to use approximations to QCD (large-$N_c$ QCD).

Among the interesting issues that need to be addressed with the above-mentioned techniques, an oustanding one is to measure the amount of CP violation provided by the Standard Model as compared with the one observed experimentally. Since its discovery more than 50 years ago, the source of CP violation still remains an open field in particle physics. The preferred scenario to test the Standard Model has been for years kaon physics. More recently, the same phenomena has been observed in B physics, opening an era of high precision measurements.

The present work is devoted to determining some of the most relevant parameters that account for CP violation in the Standard Model in a model independent way. Our framework will be restricted to kaon physics, even though our methodology can easily be extended to B physics. Our analysis relies on the use of an {\it{approximation}} to large-$N_c$ QCD (coined {\it{Minimal Hadronic Approximation}}), to deal with the non-perturbative aspects inside kaon matrix elements. Matching between short and long distances is an issue in kaon weak interactions, which our approach solves in a very natural way. A key ingredient therefore is to provide the right matching condition onto the OPE of QCD. This is precisely how scale dependencies are removed from physical calculations, and no spurious cut-off shows up. As far as we know, this is the only existing technique which allows this consistent scale and scheme-independent determination of hadronic parameters.    

The outine of the present work will be as follows: in chapter 2 we give a brief reminder of the Standard Model together with its low energy chiral realizations in the strong and electroweak sectors. 

Chapter 3 introduces the $1/N_C$ expansion and then deals with extensions of the chiral Lagrangian to higher energies through the introduction of explicit resonance fields, so-called Resonance Chiral Lagrangians. We show how quantum corrections to such a Lagrangian can be performed provided we use the power counting rule coming from large-$N_C$ QCD. We compute, as an example, the low energy coupling $L_{10}$ and test thereby the validity of the Lagrangian.

Chapter 4 is devoted to a review of kaon phenomenology and the introduction of the relevant parameters in the study of CP violation. Determination of these phenomenological parameters is the subject of chapter 5 and the main core of the present work: we compute $B_K$ and $g_{27}$ in the chiral limit and to subleading order in the $1/N_c$ expansion. Furthermore, we reassess the determination of the kaon mass difference $\Delta m_K$ and the parameter $\epsilon_K$ in the large-$N_c$ and chiral limits when corrections in inverse powers of the charm mass are taken into account. 

In chapter 6 we move to another subject and deal with the issue of quark-hadron duality violations, which arise due to the fact that the OPE is not a valid expansion in the whole complex $q^2$ plane, but breaks down (at least) in the Minkowski half real axis. Determination of the OPE coefficients using spectral sum rules is then possible up to some inherent uncertainty: the duality violating piece, whose effect is systematically ignored in all determinations of OPE coefficients. There is no theory behind duality violations and one has to resort to models to study them. Armed with a toy model of large-$N_c$ inspiration, we assess the amount of duality violation in a particular QCD two-point Green function, the $<VV-AA>$ correlator, whose OPE condensates are relevant for kaon decays, looking for the best strategy to extract the OPE coefficients reliably. Finally, we end with the conclusions.


\chapter[The Standard Model at Low Energies]{The Standard Model at Low Energies}
 The Standard Model has been on the market for the last 40 years and many good reviews on the subject exist. We will here skip the many details and content ourselves with a brief overview designed to serve as a quick reference.  For more comprehensive accounts we refer the reader to, {\it{e.g.}}, \cite{Pich:2005mk,Novaes:1999yn}.
 
The Standard Model is a Yang-Mills theory invariant under the gauge group 
\begin{equation}
SU(3)_C \times SU(2)_L \times U(1)_Y
\end{equation}
The first factor $SU(3)_C$ accounts for the strong interactions, whose discussion will be the subject of section (2.2). The remaining $SU(2)_L \times U(1)_Y$ factor is responsible for the electroweak interactions. We will first discuss the inner structure of the electroweak interactions, to move afterwards to the strong sector.  

\section{The Electroweak Sector}
The electroweak interactions merge the weak and electromagnetic interactions together under the unified $SU(2)_L\times U(1)_Y$ gauge group. This merging of the two interactions is however highly non-trivial. The notion of {\it{spontaneous symmetry breaking}}, to be introduced in short, is the mechanism by which one can single out and recover both interactions. Before we proceed, it is convenient to expand the Dirac fields for the fermions in the heliciy basis using the helicity projectors,{\footnote{Indeed, it is straightforward to check that they verify
\begin{equation}
P_L+P_R=1\qquad , \qquad P_L \cdot P_R=0 \qquad , \qquad P_{(L,R)}^2=P_{(L,R)}\nonumber
\end{equation}}}
\begin{equation}
P_L \equiv \left(\fr{1-\gamma_5}{2}\right)\quad , \quad P_R \equiv \left(\fr{1+\gamma_5}{2}\right) 
\end{equation}
and introduce the short-hand notation, to be used hereafter,
\begin{equation}\label{projectors}
\psi_L \, \equiv \, P_L\ \psi =\left(\fr{1-\gamma_5}{2}\right)\psi \quad , \quad \psi_R \, \equiv \, P_R\ \psi =\left(\fr{1+\gamma_5}{2}\right)\psi
\end{equation}
The matter content of the Standard Model consists of three families of leptons and three families of quarks, to be organized in three generations as follows
 
\begin{equation}
\left \{
\left (\begin{array}{c}
\nu_e \nonumber\\
e^{-}
\end{array}\right)_L \quad , \quad
e^-_R \quad , \quad
\left (\begin{array}{c}
u \nonumber\\
\tilde{d}
\end{array}\right)_L\quad , \quad
u_R \quad , \quad
\tilde{d}_R \right \}
\end{equation}
\begin{equation}
\left \{
\left (\begin{array}{c}
\nu_{\mu} \nonumber\\
\mu^{-}
\end{array}\right)_L \quad , \quad
\mu^-_R \quad , \quad
\left (\begin{array}{c}
c \nonumber\\
\tilde{s}
\end{array}\right)_L\quad , \quad
c_R \quad , \quad
\tilde{s}_R \right \}
\end{equation}
\begin{equation}\label{matter}
\left \{
\left (\begin{array}{c}
\nu_e \\
\tau^{-}
\end{array}\right)_L \quad , \quad
\tau^-_R \quad , \quad
\left (\begin{array}{c}
t \\
\tilde{b}
\end{array}\right)_L\quad , \quad
t_R \quad , \quad
\tilde{b}_R \right \}
\end{equation}
together with the gauge vector bosons
\begin{equation}
W_{\mu}^+\quad , \quad W_{\mu}^-\quad , \quad Z_{\mu}^0\quad , \quad A_{\mu}
\end{equation}
and the would-be intriguing Higgs scalar boson
\begin{equation}
H
\end{equation}
Equation (\ref{matter}) shows that left-handed fermions are isodoublets of $SU(2)_L$, whereas the right-handed ones are isosinglets. Defining
\begin{equation}
L_e\equiv \left (\begin{array}{c}
\nu_e \nonumber\\
e^{-}
\end{array}\right)_L \quad , \quad L_{\mu}\equiv \left (\begin{array}{c}
\nu_{\mu} \nonumber\\
\mu^{-}
\end{array}\right)_L \quad , \quad L_{\tau}\equiv \left (\begin{array}{c}
\nu_{\tau} \nonumber\\
\tau^{-}
\end{array}\right)_L \quad , \quad
\end{equation}
\begin{equation}
L_u\equiv \left (\begin{array}{c}
u \nonumber\\
\tilde{d}
\end{array}\right)_L \quad , \quad L_c\equiv \left (\begin{array}{c}
c \nonumber\\
\tilde{s}
\end{array}\right)_L \quad , \quad L_t\equiv \left (\begin{array}{c}
t \nonumber\\
\tilde{b}
\end{array}\right)_L \quad , \quad
\end{equation}
\begin{equation}
R_e\equiv e_R^{-} \quad , \quad R_{\nu}\equiv {\nu}_R^{-} \quad , \quad \cdots \quad , \quad R_u\equiv u_R \quad , \quad R_d={\tilde{d}}_R \quad , \quad \cdots
\end{equation}
we can write the free fermion sector of the Standard Model as follows
\begin{equation}\label{dirac}
\Lcal_{Dirac}=i\,\sum_k\,\bigg[\,\bar{L}_k\gamma^{\mu}{\partial}_{\mu}L_k+\bar{R}_k\gamma^{\mu}\partial_{\mu}R_k\,\bigg]
\end{equation}
where $k$ is to be understood as running over all multiplets. Interaction terms can then be read off by invoking the gauge principle. The usual trick is to promote ordinary derivatives to covariant ones, 
\begin{eqnarray}
D_{\mu}^{L}&=&\partial_{\mu}-i\,g \sum_{j=1}^3\fr{\tau^{j}}{2}\, W_{\mu}^{j}-i\,\tilde{g}\,\fr{1}{2}\ Y\,B_{\mu}\nonumber\\
D_{\mu}^{R}&=&\partial_{\mu}-i\,\tilde{g}\,\fr{1}{2}\ Y\,B_{\mu}
\end{eqnarray}
where the superscripts refer to the different chiralities. $\tau^{j}$ are the three generators of the weak isospin group $SU(2)_L$ in the form of the Pauli matrices, listed in Appendix A, and $Y$ is the weak hypercharge, the generator of the $U(1)_Y$ gauge group. $W_{\mu}^{j}$ and $B_{\mu}$ are auxiliary fields, the so-called gauge fields, whose kinematical term is
\begin{equation}\label{kin}
\Lcal_{Kin}=-\fr{1}{4}\, W_{\mu\nu}^{(a)}\, W^{\mu\nu}_{(a)}-\fr{1}{4}\, B_{\mu\nu}\, B^{\mu\nu}
\end{equation}
where the field strength tensors are defined through
\begin{eqnarray}
W_{\mu\nu}^{i}&=&\partial_{\mu}\,W_{\nu}^{i}-\partial_{\nu}\,W_{\mu}^{i}-g\,\varepsilon^{ijk}\,W_{\mu}^{j}\,W_{\nu}^{k}\nonumber\\
B_{\mu\nu}&=&\partial_{\mu}\,B_{\nu}-\partial_{\nu}\,B_{\mu}
\end{eqnarray}
Once expanded, the covariant derivatives in the Dirac Lagrangian (\ref{dirac}) will generate the interactions between quarks and leptons through the gauge bosons. With the use of Noether theorem, we can determine the (conserved) isospin and hypercharge currents, 
\begin{eqnarray}
J_{\mu}^i&=&\sum_{k}\ \bar{L}_k \gamma_{\mu}\,\fr{\tau^i}{2}\,L_k\nonumber\\
J_{\mu}^Y&=&-\sum_{k}\ \bigg[ \bar{L}_k \gamma_{\mu} L_k+2\ \bar{R}_k \gamma_{\mu} R_k\bigg]
\end{eqnarray}
Whereas $\tau^3$ is diagonal, $\tau^1$ and $\tau^2$ are antidiagonal and give rise to charged currents. The interaction Lagrangian for weak charged currents reads
\begin{eqnarray}\label{cc}
\Lcal_{CC}&=&g\ (\,J^{\mu}_1\,W_{\mu}^1+J^{\mu}_2\,W_{\mu}^2\, )\nonumber\\
&=&\fr{g}{\sqrt{2}}\ \bigg(\,J^{\mu}_{+}\ W_{\mu}^{+}+J^{\mu}_{-}\ W_{\mu}^{-}\,\bigg)
\end{eqnarray}
where in the second line diagonalisation has been performed so as to single out each charged current. The new current basis is
\begin{equation}
J^{\mu}_{\pm}=2\,(\,J^{\mu}_{1} \mp i\,J^{\mu}_2\,)
\end{equation}
with their associated gauge bosons given by
\begin{equation}
\left(\begin{array}{c}
W_{\mu}^{+}\\
{}\\
W_{\mu}^{-}\end{array}\right)=\fr{1}{\sqrt{2}}\left(\begin{array}{cc}
1 & -i\\
& \\
1 & +i \end{array}\right)\ \left(\begin{array}{c}
W_{\mu}^1\\
{}\\
W_{\mu}^2\end{array}\right)
\end{equation}
Likewise, neutral currents appear in the Standard Model in the following manner\begin{eqnarray}\label{nc}
\Lcal_{NC}&=&g\,J_3^{\mu}\,W_{\mu}^3+\fr{1}{2}\,\tilde{g}\,J_Y^{\mu}\,B_{\mu}\nonumber\\
&=& J_{\mu}^{em}A^{\mu}+\fr{g}{2\cos{\theta_W}}\,J_{\mu}^{0}\,Z^{\mu}
\end{eqnarray}
Again, in the second line we have rotated to the physical basis,
\begin{equation}
\left(\begin{array}{c}
Z_{\mu}\\
{}\\
A_{\mu}\end{array}\right)=\left(\begin{array}{cc}
\cos{\theta_W} & -\sin{\theta_W}\\
& \\
\sin{\theta_W} & \cos{\theta_W}\end{array}\right)\ \left(\begin{array}{c}
W_{\mu}^3\\
{}\\
B_{\mu}\end{array}\right)
\end{equation} 
where $\theta_W$ is the so-called {\it{electroweak mixing angle}}, defined as
\begin{equation}
\cos{\theta_W}=\fr{g}{\sqrt{g^2+\tilde{g}^2}}
\end{equation}
This way, $A_{\mu}$ can now be interpreted as the photon, since it actually couples to the electromagnetic current
\begin{eqnarray}\label{emcurrent}
J_{\mu}^{em}&=&\sum_{k}\,e\,Q_k\,\bigg[\,\bar{L}_k\gamma_{\mu}L_k+\bar{R}_k\gamma_{\mu}R_k\,\bigg]\nonumber\\&=&J_{\mu}^3+\fr{1}{2}\,J_{\mu}^Y
\end{eqnarray}
The charge of the electron $e$ can be expressed as the harmonic sum of the electroweak couplings
\begin{equation}
\fr{1}{e}=\fr{1}{g}+\fr{1}{\tilde{g}}
\end{equation}
and $eQ_k$ in (\ref{emcurrent}) is the electric charge in units of the electron charge. The neutral current coupled to the $Z_{\mu}$ reads
\begin{equation}
J_{\mu}^{0}=\sum_k\ \bar{\psi_k}\,\gamma_{\mu}\,(\,g_V^k-g_A^k\,\gamma_5\,)\,\psi_k
\end{equation}
\begin{equation}
g_V^k=T_3^k-2\,Q_k\,\sin^2{\theta_W},\qquad g_A^k=T_3^{k}
\end{equation}
where values of the generators for different particles can be extracted from table (2.1). Note that equation (\ref{emcurrent}) assigns the electrical charge to the known particles from knowledge of their weak isospin and weak hypercharge. In terms of Noether charges, it leads to the well-known Gell-Mann-Nishijima relation
\begin{equation}
Q=T^3+\fr{1}{2}\,Y
\end{equation} 
Gathering it all, the Standard Model Lagrangian can be expressed in the concise form
\begin{equation}\label{smlag}
\Lcal_{SM}=\Lcal_{Kin}+\Lcal_{D}+\Lcal_{NC}+\Lcal_{CC}
\end{equation}
with terms given in (\ref{kin}), (\ref{dirac}), (\ref{nc}) and (\ref{cc}), respectively.
\begin{table}
\begin{center}
\begin{tabular}{|c||c|c|c|c|c|c|c|}
\hline
 & $\nu_e$ & $e^-_L$ & $e^-_R$ & $u_L$ & $\tilde{d}_L$ & $u_R$ &$\tilde{d}_R$ \\
\hline \hline
$T$ & 1/2 & 1/2 & 0 & 1/2 & 1/2 & 0 & 0\\
\hline
$T_3$ & 1/2 & -1/2 & 0 & 1/2 & -1/2 & 0 & 0\\
\hline
$Y$ & -1 & -1 & -2 & 1/3 & 1/3 & 4/3 & -2/3\\
\hline
$Q$ & 0 & -1 & -1 & 2/3 & -1/3 & 2/3 & -1/3\\
\hline
\end{tabular}
\end{center}
\caption{{\it{Quantum numbers of the particle content of the Standard Model. For simplicity, only the first generation is shown.}}}\label{tab}
\end{table}
We have not yet talked about mass in the Standard Model. As it stands in (\ref{smlag}), a Yang-Mills theory does not allow a mass term for the gauge bosons, otherwise the gauge symmetry would be broken. This is rather annoying, for we know that all gauge bosons have mass with the exception of the photon, which remains massless. A way out is provided by the Higgs-Kibble mechanism\footnote{See, {\it{e.g.}}, \cite{Kaku}.}.
\subsection{Spontaneous Symmetry Breaking and the Higgs-Kibble mechanism}
Generation of masses in the Standard model is driven by the {\it{spontaneous breaking}} of the gauge group $SU(2)_L\times U(1)_Y$, following the pattern
\begin{equation}
  SU(2)_L \times U(1)_Y\  \stackrel{SSB}{\longrightarrow}\ U(1)_{em}
\end{equation}
Symmetries in quantum field theory can be realized in two ways, depending on whether the Noether charges associated to the symmetries annihilate the vacuum or not. The latter possibility is termed {\it{spontaneous symmetry breaking}}.  
Formally, the simplest way of parametrising our ignorance about the vacuum structure is to postulate the existence of a scalar particle, called the Higgs boson, arranged as a complex isodoublet
\begin{equation}
\Phi=\left( \begin{array}{c}
\phi^+\\
\phi^0
\end{array}\right)
\end{equation} 
and enlarge the Standard Model Lagrangian with an extra piece
\begin{equation}\label{higgsterm}
\Lcal_H=\fr{1}{2}\ (D_{\mu}\Phi)^{\dagger}D^{\mu}\Phi-\mu^2\ \Phi^{\dagger}\Phi+\lambda\ (\Phi^{\dagger}\Phi)^2
\end{equation}
The wrong sign for the mass term shifts the minimum away from $\Phi=0$ and makes the field acquire a vacuum expectation value. Actually, there is a multiplicity of degenerate minima. In order to apply perturbation theory consistently, we have to single out one of the infinitely many degenerate minima. Typically,
\begin{equation}
\langle \Phi \rangle_0 =\left(\begin{array}{c}
 0\\
\fr{v}{\sqrt{2}}
\end{array}\right)\quad , \quad v=\sqrt{\fr{-\mu^2}{\lambda}} 
\end{equation}
The field $\phi$ has to be shifted accordingly for perturbation theory to be valid. This can be done as follows
\begin{equation}
\phi=\fr{v+H(x)}{\sqrt{2}}\ {\textrm{exp}}\,\left[\,i\,\fr{\tau^k}{2}\,\fr{\chi_k (x)}{v}\,\right]\left(\begin{array}{c}0\\
1
\end{array}\right)
\end{equation}
Due to gauge freedom, the $\chi$ fields can be removed from the theory. Their degrees of freedom are then converted to longitudinal modes of the gauge bosons, {\it{i.e.}}, mass terms.
Plugging the expression above to the Higgs Lagrangian (\ref{higgsterm}) we get
\begin{eqnarray}\label{higgslag}
\Lcal_{H}&=&\fr{1}{2}\ (\partial_{\mu}\Phi)^{\dagger}\,\partial^{\mu}\Phi+\fr{g^2}{4}\,(\,v+H\,)^2\left(W_{\mu}^{+}\,W^{\mu\, -}+\fr{1}{2\cos^2{\theta_W}}\ Z_{\mu}Z^{\mu}\right)-\nonumber\\
&&-\mu^2\,\fr{(\,v+H\,)^2}{2}-\lambda\,\fr{(\,v+H\,)^4}{4}
\end{eqnarray}
from which the masses of the gauge bosons can be readily determined to be
\begin{equation}
m_H^2=-2\mu^2=2\lambda v^2
\end{equation}
\begin{equation}
m_W^2=\fr{v^2}{4}\,g^2\sim (80 \, {\mathrm{GeV}})^2
\end{equation}
\begin{equation}
m_Z^2=\fr{v^2}{4}\,\bigg(g^2+\tilde{g}^2\bigg)\sim  (90 \, {\mathrm{GeV}})^2
\end{equation}
This means that the existence of a non-trivial vacuum, to which the $W_{\mu}^+$, $W_{\mu}^-$ and $Z_{\mu}^0$ gauge bosons are not transparent, makes them acquire a mass, whereas the photon remains unaffected by the vacuum and thus massless. The remaining terms in (\ref{higgslag}) are the couplings of the Higgs boson to itself and to the gauge bosons through triple and quartic vertices. 

Explicit fermion mass terms in the Standard Model Lagrangian are also forbidden by gauge symmetry. Fortunately, the Higgs mechanism allows fermion masses to be accomodated into the Lagrangian as the Yukawa couplings to the Higgs field.
The picture, beautiful as it is from a theoretical point of view, is nonetheless rather disappointing on the phenomenological side, because the only thing we know experimentally is the vacuum expectation value,
\begin{equation}
v=\left(\sqrt{2}\ G_F\right)^{\fr{1}{2}}\sim 246 \,\mathrm{GeV} 
\end{equation}
from which the gauge boson masses can be determined. However, neither $\mu$ nor $\lambda$ are known, which means that there is {\it{no prediction}} for the Higgs mass. Hopefully, the LHC will soon dilucidate if vacuum effects can be cast in the form of a Higgs field or more complicated structures are needed.

\subsection{Quark Mixing and CP Violation}
When introducing the quark doublets in (\ref{matter}) we denoted the lower quarks with a tilde without further explanation. The reason for this is that quark fields are not simultaneous eigenstates of the strong and electroweak interactions. This is because the electroweak interactions do not respect the global $SU(3)_F$ flavour symmetry, while strong interactions do, and some ambiguity arises as to the definition of quark fields. Therefore, they can in principle couple as mixtures of these strong eigenstates. It is customary to shift this mixing to the lower quarks in each multiplet, something which can always be done with a proper redefinition of the quark fields. Therefore, the electroweak quark fields are rotated with respect to the strong quark fields as
\begin{equation}
\left(\begin{array}{c}
\tilde{d}\\
\tilde{s}\\
\tilde{b}\end{array}\right)=V_{CKM}\left(\begin{array}{c}
d\\
s\\
b\end{array}\right)
\end{equation}
where $V_{CKM}$ is a unitary matrix known as the {\it{Cabbibo-Kobayashi-Maskawa}} matrix, which can be parameterised with three Euler angles $\theta_1$, $\theta_2$, $\theta_3$ and a phase $\delta$ as follows
\begin{eqnarray}\label{CKM}
V_{CKM}& \doteq &\left(\begin{array}{ccc}
V_{ud} & V_{us} & V_{ub}\\
V_{cd} & V_{cs} & V_{cb}\\
V_{td} & V_{ts} & V_{tb}
\end{array}\right)\nonumber\\
&=&\left(\begin{array}{ccc}
1 & 0 & 0\\
0 & \cos{\theta_2} & \sin{\theta_2}\\
0 & -\sin{\theta_2} & \cos{\theta_2}
\end{array}\right)
\left(\begin{array}{ccc}
\cos{\theta_1} & \sin{\theta_1} & 0\\
-\sin{\theta_1} & \cos{\theta_1} & 0\\
0 & 0 & 1
\end{array}\right)\cdot \nonumber\\
& & \qquad \qquad \qquad \qquad \qquad 
\cdot \left(\begin{array}{ccc}
1 & 0 & 0\\
0 & 1 & 0\\
0 & 0 & e^{i\delta}
\end{array}\right)
\left(\begin{array}{ccc}
1 & 0 & 0\\
0 & \cos{\theta_3} & \sin{\theta_3}\\
0 & -\sin{\theta_3} & \cos{\theta_3}
\end{array}\right)
\end{eqnarray}
Due to the existence of a relative phase between quark fields, the $V_{CKM}$ matrix picks an imaginary part, which is the source of CP violation in the Standard Model\footnote{There is also another source of CP violation, the strong CP term in the QCD Lagrangian, to be discussed in the next section. Its impact is nonetheless negligible, and quark mixing bears the bulk of CP violation in the Standard Model.}. This is unlike what happens if there only existed two families of quarks. The Euler angles are then reduced to just one angle, the well-known $Cabbibo$ angle $\theta_C$. This mixing angle is responsible for the suppression of the bothersome flavour-changing neutral currents (FCNC) through the {\it{Gell-Mann-Illiopoulos-Maiani}} (GIM) mechanism\footnote{At the time of its introduction, only the three light quarks had been postulated. The GIM mechanism spurred the introduction of the fourth (charm) quark, with which the orthogonality of the Cabbibo matrix rules out any flavour-changing neutral current transition. The addition of a third family does not spoil this statement.}. The most important aspect of the two-family model is the fact that the imaginary phase in (\ref{CKM}) becomes global and can be factored out to render a self-adjoint matrix, thus forbidding any violation of CP. Therefore, the presence of the third family has a pivotal role in the origin of CP violation in the Standard Model. The CKM matrix structure (\ref{CKM}) can be simplified recalling that $\theta_2$ and $\theta_3$ are small angles, and to a very good approximation\cite{Buras:98}
\begin{equation}
\sin{\theta_1}\simeq |V_{us}|\, , \quad \sin{\theta_2}\simeq |V_{ub}|\, , \quad \sin{\theta_3}\simeq |V_{cb}|\, , \quad \delta
\end{equation}
It is customary, however, to employ the so-called Wolfenstein parameterization, which is an expansion in the parameter $\lambda\doteq |V_{us}|$
\begin{equation}
V_{CKM}=\left(\begin{array}{ccc}
1-\fr{\lambda^2}{2} & \lambda & A\lambda^3(\rho-i\eta)\\
&&\\
-\lambda & 1-\fr{\lambda^2}{2} & A\lambda^2\\
&&\\
 A\lambda^3(1-\rho-i\eta) & -A\lambda^2 & 1
\end{array}\right)
+{\cal{O}}(\lambda^4)
\end{equation}
Unitarity of the Cabbibo-Kobayashi-Maskawa matrix results in a set of unitarity relations that bind the $CKM$ matrix elements, {\it{e.g.}},
\begin{equation}\label{unitarity}
V_{ud}\ V_{ub}^*+V_{cd}\ V_{cb}^*+V_{td}\ V_{tb}^*=0
\end{equation}

\begin{figure}
\renewcommand{\captionfont}{\small \it}
\renewcommand{\captionlabelfont}{\small \it}
\centering
\psfrag{A}{$\gamma$}
\psfrag{B}{$\alpha$}
\psfrag{C}{$\beta$}
\psfrag{D}{$1-{\bar{\rho}}-i{\bar{\eta}}$}
\psfrag{G}{${\bar{\rho}}+i{\bar{\eta}}$}
\includegraphics[width=2.1in]{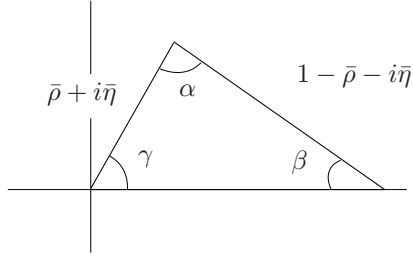}
\caption{The unitarity triangle.}\label{triangle}
\end{figure}

The above relation can be visualized geometrically in terms of the so-called unitarity triangle\footnote{ Out of the number of possible triangles, the one coming from (\ref{unitarity}) is the only one where all sides are of the same order in $\lambda$. This is why people usually refer to it as {\it{the}} unitarity triangle.} of figure (\ref{triangle}). The amount of CP violation is then proportional to the area of the triangle. The strategy in phenomenology of K and B physics is to overconstrain the triangle to check the unitarity of the CKM matrix inside the Standard Model. Should inconsistencies be found, this would signal at physics beyond the Standard Model. 

After this aside to discuss quark mixing in the Standard Model, we complete our description of the Standard Model with the strong interactions.

\section{The Strong Sector}

So far, we have focussed on the Electroweak Interactions. Strictly speaking, one cannot speak of a unification of the Weak Interactions and the Electromagnetic Interactions, in the sense that we do not provide a single coupling constant. However, the picture introduced previously showed a deep intertwining between the two through the Higgs-Kibble mechanism. Likewise, the Strong Interactions could be seen as another patch of the Standard Model, since it comes with its own coupling constant. However, there is a deep non-trivial consistency relation between the three interactions in the cancellation of axial anomalies. Anomalies show up as a manifestation of a classical symmetry which does not survive quantization. In particular, anomalous triangle diagrams involving electroweak gauge bosons could spoil renormalization of the Standard Model. This potentially dangerous contributions are cancelled due to a subtle combined action of quarks and leptons in each of the three generations, provided that quarks appear with three different colours, $N_C=3$. It is this inner consistency and self-dependence that make sense of the word Standard Model.     

Strong interactions\footnote{There is a vast literature on the subject. We refer to \cite{Smilga,Pich:1995ua} for further details.} bind quarks inside the atomic nuclei due to the mediation of gluons, the gauge bosons of the color group. The interaction is described by a Yang-Mills quantum field theory under the non-abelian gauge group $SU(3)_C$ of colour. The Lagrangian of the theory reads,
\begin{equation}\label{qcdlagrangian}
\Lcal_{QCD}=-\fr{1}{4}\ G_{\mu\nu}^{(a)}\,G^{\mu\nu}_{(a)}
+i\,\sum_{q=1}^6\ ({\bar{\psi}}_q)_i\,\bigg(\,\gamma^{\mu}(D_{\mu})^i_j-m_q\,\delta^i_j\,\bigg)\, (\psi_q)^j+\fr{\theta}{32\pi^2}\ G_{\mu\nu}^{(a)}\,{\widetilde G}^{\mu\nu}_{(a)}
\end{equation}
where the second term is the Dirac Lagrangian for the six quark fields, and $D_{\mu}$ is the covariant derivative for the group $SU(3)_C$, defined as
\begin{equation}
(D_{\mu})^i_j=\delta^i_j\ \partial_{\mu}+i\,g_s\,A_{\mu} 
\end{equation}
where $A_{\mu}$ is the gluon field,
\begin{equation}
A_{\mu}=\sum_a \fr{(\lambda^a)^i_j}{2}\,A_{\mu}^{(a)}
\end{equation}
and $\lambda^a$ the Gell-Mann matrices (see appendix A). The first term in (\ref{qcdlagrangian}) is the gluon kinematical term written in terms of the gluon strength field tensor, to wit
\begin{equation}
G_{\mu\nu}^{(a)}=\partial_{\mu}A_{\nu}^{(a)}-\partial_{\nu}A_{\mu}^{(a)}- g_s\,f_{abc}\,A_{\mu}^{(b)} A_{\nu}^{(c)}
\end{equation}
where $g_s$ is the strong coupling constant and $f_{abc}$ the SU(3) structure constants, whose expressions are listed in appendix A. From a mathematical point of view, gluons belong to the adjoint representation of $SU(3)$, while quarks sit in the fundamental representation. From a physical viewpoint, this means that quarks come as triplets of colour, whereas gluons are combinations of colour and anticolour and their number is fixed by the group structure to be $N_c^2-1$. Finally, the third term in (\ref{qcdlagrangian}) is a source of CP violation allowed by symmetry arguments, where the dual field strength tensor ${\widetilde{G}}^{\mu\nu}$ is defined as
\begin{equation}
\widetilde{G}^{\mu\nu}=\fr{1}{2}\ \varepsilon^{\mu\nu\rho\lambda}\,G_{\rho\lambda}\, , \qquad \varepsilon^{0123}=+1
\end{equation}
The parameter $\theta$ is related to the neutron anomalous magnetic moment and happens to be extremely small. Present bounds\footnote{See, for instance, \cite{Smilga}.} give $\theta\lesssim 10^{-10}$. This smallness constitutes the so-called strong CP problem, which still remains an open issue in particle physics.
  
Contrary to what happens in QED, the strong coupling constant is small at high energies. In other words, at high energies quarks are loosely bound and behave as free particles, thus allowing perturbation theory techniques. This crucial observation, coined {\it{asymptotic freedom}}, has been awarded last year's Nobel Prize in Physics, and it explains, in particular, the successes of the parton model. But at low energies, quarks become more and more tightly bound, {\it{i.e.}}, the strong coupling constants increases its value and we enter a non-perturbative regime. This is known as {\it{infrared slavery}} and leads to the notion of {\it{confinement}}. Due to its non-abelian nature, gluons not only interact with the quark families, but they also interact with each other, leaving a very intrincate picture of the strong interactions, which is far from being understood in detail. Confinement makes the quark-gluon picture transform to the hadronic picture we observe in particle accelerators. This leads to a misfortunate situation: we have been able to write down the QCD Lagrangian in terms of quarks and gluons but we do not know how to evolve it to its dual description in terms of the asymptotic hadronic states. Fortunately, effective field theories and symmetry arguments provide a way to progress in that direction. This goes under the name of Chiral Perturbation Theory and will be the subject of the remaining parts of the chapter.

\section{Chiral Perturbation Theory}

\subsection{Effective Field Theories of the Standard Model}

Consider a toy QFT in which, for simplicity, the matter content is restricted to one heavy field $\Psi$ and one light field $\psi$. For concreteness, its Lagrangian would be\footnote{For general reviews on the subject, we refer to \cite{Pich:eff}-\cite{Lepage}.}

\begin{equation}
\Lcal\ (\psi,\partial_{\mu} \psi;\,\Psi, \partial_{\mu}\Psi)=\sum_k \, g_k\,\, {\cal{O}}_k(\psi, \Psi)
\end{equation}
At energy scales below the heavy mass threshold, only $\psi$ is dynamical, while $\Psi$ is frozen. At the level of the Lagrangian, this can be implemented by integrating out the heavy degree of freedom in the path integral sense,
\begin{eqnarray}\label{genfunc}
Z & = &
\int\mathcal{D}\psi\ \mathcal{D}\Psi\,\,{\mathrm{exp}}\left[i\int\,dx^4\,\Lcal\ (\psi,\partial_{\mu} \psi;\,\Psi, \partial_{\mu}\Psi)\right]\nonumber\\
& = & \int\mathcal{D}\psi\,\,{\mathrm{exp}}\left[i\int\,dx^4\,\Lcal_{eff}\,(\psi,\partial_{\mu} \psi)\right]
\end{eqnarray}
The generating functional (\ref{genfunc}) is obviously left invariant, but this is not so for the Lagrangian. Formally, what one has is
\begin{equation}\label{inf}
\Lcal_{eff}(\psi)=\sum_k \, c_k\,(m_{\Psi})\ {\cal{O}}_k\,(\psi,\partial_{\mu}\psi)
\end{equation}
which can be thought of as a series expansion in inverse powers of the heavy mass. All the information of the heavy degrees of freedom is encapsulated in the new coupling constants $c_k(m_{\Psi})$, whereas the operators are made up with the (dynamical) light field alone\footnote{See, for instance, \cite{Appel}. For a more general discussion, we refer to \cite{John}.}. Whereas symmetry requirements suffice to constrain the form of the (infinitely many) operators, the low-energy couplings are to be determined through a matching condition to the original theory. This ensures that physics is the same in the original and the effective Lagrangian as long as one stays below the heavy mass threshold. The advantage of working with an effective field theory is that the integration of non-dynamical degrees of freedom cleans calculational efforts from unnecessary complications.

This quite simple idea can be shown to be consistent once it is provided with a criteria to organize the (infinite) operators that come out in (\ref{inf}). This is what is known as a {\it{power counting rule}}. Predictability is then ensured even though the theory is intrinsically non-renormalizable: with a proper renormalization procedure, all ultraviolet divergences can be absorbed in the couplings of higher dimensional operators in a controlled manner. In an effective field theory, renormalizability is ensured order by order.

In the case of QCD, the mass gap between the lowest-lying pion octet and the first resonances of the spectrum ($\rho(770)$, $a_1(1260)$, ...) induced by chiral symmetry is of the outmost importance, since it allows an effective field theory treatment. There is, however, an additional sublety to the example shown above: at low energies the relevant degrees of freedom change due to confinement, and one is forced to change language and speak in terms of mesons instead of the fundamental quark and gluon fields. Knowing the relation between the two sets of variables would be tantamount to solving QCD: we would know how quarks and gluons assemble to form hadrons, something which seems quite out of reach by now. Fortunately, symmetry requirements alone provide a great deal of information.
\subsection{Chiral Symmetry}
Out of the six quark flavours in nature, at energies below the charm mass threshold only three of them are dynamically active, namely $u,d,s$, while the remaining decouple\footnote{For more comprehensive accounts of chiral symmetry, we refer to \cite{Ecker}-\cite{Coll}.}. This is motivated by the separation of scales in the quark mass spectrum. Therefore, at low energies, only the light quarks are the relevant degrees of freedom. Using the projector basis of (\ref{projectors}) we can split the Dirac term of the QCD Lagrangian as follows
\begin{eqnarray}
{\bar{\psi}}\,\gamma^{\mu}D_{\mu}\,\psi&=&{\bar{\psi}}_L\,\gamma^{\mu}D_{\mu}\,\psi_L+{\bar{\psi}}_R\,\gamma^{\mu}D_{\mu}\,\psi_R\nonumber\\
m\,{\bar{\psi}}\psi &=& m\,{\bar{\psi}}_R\,\psi_L+m\,{\bar{\psi}}_L\,\psi_R
\end{eqnarray}
and rewrite the QCD Lagrangian in the following fashion
\begin{equation}
\Lcal_{QCD}=-\fr{1}{4}\ G_{\mu\nu}^{(a)}\, G^{\mu\nu}_{(a)}+ i\,\sum_{q=1}^3\, \bigg [\,{\bar{q}}_L\,\gamma^{\mu}D_{\mu}\,q_L+\,{\bar{q}}_R\,\gamma^{\mu}D_{\mu}\,q_R-m_q\,({\bar{q}}_R\,q_L+{\bar{q}}_L\,q_R) \bigg]
\end{equation}
The first terms can be shown to enjoy a {\it{global}} $U(3)_R^f \times U(3)_L^f$ flavour symmetry, since left-handedness and right-handedness do not mix. This symmetry is explicitly broken by the mass term, which connects right and left chiralities. However, since the light quark masses are small, this breaking is very soft and one can treat the mass term as a perturbation. Therefore, to a very good approximation, the QCD Lagrangian for three flavours is chiral invariant. This $U(3)_R^f \times U(3)_L^f$ can be decomposed into the subgroup factors $U(1)_V \times SU(3)_R^f \times SU(3)_L^f$ times an axial coset $U(1)_A$.

According to Noether's theorem, there should exist 8 conserved currents for each handedness, $Q_L$ and $Q_R$, together with $Q_V$ and $Q_A$. $Q_V$ expresses nothing but baryon number conservation and $Q_A$ was long ago shown to be broken at the quantum level due to the chiral anomaly.  We are thus left with the 16 generators of the so-called chiral group $SU(3)_R^f \times SU(3)_L^f$. Continuous symmetries, however, can be realized in two ways, depending on its response to the presence of the vacuum. We already commented on that when discussing the electroweak sector. If the vacuum is symmetric, then the currents annihilate it
\begin{equation}
Q_i\,|\,0>\ =\ 0
\end{equation}
and, following Coleman's theorem, there should exist degenerate parity multiplets in the spectrum. On the other hand, if the operator does not annihilate the vacuum
\begin{equation}\label{Nambu}
Q_i\,|\,0>\ \neq\ 0
\end{equation}  
the symmetry is said to be {\it{spontaneously broken}}. The multiplets are no longer degenerate and, according to Goldstone's theorem, a set of massless, spinless particles, as many as broken symmetries, appear in the spectrum. Furthermore, their parity and internal quantum numbers are the same as those of the broken generators. 

We happen to live in an asymmetric vacuum, as pointed at by the experimental fact that there are no parity multiplets in the QCD spectrum: axial-vector multiplets have higher energies than their vector partners. What remains to dilucidate is the precise pattern of symmetry breaking.
Starting from the chiral group $SU(3)_R \times SU(3)_L$, Vafa and Witten showed that the subgroup $SU(3)_V$ could not be spontaneously broken by the QCD vaccum \cite{Vafa}. The remaining coset, $SU(3)_A$, should therefore be spontaneously broken, so that the accepted pattern of symmetry breaking in QCD is
\begin{equation}
G\,\equiv\, SU(3)_R \times SU(3)_L\ \stackrel{SSB}{\longrightarrow}\ SU(3)_V\,\equiv\, H
\end{equation}
The pion\footnote{Under pion multiplet we are actually referring to the {\it{full}} multiplet, which consists of the $\pi$ states but also of the $K$ and $\eta$ particles.} multiplet is interpreted as the 8 Goldstone bosons of the theory, one for every $SU(3)_A$ generator. This explains why pions are lighter than they should be according to their quark content: chiral symmetry protects them from acquiring mass.

An immediate consequence of (\ref{Nambu}) is that 
\begin{equation}
\langle\, 0\,|\,[\,Q,{\cal{O}}\,]\,|\,0\,\rangle\ \neq\ 0
\end{equation}
where ${\cal{O}}$ is an operator which does not commute with Q. This quantity, signalling at spontaneous symmetry breaking, is commonly called an {\it{order parameter}}. Clearly, according to our previous discussion, ${\cal{O}}$ must be a pseudoscalar operator. The lowest dimension one is ${\cal{O}}^a={\bar{q}}\gamma_5\lambda^a\,q$, which yields
\begin{equation}
\langle\, 0\,|\,[\,Q_A^a,{\cal{O}}^b\,]\,|\,0\,\rangle =-\fr{1}{2}\ \langle\,0\,|\,{\bar{q}}\, \{\lambda^a,\lambda^b\} \,q\,|\,0\,\rangle=-\fr{2}{3}\ \delta^{ab}\,\langle\,0\,|\,{\bar{q}}\,q\,|\,0\,\rangle
\end{equation}
which means that the quark condensate is the natural order parameter of chiral symmetry breaking\footnote{For an alternative view, see \cite{Stern}.}. 
Our purpose is to find out the effective theory of the strong interactions in terms of the dynamics of the pseudo-Goldstone bosons. From a group point of view, they belong to the coset $G/H$. Let us parametrize them as $\phi^a$ under the representation 
\begin{equation}
\xi(\phi)=\bigg(\,\xi_L(\phi),\xi_R(\phi)\,\bigg)
\end{equation}
It can be readily shown that $\xi(\phi)$ satisfies the following transformation rule under a chiral rotation
\begin{eqnarray}
\xi_L(\phi)\stackrel{G}{\longrightarrow} g_L\ \xi_L(\phi)\ h^{\dagger}(\phi,g)\, &,& \xi_R(\phi)\stackrel{G}{\longrightarrow} g_R\ \xi_R(\phi)\ h^{\dagger}(\phi,g)\\
&&\nonumber\\
g=(\,g_L,g_R\,)\,\,\, \in\,\,\, G\, , \qquad &h(\phi,g)\,\,\, \in\,\,\, H\,& , \qquad  \xi(\phi)=\bigg(\,\xi_L(\phi),\xi_R(\phi)\,\bigg)\,\,\, \in\,\,\, G/H\nonumber
\end{eqnarray}
For simplicity it is convenient to work with the combination $U(\phi)=\xi_R(\phi)\,\xi_L^{\dagger}(\phi)$. The transformation rule is then independent of $h$, to yield
\begin{equation}
U(\phi)\stackrel{G}{\longrightarrow} g_R\ U(\phi)\ g_L^{\dagger}
\end{equation}
Obviously, physical results will not depend on the chosen representation. However, linear representations are known to lead to the appearance of extra particles. Since we only want to describe Goldstone boson dynamics, the only requirement we impose is that $U$ be a non-linear realization. The conventional choice is to take $\xi_R(\phi)=\xi_L^{\dagger}(\phi)=u(\phi)$ and 
\begin{equation}
U(\phi)=u(\phi)^2={\textrm{exp}}\,\left\{\fr{i}{F_0}\sum_a\,{\phi_a\,\lambda^a}\right\}
\end{equation}
where the Goldstone bosons are collected in a $SU(3)$ multiplet as follows
\begin{equation}
\phi_a\frac{\lambda^a}{\sqrt{2}}={\bf{\Phi}}(x)=\left(\begin{array}{ccc}
\frac{1}{\sqrt{2}}\pi^0+\frac{1}{\sqrt{6}}\eta_8 & \pi^{+} &
K^{+}\\
&&\\
\pi^{-} & -\frac{1}{\sqrt{2}}\pi^0+\frac{1}{\sqrt{6}}\eta_8 &
K^{0}\\
&&\\
K^{-} & \bar{K}^{0} & -\frac{2}{\sqrt{6}}\eta_8
\end{array}\right)
\end{equation}
Armed with all these definitions, we are ready to write down the Chiral Lagrangian, understood as the most general chiral-symmetric Lagrangian in terms of $U$ and its derivatives. In terms of the effective action, we have to solve for the following   
\begin{eqnarray}\label{effaction}
Z & = &
\int\mathcal{D}q\,\mathcal{D}\bar{q}\,\mathcal{D}{G}_{\mu}\,{\mathrm{exp}}\bigg[i\int\,dx^4\,\mathcal{L}_{QCD}^0\bigg]\nonumber\\
& = & \int\mathcal{D}U\,{\mathrm{exp}}\bigg[i\int\,dx^4\,\mathcal{L}_{eff}(U,D_{\mu}U)\bigg]
\end{eqnarray}
This was the method employed by Gasser and Leutwyler \cite{Gass,Gass1} to unveil the form of the chiral lagrangian order by order in the mass and momentum expansion. Since we eventually would like to compute QCD Green functions, we have to make sure that chiral Ward identities are fulfilled. This can be accomplished most easily with the inclusion of external sources in the QCD Lagrangian and afterwards imposing local gauge invariance. The QCD Lagrangian with the addition of these auxiliary fields now looks like
\begin{eqnarray}
{\cal{L}}_{QCD}&=&{\cal{L}}_{QCD}^0+ J_if^i\nonumber\\
 &=&{\cal{L}}_{QCD}^0+\bar{q}\,\gamma^{\mu}(v_{\mu}+\gamma^5a_{\mu})\,q-\bar{q}\,(s-i\gamma^5p)\,q
\end{eqnarray}
where $v_{\mu}$, $a_{\mu}$, $s$ and $p$ stand for vector, axial-vector, scalar and pseudoscalar external sources, respectively, defined to be the following $SU(3)$ elements, 
\begin{equation}\label{extsour}
v_{\mu}=\sum_a\, \fr{\lambda_a}{2}\ v_{\mu}^{(a)}\, , \qquad a_{\mu}=\sum_a\, \fr{\lambda_a}{2}\ a_{\mu}^{(a)}\, , \qquad s=s_0+\sum_a\, \lambda_a\,s^{(a)}\, , \qquad p=p_0+\sum_a\, \lambda_a\,p^{(a)}\, \qquad 
\end{equation} 
The QCD Green functions are then computed as derivatives of the generating functional with respect to the external fields around the point
\begin{equation}
v_{\mu}=a_{\mu}=p=0\, , \qquad s={\textrm{diag}}\,(m_u,m_d,m_s) 
\end{equation}
Imposing {\it{local}} chiral symmetry induces the following transformation rules for the external fields
\begin{eqnarray}
{\hat{q}}\,({\hat{x}})&=&(g_R\,q_R+g_L\,q_L)\,(x)\nonumber\\
{\hat{r}}_{\mu}\,({\hat{x}})\equiv ({\hat{v}}_{\mu}+{\hat{a}}_{\mu})\,({\hat{x}})&=&(g_R\,r_{\mu}\,g_R^{\dagger}+ig_R\,\partial_{\mu}g_R^{\dagger})\,(x)\nonumber\\
{\hat{l}}_{\mu}\,({\hat{x}})\equiv ({\hat{v}}_{\mu}-{\hat{a}}_{\mu})\,({\hat{x}})&=&(g_L\,l_{\mu}\,g_L^{\dagger}+ig_L\,\partial_{\mu}g_L^{\dagger})\,(x)\nonumber\\
({\hat{s}}+i\,{\hat{p}})\,({\hat{x}})&=&(g_R\,[\,s+i\,p\,]\,g_L^{\dagger})\,(x)\nonumber\\
({\hat{s}}-i\,{\hat{p}})\,({\hat{x}})&=&(g_L\,[\,s-i\,p\,]\,g_R^{\dagger})\,(x)
\end{eqnarray}
In principle, external sources are formal fields whose importance is to enforce Ward identities. However, they can also be thought of as background fields when one switches on the electroweak interactions. Then, the external sources (\ref{extsour}) can be identified with physical fields via
\begin{eqnarray}
r_{\mu}&=&-e\,Q\,(A_{\mu}-\tan{\theta_W}\,Z_{\mu})\nonumber\\
l_{\mu}&=&-e\,Q\,A_{\mu}-\fr{e}{\sin{\theta_W}\cos{\theta_W}}\ \left(-Q\sin^2{\theta_W}+Q-\fr{1}{6}{\mathrm{1}}\right)\,Z_{\mu}-\nonumber\\
&&\qquad \qquad \qquad  -\fr{e}{\sqrt{2}\sin{\theta_W}}\ \left[\,Q_L^{(+)}\,W_{\mu}^{(+)}+\,Q_L^{(-)}\,W_{\mu}^{(-)}\,\right]\nonumber\\
s+ip&=&\left(\begin{array}{ccc}
m_u & 0 & 0\\
0 & m_d & 0\\
0 & 0 & m_s
\end{array}\right)\equiv {\cal{M}}
\end{eqnarray}
where $Q$ is the matrix of the electric charge of quarks
\begin{equation}
Q={\mathrm{diag}}\left(\fr{2}{3},-\fr{1}{3},-\fr{1}{3}\right)
\end{equation}
and $Q_L$ contains the flavour changing Kobayashi-Maskawa matrix elements, to wit
\begin{equation}
Q_L^{(+)}=\left(\begin{array}{ccc}
0 & V_{ud} & V_{us}\\
0 & 0 & 0\\
0 & 0 & 0
\end{array}\right)\, , \qquad 
Q_L^{(-)}=\left(\begin{array}{ccc}
0 & 0 & 0\\
V_{ud}^* & 0 & 0\\
V_{us}^* & 0 & 0
\end{array}\right)
\end{equation}
\subsection{The Chiral Expansion}
The effective action (\ref{effaction}) can be expanded in powers of momenta and masses, once power counting rules are provided for all parameters,
\begin{eqnarray}
U & \qquad & {\mathcal{O}}(p^0)\nonumber\\
v_{\mu}\ ,\ a_{\mu} & \qquad & {\mathcal{O}}(p^1)\nonumber\\
s\ ,\ p & \qquad & {\mathcal{O}}(p^2)
\end{eqnarray}
At lowest order, ${\cal{O}}(p^2)$, we obtain 
\begin{equation}
\mathcal{L}_2=\frac{F_0^2}{4}\ \langle\,
D_{\mu}U^{\dagger}D^{\mu}U+U^{\dagger}\,\chi+\chi^{\dagger}\,U\,\rangle
\end{equation}
where the covariant derivatives are defined, as usual, by
\begin{eqnarray}
D_{\mu}U&=&\partial_{\mu}U-i\,r_{\mu}U+i\,U\,l_{\mu}\nonumber\\
D_{\mu}U^{\dagger}&=&\partial_{\mu}U+i\,U^{\dagger}r_{\mu}-i\,l_{\mu}U^{\dagger}
\end{eqnarray}
and the matrix $\chi$ is usually parametrized as
\begin{equation}
\chi=2\,B_0\,(s+i\,p)
\end{equation}
Thus, at leading order in the chiral expansion we are left with only two low-energy constants, $F_0$ and $B_0$, which can be related to QCD parameters through a matching procedure, to wit
\begin{equation}
F_0=-i\,\fr{p_{\mu}}{\sqrt{2}p^2}\,\langle\, 0\,|\,\fr{\delta  \Lcal_2}{\delta a_{\mu}}\,|\,\pi^+(p)\,\rangle\ =\ f_{\pi}\sim 93.3\ {\textrm{MeV}}
\end{equation}
\begin{equation}
B_0=\fr{1}{F_0^2}\,\langle\, 0\,|\,\fr{\delta  \Lcal_2}{\delta s}\,|\,0\,\rangle\ =\ -\fr{1}{F_0^2}\,\langle\, 0\,|\,{\bar{q}}\,q\,|\,0\,\rangle\ , \qquad \langle\, 0\,|\,{\bar{q}}\,q\,|\,0\,\rangle(2\,{\textrm{GeV}})\sim -[(\ 280 \pm 30)\ {\textrm{MeV}}]^3 
\end{equation}
$F_0$ can therefore be identified with the pion decay constant, and the parameter $B_0$ is proportional to the quark condensate, which takes into account the effect of non-vanishing quark masses.
This very first term was already written down by Weinberg \cite{Weinberg}, and to lowest order it reproduces the current algebra results. At ${\cal{O}}(p^4)$, one has ten low energy couplings $L_i$ plus two contact terms $H_i$, to wit \cite{Gass1}
\begin{eqnarray}\label{chiralexp4}
\mathcal{L}_4 & = & L_1\,\langle
D_{\mu}{U}^{\dagger}D^{\mu}U\rangle^2+L_2\,\langle
D_{\mu}{U}^{\dagger}D_{\nu}U\rangle\,\langle
D^{\mu}{U}^{\dagger}D^{\nu}U\rangle+\nonumber\\ & + &
L_3\,\langle D_{\mu}{U}^{\dagger}D^{\mu}U\,
D_{\nu}{U}^{\dagger}D^{\nu}U\rangle+L_4\,\langle
D_{\mu}{U}^{\dagger}D^{\mu}U\rangle\,\langle{U}^{\dagger}\chi+{\chi}^{\dagger}U\rangle+\nonumber\\
& + & L_5\,\langle
D_{\mu}{U}^{\dagger}D^{\mu}U\,({U}^{\dagger}\chi+{\chi}^{\dagger}U)\rangle+L_6\,\langle\,{U}^{\dagger}\chi+{\chi}^{\dagger}U\rangle^2+\nonumber\\
& + &
L_7\,\langle\,{U}^{\dagger}\chi-{\chi}^{\dagger}U\rangle^2+L_8\,\langle\,{\chi}^{\dagger}U{\chi}^{\dagger}U+{U}^{\dagger}\chi{U}^{\dagger}\chi\rangle-\nonumber\\
& - & i\,L_9\,\langle
F_R^{\mu\nu}D_{\mu}UD_{\nu}{U}^{\dagger}+F_L^{\mu\nu}D_{\mu}{U}^{\dagger}D_{\nu}U\rangle+L_{10}\,\langle\,{U}^{\dagger}F_R^{\mu\nu}UF_{L\mu\nu}\rangle+\nonumber\\
  & + & H_1\,\langle
  F_{R\mu\nu}F_R^{\mu\nu}+F_{L\mu\nu}F_L^{\mu\nu}\rangle+H_2\,\langle{\chi}^{\dagger}\chi\rangle
\end{eqnarray}
where the chiral field strengths are defined as follows,
\begin{equation}
F_L^{\mu\nu}=\partial^{\mu}l^{\nu}-\partial^{\nu}l^{\mu}-i\,[l^{\mu},l^{\nu}];\qquad
F_R^{\mu\nu}=\partial^{\mu}r^{\nu}-\partial^{\nu}r^{\mu}-i\,[r^{\mu},r^{\nu}]
\end{equation}
As already emphasised, effective field theories have to be provided with a {\it{power counting rule}}. At ${\mathcal{O}}(p^d)$, the diagrams that contribute are dictated by
\begin{equation}
d=2+\sum_n\ N_n\ (n-2)+2\ N_L\, , \qquad n\ =\ 2,4,6,\dots
\end{equation}
where $N_n$ is the number of vertices coming from ${\mathcal{O}}(p^n)$ operators, and $N_L$ is the number of loops.
However, the drastic increase in the number of operators as one goes to higher orders in the chiral expansion, together with the poorly-known associated low-energy couplings, makes one stop already at ${\mathcal{O}}(p^4)$ and only in certain observables to go to ${\mathcal{O}}(p^6)$.

As stated in our discussion of effective field theories, the low energy couplings $L_i$ should absorb the divergences of the theory at ${\mathcal{O}}(p^4)$ to guarantee the renormalisability of the theory up to that order. Using dimensional regularization and the Gasser-Leutwyler renormalization scheme \cite{Gass1},
\begin{eqnarray}
L_i&=&L_i^r(\mu)+\frac{\Gamma_i}{32\pi^2}\left(\frac{1}{\epsilon}-\log{4\pi}+\gamma-1+\log{\mu^2}\right)\nonumber\\
H_i&=&H_i^r(\mu)+\frac{\tilde{\Gamma}_i}{32\pi^2}\left(\frac{1}{\epsilon}-\log{4\pi}+\gamma-1+\log{\mu^2}\right)
\end{eqnarray}
and evolution under the renormalization group is then given by
\begin{equation}
L_i^r(\mu_2)=L_i^r(\mu_1)+\frac{\Gamma_i}{16\pi^2}\log{\left(\frac{\mu_1}{\mu_2}\right)}
\end{equation}
where $\Gamma_i$ take the values listed below
\begin{equation}
\begin{array}{cccccccc}
\Gamma_1=\frac{3}{32}&\qquad&\Gamma_5=\frac{3}{8}&\qquad&\Gamma_9=\frac{1}{4}\\
&&&&\\
\Gamma_2=\frac{3}{16}&\qquad&\Gamma_6=\frac{11}{144}&\qquad&\Gamma_{10}=-\frac{1}{4}\\
&&&&\\
\Gamma_3=0&\qquad&\Gamma_7=0&\qquad&\tilde{\Gamma_1}=-\frac{1}{8}\\
&&&&\\
\Gamma_4=\frac{1}{8}&\qquad&\Gamma_8=\frac{5}{48}&\qquad&\tilde{\Gamma_2}=\frac{5}{24}
\end{array}
\end{equation}
Since we ignore the details of the underlying theory (QCD), the matching procedure to determine the low-energy couplings cannot be performed in an analytical way, and one has to resort to approximations or experimental data, when available. Their values, as determined from experiment, are listed in table (2.2).
\begin{table}\label{li}
\begin{center}
\begin{tabular}{|c|c|c|}
\hline
$L_i$ & Experimental value ($10^3 \cdot L_i(M_{\rho})$) & source\\
\hline \hline
$L_1$ & $0.7\pm0.3$ & $K_{e4}$, $\pi\pi\rightarrow \pi\pi$ \\
\hline
$L_2$ & $1.4\pm0.3$ & $K_{e4}$, $\pi\pi\rightarrow \pi\pi$ \\
\hline
$L_3$ & $-3.5\pm1.1$ & $K_{e4}$, $\pi\pi\rightarrow \pi\pi$ \\
\hline
$L_4$ & $-0.3\pm0.5$ & Zweig rule \\
\hline
$L_5$ & $1.4\pm0.5$ & $F_K/F_{\pi}$ \\
\hline
$L_6$ & $-0.2\pm0.3$ & Zweig rule \\
\hline
$L_7$ & $-0.4\pm0.15$ & Gell-Mann-Okubo, $L_5$, $L_8$ \\
\hline
$L_8$ & $0.9\pm0.3$ & $M_{K^0}-M_{K^+}$, $L_5$ \\
\hline
$L_9$ & $6.78\pm0.15$ & $\langle r^2\rangle_V^{\pi}$ \\
\hline
$L_{10}$ & $-5.13\pm0.19$ & $\pi\rightarrow e\nu\gamma$ \\
\hline
\end{tabular}
\end{center}
\caption{{\it{Values of the low-energy coupling constants appearing in $\Lcal_4$ (\ref{chiralexp4}) \cite{Sui1,Perrot,Gol,Davier:1} and references therein. See also \cite{Amoros:2000,Anan:2001}.}}}
\end{table}
So far, we have stressed that one key ingredient for building up chiral perturbation theory is the existence of a mass gap. However, we have not yet identified the expansion parameter, {\it{i.e.}}, the scale $\Lambda_{\chi}$ which makes the expansion in momenta and masses
\begin{equation}
\fr{p^2}{\Lambda_{\chi}}\ll 1 \quad , \quad \fr{m_i}{\Lambda_{\chi}}\ll 1
\end{equation}
meaningful. According to our discussion of effective field theories, the natural order of $\Lambda_{\chi}$ is that of the last integrated-out degrees of freedom, {\it{i.e.}}, the first hadronic multiplet: $\rho(770)$, $a_1(1260)$, ... Therefore, one expects
\begin{equation}
\Lambda_{\chi}\simeq 1\, {\mathrm{GeV}}
\end{equation}
This means that, even though the light quark masses are rather different
\begin{equation}
\fr{m_u}{m_d}\ \sim\  0.5\ , \qquad \fr{m_s}{m_d}\ \sim\ 20 
\end{equation}
convergence, at least for two flavours, is good
\begin{equation}
\fr{m_u}{\Lambda_{\chi}}\ \simeq\ \fr{m_d}{\Lambda_{\chi}}\ \lesssim\ \fr{m_s}{\Lambda_{\chi}}
\end{equation} 
Above that scale $\Lambda_{\chi}$, new fields can become dynamical and have to be included in the form of new operators. Such an attempt was done in \cite{Sui1}, which proposed a Lagrangian made of the first resonance multiplets in the vector, axial, scalar and pseudoscalar channels. We will discuss this approach in detail in the next chapter. 

The Chiral Lagrangian terms of (\ref{chiralexp4}) have an additional {\it{even parity}} symmetry which QCD does not possess. This is so because in going from the first line to the second in (\ref{effaction}) attention has to be paid to anomalies, which manifest themselves in the integration measure jacobian.  Once this is taken into account, one sees that the chiral anomaly is an ${\cal{O}}(p^4)$ effect, known as the {\it{Wess-Zumino-Witten anomalous term}} \cite{Wess}-\cite{Witten:ano}. Following the conventions of \cite{Pich:eff}, it can be cast in the following fashion
\begin{eqnarray}\label{WZW}
S\,[U,l,r]_{WZW}&=&-\frac{iN_C}{240 {\pi}^2} \int d {\sigma}^{ijklm}\, \langle\,  {\Sigma}_i^L\, {\Sigma}_j^L\, {\Sigma}_k^L\, {\Sigma}_l^L\, {\Sigma}_m^L\, \rangle-\nonumber\\
&&-\frac{iN_C}{48\pi^2}\int\,\,d^4x\,\,\varepsilon_{\mu\nu\alpha\beta}\ \bigg[\,W(U,l,r)^{\mu\nu\alpha\beta}-W({\bf{1}},l,r)^{\mu\nu\alpha\beta}\,\bigg]
\end{eqnarray}
where
\begin{equation}
\Sigma_{\mu}^L=U^{\dagger}\, \partial_{\mu}U,\qquad \Sigma_{\mu}^R=U\, \partial_{\mu}U^{\dagger}
\end{equation}
The first line in (\ref{WZW}) is only local in five dimensions and contains information about odd-parity processes purely among Goldstone bosons. The second term contains the couplings to external sources that guarantee chiral invariance. The tensor $W_{\mu\nu\alpha\beta}$ in the second line reads
\begin{eqnarray}
W(U,l,r)_{\mu\nu\alpha\beta} &=& \langle\,Ul_{\mu}l_{\nu}l_{\alpha}U^{\dagger}r_{\beta}+\frac{1}{4}\,Ul_{\mu}U^{\dagger}r_{\nu}Ul_{\alpha}U^{\dagger}r_{\beta}+iU\partial_{\mu}l_{\nu}l_{\alpha}U^{\dagger}r_{\beta}
+ i\partial_{\mu}r_{\nu}Ul_{\alpha}U^{\dagger}r_{\beta}-\nonumber\\
&-& i\Sigma_{\mu}^Ll_{\nu}U^{\dagger}r_{\alpha}Ul_{\beta}+\Sigma_{\mu}^LU^{\dagger}\partial_{\nu}r_{\alpha}Ul_{\beta}-\Sigma_{\mu}^L\Sigma_{\nu}^LU^{\dagger}r_{\alpha}Ul_{\beta}+\Sigma_{\mu}^Ll_{\nu}\partial_{\alpha}l_{\beta}+\nonumber\\
&+&\Sigma_{\mu}^L\partial_{\nu}l_{\alpha}l_{\beta}- i\Sigma_{\mu}^Ll_{\nu}l_{\alpha}l_{\beta}+\frac{1}{2}\,\Sigma_{\mu}^Ll_{\nu}\Sigma_{\alpha}^Ll_{\beta}-i\Sigma_{\mu}^L\Sigma_{\nu}^L\Sigma_{\alpha}^Ll_{\beta}\rangle-(L \longleftrightarrow R)\nonumber\\
&&
\end{eqnarray}
where the left-right exchange in the last line amounts to do the following substitutions
\begin{equation}
U \longleftrightarrow U^{\dagger},\qquad l_{\mu} \longleftrightarrow r_{\mu}, \qquad \Sigma_{\mu}^L \longleftrightarrow \Sigma_{\mu}^R
\end{equation}

\subsection{Chiral Electroweak Lagrangians}

Having discussed in the previous section the low energy effective lagrangian of the strong interactions as an expansion in powers of the light quark masses and external momenta, we can now apply the same procedure to account for the electroweak interactions at low energies. We will consider them as perturbations to be included in the strong chiral lagrangian.

At leading order, the electromagnetic corrections can be summarized in the following operator
\begin{equation}
\Lcal_2=e^2\,F_0^4\ \widehat{c}_2\, \langle {\cal{Q}}_L{\cal{Q}}_R \rangle\,+{\mathcal{O}}(e^4;e^2\,p^2)
\end{equation}
where $\widehat{c}_2$ is the resulting low energy coupling, to be determined from experiment, and the following notations are adopted
\begin{equation}
{\cal{Q}}_L={\cal{Q}}_R^{\dagger}=u\,Q\,u^{\dagger}\, ,\qquad 
Q={\rm{diag}}\left(\fr{2}{3},-\fr{1}{3},-\fr{1}{3}\right)
\end{equation}
As for the electroweak chiral lagrangian, we will be concerned with non-leptonic $\Delta S=1$ and $\Delta S=2$ processes, which are the relevant ones in the study of kaon physics. The procedure again relies on the method of external sources. Define a scalar source $\lambda$ to transform like
\begin{equation}
{\hat{\lambda}}=g_L\,\lambda\,g_L^{\dagger}
\end{equation}
Chiral operators linear in $\lambda$ are related to ${\cal{O}}(G_F)$ processes, while chiral operators quadratic in $\lambda$ account for ${\cal{O}}(G_F^2)$ processes. Fixing the proper flavour content of $\lambda$ to be
\begin{equation}
\lambda=\fr{1}{2}\,\bigg[\lambda_6-i\,\lambda_7\bigg]\equiv \lambda_{23}
\end{equation} 
the nonleptonic $\Delta S=1,2$ chiral Lagrangians can be determined. At leading order, $\Delta S=1$ transitions are described by the following Lagrangian\footnote{See, for instance, \cite{Ambrosio} and \cite{PichdeR} and references therein to the original literature.}
\begin{equation}\label{deltas=1}
\Lcal_2^{\Delta S=1}=\fr{G_F}{\sqrt{2}}\,\lambda_u\,F_0^4\, \bigg[g_8\,\Lcal_8^{(2)}+g_{27}\,\Lcal_{27}^{(2)}+e^2\, F_0^2 \,g_{ew}\,\Lcal_8^{(0)}\bigg]+{\mathrm{h.c}}+{\mathcal{O}}(p^4;e^2\,p^2)
\end{equation}
where $G_F$ and $F_0$ are factored out of the coupling constants $g_8$, $g_{27}$ and $g_{ew}$. $\Lcal_8^{(2)}$, $\Lcal_8^{(0)}$ and $\Lcal_{27}^{(2)}$ collect the operators which transform as $(8_L,1_R)$, $(8_L,8_R)$ and $(27_L,1_R)$, respectively, under the chiral group. Superscripts stand for their power counting in powers of momenta. Thus, $\Lcal_8^{(2)}$ and $\Lcal_{27}^{(2)}$ are ${\cal{O}}(p^2)$, while $\Lcal_8^{(0)}$ is ${\cal{O}}(p^0)$. Their expressions are
\begin{eqnarray}
\Lcal_8^{(2)}&=&\langle\,D_{\mu}U^{\dagger}\,D^{\mu}U\,\rangle_{23}\nonumber\\
\Lcal_{27}^{(2)}&=&\langle \,U^{\dagger}\,D_{\mu}U\,\rangle_{23}\,\langle \,U^{\dagger}\,D^{\mu}U\,\rangle_{11}+\fr{2}{3}\,\langle \,U^{\dagger}\,D_{\mu}U\,\rangle_{21}\,\langle \,U^{\dagger}\,D^{\mu}U\,\rangle_{13}\nonumber\\
\Lcal_8^{(0)}&=& \langle U\,\lambda_{32}\,U^{\dagger}\,Q \rangle
\end{eqnarray}
where the subscripts under the traces stand for the flavour content. 

As for $\Delta S=2$ transitions, the effective Lagrangian is 
\begin{equation}
\Lcal_2^{\Delta S=2}=\fr{G_F^2}{16\pi^2}\,F_0^4\,\Lambda^2_{S=2}\,{\mathrm{Tr}}\bigg[\,\lambda_{32}\,(D^{\mu}U^{\dagger})\,U\,\lambda_{32}\,(D_{\mu}U^{\dagger})\,U\,\bigg]+{\mathrm{h.c.}}+{\mathcal{O}}(p^4;e^2\,p^2)
\end{equation}
where, again, $G_F$, $F_0$ and the loop factor are kept explicit. $\Lambda^2_{S=2}$ is the low energy coupling encoding the physics of the heavy degrees of freedom which have been integrated out from the Standard Model Lagrangian. Dealing, as we do, with effective field theories, this coupling plays the pivotal role in the description of the $K^0-{\bar{K}}^0$ mixing. We postpone its determination, which will be addressed in chapter 5.

Contributions beyond leading order have also been computed for $\Delta S=1$ transitions. However, similarly to what happens in the strong sector, the number of low energy couplings increases dramatically, to the point that the number of couplings exceeds the experimental skill. 


\chapter[Chiral Lagrangians in the Resonance Region]{Chiral Lagrangians in the Resonance Region}

\section{Motivation}
In the previous chapter we have discussed chiral symmetry and its implementation to study the low energy dynamics of the pion octet fields. We estimated the radius of convergence of the chiral expansion to be $\Lambda_{\chi}\sim 1\,{\mathrm{GeV}}$, which lies close to the first resonance multiplet. A natural extension would then be to further exploit chiral symmetry by incorporating these higher energy excitation states in a (chiral) Lagrangian, thus pushing $\Lambda_{\chi}$ above the $1\,{\mathrm{GeV}}$ threshold.   

One such Lagrangian would then allow a prediction of the low energy parameters of the strong interactions ($L_i$ defined in chapter 2) in terms of the masses and decay constants of the included resonances. Out of the infinitely many resonances one can include, phenomenological evidence seems to favour the lowest lying multiplets as having more specific weight in the $L_i$ than the remaining hadronic multiplets. If one splits the contribution from the hadronic spectrum to the $L_i$ as
\begin{equation}\label{lidef}
L_i\ (\mu)\ =\ L_i^R\ +\ {\widehat{L}}_i\ (\mu)
\end{equation}
where the first term corresponds to the contribution from the lowest lying multiplets and the remaining piece gathers the remaining contribution from the hadronic spectrum, this {\it{lowest meson dominance}} suggests that the first term basically {\it{saturates}} the previous equation. This was the philosophy behind one such construction of a resonance chiral Lagrangian, already proposed in the late 1980's \cite{Sui1}{\footnote{There have been other attempts to describe the dynamics of the lowest energy resonances. See for instance \cite{meissner,bando}.}}, which succeeded in reproducing the above-mentioned resonance saturation and yielded expressions for the $L_i$ in terms of a finite amount of resonance parameters.

Later developments showed that that Lagrangian could actually be interpreted as an approximation to large--$N_C$ QCD \cite{Perrot,Gol}. This has many interesting implications. For instance, the predictions of the $L_i$, once supplied with QCD short distance information, are constraining enough to provide a set of relations between the different parameters $L_i$. To the degree that the resonance Lagrangian is an approximation to large-$N_C$ QCD, those relations can be envisaged as manifestations of a hypothetical underlying symmetry of large-$N_C$ QCD itself. However, for our purposes we will be mainly interested in the fact that once embedded in a large-$N_C$ framework, the resonance Lagrangian inherits a power counting (that of large-$N_C$) which enables a consistent computation of quantum effects. This turns out to be important, because it allows one to check whether the alleged resonance saturation still holds at the quantum level or, on the contrary, the resonance Lagrangian needs more structure to reach that goal.

 Bearing this point in mind, it is convenient, before we proceed, to introduce the basics of the $1/N_C$ expansion.

\section{The Expansion in the Number of Colours}

\subsection{Motivation}

Based on ideas of Condensed Matter Physics, 't Hooft \cite{thooft} proposed to use the number of colours in QCD as an expansion parameter around $N_C\rightarrow \infty$. The $1/N_C$ expansion is therefore a perturbative approach to QCD, not in the strong coupling constant but in a topological parameter, namely the index of the colour gauge group (actually, its inverse). As such, since the expansion parameter does not depend on energy scales, the approximation is valid over the whole range of energies.

Somewhat counterintuitively, increasing the number of colours leads to a much simpler description of the strong interactions. However, this is only true in some sense. We {\it{do not}} know how to solve for Large--$N_C$ QCD, and presumably this is not easier than actually solving QCD itself, in the sense that there are more dynamical variables in Large--$N_C$ QCD than in real QCD. All we know is that the increase in number of colours allows a much simpler treatment of the gross features of the theory. In some sense, having $N_C$ arbitrarily large sheds some light on generic features of the theory shadowed by the poor perspective of having $N_C=3$. The natural question to ask is whether Large--$N_C$ QCD has any resemblance to the real $N_C=3$ world. The answer to this question relies heavily on the skill of the theory to pass the phenomenological test. Interestingly, a great deal of phenomena find a natural explanation in the large-$N_C$ framework (sometimes it is the only explanation available). Just to mention some of them: 
\begin{itemize}
\item It provides a rationale to understand the Zweig rule.
\item It is compatible with Regge phenomenology and, at least in two dimensions, reproduces Regge trajectories.
\item It favours two-body meson decays, as it is experimentally observed.
\item Under quite general hypothesis, it can be proven that chiral symmetry breaking follows the same pattern as that observed in real QCD \cite{Coleman}.     
\end{itemize}
\subsection{Large--$N_C$ QCD}
The Lagrangian of the theory stands as in $N_C=3$ QCD (after all, it is still a Yang-Mills Lagrangian) except for the fact that we have changed the gauge group\footnote{For general accounts of the $1/N_C$ expansion, we refer to \cite{Manohar,Lebed}.}. That means that each quark field, since they sit in the fundamental representation, appears as an $N_C$-plet. Gluons, on the contrary, live in the adjoint representation and enlarge their number to $N_C^2-1$. In practice, it is commonplace to approximate this to $N_C^2$; in other words, we are skipping the tracelessness constraint (we are taking $U(N_C)$ instead of $SU(N_C)$). The missing gluon is numerically unimportant at sufficiently large $N_C$. Besides, it can be shown that the abelian factor is indeed suppressed at large $N_C$. Notice that the theory we are looking at differs from $N_C=3$ QCD in that there exist far more gluons than quarks (the former scale with $N_C^2$ while the latter only with $N_C$).

Our purpose will be to show the topological ordering of diagrams induced by the large--$N_C$ power counting scheme. For clarity, it is convenient to use {\it{'t Hooft double-line notation}}.
\subsection{The Double-Line Notation and Planarity}
Unlike the perturbative approaches we are used to, in which we expand in powers of a coupling constant, we have to change our strategy. In perturbative QED or QCD there is only one coupling constant which shows up to couple fermions and antifermions. That is why Feynman diagrams are so useful to organize calculations in powers of the coupling constant: you only need to count the number of vertices.

In the $1/N_C$ expansion, we need to keep track not only of the vertices of the theory (we will show later on that the coupling constant at large--$N_C$ is colour-dependent) but also of colour flow inside the diagram. We would like to have a pictorical approach to be able to determine in an easy way the scaling of physical amplitudes with the colour factor. The double-line notation introduced by 't Hooft makes it transparent to extract the colour scaling of a given amplitude altering only slightly the Feynman diagrams we are used to. Consider the following pictorical recipee: represent every quark line by a colour line, right-faced arrow meaning colour flow, left-faced arrow meaning anticolour flow. This does not change much the Feynman picture. When it comes to gluons, however, we have to interpret their colour indices as a system of colour and anticolour line. Figure (3.1) shows some examples of converting Feynman diagrams to double-line diagrams.
\begin{figure}\label{dou}
\renewcommand{\captionfont}{\small \it}
\renewcommand{\captionlabelfont}{\small \it}
\centering
\includegraphics[width=2.0in]{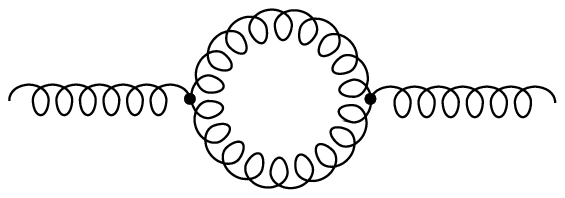}\qquad \qquad
\includegraphics[width=2.0in]{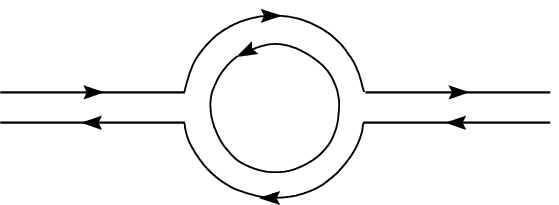}\\
\includegraphics[width=1.4in]{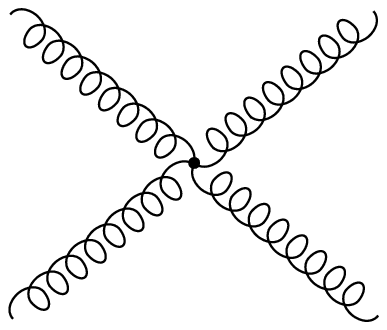}\qquad \qquad \qquad
\includegraphics[width=1.5in]{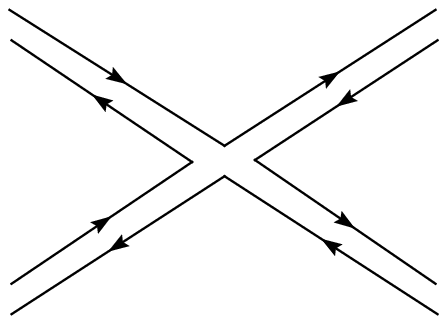}
\caption{Feynman diagrams with its counterparts in the double line notation introduced by 't Hooft.}
\end{figure}

We shall next prove that the expansion in inverse powers of the number of colours is equivalent to a topological expansion, where the planar topologies are the leading ones.
Let us characterize a diagram by its vertices (V), propagators (P) and colour loops (L). Then every diagram is holomorphic to a polyhedrum of L faces, V vertexs and P faces. Once this identifications are made, it can be shown that the equivalence classes of diagrams in $1/N_C$ power counting are holomorphic to the equivalence topological classes of polyhedra. Therefore, diagrams scale as
\begin{equation}\label{scalinginch2}
{\cal{M}} \sim g_s^{-2\chi_E}\,(g_s^2N_C)^C 
\end{equation}
where $\chi_E$ is a topological invariant called the Euler characteristic. That leaves the {\it{only}} non-trivial scaling that renders the theory finite to be
\begin{equation}
g_s \sim (N_C)^{-\fr{1}{2}}
\end{equation} 
which means that indeed the $1/N_C$ expansion is purely topological.
This leads to a set of {\it{selection rules}}, which can be summarized as follows:
\begin{itemize}
\item The leading diagrams are the subset of {\it{planar}} diagrams which minimize the internal quark loops and whose arbitrary number of internal gluon lines do not cross except at interacting vertices. Quark colour-flow lines have to be external. 
\item Most commonly, a Green function is made of a product of currents, {\it{i.e.}}, of quark bilinears of the form ${\bar{\psi}}_i\Gamma^i_j\psi_j$, where $\Gamma$ is a Dirac basis element. Then, {\it{planar}} diagrams are those that contain a single quark loop sitting in the border. {\it{Non-planar}} diagrams are suppressed by $N_C^{-2}$. The introduction of an internal quark loop supplies a $N_C^{-1}$ suppression factor. 
\end{itemize}
In other words, planar diagrams are those in which the colour factors compensate for the vertex factors. This explains why one can supply extra gluons in a planar diagram without altering its $N_C$ scaling, while extra quarks cannot compensate vertices with combinatoric colour factors.

\subsection{Mesons in Large--$N_C$ QCD: selection rules}
In the previous section we have found that QCD with a large number of colours provides a power-counting scheme, in which the so-called {\it{planar}} diagrams are the leading ones in powers of $N_C$. However, as it happens with $N_C=3$ QCD, nature forces us to deal with mesons instead of quarks and gluons. Therefore, we eventually want to translate the above statements into the meson picture of QCD. Surprisingly enough, this will provide us with far-reaching results\footnote{We will follow the excellent review by Witten \cite{Witten:baryons}.}.

First and foremost, we have to assume that large--$N_C$ QCD is confining. This has to be imposed, otherwise it would be sterile to try to make any contact between QCD and large--$N_C$ QCD. Confinement therefore is a hypothesis we make at the very beginning, not an output of large--$N_C$ QCD.
\begin{figure}\label{piful}
\renewcommand{\captionfont}{\small \it}
\renewcommand{\captionlabelfont}{\small \it}
\centering
\includegraphics[width=2.3in]{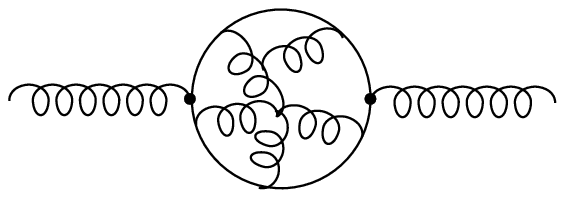}\qquad \quad
\includegraphics[width=1.7in]{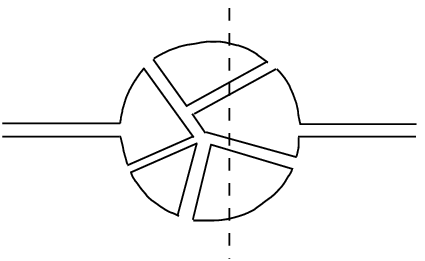}
\caption{A representative Feynman diagram contributing to the two-point function defined in (\ref{def}) together with its counterpart in double line notation. The cut illustrates the fact that in planar diagrams only one-particle intermediate states are allowed.}
\end{figure}

We will consider QCD two-point correlators as our starting point. In the following chapters we will need to consider three-point correlators and even four-point correlators. We will discuss its detailed large--$N_C$ behaviour in due course. Consider
\begin{equation}\label{def}
\Pi_{\mu\nu}(q^2)=i\int\, d^4x\, e^{i\,q\cdot x}\ \langle\, 0\,|\,T\{ j_{\mu}(x)j_{\nu}(0)\}\,|\,0\,\rangle
\end{equation}
where $j_{\mu}$, $j_{\nu}$ are quark bilinears. Lorentz symmetry tells us that
\begin{equation}
\Pi_{\mu\nu}(q^2)=(q^2g_{\mu\nu}-q_{\mu}q_{\nu})\, \Pi(q^2)
\end{equation}
Using the {\it{selection rules}} of the previous section, we know that planar diagrams will be those with no internal quark lines and arbitrary gluons. There are infinitely many such diagrams, and we do not know yet how to resum them in an intelligent manner. However, once the diagrams are written in double-line notation, we can look for intermediate states by cutting the diagram (see figure (3.2)). Then one realizes that there is only one colour contraction, no matter how many gluons plague the diagram, all making up a single $q{\bar{q}}$ meson state. Summing all the {\it{planar}} diagrams would be equivalent to determine the large--$N_C$ $q{\bar{q}}$ meson wave function in terms of quarks and gluons. This we do not know, but the crucial point to stress is that intermediate states are one-particle states, whatever its precise quark and gluon content. Multiparticle cuts are suppressed, {\it{i.e.}}, they show up in non-planar diagrams, because internal quark lines are needed.

Therefore, in full generality,
\begin{equation}\label{examtwo}
\Pi(q^2)=\sum_n\, \fr{F_n^2}{q^2-m_n^2}
\end{equation} 
So far we have only discussed the analytical structure of the Green function, but indeed much information can be inferred for mesons themselves. The expression above has to match onto the parton model logarithm, and this requires the sum to be infinite. On the other hand, the right-hand side of (\ref{examtwo}) picks no imaginary part, so that the mesons are stable: their decay width is $1/N_C$ suppressed.
\begin{figure}\label{gfunctions}
\renewcommand{\captionfont}{\small \it}
\renewcommand{\captionlabelfont}{\small \it}
\centering
$\langle J\,J\rangle$\ = $\sum\qquad$
\includegraphics[width=0.5in]{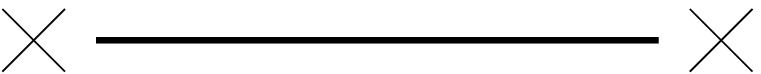}\\
\vskip 1cm
$\langle J\,J\,J\rangle$\ = $\sum \quad \Bigg[\quad$
\includegraphics[width=0.5in]{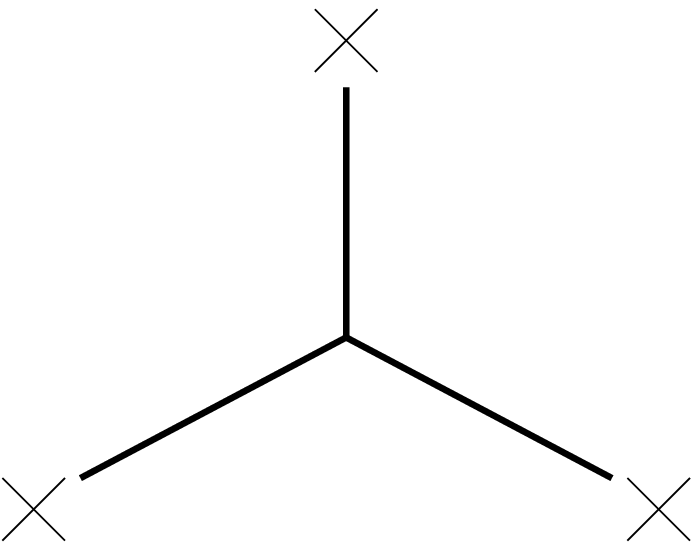}\qquad
$+\qquad$
\includegraphics[width=0.5in]{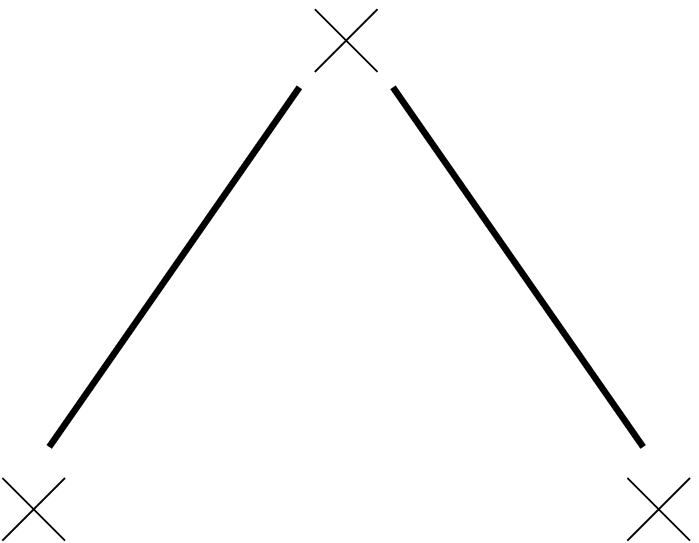}\qquad \Bigg]\\
\vskip 1cm
$\langle J\,J\,J\,J\rangle$\ = $\sum \quad \Bigg[\quad$
\includegraphics[width=0.5in]{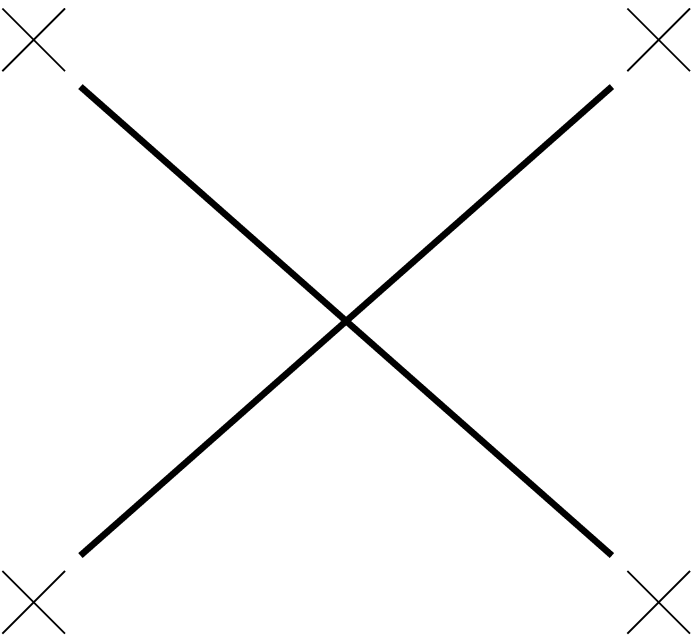}\qquad
$+\qquad$
\includegraphics[width=0.5in]{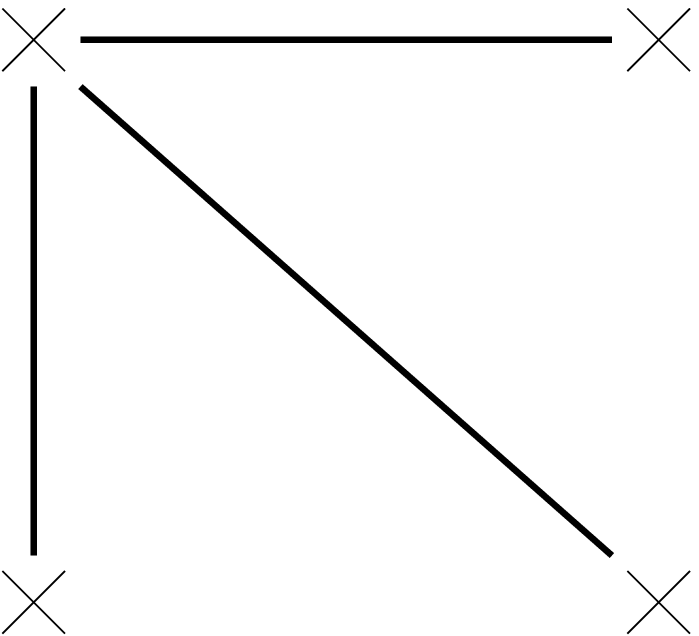}\qquad
+\qquad
\includegraphics[width=0.5in]{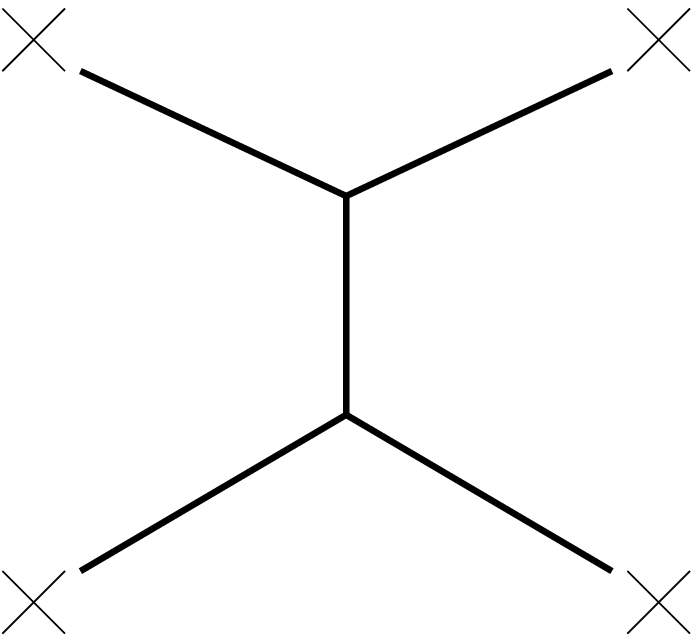}\qquad
+\qquad
\includegraphics[width=0.5in]{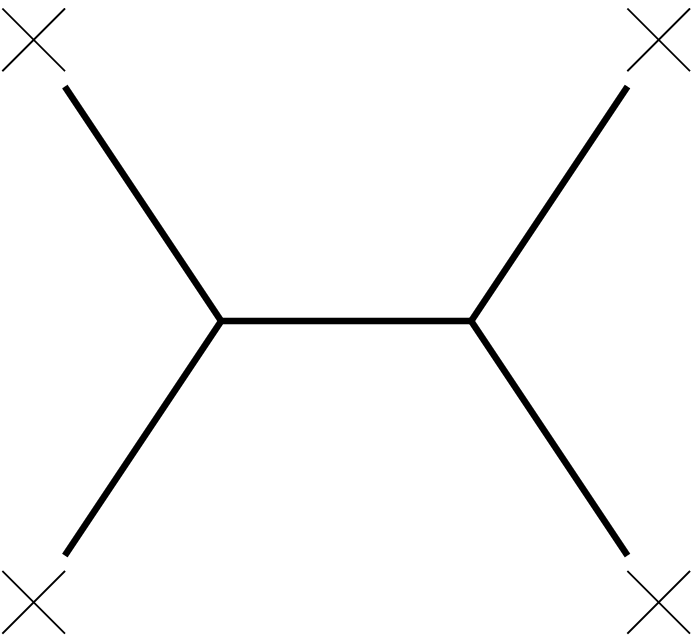}\qquad \Bigg]
\caption{Schematic decomposition of two-point, three-point and four-point Green functions in the large-$N_C$ limit. For instance, the last line shows the four-meson vertex, three-meson decay and two meson scattering in the $t$ and $s$ channels.}
\end{figure}

This is as far as we can get with two-point functions: two-point functions provide information on the propagator of the mesons. To get further, one should consider three-point and four-point functions. Their large--$N_C$ structure is depicted in figure (3.3). One can then show that indeed the singularity structure is restricted to poles: single poles, double poles, triple poles,... but no cuts. Additionaly, a detailed, though quite simple, study of three point correlators allow for a determination of the $N_C$ scaling of meson decay constants, to wit
\begin{equation}
F_n\qquad \qquad {\mathcal{O}}(\sqrt{N_C})
\end{equation}
This is so because a three point function has the vertex of a meson decaying into two mesons. In an analogous manner, four-point correlators have information on meson-meson scattering. A similar analysis thereupon shows that meson scattering is $1/N_C$ suppressed altogether.  

We have ended up with a very nice picture of large--$N_C$ QCD. Assuming confinement and planarity, the theory is made of infinitely many stable non-interacting mesons. This picture has some flavour of a semiclassical theory of hadrons, and indeed some efforts have been done towards going further in this very suggesting direction. 

\subsection{Two-dimensional Large--$N_C$ QCD: the {\it{'t Hooft model}}}
When introducing the foundations of large--$N_C$ QCD, we have stressed the fact that even though there are some simplifications, these are quite akin to many-body systems, where gross features can be dealt with while the dynamical details are in general much more complicated.

There is, however, an exception, and that is QCD in two dimensions, in short QCD$_{(2d)}$. 't Hooft showed \cite{thooft:2} that an analytical solution existed to solve the spectrum of resonances in terms of quark degrees of freedom in an explicit way. Also, unlike large--$N_C$ QCD$_{(4d)}$, in the 't Hooft model confinement is no longer an assumption but an explicit feature of the theory. This is not that surprising if one notices that in 2 dimensions the gluon propagator behaves in position space as
\begin{equation}
V(r)\sim r
\end{equation}
which indeed is confining. We will not delve here into technicalities and instead give a brief account of the most relevant results.

The reason for the solvability of the model in the {\it{planar}} limit is the fact that the reduction in dimensionality allows one to choose a gauge in which gluonic self-interacting vertices simply cancel. One is then left with two diagram topologies, {\it{rainbow}} diagrams and {\it{ladder}} diagrams, which can be resummed to yield an eigenvalue equation, the so-called 't Hooft equation \cite{thooft:2}, which in the notation of \cite{Gross} reads\footnote{See also \cite{Einh}.}:
\begin{equation}
M_n^2\, \phi_n(x)=\fr{m_q^2}{x\,(1-x)}\,\phi_n(x)-m_0^2\int dy\, \fr{\phi_n(y)}{(x-y)^2}\end{equation}
where $x$ is the momentum fraction carried by the quark $q$ inside the hadron. This equation shows that the resulting spectrum is discreet, labeled by index $n$, with masses piled up in a tower with characteristic mass parameter
\begin{equation}
m_0^2=\fr{g^2 N_C}{\pi} 
\end{equation}
This tower is asymptotically linear, {\it{i.e.}}, at large excitation number $n$, masses behave as
\begin{equation}\label{scaling}
m_n^2\sim \pi^2 m_0^2\, n
\end{equation}
and hadronic wave functions take the rather simple expression
\begin{equation}
\phi_n(x)\sim\sqrt{2}\,\sin{(n\pi x)}
\end{equation}
Additionally, one can show that the {\it{operator product expansion}} of two-point functions is an asymptotic series, as signalled by the factorial scaling of the Wilson coefficients, which is of the form \cite{Zhit}
\begin{equation}
c_{\,4n}\sim (-1)^{n-1}\left(\fr{g^2\,N_C}{2}\right)^{2n}(2n-1)!
\end{equation}
We do not know how this whole picture changes when one is considering QCD$_{(4d)}$. Certainly it may very well be that none of them survive in 4 dimensions, but at least some of them seem to be independent of dimensionality. For instance, Regge phenomenology hints at an asymptotic mass scaling of the form of (\ref{scaling}). Also, several considerations tend to catalog the OPE in QCD as an asymptotic series, a feature that QCD$_{(2d)}$ shares. Therefore, at least some qualitative features seem to survive.

\subsection{Large-$N_C$ $\chi$PT}
Being $\chi$PT an effective field theory of the strong interactions, it is also possible to implement there large-$N_c$ methods in what has been termed {\it{large-$N_c$ $\chi$PT}} \cite{Kaiser,Leut}. The mass difference between the octet and singlet elements of a multiplet is a $1/N_c$-suppressed effect, meaning that in the large-$N_c$ limit octet and singlet fields are degenerate and should be merged in a nonet \cite{Witten:79}, an element of the enlarged $U(N_C)_L\times U(N_C)_R$ chiral group. The procedure to follow in the chiral limit will be the introduction of an additional diagonal Gell-Mann matrix
\begin{equation}
\lambda^{(0)}=\sqrt{\fr{2}{3}}\ {\mathbf{1}}
\end{equation}
with the following algebra
\begin{equation}
f^{ab0}=0\, , \qquad d^{ab0}=\sqrt{\fr{2}{3}}\ \delta^{ab0}
\end{equation}
The matrix collecting Goldstone bosons gets therefore modified
\begin{equation}
\phi_a\frac{\lambda^a}{\sqrt{2}}={\bf{\Phi}}(x)=\left(\begin{array}{ccc}
\frac{1}{\sqrt{2}}\pi^0+\frac{1}{\sqrt{6}}\eta_8+\sqrt{\fr{2}{3}}\eta_1 & \pi^{+} &
K^{+}\\
\pi^{-} & -\frac{1}{\sqrt{2}}\pi^0+\frac{1}{\sqrt{6}}\eta_8+\sqrt{\fr{2}{3}}\eta_1 &
K^{0}\\
K^{-} & \bar{K}^{0} & -\frac{2}{\sqrt{6}}\eta_8+\sqrt{\fr{2}{3}}\eta_1
\end{array}\right)\\
\end{equation}
but the chiral Lagrangian stays the same. The power counting induced by the $1/N_c$-expansion can be inferred once the $N_c$ dependence of the low energy parameters are specified, {\it{i.e.}},
\begin{equation}
F_0 \qquad \qquad \qquad {\mathcal{O}}(\sqrt{N_c})
\end{equation}
since it is a decay constant and
\begin{equation}
B_0 \qquad \qquad \qquad  {\mathcal{O}}(1)
\end{equation}
since it is proportional to a mass term. A quite more involved analysis also leads to the following, 
\begin{eqnarray}
L_1,\ L_2,\ L_3,\ L_5,\ L_8,\ L_9,\ L_{10},\ H_1,\ H_2&\qquad & {\mathcal{O}}(N_c)\nonumber\\
L_4,\ L_6,\ L_7,\ 2L_1-L_2&\qquad & {\mathcal{O}}(1)
\end{eqnarray}
a hierarchy which is in agreement with experimental data.
It should be noted that the scaling above is specific of the $U(N_C)_L\times U(N_C)_R$ colour group. For large $N_C$, we have shown that this is basically equivalent to $SU(N_C)_L\times SU(N_C)_R$, but in real $N_C=3$ QCD, this is no longer an allowed approximation. The $\eta_1$ particle decouples from the pion-kaon octet and becomes massive due to the chiral anomaly. $L_7$ receives contributions from $\eta_1$, and this means that when comparing with experimental data, $L_7$ has a special status, since its large-$N_C$ version is ill-defined \cite{Peris1}. Leaving this subtlety aside, we can immediately verify that the scale of chiral symmetry breaking $\Lambda_{\chi}$ introduced in the previous chapter scales as
\begin{equation}
\Lambda_{\chi}\ \sim {\mathcal{O}}(N_C)
\end{equation} 
which means that chiral corrections are suppressed by $1/N_c$ powers.
\section{The Minimal Hadronic Approximation}
We have seen in previous sections that large--$N_C$ QCD is able to constrain the analytical form of QCD Green functions to be meromorphic functions. For a typical two-point correlator,
\begin{equation}\label{large}
\Pi(Q^2)=\sum_{n=1}^{\infty}\,\fr{F_n^2}{q^2-m_n^2}
\end{equation}
\begin{figure}\label{MHAeuclidean}
\renewcommand{\captionfont}{\small \it}
\renewcommand{\captionlabelfont}{\small \it}
\centering
\includegraphics[width=2.5in]{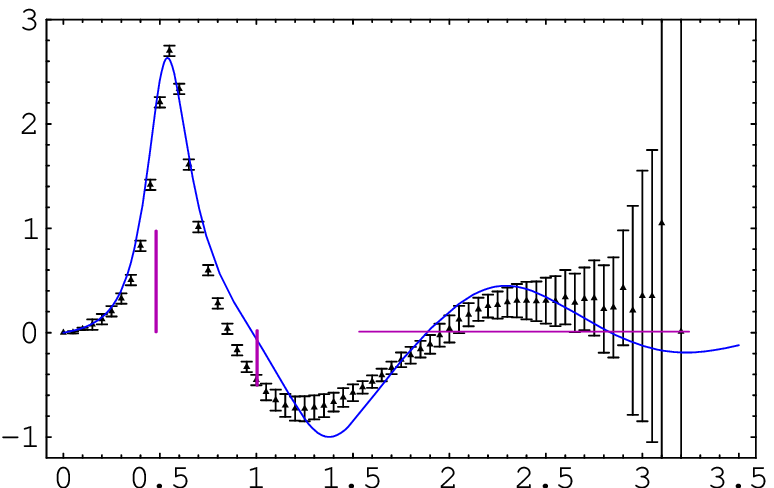}
\hspace{1cm}
\includegraphics[width=2.5in]{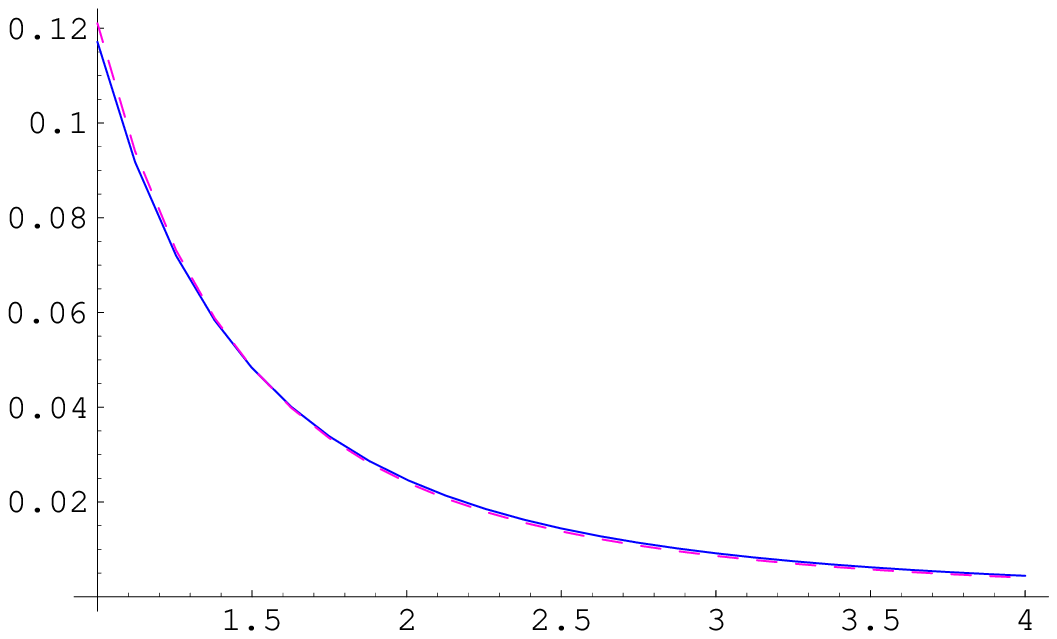}
\caption{The MHA to the two-point $<$VV-AA$>$ correlator as compared to experimental data from ALEPH and OPAL (left) and to a model to be defined in chapter 6 (solid blue curve). Comparison is done in the Minkowski (${\textrm{Re}}\,q^2\geq\,0$) and Euclidean (${\textrm{Re}}\,q^2\leq\,0$) half planes (left and right figures, respectively). Notice that, even though the spectral functions are drastically different, their analytic extensions to the Euclidean fit impressively.}
\end{figure}
where the sum extends to infinity, in order to reproduce the parton model logarithm. If we knew the solution to large--$N_C$ QCD, then we could derive analytically the infinite number of poles and residues of the Green function. Since this is not at hand, equation (\ref{large}) is of not much phenomenological use. A much useful approximation, which has been coined the {\it{Minimal Hadronic Approximation}} \cite{knecht:mha,MIT}, consists in truncating the series of (\ref{large}) to retain only a finite number of terms. The poles and residues are then to be determined by matching onto the OPE to get the right short distances and to $\chi$PT to reproduce the long distance behaviour. From a mathematical point of view, such interpolating functions come under the name of {\it{rationale approximants}}. 

For the subset of Green function which are order parameter of spontaneous chiral symmetry breaking (S$\chi$SB) such an approximation should work better than with conventional Green functions (see figure (3.4)). Order parameters of S$\chi$SB receive no contribution from perturbative QCD and are therefore expected to converge more rapidly at high energies (sometimes they are termed superconvergent correlators). This suggests that a small number of resonances should be enough to reproduce the behaviour of the Green function. In other words, that we do not expect dramatic changes as more terms are considered in (\ref{large}), but a soft convergence. Another fact supporting this point is that the region of intermediate energies is rather narrow: $\chi$PT is valid at about 1 GeV and common lore sets the onset of the perturbative regime below 3 GeV. Not much room is left in between, and it is not natural to think that a bump will show up in the Euclidean region.  Classical vector meson dominance also relied on some of this points and proved to be phenomenologically successful. Therefore, the approximation consists in truncating (\ref{large}) to 
\begin{equation}\label{trunc}
\Pi(Q^2)=\sum_{n=1}^{\mathcal{N}}\,\fr{F_n^2}{q^2-m_n^2}=A_N\, \prod_{i=1}^P\left[\fr{1}{q^2+m_i^2}\right]\,\prod_{j=1}^Z \bigg[q^2-\sigma_j^2\bigg]
\end{equation}
where in the second equality we have rearranged the finite sum as a product of poles and zeros. Such an approach is called an approximation because one can in principle add more and more resonances in the equation above to improve. Each new resonance in (\ref{trunc}) has to be included in a way compatible with the short distance and long distance behaviour of the Green function. Therefore, there is some freedom in choosing the constraints coming from high and low energies. However, one has to ensure that at least the leading OPE constraint is fulfilled\footnote{Actually, this will turn out to be the crucial ingredient to obtain the right matching between short and long distances in nonleptonic weak interactions (see chapter 5).}. This imposes a lower bound for the number of resonances to be included, which can be cast in the form of a theorem \cite{MIT}
\begin{equation}
{\cal{Z}}-{\cal{P}}=-p_{OPE}
\end{equation}
where ${\cal{N}}$ is the number of zeros and ${\cal{P}}$ that of poles in the Green function, whereas $p_{OPE}$ is the leading fall-off power ($1/(Q^2)^{p_{OPE}}$) of the Operator Product Expansion. Since ${\mathcal{N}}\geq 0$, this leads to
\begin{equation}
{\mathcal{P}}\geq p_{OPE}
\end{equation} 
and so fixes the minimal number of resonances to be considered.

\section{A chiral Lagrangian with Resonance fields}
We now turn to our initial goal of including resonances in a chiral Lagrangian. The building of an effective Lagrangian out of Goldstone bosons and resonance excitation fields in a chiral-invariant way can be achieved once the resonance fields are embodied with a chiral representation. Consider $R$ and $R_1$ as the octet and singlet components of a resonance multiplet. Their transformation under the chiral group has to be of the form
\begin{equation}\label{trans}
R\rightarrow h\,R\,h^{\dagger} \, , \qquad R_1\rightarrow R_1\, , \qquad h\,\in SU(3)_V
\end{equation}
This suggests, by analogy with the Goldstone fields, to collect the multiplets as
\begin{equation}
V_{\mu\nu}(x)=V_{\mu\nu}^a\frac{\lambda_a}{\sqrt{2}}=\left(\begin{array}{ccc}
\frac{1}{\sqrt{2}}\rho^0+\frac{1}{\sqrt{6}}\omega_8 & \rho^{+} &
K^{*+}\\
\rho^{-} & -\frac{1}{\sqrt{2}}\rho^0+\frac{1}{\sqrt{6}}\omega_8 &
K^{*0}\\
K^{*-} & \bar{K}^{*0} & -\frac{2}{\sqrt{6}}\omega_8
\end{array}\right)_{\mu\nu}
\end{equation}  
for the vector channel and similarly for the axial, scalar and pseudoscalar channels. The transformation rules (\ref{trans}) and chiral invariance fix the interactions between the Goldstone and Resonance states.

However, in our discussion of effective field theories we already stressed the fact that they consist of an infinite number of operators. The rigorous way to proceed is to identify a power counting to systematically order them according to their relevance. Hence, in the chiral lagrangian we saw that dimensional power counting provides a way to truncate the chiral expansion consistently. Unfortunately, no such power counting argument exists for the present case. As a starting approximation, one could consider the following Lagrangian \cite{Sui1}
\begin{eqnarray}\label{Ecker}
\Lcal_2[V(1^{--})]&=&\sum_i\,\left \{ \frac{F_{V_i}}{2\sqrt{2}}\,\langle\,V^i_{\mu\nu}f_{+}^{\mu\nu}\rangle\,+\,i\frac{G_{V_i}}{\sqrt{2}}\langle\,
V^i_{\mu\nu}u^{\mu}u^{\nu}\rangle\right \}\nonumber\\
\Lcal_2[A(1^{++})]&=&\sum_i\, \frac{F_{A_i}}{2\sqrt{2}}\,\langle A^i_{\mu\nu}f_{-}^{\mu\nu}\rangle\nonumber\\
\Lcal_2[S(0^{++})]&=&\sum_i\, \left\{ c_{d_i} \,\langle S_i u_{\mu} u^\mu \rangle\,
+\,c_{m_i}\,\langle S_i \chi_+\rangle + \tilde{c}_{d_i} \, S_{1i}\,
\langle u_{\mu} u^\mu \rangle\,+\tilde{c}_{m_i}\,S_{1i} \langle \chi_+\rangle\right\}\nonumber\\
\Lcal_2[P(1^{-+})]&=&\sum_i\, \left\{id_{m_i}\langle P_i\chi_-\rangle+i\tilde{d}_{m_i} P_{1i}\langle\chi_+\rangle\right\}
\end{eqnarray}
where
\begin{equation}
u_{\mu}\equiv i u^{\dagger}D_{\mu}U\,u^{\dagger}\, , \qquad f_{\pm}^{\mu\nu}\equiv u\,F_L^{\mu\nu}\,u^{\dagger}\pm u^{\dagger}F_R^{\mu\nu}u \, , \qquad \chi_{\pm}\equiv u^{\dagger}\chi u^{\dagger}\pm u\chi^{\dagger}u
\end{equation}
and the sum runs up to infinity to include each hadronic state. The previous Lagrangian is indeed chiral invariant, but only includes linear operators in the resonance fields. Without any power counting behind, we have to consider them as an ansatz and test it phenomenologically. Integration of the whole tower of hadronic states would result in a determination of the low-energy couplings $L_i$ of the original chiral Lagrangian in terms of hadronic parameters (masses and decay constants). Dimensional analysis leads to a straightforward estimation \cite{pich:poly}
\begin{equation}\label{est}
L_k\sim \sum_i\fr{F_{R_i}^2}{M_{R_i}^2}
\end{equation}
 For instance,
\begin{equation}
L_{10}=\sum_i\left\{ \fr{F_{A_i}^2}{4M_{A_i}^2}-  \fr{F_{V_i}^2}{4M_{V_i}^2}\right\}
\end{equation}
showing that the higher the resonance states the fewer the impact on the low energy couplings. This is supported by the phenomenologically favoured {\it{vector meson dominance}}. A truncation of the sums in (\ref{Ecker}) to include the lowest lying multiplet for each channel thus seems to be favoured experimentally. This so-called {\it{lowest meson dominance}} can then be viewed as a natural extension of the {\it{vector meson dominance}}\footnote{See, {\it{e.g.}}, \cite{Sakurai}.}. A step forward was later on provided by \cite{Perrot,Gol}, which embedded the previous Lagrangian in a $1/N_C$ framework. As it stands, the Lagrangian (\ref{Ecker}) indeed contains an infinite number of narrow-width resonance states, as the large-$N_C$ limit demands\footnote{Obviously, in the large-$N_C$ limit the multiplets have nine components, as we have seen previously, which means that the distinction between singlet couplings and octet couplings in the scalar and pseudoscalar sector is somewhat artificial, since one expects them to be one and the same. However, taking into account that the $1/N_C$ expansion works, in general, worse precisely for these channels, it is phenomenologically advisable to split them apart. If the $1/N_C$ expansion happens to be a good approximation, this will show up in a nearly degenerate couplings. We therefore prefer to check if the splitting was unnecessary {\it{a posteriori}}, based on phenomenological grounds.}. $1/N_C$ counting then yields the following behaviour for the parameters in (\ref{Ecker})
\begin{eqnarray}
F_V,\,F_A,\,G_V,\,c_d,\,\tilde{c}_d\qquad &\sim& {\cal{O}}(\sqrt{N_c})\nonumber\\
M_V,\,M_A,\,M_S,\,M_{S_1}\qquad &\sim& {\cal{O}}(N_c^0)
\end{eqnarray}
Truncation of the sums then amounts to be working in the already mentioned {\it{minimal hadronic approximation}}. The importance of endowing the resonance chiral Lagrangian with a large-$N_C$ framework is thus to give a rationale, namely the MHA, for the otherwise purely phenomenological truncation of the infinite sums. In the following, it will prove convenient to adopt the MHA point of view. This eventually would allow us to compute quantum corrections ({\it{i.e.}}, $1/N_C$ corrections) with the Lagrangian (\ref{Ecker}). However, we want to emphasise that large-$N_C$ is able to provide a power-counting rule for quantum corrections once the Lagrangian is given, but it does not yield a power-counting criteria to build the Lagrangian (\ref{Ecker}). The issue of whether (\ref{Ecker}) gets any close to the large-$N_C$ QCD Lagrangian or, on the contrary, fails, remains unanswered. All we know is that, at least for certain two-point Green functions, (\ref{Ecker}) agreement with QCD is provided by imposing QCD short distances constraints, as we will see later on, but there are indications that for certain three-point Green functions this agreement ceases to hold.

Enforcement of local chiral symmetry on the truncated version of (\ref{Ecker}) requires the definition of a covariant derivative acting upon the resonance fields
\begin{equation}
\nabla_{\mu}R_i=\partial_{\mu}R_i+[\,\Gamma_{\mu},r_i\,]
\end{equation}
where the connection $\Gamma_{\mu}$ is defined as
\begin{equation}
\Gamma_{\mu}=\frac{1}{2}\bigg\{u^{\dagger}[\partial_{\mu}-ir_{\mu}]\,u+u\,[\partial_{\mu}-il_{\mu}]\,u^{\dagger}\bigg\}
\end{equation}
leading to the kinetic terms
\begin{eqnarray}\label{kinterm}
\Lcal_{kin}(V,A,S,P)&=&{\sum_{R=V,A}}\left[-\frac{1}{2}\langle {\nabla}^\lambda
R_{\lambda\mu}{\nabla}_\nu
R^{\nu\mu}- \frac{1}{2} M_R^2
R_{\mu\nu}R^{\mu\nu}\rangle\,\right]-\nonumber\\
&-&\sum_{R=V,A}\left[\frac{1}{2}\partial^{\lambda}(R_1)_{\lambda\mu}\partial_{\nu}(R_1)^{\nu\mu}\,+\,\frac{M_{R_1}^2}{4}(R_1)_{\mu\nu}(R_1)^{\mu\nu}\right]+\nonumber\\
&+&{\sum_{R=S,P}}\left[\frac{1}{2}\, \langle
\nabla^\mu R \nabla_\mu R - M_R^2 R^2\rangle +\, \frac{1}{2}\, \left(
\partial^\mu R_1 \partial_\mu R_1 - M_{R_1}^2 R_1^2\right)\right]\nonumber\\
&&
\end{eqnarray}
The use of a tensorial antisymmetric representation for the resonance multiplets is especially convenient when dealing with gauge fields. A comprehensive treatment of tensorial representation of vectorial fields is given in \cite{Sui1,JJ}. For instance, the propagator is
\begin{equation}
\Delta_{\mu\nu\rho\lambda}=\fr{i}{M^2-k^2-i\epsilon}\left\{\ g_{\mu\rho}g_{\nu\lambda}\left(1-\fr{k^2}{M^2}\right)+g_{\mu\rho}\fr{k_{\nu}k_{\lambda}}{M^2}-g_{\nu\lambda}\fr{k_{\mu}k_{\rho}}{M^2}-(\mu\leftrightarrow \nu)\right \}
\end{equation}
 Other commonly used representations include the Yang-Mills representation (see \cite{meissner} and references therein) or the hidden symmetry representation \cite{bando}. Obviously some ambiguities arise between them, but they were shown to disappear once the right QCD behaviour is imposed on certain two-point Green functions \cite{Sui2}. A general proof of the equivalence between them was given in \cite{Bijnens:1995ii}, showing explicitly that they are linked through different redefinitions of the same Lagrangian.
 
We can now integrate the newly introduced resonance fields in (\ref{Ecker}) and (\ref{kinterm}), something which leads to a determination of the low energy couplings, to wit \cite{Sui1},

\begin{equation}\label{predinch2}
\begin{array}{lcccccccccccc}
L_1=&\frac{G_V^2}{8M_V^2}&+&
&-&\frac{c_d^2}{6M_S^2}&+&\frac{{\tilde{c}}_d^2}{2M_{S_1}^2}&+& &+& &\\
&&&&&&&&&&&&\\
L_2=&\frac{G_V^2}{4M_V^2}&+& &+& &+& &+& &+& &\\
&&&&&&&&&&&&\\
L_3=&-\frac{3G_V^2}{4M_V^2}&+& &+&\frac{c_d^2}{2M_S^2}&+& &+& &+& &\\
&&&&&&&&&&&&\\
L_4=& &+&
&-&\frac{c_dc_m}{3M_S^2}&+&\frac{\tilde{c}_d\tilde{c}_m}{M_{S_1}^2}&+&
&+& &\\
&&&&&&&&&&&&\\
L_5=& &+& &+&\frac{c_dc_m}{M_S^2}&+& &+& &+& &\\
&&&&&&&&&&&&\\
L_6=& &+& &-&\frac{c_m^2}{6M_S^2}&+&\frac{\tilde{c}_m^2}{2M_{S_1}^2}&+&
&+& &\\
&&&&&&&&&&&&\\
L_7=& &+& &+&
&+& &+&\frac{d_m^2}{6M_P^2}&-&\frac{\tilde{d}_m^2}{2M_{{\eta}_1}^2}&\\
&&&&&&&&&&&&\\
L_8=& &+& &+&\frac{c_m^2}{2M_S^2}&+& &-&\frac{d_m^2}{2M_P^2}&+& &\\
&&&&&&&&&&&&\\
L_9=&\frac{F_VG_V}{2M_V^2}&+& &+& &+& &+& &+& &\\
&&&&&&&&&&&&\\
L_{10}=&-\frac{F_V^2}{4M_V^2}&+&\frac{F_A^2}{4M_A^2}&+& &+& &+& &+& &
\end{array}
\end{equation}
With the previous determination of the low energy couplings the above-mentioned argument in favour of lowest-lying resonances can be tested. Experimental values as compared with predictions are summarized in table (\ref{predexp}).
\begin{table}
\begin{center}
\begin{tabular}{|c|c|c|c|c|c|c|c|c|}
\hline
$L_i$ & Experimental value $(\ \mu\ =\ m_{\rho})$ &\ $V$\ &\ $A$\ &\ $S$\ &\ $S_1$\ &\ $\eta_1$\ & Prediction \\
\hline \hline
$L_1$ &     $0.7\pm0.3$ &   $0.6$ & $0$ & $-0.2$ & $0.2$ & $0$ & $0.6$       \\
\hline
$L_2$ &     $1.3\pm0.7$ &   $1.2$ & $0$ & $0$ & $0$ & $0$ & $1.2$      \\
\hline
$L_3$ &            $-4.4\pm2.5$ & $-3.6$ & $0$ & $0.6$ & $0$ & $0$ & $-3.0$         \\
\hline
$L_4$ &             $-0.3\pm0.5$ & $0$ & $0$ & $-0.5$ & $0.5$ & $0$ & $0.0$        \\
\hline
$L_5$ &        $1.4\pm 0.5$ & $0$ & $0$ & $1.4$ & $0$ & $0$ & $1.4$    \\
\hline
$L_6$ &         $-0.2\pm0.3$ & $0$ & $0$ & $-0.3$ & $0.3$ & $0$ & $0.0$       \\
\hline
$L_7$ &          $-0.4\pm0.15$ & $0$ & $0$ & $0$ & $0$ & $-0.3$ & $-0.3$       \\
\hline
$L_8$ &           $0.9\pm0.3$  & $0$ & $0$ & $0.9$ & $0$ & $0$ & $0.9$     \\
\hline
$L_9$ &            $6.9\pm 0.2$ & $6.9$ & $0$ & $0$ & $0$ & $0$ & $6.9$         \\
\hline
$L_{10}$ &          $-5.2\pm 0.3$ & $-10.0$ & $4.0$ & $0$ & $0$ & $0$ &   $-6.0$       \\
\hline

\end{tabular}
\end{center}
\caption{{\it{Predictions for the low energy couplings $L_i$ from the expressions in (\ref{predinch2}) versus experimental values. Masses and decay constants are estimated phenomenologically with the values given in \cite{Sui1}.}}}\label{predexp}
\end{table}
In the light of the results, there seems to be an amazing agreement, signalling at the fact that the lowest lying multiplets are enough to account for the low energy couplings of QCD, a phenomenon coined thereafter {\it{resonance saturation}}.
A more careful approach would be to reduce the phenomenological input and at the same time make the theory resemble QCD to a higher extent. For instance, one could demand the right ultraviolet behaviour of certain two-point Green functions, {\it{e.g.}}, the pion electromagnetic form factor $F_V\,(q^2)$, the axial form factor $G_A\,(q^2)$ in $\pi\rightarrow e\ \nu_e\ \gamma$ decay, the $VV-AA$ two-point function $\Pi^{LR}\,(q^2)$ and the $SS-PP$ two-point function $\Pi^{SP}\,(q^2)$,  
\begin{eqnarray}\label{vecfor}
F_V(q^2)&=&1+\sum_i\frac{F_{V_i}G_{V_i}}{F_{0}^2}\frac{q^2}{M_{V_i}^2-q^2}\\
G_A(q^2)&=&\sum_i\left[\frac{2F_{V_i}G_{V_i}-F_{V_i}^2}{M_{V_i}^2}+\frac{F_{A_i}^2}{M_{A_i}^2-q^2}\right]\label{axfor}\\
\Pi^{LR}(q^2)&=&\ \fr{F_{0}^2}{q^2}\ +\ \sum_i\ \fr{F_{V_i}^2}{M_{V_i}^2-q^2}\ -\  \sum_i\ \fr{F_{A_i}^2}{M_{A_i}^2-q^2}\label{pifor}\\
\Pi^{SP}(q^2)&=&16B_0^2\left[\sum_i\ \fr{c_{m_i}^2}{M_{S_i}^2-q^2}\ -\ \sum_i\ \fr{d_{m_i}^2}{M_{P_i}^2-q^2}\ +\fr{F_{0}^2}{8\ q^2}\right]\label{scafor}
\end{eqnarray}
Truncating the previous expressions to the first multiplet and comparing with QCD results in the following set of {\it{matching}} conditions \cite{Sui2}\footnote{Recall that this is actually a big jump. We found a convincing argument to truncate the hadronic tower at {\it{low energies}}, namely, that higher mass resonances yield a (in principle) small contribution to the low energy couplings $L_i$ (see (\ref{est})). However, what it is implicitly assumed now in the matching equations (\ref{formfac1}-\ref{scalar2}) is that they also saturate the {\it{high energies}}.}  
\begin{eqnarray}\label{formfac1}
F_V\ G_V&=&F_0^2 \\
\fr{2F_V\ G_V-F_V^2}{M_V^2}&=&0\label{formfac2}
\end{eqnarray}
together with \cite{wein:sr}
\begin{eqnarray}\label{firstWSR}
F_V^2-F_A^2&=&F_0^2 \\
F_V^2\,M_V^2-F_A^2\,M_A^2&=&0\label{secondWSR}
\end{eqnarray}
and with \cite{Gol}
\begin{eqnarray}\label{vect}
F_V^2\ M_V^4-F_A^2\ M_A^4&=&-\fr{3072}{25}\pi^4\, F_0^6\\
c_m^2-d_m^2&=&\fr{F_0^2}{8}\label{scalar1}\\
c_m^2\ M_S^2-d_m^2\ M_P^2&=&\fr{3\pi\alpha_s}{4}\ F_0^4\label{scalar2}
\end{eqnarray}
where the last two equations are the akin Weinberg sum rules for the scalar-pseudoscalar sector. Combining (\ref{formfac1})-(\ref{firstWSR}) results in the following relations \cite{Gol} for the couplings in terms of $F_0$
\begin{equation}\label{decay}
F_V\,=\,2G_V\,=\,\sqrt{2}F_A\,=\,\sqrt{2}F_{0}
\end{equation}
The above relation, together with (\ref{secondWSR}), leads to the following expression for the masses
\begin{equation}\label{masses}
M_V\,=\,\fr{M_A}{\sqrt{2}}\,=\,4\pi F_{0}\left(\frac{\sqrt{6}}{5}\right)^{1/2}\end{equation}
where in the second equality use has been made of (\ref{vect}). Therefore, we are able to express couplings and masses in terms of $F_{0}^2$ alone. Additionally, this means that there appears a parameter-free {\it{prediction}} of the low energy couplings, to wit   
\begin{equation}\label{predictions}
6\,L_1\,=\,3\,L_2\,=\,-\frac{8}{7}\,L_3\,=\,4\,L_5\,=\,8\,L_8\,=\,\frac{3}{4}\,L_9\,=\,-L_{10}\,=\,\frac{3}{8}\frac{f_{\pi}^2}{M_V^2}\,=\,\frac{15}{8\sqrt{6}}\frac{1}{16\pi^2}
\end{equation}
where (\ref{scalar1}) and (\ref{scalar2}) were used in the prediction of the $L_5$ and $L_8$ couplings\footnote{The analysis leading to the relations for $L_5$ and $L_8$ in (\ref{predictions}) actually requires more input than just (\ref{scalar1}) and (\ref{scalar2}). We refer to \cite{Perrot} and \cite{Gol} for details.}.
The set of relations found in the previous equation among the low energy couplings of the strong interactions, far from being a simple curiosity, seem to point at a more profound structure of the large-$N_c$ limit of the strong interactions, as we commented on earlier.
 To illustrate this point clearlier, it is worth making an aside and consider an analogy with the determination of the $\rho$ parameter in the Standard Model.

The $\rho$ parameter measures the ratio between the neutral and charged currents in the Standard Model. Its expression at tree level reads\footnote{See, for instance, \cite{cheng}.}
\begin{equation}
\rho=\frac{1}{\cos^2{\theta_W}}\frac{M_W^2}{M_Z^2}
\end{equation}
whose analog in our study would be the parameter $L_{10}$ at tree level
\begin{equation}
L_{10}=\fr{F_A^2}{4\,M_A^2}-\fr{F_V^2}{4\,M_V^2}
\end{equation}
We already discussed in chapter 2 that the dynamical symmetry of the Standard Model is an $SU(2)_L\times U(1)_Y$. However, there also exists a global $SU(2)_L\times SU(2)_R$ symmetry, coined {\it{custodial symmetry}}\footnote{To be precise, it is $SU(2)_R$ which is referred to as custodial symmetry.}, which yields the prediction
\begin{equation}
\rho=1
\end{equation}
and the parameter is symmetry-protected. Much in the same fashion, enforcement of this would-be symmetry of the large-$N_C$ limit of QCD leads to
\begin{equation}\label{pred}
L_{10}=-\frac{1}{4}\left(\frac{15}{32\pi^2\sqrt{6}}\right)
\end{equation}
For simplicity, we will give a detailed accound of the previous statements for the $\rho$ parameter in the linear sigma model. Consider the Lagrangian
\begin{eqnarray}\label{sigma}
\mathcal{L}&=&\frac{1}{2} \left(\partial_{\mu} \sigma\right)^2 -
  \frac{m_{\sigma}^2}{2} \sigma^2
  + \left(\frac{\sigma^2}{4}+ f
\frac{\sigma}{2} \right)
  \mathrm{Tr}D_{\mu}U D^{\mu}U^{\dag}-\nonumber \\
  &-&  \frac{1}{\rho}\ \frac{f^2}{2}
  \left(\mathrm{Tr}\frac{\tau_{3}}{2} U^{\dag} D_{\mu}U\right)^2
  -  \frac{f^2}{2}\sum_{a=1,2}
  \left(\mathrm{Tr}\frac{\tau_{a}}{2} U^{\dag} D_{\mu}U \right)^2+
  SSB
\end{eqnarray}
where $U$ is the $SU(2)$ Goldstone matrix,
\begin{equation}
U={\textrm{exp}}\,\bigg(i\fr{\pi^a\tau_a}{f}\bigg)
\end{equation}
and we have gauged $U(1)_Y$ out of $SU(2)\times U(1)$, so that the covariant derivative reads\footnote{See, {\it{e.g.}}, chapter 2 of \cite{Geo:book}.}
\begin{equation}
D_{\mu}U=\partial_{\mu}U+ig'\frac{\tau_3}{2}U
\end{equation}
$SSB$ in the last line stands for the spontaneous symmetry breaking terms, which are responsible for the vacuum expectation value $f$. The previous Lagrangian is therefore invariant under $SU(2)_{global}\times U(1)_{local}$. We can now enforce the right high-energy behaviour of certain Green functions, much in the same fashion as we did in (\ref{vecfor})-(\ref{scafor}). We consider the matrix element for pion scattering \cite{Chanowitz}
\begin{equation}\label{chan}
 \mathcal{M}(\pi^+\pi^-\rightarrow \pi_3\pi_3) = \frac{1}{ f^2}\left(
  \frac{s}{\rho} - \frac{s^2}{s-m_{\sigma}^2}\right) 
\end{equation}
If we require unitarity to be preserved, then $\rho=1$, and as a by-product this leads to an extra custodial $SU(2)$ symmetry of (\ref{sigma}). We can turn the argument round by saying that the custodial symmetry ensures the condition $\rho=1$ to hold. Recall that this is akin to what happens with the set of constraints (\ref{formfac1})-(\ref{scalar2}), which eventually lead to (\ref{pred}).

If we now integrate out the $\sigma$ particle in (\ref{sigma}), in a similar fashion as what we did earlier on with the resonances in (\ref{Ecker}) and (\ref{kinterm}), we find the effective theory for the Goldstone bosons
\begin{equation}
 \mathcal{L}= - \frac{1}{\rho(\mu)}\ \frac{f^2}{2}
  \left(\mathrm{Tr}\frac{\tau_{3}}{2} U^{\dag} D_{\mu}U \right)^2 -
\frac{f^2}{2}\sum_{a=1,2}
  \left(\mathrm{Tr}\frac{\tau_{a}}{2} U^{\dag} D_{\mu}U \right)^2\ ,
\end{equation}
which yields a prediction for the $\rho$ parameter to be
\begin{equation}
  \rho(\mu)= 1 - \frac{3}{4}\frac{g'^2}{16 \pi^2} \log
  \frac{m_{\sigma}^2}{\mu^2} 
\end{equation}
where the first term is the prediction one gets from imposing the custodial $SU(2)$ symmetry, while the second term comes from the integration of the $\sigma$ resonance. By the same token, integration of resonances in (\ref{Ecker}) should lead to a prediction for $L_{10}$ of the form 
\begin{equation}
L_{10}(\mu)=-\frac{1}{4}\left(\frac{15}{32\pi^2\sqrt{6}}\right)-\frac{1}{64\pi^2}\log{\frac{\Lambda_{\chi}}{\mu}}
\end{equation}
where, if resonance saturation holds in QCD, $\Lambda_{\chi}$ would be a function of the resonance masses and couplings close to the scale of the integrated resonances, {\it{i.e.}}, $\Lambda_{\chi}\,\sim\, m_{\rho}\,\sim\,1$ GeV. For the sake of clarity, table (\ref{tauparall}) summarizes the parallelism between the determination of $L_{10}$ in the resonance chiral Lagrangian and the $\rho$ parameter in the linear sigma model.  
\begin{table}
\begin{center}
\begin{tabular}{|c|}\hline
  $\rho(\mu)\leftrightarrow L_{10}(\mu)$ \\\hline
  $g'\leftrightarrow 1/N_c$ \\\hline
   $\sigma\leftrightarrow \{V,A,S,S_1\}$ \\\hline
  Custodial Symmetry $SU(2)$ $ \leftrightarrow$
  Symmetry of QCD($N_c\rightarrow\infty$)\\ \hline
\end{tabular}
\caption{{\it{Analogies between the $\rho$ parameter in the linear sigma model and the parameter $L_{10}$ in the Resonance Chiral Lagrangian, understood as an approximation to large-$N_C$ QCD.}}}\label{tauparall}
\end{center}
\end{table}
Another analogy can also be drawn with the well-known $SU(5)$ grand unification group, where the seemingly unrelated couplings of the electromagnetic, weak and strong interactions (the analogs to the $L_i$ Gasser-Leutwyler couplings) happen to converge at sufficiently high energies, {\it{i.e.}},
\begin{equation}
g_{SU(3)_C}\,(M_{GUT})\ \sim\ g_{SU(2)_W}\,(M_{GUT})\ \sim\ g_{U(1)_Y}\,(M_{GUT})
\end{equation}
which provides a prediction for the electroweak mixing angle $\theta_W$, to wit
\begin{equation}
\sin^2\ \theta_W\ (\mu)\ =\ \fr{3}{8}-\fr{55}{24}\, \fr{\alpha}{\pi}\ \log{\fr{m_{GUT}}{\mu}}
\end{equation}
The analogy would then be best illustrated as follows
\vskip 0.6cm
\begin{center}
\begin{tabular}{|c|c|}
\hline
\multicolumn{2}{|c|}{$\mathbf{SU(5)}$} \\
\hline
$\sin^2{\theta_W}=\frac{3}{8}$ &
$\alpha_{SU(3)}(M_{GUT})=\alpha_{SU(2)}(M_{GUT})=\alpha_{U(1)}(M_{GUT})$\\
\hline
\end{tabular}
\end{center}
\vspace*{0.3 cm}
\begin{center}
\begin{tabular}{|c|c|}
\hline
\multicolumn{2}{|c|}{\bf{Hypothetical Symmetry of Large--$N_c$}} \\
\hline
$L_{10}=-\frac{15}{8\sqrt{6}}\frac{1}{16\pi^2}$ & $6L_1=3L_2=-\frac{8}{7}L_3=\frac{3}{4}L_9=-L_{10}=\frac{3}{8}\frac{f_{\pi}^2}{M_V^2}=\frac{15}{8\sqrt{6}}\frac{1}{16\pi^2}$\\
\hline
\end{tabular}
\end{center}
\vspace*{0.2 cm}
As stressed above, the fact that the lagrangian can be endowed with a large-$N_c$ behaviour has many advantages. First and foremost, it provides a consistent power counting rule, therefore allowing a consistent computation of quantum corrections. Obviously, the {\it{minimal hadronic approximation}} we are using throughout is not large-$N_c$ QCD, but it can be systematically improved towards that limit with the addition of more resonance states. In the following we will concentrate on a prediction for $L_{10}$ beyond leading order, as an example of how the adoption of the power counting supplied by large-$N_C$ QCD enables one to treat the Lagrangian (\ref{Ecker}) at the quantum level.
\section{Prediction for $L_{10}$ beyond leading order}
Consider the following two-point Green function\footnote{We will follow closely \cite{Cata:1}.}
\begin{equation}
\Pi_{LR}^{\mu\nu}(q)\,\delta_{ab}=
 2i\int d^4 x\,e^{iq\cdot x}\langle 0\vert
\mbox{\rm T}\left(L^{\mu}_a(x)R^{\nu}_b(0)^{\dagger} \right)\vert
0\rangle
\end{equation}
where $R^{\nu}_b$ and $L^{\mu}_a$ are the QCD currents
\begin{equation}
  R_a^{\mu}=
\bar{q}(x)\gamma^{\mu}\ \frac{\lambda_a}{\sqrt{2}}\
\frac{(1+\gamma_5)}{2}\ q(x)\, , \qquad  L_a^{\mu}=
\bar{q}(x)\gamma^{\mu}\ \frac{\lambda_a}{\sqrt{2}}\
\frac{(1-\gamma_5)}{2}\ q(x)\, 
\end{equation}
Lorentz symmetry constraints the tensorial structure to consist of just one form factor
\begin{equation}
\Pi_{LR}^{\mu\nu}(q^2)=(g^{\mu\nu}q^2-q^{\mu}q^{\nu})\,\Pi_{LR}(q^2)
\end{equation}
where the low energy expansion of $\Pi^{LR}(q^2)$ yields
\begin{equation}
\Pi_{LR}(Q^2)=-\frac{F_{0}^2}{Q^2}+4L_{10}+{\cal{O}}(Q^2)
\end{equation}
In other words, it provides a definition for the low energy coupling $L_{10}$, to wit
\begin{equation}\label{l10def}
L_{10}=-\fr{1}{4}\,\fr{d}{dq^2}\, \bigg( \, q^2\ \Pi^{LR}(q^2) \bigg) \bigg|_{q^2=0}
\end{equation}
The previous result is just an example of the very general fact that low energy couplings of the strong interactions can be defined as coefficients of the Taylor expansion of QCD Green functions\footnote{A completely different thing happens for low energy couplings in the electroweak sector, where they are expressible in terms of integrals of Green functions ({\it{cf.}} chapter 5).}.
Our strategy hereafter would be to compute the contributions to $\Pi^{LR}(q^2)$ both in the chiral lagrangian and in the resonance chiral lagrangian. The chiral lagrangian, up to $O(p^4)$ and in the chiral limit reads
\begin{eqnarray}
\Lcal_{\chi PT}&=&\frac{F_0^2}{4}\langle
D_{\mu}U^{\dagger}D^{\mu}U\rangle+ L_1\,\langle
D_{\mu}{U}^{\dagger}D^{\mu}U\rangle^2\nonumber\\
&+&L_2\,\langle
D_{\mu}{U}^{\dagger}D_{\nu}U\rangle\langle
D^{\mu}{U}^{\dagger}D^{\nu}U\rangle+
L_3\,\langle D_{\mu}{U}^{\dagger}D^{\mu}U
D_{\nu}{U}^{\dagger}D^{\nu}U\rangle-\nonumber\\
& - & iL_9\,\langle
F_R^{\mu\nu}D_{\mu}UD_{\nu}{U}^{\dagger}+F_L^{\mu\nu}D_{\mu}{U}^{\dagger}D_{\nu}U\rangle+L_{10}\,\langle{U}^{\dagger}F_R^{\mu\nu}UF_{L\mu\nu}\rangle
\end{eqnarray}
where we omitted the WZW term, which plays no role in the determination of two-point functions. By contrast, the resonance chiral lagrangian in the chiral limit ($\chi\rightarrow 0$) reads
\begin{eqnarray}
\Lcal_{kin}(V,A,S,P)&=&{\sum_{R=V,A}}\left[-\frac{1}{2}\langle {\nabla}^\lambda
R_{\lambda\mu}{\nabla}_\nu
R^{\nu\mu}- \frac{1}{2} M_R^2
R_{\mu\nu}R^{\mu\nu}\rangle\,\right]+\nonumber\\
&+&\left[\frac{1}{2}\, \langle
\nabla^\mu S \nabla_\mu S - M_S^2 S^2\rangle +\, \frac{1}{2}\, \left(
\partial^\mu S_1 \partial_\mu S_1 - M_{S_1}^2 S_1^2\right)\right]+\nonumber\\
&+&\left \{ \frac{F_{V}}{2\sqrt{2}}\,\langle\,V_{\mu\nu}f_{+}^{\mu\nu}\rangle\,+\,i\frac{G_{V}}{\sqrt{2}}\langle\,
V_{\mu\nu}u^{\mu}u^{\nu}\rangle\right \}+\nonumber\\
&+& \frac{F_{A}}{2\sqrt{2}}\,\langle A_{\mu\nu}f_{-}^{\mu\nu}\rangle+\nonumber\\
&+&\bigg\{ c_{d} \,\langle S u_{\mu} u^\mu \rangle\,
 + \tilde{c}_{d} \, S_{1}\,
\langle u_{\mu} u^\mu \rangle\bigg\}+\nonumber\\
&+&\frac{F_0^2}{4}\langle
D_{\mu}U^{\dagger}D^{\mu}U\rangle+ {\widehat{L}}_1\langle
D_{\mu}{U}^{\dagger}D^{\mu}U\rangle^2+\cdots +{\widehat{L}}_{10}\langle{U}^{\dagger}F_R^{\mu\nu}UF_{L\mu\nu}\rangle \nonumber\\
&&
\end{eqnarray}
where $L_i$ and ${\widehat{L}}_i$ are as defined in (\ref{lidef}). The previous lagrangian lacks the pseudoscalar channel, since it does not couple either to $L_{\mu}$ nor to $R_{\mu}$ currents, as it can be checked by inspection. The same reasoning justifies the absence of vector and axial vector singlet states.
\begin{figure}
\renewcommand{\captionfont}{\small \it}
\renewcommand{\captionlabelfont}{\small \it}
\centering
\psfrag{A}{$R_{\mu}$}
\psfrag{B}{$L_{\nu}\,\,\, \equiv \,$}
\psfrag{C}{$L_{10}$}
\includegraphics[width=1.4in]{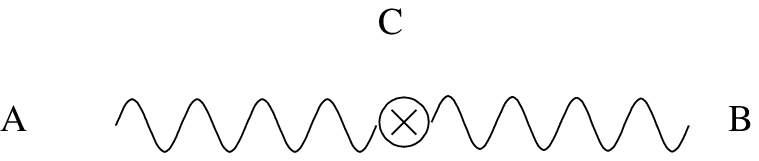}
\hspace{1cm}
\psfrag{A}{$R_{\mu}$}
\psfrag{B}{$L_{\nu}\,\,\, +\,$}
\psfrag{C}{$V_{\mu\nu},A_{\mu\nu}$}
\includegraphics[width=1.4in]{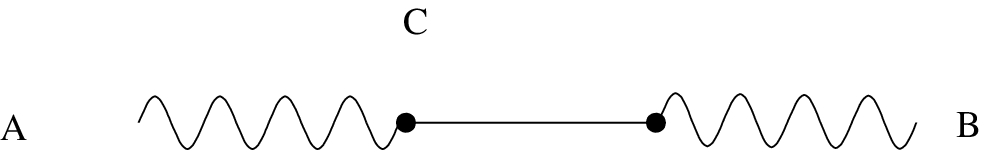}
\hspace{0.8cm}
\psfrag{A}{$R_{\mu}$}
\psfrag{B}{$L_{\nu}$}
\psfrag{C}{${\widehat{L}}_{10}$}
\includegraphics[width=1.4in]{point.eps}
\caption{Matching condition for (\ref{l10defff}) at leading order, the left-hand side diagram coming from the chiral Lagrangian and the right-hand side ones from the resonance chiral Lagrangian.}\label{dotl10}
\end{figure}
Computation of (\ref{l10def}) at leading order is straightforward and leads to the matching condition 
\begin{equation}\label{l10defff}
L_{10}=\frac{F_A^2}{4M_A^2}-\frac{F_V^2}{4M_V^2}+{\widehat{L}}_{10}=-\frac{1}{4}\left(\frac{15}{32\pi^2\sqrt{6}}\right)
+{\widehat{L}}_{10} 
\end{equation}
where (\ref{decay}) and (\ref{masses}) were used in the second equality. This matching condition is depicted in figure (3.5).
Next to leading order corrections are also relatively easy to compute in the chiral lagrangian side. We shall use the standard dimensional regularisation with the renormalization scheme chosen in the foundational papers of Gasser and Leutwyler. This leads to
\begin{equation}\label{matching1}
-\fr{d}{dq^2}\, \bigg( \, q^2\, \Pi^{LR}_{\chi PT}(q^2) \bigg) \bigg|_{q^2=0}=4L_{10}(\mu)-\frac{1}{32\pi^2}\left(\frac{5}{3}-\log{\frac{-q^2}{\mu^2}}\right)
\end{equation}
where the second term corresponds to the pion loop depicted in figure (\ref{resonloop}).

Its counterpart with the resonance chiral lagrangian requires a bit more effort. Diagrams to be taken into account are listed in figure \ref{resonloop}. Their contributions render {\footnote{To be consistent, we will all along stick to the Gasser-Leutwyler renormalization scheme introduced in chapter 2.}}
\begin{eqnarray}\label{matching2}
-\fr{d}{dq^2}\, \bigg( \, q^2\, \Pi^{LR}_{R \chi}(q^2) \bigg) \bigg|_{q^2=0}&=&\frac{3}{2}\frac{1}{(4\pi)^2}\left(\frac{F_V}{f_\pi}\right)^2\left(\frac{1}{2}-
\log\frac{M_V^2}{\mu^2}\right)+\frac{3}{2}\frac{1}{(4\pi)^2}\left(-\frac{1}{3}-\log\frac{M_V^2}{\mu^2}\right)-\nonumber\\
&\!\!\!\!\!\!\!\!\!\!\!\!\!\!\!\!\!\!\!\!\!\!\!\!\!\!-&\!\!\!\!\!\!\!\!\!\!\!\!\!\!\!\!\frac{5}{(4\pi)^2}\left(\frac{G_V}{f_\pi}\right)^2
\left(-\frac{17}{30}-\log\frac{M_V^2}{\mu^2}\right)-\frac{3}{2}\frac{1}{(4\pi)^2}\left(\frac{F_A}{f_\pi}\right)^2\left(\frac{1}{2}-
\log\frac{M_A^2}{\mu^2}\right)+\nonumber\\
&\!\!\!\!\!\!\!\!\!\!\!\!\!\!\!\!\!\!\!\!\!\!\!\!\!\!+&\!\!\!\!\!\!\!\!\!\!\!\!\!\!\!\!\frac{3}{2}\frac{1}{(4\pi)^2}\left(-\frac{1}{3}-\log\frac{M_A^2}{\mu^2}\right)-\frac{4}{3}\frac{1}{(4\pi)^2}\left(\frac{{\tilde{c}}_d}{f_{\pi}}\right)^2
\left(\frac{1}{6}+\log\frac{M_{S_1}^2}{\mu^2}\right)-\nonumber\\
&\!\!\!\!\!\!\!\!\!\!\!\!\!\!\!\!\!\!\!\!\!\!\!\!\!\!-&\!\!\!\!\!\!\!\!\!\!\!\!\!\!\!\!\frac{10}{9}\frac{1}{(4\pi)^2}
\left(\frac{c_d}{f_{\pi}}\right)^2
\left(\frac{1}{6}+\log\frac{M_S^2}{\mu^2}\right)+\frac{1}{2}\frac{1}{(4\pi)^2}
\left(1+\log\frac{M_S^2}{\mu^2}\right)-\nonumber\\
&\!\!\!\!\!\!\!\!\!\!\!\!\!\!\!\!\!\!\!\!\!\!\!\!\!\!-&\!\!\!\!\!\!\!\!\!\!\!\!\!\!\!\!\frac{4}{9}\frac{1}{(4\pi)^2}\left(\frac{c_d}{f_{\pi}}\right)^2
\left[\frac{1}{6}+\log\frac{M_S^2}{\mu^2}+ 2B + 2B^2
    - (2 B^3 + 3 B^2)
    \log\frac{M_S^2}{M_{\eta_{1}}^2}\right]+\nonumber\\
&\!\!\!\!\!\!\!\!\!\!\!\!\!\!\!\!\!\!\!\!\!\!\!\!\!\!+&\!\!\!\!\!\!\!\!\!\!\!\!\!\!\!\!4{\widehat{L}}_{10}(\mu)-\frac{1}{32\pi^2}\left(\frac{5}{3}-\log{\frac{-q^2}{\mu^2}}\right)
\end{eqnarray}
where $B$ is the mass ratio
\begin{equation}
B=\frac{M_{\eta_{1}}^2}{(M_S^2-M_{\eta_{1}}^2)}
\end{equation}
The first important point to stress is the vectorial dominance in the $\beta$-function of the $L_{10}$ coupling. Indeed, using (\ref{decay}) the logarithmic dependence on the axial channel identically vanishes.
\begin{equation}
\bigg[-\frac{3}{2}\frac{1}{(4\pi)^2}\left(\frac{F_A}{f_\pi}\right)^2\left(\frac{1}{2}-
\log\frac{M_A^2}{\mu^2}\right)+\frac{3}{2}\frac{1}{(4\pi)^2}\left(-\frac{1}{3}-\log\frac{M_A^2}{\mu^2}\right)\bigg]\bigg|_{F_A=f_{\pi}}=-\fr{5}{4}\fr{1}{(4\pi)^2}
\end{equation}
This again is intriguing: the relations between the parameters (\ref{predictions}) suffices to ensure this vector dominance. Also the scalar sector shows some degree of cancellation in the interplay between the singlet and the octet. It seems that somehow the hypothetical symmetry behind (\ref{predictions}) tends to protect the {\it{vector meson dominance}} even at the one-loop level. 

It is also worth mentioning that the previous equation is very insensitive to the $\eta_1$ mass. Recall that the $\eta_1$ mass arises as a $1/N_C$ effect, therefore decoupling from the octet away from the strict large-$N_C$ limit. Should large-$N_C$ be a good approximation to the $N_C=3$ real world, one necessary condition is precisely this mild dependence on the $\eta_1$ particle. Therefore, we find this a very appealing feature, pointing at a smooth transition between the large-$N_C$ limit and $N_C=3$ QCD.  
\begin{figure}
\renewcommand{\captionfont}{\small \it}
\renewcommand{\captionlabelfont}{\small \it}
\centering
\psfrag{A}{$R_{\mu}$}
\psfrag{B}{$L_{\nu}$}
\psfrag{C}{$V_{\mu\nu},A_{\mu\nu},S$}
\psfrag{D}{$V_{\mu\nu},A_{\mu\nu},S$}
\includegraphics[width=2.1in]{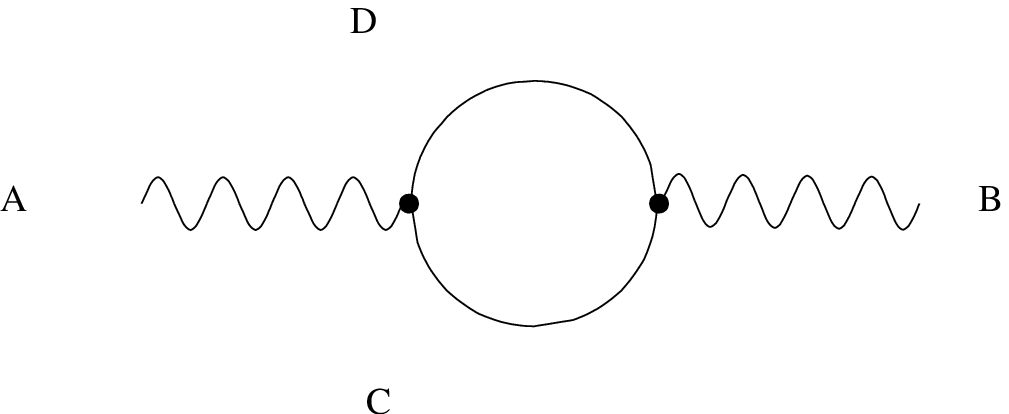}
\hskip 1 cm
\psfrag{C}{$\pi$}
\psfrag{D}{$V_{\mu\nu},A_{\mu\nu},S,S_1$}
\includegraphics[width=2.0in]{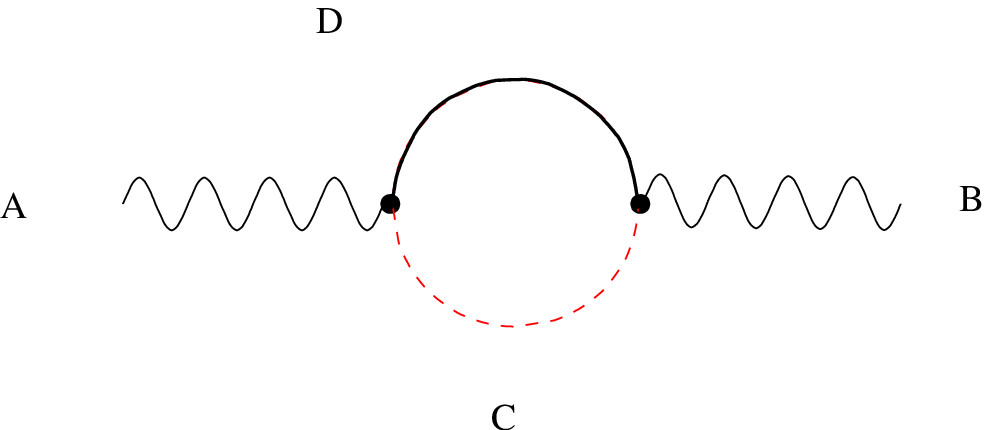}\\
\vskip 1 cm
\psfrag{C}{$\pi$}
\psfrag{D}{$V_{\mu\nu}$}
\includegraphics[width=2.1in]{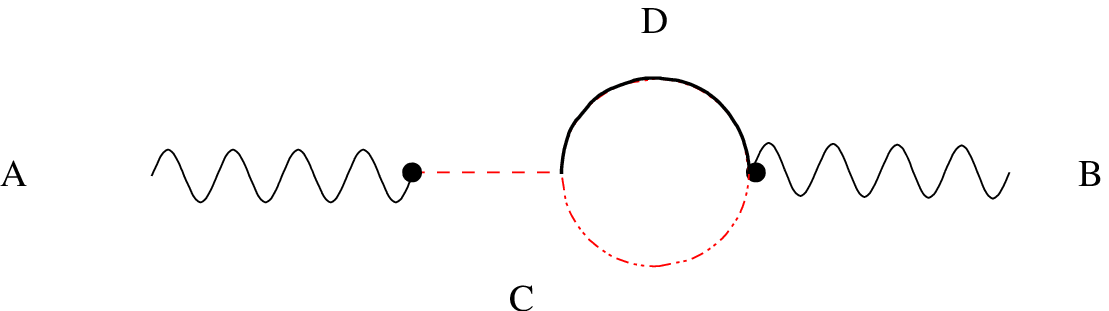}
\hskip 1 cm
\psfrag{C}{$\pi$}
\psfrag{D}{$V_{\mu\nu}$}
\includegraphics[width=2.1in]{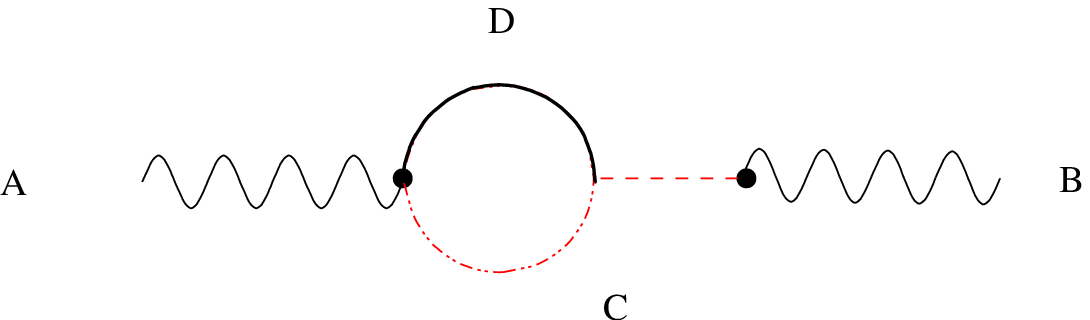}
\vskip 0.7cm
\psfrag{A}{$R_{\mu}$}
\psfrag{B}{$L_{\nu}$}
\psfrag{C}{$\pi$}
\psfrag{D}{$\pi$}
\includegraphics[width=2.1in]{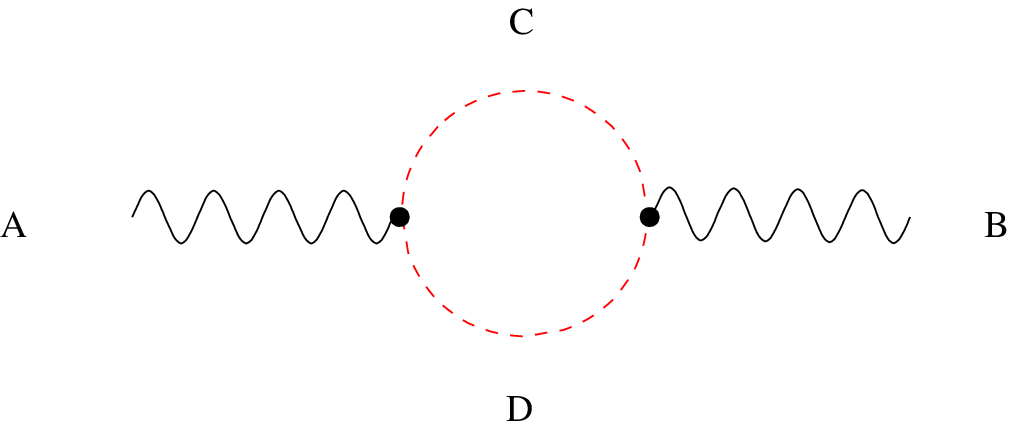}
\caption{Next to leading order contribution to (\ref{l10def}) stemming from the Chiral Lagrangian and the Resonance Chiral Lagrangian. The pion loop is reproduced by both Lagrangians and identically cancels in the matching condition (see main text).}\label{resonloop}
\end{figure}

Equating (\ref{matching1}) and (\ref{matching2}) yields the determination for $L_{10}$ up to next to leading order, 
\begin{eqnarray}\label{resultinch2}
4\ L_{10}^r(\mu)&=&-\frac{15}{32\pi^2\sqrt{6}}\nonumber\\
&&-\frac{3}{2}\frac{1}{(4\pi)^2}\left(\frac{F_A}{f_\pi}\right)^2\left(\frac{1}{2}-
\log\frac{M_A^2}{\mu^2}\right)+
\frac{3}{2}\frac{1}{(4\pi)^2}\left(\frac{F_V}{f_\pi}\right)^2\left(\frac{1}{2}-
\log\frac{M_V^2}{\mu^2}\right)\nonumber\\
&&-\frac{5}{(4\pi)^2}\left(\frac{G_V}{f_\pi}\right)^2
\left(-\frac{17}{30}-\log\frac{M_V^2}{\mu^2}\right)\nonumber\\
&&+\frac{3}{2}\frac{1}{(4\pi)^2}\left(-\frac{1}{3}-\log\frac{M_A^2}{\mu^2}\right)
+\frac{3}{2}\frac{1}{(4\pi)^2}\left(-\frac{1}{3}-\log\frac{M_V^2}{\mu^2}\right)
\nonumber\\
&&
-\frac{4}{3}\left(\frac{{\tilde{c}}_d}{f_{\pi}}\right)^2\frac{1}{(4\pi)^2}
\left(\frac{1}{6}+\log\frac{M_{S_1}^2}{\mu^2}\right)-\frac{10}{9}\frac{1}{(4\pi)^2}
\left(\frac{c_d}{f_{\pi}}\right)^2
\left(\frac{1}{6}+\log\frac{M_S^2}{\mu^2}\right)\nonumber\\
&&+\frac{1}{2}\frac{1}{(4\pi)^2}
\left(1+\log\frac{M_S^2}{\mu^2}\right)-\nonumber\\
&&-\frac{4}{9}\frac{1}{(4\pi)^2}\left(\frac{c_d}{f_{\pi}}\right)^2
\left[\frac{1}{6}+\log\frac{M_S^2}{\mu^2}+ 2B + 2B^2 - (2 B^3 + 3 B^2) \log\frac{M_S^2}{M_{\eta_{1}}^2}\right]\nonumber\\
&&+ 4\ ({\widehat{L}}_{10})^r(\mu) 
\end{eqnarray} 
We are now in a position to assess whether the {\it{resonance saturation}} survives at the quantum level, {\it{i.e.}}, whether the contribution from the higher mass multiplets to $L_{10}$, which we noted as ${\widehat{L}}_{10}$, can be dropped from
\begin{equation}
L_{10}(\mu)=L_{10}^R(\mu)+{\widehat{L}}_{10}(\mu)
\end{equation}
It turns out that one can not get rid of this term (see \cite{Cata:1} for details) or, in other words, the integration of the resonances does not predict the right evolution for $L_{10}$ under the renormalization group equation, as dictated by chiral perturbation theory
\begin{equation}
L_{10}(\mu)=L_{10}(M_{\rho})-\fr{1}{64\pi^2}\,\log{\fr{M_{\rho}}{\mu}}
\end{equation}
Our conclusion is that the Lagrangian of \cite{Sui1}, despite yielding successful phenomenological results at tree level, fails at the quantum level. We are not the first ones to claim that that Lagrangian is incomplete\footnote{See, {\it{e.g.}}, \cite{Mouss:incom,Knecht:Ny,Rene:2002}.}. The message we wanted to convey in \cite{Cata:1} is that one can go beyond tree level with such resonance Lagrangians, because the large-$N_C$ framework provides an expansion parameter with which to do quantum corrections in a consistent way, namely $1/N_C$.

In the past three years there have been considerable efforts to improve on the Lagrangian of \cite{Sui1}, by adding more operators and looking for agreement with QCD short distances of certain two and three-point Green functions at the quantum level in the $1/N_C$ expansion\footnote{See, {\it{e.g.}}, \cite{Cir:2005,Rosell} and references therein.}.  However, as already pointed out, the absence of a power counting rule to tell which terms have the bigger impact makes it difficult to make further progress in that direction\footnote{Recall that we are actually attempting to model the large-$N_C$ Lagrangian of QCD. Therefore, large-$N_C$ power counting rules can only tell us how to go to the quantum realm once the Lagrangian is given. For this ambitious task we would need to know much than we nowadays do about the $1/N_C$ expansion itself.}: the new operators included {\it{ad hoc}} to improve on certain Green functions might as well spoil other Green functions not yet considered. 

From this point of view, we think that we can provide one of the simplest starting tests {\it{any}} resonance Lagrangian has to pass. The addition of more operators has to render, among other things, the right running for $L_{10}(\mu)$. This is a necessary condition to eventually yield a finite prediction for $\Pi^{LR} (q^2)$, one of the simplest QCD two-point functions. Should we have that Green function under control, we could then move to more complicated ones. However, to the best of our knowledge, no prediction exists yet even for $L_{10}(\mu)$. 


\chapter[Hadronic Matrix Elements of Kaons]{Hadronic Matrix Elements of Kaons}

In chapter 2 we saw that the CKM quark mixing matrix had a non-factorizable phase signalling at CP violation in the Standard Model. In this chapter we will show how this phenomenon manifests itself at the meson level in neutral particle-antiparticle systems, the paradigmatic one being the $K^0-{\bar{K}}^0$ system, to which we will devote our attention\footnote{We will follow the treatments given in \cite{John} and \cite{rafael:kaon}.}. Two mechanisms combine as sources of CP violation: mixing of $K^0-{\bar{K}}^0$ and kaon decay. The first part of the chapter is oriented to characterize these effects in terms of a few phenomenological parameters, to be determined later on.
In chapter 2 we already saw that the Standard Model at low energies admitted an expansion in powers of momenta with the Goldstone octet as dynamical fields. This will be the appropriate tool to be employed all through our analysis.

\section{Phenomenology of kaon CP-Violation}

The neutral kaons that appear in the chiral lagrangian are the ones predicted by the Eightfold Way of Gell-Mann and Ne'eman, namely
\begin{equation}
K^0=\bar{s}d\, ,\quad {\bar{K}}^0=\bar{d}s
\end{equation}
They are eigenstates of strangeness, but they have no well-defined CP parity. They transform in the following way\footnote{Actually, in full generality they are related by a phase $\xi_K$, 
\begin{equation}
CP\,|K^0\rangle=\xi_K\,|{\bar{K}}^0\rangle\nonumber
\end{equation}
which we choose to be $\xi_K=-1$ for convenience. Physical quantities should be independent of the phase choice.}  
\begin{equation}
CP\,|\,K^0\rangle=-|\,{\bar{K}}^0\rangle
\end{equation}
In the absence of electroweak interactions, neutral kaons are stable and they behave as distinct states, since they are protected by their different strangeness quantum number,{\it{i.e.}}, no mixing nor decay is to be expected, and since they are antiparticles, CPT guarantees that they have degenerate masses. However, once the electroweak interaction is switched on, they are allowed to decay weakly into two-pion and three-pion states. Strangeness is no longer a conserved quantum number and, owing to the well-defined CP parity of the decay modes, it is convenient to build the CP conserving states,
\begin{equation}
K_1=\fr{1}{\sqrt{2}}\bigg(K^0-\bar{K}^0\bigg),\quad K_2=\fr{1}{\sqrt{2}}\bigg(K^0+\bar{K}^0\bigg)
\end{equation}
whence,
\begin{equation}
CP\,|\,K_1\rangle=+\,|K_1\rangle\, ,\quad CP\,|\,K_2\rangle=-\,|K_2\rangle
\end{equation}
Hence, were CP conserved in the neutral kaon system, $K_1$ should decay exclusively into two-pion states and $K_2$ into three-pion states. The observed difference in lifetimes
\begin{equation}
\Gamma_L^{-1}\simeq 10^3\,\Gamma_S^{-1}
\end{equation}
is what one would expect by na{\"i}ve phase space arguments.
However, these states are not the physical ones, because even CP is violated by the electroweak interactions. Quite generally, we can write the physical kaon basis as
\begin{eqnarray}
|\,K_L\rangle&=& \fr{1}{\sqrt{|p^2|+|q^2|}}\bigg(p\,|K^0\rangle+q\,|\bar{K}^0\rangle\bigg)\nonumber\\
|\,K_S\rangle&=& \fr{1}{\sqrt{|p^2|+|q^2|}}\bigg(p\,|K^0\rangle-q\,|\bar{K}^0\rangle\bigg)
\end{eqnarray}
Departure from CP conservation is signalled by the parameter $\tilde{\epsilon}$, defined as
\begin{equation}
\tilde{\epsilon}=\fr{p-q}{p+q}
\end{equation}
in terms of which, the physical kaons look like
\begin{eqnarray}
|\,K_L\rangle&=& \fr{1}{\sqrt{1+|\tilde{\epsilon}|^2}}\bigg(|K_2\rangle+\tilde{\epsilon}\,|\bar{K}_1\rangle\bigg)\nonumber\\
|\,K_S\rangle&=& \fr{1}{\sqrt{1+|\tilde{\epsilon}|^2}}\bigg(|K_1\rangle+\tilde{\epsilon}\,|\bar{K}_2\rangle\bigg)
\end{eqnarray}
\subsection{$K^0-{\bar{K}}^0$ mixing}
If we define the wavefunction describing the kaon system evolution as
\begin{equation}
|\Psi(t)\rangle\equiv c_1(t)\,|K^0\rangle+c_2(t)\,|{\bar{K}}^0\rangle
\end{equation}
elementary quantum mechanics of two-state systems renders the following Schr{\"o}dinger equation
\begin{equation}
i\fr{d}{dt}|\Psi(t)\rangle={\mathcal{M}}\,|\Psi(t)\rangle
\end{equation}
where ${\mathcal{M}}$ stands for the Hamiltonian,
\begin{equation}
{\cal{M}}=\left(\begin{array}{cc}
\langle K^0|H_{SM}|K^0\rangle & \langle K^0|H_{SM}|\bar{K}^0\rangle\nonumber\\
\langle {\bar{K}}^0|H_{SM}|K^0\rangle & \langle \bar{K}^0|H_{SM}|\bar{K}^0\rangle
\end{array}\right)
\end{equation}
whose entries can be parametrized as follows 
\begin{equation}
{\cal{M}}_{ij}=M_{ij}-\fr{i}{2}\Gamma_{ij}
\end{equation}
The imaginary part in the equation above accounts for the fact that kaons actually decay.
Conventional quantum mechanics perturbation theory then yields
\begin{equation}\label{pert}
{\cal{M}}_{ij}=m_K\,\delta_{ij}+\fr{1}{2 m_K}\langle K_i^0|H_{eff}^{\Delta S=2}|K_j^0\rangle+\fr{1}{2 m_K}\sum_n\fr{\langle K_i^0|H_{eff}^{\Delta S=1}| n\rangle \langle n|H_{eff}^{\Delta S=1}|K_j^0\rangle}{m_K - E_n + i \epsilon}
\end{equation}
which provides a link between kaon masses and kaon decay widths in terms of the dynamics of the Standard Model.
CPT invariance, which is a consequence of any quantum field theory, further constrains the mass matrix elements \cite{MIT},
\begin{equation}
{\cal{M}}_{11}={\cal{M}}_{22} \, , \quad {\cal{M}}_{12}=({\cal{M}}_{21})^*
\end{equation}  
The diagonal terms $M_{ij}$ are just the masses of the neutral kaons, whereas the off-diagonal entries account for kaon mixing. In terms of kaon masses and decay widths the parameter $\tilde{\epsilon}$ can then be written as
\begin{equation}\label{defeps}
\tilde{\epsilon}=\fr{p-q}{p+q}=\fr{i}{2}\,\fr{{\mathrm{Im}}M_{12}-\fr{i}{2}\,{\mathrm{Im}}\Gamma_{12}}{{\mathrm{Re}}M_{12}-\fr{i}{2}\,{\mathrm{Re}}\Gamma_{12}}
\end{equation}
This in particular means that if kaons were stable nonetheless they would mix.{\footnote{There is increasing evidence that neutrinos indeed follow this pattern of mixing without decaying.}} This suggests to study kaon decay and kaon mixing separately. 
 
For kaon mixing, our main concern will be the kaon mass difference, to be defined as
\begin{equation}
\Delta m_K\equiv m_L-m_S
\end{equation}
A close look at equation (\ref{pert}) reveals that the kaon mass difference receives contributions from the $\Delta S=2$ sector of the Standard Model, which connects both neutral kaons through the well-known box diagrams, depicted in figure (\ref{box}). This contribution leads to the $\Delta S=2$ Hamiltonian we mentioned at the end of chapter 2, whose leading order in inverse powers of the masses of the integrated out particles is
\begin{equation}
\Lcal_2^{\Delta S=2}=-\fr{G_F^2m_W^2}{4\pi^2}\,\bigg[\lambda_c^2 F_1+\lambda_t^2F_2+2\lambda_c\lambda_t F_3\bigg]\,c_i(\mu)\,{Q}_i(\mu)
\end{equation}
where
\begin{equation}
\lambda_c=V_{cd}V_{cs}^*\, , \qquad \lambda_t=V_{td}V_{ts}^*
\end{equation}
and
\begin{equation}
Q=\bar{s}_L\gamma^{\mu}d_L \bar{s}_L\gamma_{\mu}d_L
\end{equation}
$F_1,F_2,F_3$ account for the electroweak and strong loop corrections, to be defined in the next chapter.
Conventionally, one defines the bag parameter $B_K$ in the following manner
\begin{equation}
\left< {\bar{K}}^0 |\,Q (0)|K^0\right> \doteq \fr{4}{3}F_K^2m_K^2B_K(\mu^2)
\end{equation} 
such that it becomes the parameter which governs kaon mixing. The determination of the $B_K$ parameter is the subject of the next section. As it is clear from (\ref{pert}), though, $B_K$ alone does not provide the whole kaon mass difference. There are also long-distance contributions in the second term in (\ref{pert}), with which we will deal in the next chapter. Furthermore, the last term in (\ref{pert}) is a genuinely long-distance contribution. If we restrict the sum over intermediate states to be saturated by the $\pi\pi$ exchange, which is a good approximation taking into account that it is this the dominant kaon decay mode, the long-distance contributions can be cast in the form of $\bigg[(\Delta S=1)\times(\Delta S=1)\bigg]$ pieces, as it appears in (\ref{pert}). There is some rationale in neglecting these $\bigg[(\Delta S=1)\times(\Delta S=1)\bigg]$ contributions \cite{John}, but no estimation of these terms has been performed so far.
  
\subsection{Kaon non-leptonic weak decays}
So far we have only remarked the fact that the electroweak interactions do not conserve CP parity, but we have not assessed yet to what extent. Were CP a good quantum number, then the aforementioned CP basis $K_1$, $K_2$ would be the physically observed basis, $K_S$, $K_L$. The experimental fact that the long-lived kaon indeed decays into a two-pion state rules out this possibility and it is seen as evidence for CP violation. To evaluate the size of this effect, several observables can be constructed, for instance
\begin{eqnarray}
\eta_{+-}&=&\fr{\langle \pi^+\pi^-|\,T\,|K_L\rangle}{\langle \pi^{+}\pi^{-}|\,T\,|K_S\rangle}\nonumber\\
\eta_{00}&=&\fr{\langle \pi^0\pi^0|\,T\,|K_L\rangle}{\langle \pi^0\pi^0|\,T\,|K_S\rangle}
\end{eqnarray} 
It is however convenient, in view of the phenomenological treatments to come, to express any observable in terms of well-defined isospin states{\footnote{Actually, there is a deeper reason: CPT invariance combined with Watson's theorem relate both neutral kaons to the same isospin final state. For a proof, we refer to {\cite{MIT}}.}}. Bose symmetry forbids $I=1$ transitions, so that we are left with $I=0$ and $I=2$ states, 
\begin{eqnarray}
|\pi^+\pi^- \rangle &=&\sqrt{\fr{2}{3}}\ |(\pi^+\pi^-)_0\rangle+\sqrt{\fr{1}{3}}\ |(\pi^+\pi^-)_2\rangle\nonumber\\
|\pi^0\pi^0 \rangle &=&-\sqrt{\fr{1}{3}}\ |(\pi^0\pi^0)_0\rangle+\sqrt{\fr{2}{3}}\ |(\pi^0\pi^0)_2\rangle
\end{eqnarray}
We define
\begin{equation}\label{definitions}
\langle (\pi\pi)_I|\,T\,|K^0\rangle=i\,A_I\,e^{i\delta_I}\quad , \quad\langle (\pi\pi)_I|\,T\,|{\bar{K}}^0\rangle=-i\,A_I^*\,e^{i\delta_I} 
\end{equation}
The natural ratios one can build thereof are
\begin{equation}
\zeta \equiv \fr{{\cal{A}}[K_L\rightarrow (\pi\pi)_{I=2}]}{{\cal{A}}[K_S\rightarrow (\pi\pi)_{I=0}]}
\end{equation}
\begin{equation}
\omega \equiv \fr{{\cal{A}}[K_S\rightarrow (\pi\pi)_{I=2}]}{{\cal{A}}[K_S\rightarrow (\pi\pi)_{I=0}]}
\end{equation}
\begin{equation}\label{epsilon}
\varepsilon_K \equiv \fr{{\cal{A}}[K_L\rightarrow (\pi\pi)_{I=0}]}{{\cal{A}}[K_S\rightarrow (\pi\pi)_{I=0}]}
\end{equation}
Nonetheless, it is commonplace to use the following basis instead
\begin{equation}
\varepsilon_K
\end{equation}
\begin{equation}
\omega
\end{equation}
\begin{equation}
\varepsilon_K^\prime=\fr{1}{\sqrt{2}}\bigg(\zeta-\varepsilon_K\cdot \omega\bigg)
\end{equation}
which relate to the physically measurable quantities $\eta_{+-}$ and $\eta_{00}$ as follows
\begin{eqnarray}
\eta_{+-}&=&\varepsilon_K+\varepsilon_K^\prime\,\fr{1}{1+\fr{1}{\sqrt{2}}\omega}\nonumber\\
\eta_{00}&=&\varepsilon_K-2\,\varepsilon_K^\prime\,\fr{1}{1-\sqrt{2}\omega}
\end{eqnarray}
Plugging (\ref{definitions}) in (\ref{epsilon}) we get
\begin{equation}
\varepsilon_K=\fr{{\tilde{\epsilon}}\ {\textrm{Re}}A_0+i\,{\mathrm{Im}}A_0}{{\textrm{Re}}A_0+i\,{\tilde{\epsilon}}\,{\mathrm{Im}}A_0}
\end{equation}
and
\begin{equation}
\varepsilon_K^\prime=\fr{i}{\sqrt{2}}(1-\tilde{\epsilon}^2)\fr{{\mathrm{Im}}A_2{\mathrm{Re}}A_0-{\mathrm{Im}}A_0{\mathrm{Re}}A_2}{({\mathrm{Re}}A_0+i\,\tilde{\epsilon}\,{\mathrm{Im}}A_0)^2}\,e^{i(\delta_2-\delta_0)}
\end{equation}
The previous expressions can be simplified using the following approximations, supported by phenomenology
\begin{equation}
\tilde{\epsilon}^2\ll 1\, , \quad \qquad \tilde{\epsilon}\,{\mathrm{Im}}A_0\ll {\mathrm{Re}}A_0\,\qquad 
\end{equation}
which yield
\begin{eqnarray}\label{approx}
\varepsilon_K&\sim&\tilde{\epsilon}+i\,\fr{{\mathrm{Im}}A_0}{{\mathrm{Re}}A_0}\nonumber\\
\varepsilon_K^\prime&\sim&\frac{1}{\sqrt{2}}\,\fr{{\mathrm{Re}} A_2}{{\mathrm{Re}} A_0} \left (\fr{{\mathrm{Im}} A_2}{{\mathrm{Re}} A_2}- \fr{{\mathrm{Im}} A_0}{{\mathrm{Re}} A_0} \right )\,e^{i(\delta_2-\delta_0+\fr{\pi}{2})}
\end{eqnarray}
Our expression for $\varepsilon_K$ can be further simplified if we make the additional approximations
\begin{displaymath}
{\mathrm{Re}}M_{12}\sim \fr{\Delta m_K}{2}\sim \fr{\Gamma_S}{4}\, , \quad \qquad {\mathrm{Re}}\Gamma_{12}\sim -\fr{\Delta \Gamma}{2}\sim \fr{\Gamma_S}{4}
\end{displaymath}
\begin{equation}
\Gamma_{12}\sim -({\mathrm{Re}}A_0+i\,{\mathrm{Im}}A_0)^2
\end{equation}
where in the last one we consider the width to be dominated by the $\pi\pi$ decay channel. Implementing the previous conditions onto the definition for ${\tilde{\epsilon}}$ (\ref{defeps}) results in 
\begin{equation}
\tilde{\epsilon}\sim \fr{1}{1+i}\left (i\,\fr{{\mathrm{Im}}M_{12}}{\Delta m_K}+\fr{{\mathrm{Im}}A_0}{{\mathrm{Re}}A_0}\right)
\end{equation}
from where, plugging the previous equation in (\ref{approx}), we end up with our final expression for $\varepsilon_K$
\begin{equation}
\varepsilon_K=\fr{1}{\sqrt{2}}\left(\fr{{\textrm{Im}}M_{12}}{\Delta m_K}+\fr{{\textrm{Im}} A_0}{{\textrm{Re}} A_0}\right)\,e^{i\fr{\pi}{4}}
\end{equation} 
From the previous equations we can readily see that $\epsilon_K$ probes both kaon mixing and kaon decay, whereas $\varepsilon_K^\prime$ is sensitive to direct CP violation in the form of the interference of the different isospin decay modes.  

\section{Effective Lagrangians for kaon processes}
Having discussed the basic formalism for the phenomenology of kaon physics, we turn our attention to its description in the framework of the Standard Model. In chapter 2 we introduced the low energy description of the Standard Model. We will focus on strangeness changing processes, $\Delta S=1$ and $\Delta S=2$, which account for kaon decay and kaon mixing, respectively. 
\subsection{$\Delta S=1$ transitions} 
$\Delta S=1$ are mediated in the Standard Model through the exchange of virtual W particles. Nevertheless, we can make use of effective field theory techniques to simplify our analysis. Operators in the flavour changing charged sector of the Standard Model giving rise to $\Delta S=1$ transitions appear as the convolution of two currents with a $W$. We can therefore integrate the $W$ particle, which results in a series expansion, the first term of which (the lowest dimension operator) is equivalent to shrinking the propagator to a point-like vertex. We are then left with a product of two currents, and we recover Fermi theory of the weak interactions. The remaining terms of the expansion are higher-dimensional local operators which conform the Operator Product Expansion, a topic to be discussed further in the last chapter. We end up in this very first stage with a set of local operators whose Wilson coefficients contain inverse powers of the W-mass. 

We then have to run the operators down to lower energies until we find the next particle to be integrated out. The evolution of the operators and Wilson coefficients in the energy gap between two particles is dictated by the renormalization group analysis. Heavy quarks can be integrated following these steps: at a particle mass threshold one integrates out the particle form the effective action, and evolves down the resulting operators with renormalization group techniques. Proceeding this way one can end up with an effective Lagrangian for strangeness changing processes of the form
\begin{equation}
\Lcal_2^{\Delta S=1}=-\fr{G_F}{\sqrt{2}}\,\lambda_u \sum_{i=1}^{10}\, c_i(\mu)\,Q_i(\mu)
\end{equation}
where
\begin{equation}
\lambda_u=V_{ud}V_{us}^*
\end{equation}
are the CKM matrix elements defined in chapter 2 and $Q_i$ are the leading operators ({\it{i.e.}}, the lowest dimensional ones) made up with the left-over light quark fields.

The method sketched above is not free of subleties: renormalization evolution equations take into account the behaviour of the operators under the strong and electromagnetic interactions. This not only affects the Wilson coefficients, but quite often generates new operators, which mix between each other under the renormalization group equation flow. So, unless the operators are {\it{multiplicatively renormalizable}}, new operators will appear\footnote{For an exhaustive review, see \cite{Buras:98}.}. That is why we started from one single operator in the lagrangian of the Standard Model and end up with ten, to wit  
\begin{equation}\label{S1}
Q_1=4\,(\bar{s}_L\gamma^{\mu}d_L)(\bar{u}_L\gamma_{\mu}u_L)\, , \quad
Q_2=4\,(\bar{s}_L\gamma^{\mu}u_L)(\bar{u}_L\gamma_{\mu}d_L)
\end{equation}
which are the so-called {\it{current-current}} operators. Furthermore, there are the so-called {\it{penguin}} operators
\begin{eqnarray}\label{penguin}
Q_3=4\,(\bar{s}_L\gamma^{\mu}d_L)\sum_q\,(\bar{q}_L\gamma_{\mu}q_L)&&
Q_4=4\,\sum_q\,(\bar{s}_L\gamma^{\mu}q_L)(\bar{q}_L\gamma_{\mu}d_L)\nonumber\\
Q_5=4\,(\bar{s}_L\gamma^{\mu}d_L)\sum_q\,(\bar{q}_R\gamma_{\mu}q_R)&&
Q_6=-8\sum_q\,(\bar{s}_L q_R)(\bar{q}_R d_L)
\end{eqnarray}
and also, with inclusion of the electromagnetic corrections, there appear the following {\it{electroweak penguin}} operators
\begin{eqnarray}\label{emS1}
Q_7=6\,(\bar{s}_L\gamma^{\mu}d_L)\sum_q e_q\,(\bar{q}_R\gamma_{\mu}q_R)&&
Q_8=4\,\sum_q e_q\,(\bar{s}_L\gamma^{\mu}q_R)(\bar{q}_R\gamma_{\mu}d_L)\nonumber\\
Q_9=6\,(\bar{s}_L\gamma^{\mu}d_L)\sum_q e_q\,(\bar{q}_R\gamma_{\mu}q_R)&&
Q_{10}=-12\sum_q e_q\,(\bar{s}_L\gamma^{\mu} q_L)(\bar{q}_L\gamma_{\mu} d_L)
\end{eqnarray}
Before turning to our low energy chiral description we have to identify the symmetries of the lagrangian, which will be our guiding line. The procedure we have sketched for integrating out particles breaks down as soon as one goes down to energies were perturbative analyses are no longer valid. We have to change our description in terms of quarks and gluons for one in terms of hadrons and connect them through symmetry arguments. As stated previously when discussing Chiral Lagrangians in the electroweak sector, the previous basis (\ref{S1}), (\ref{penguin}) and (\ref{emS1}) transforms under the chiral group as $(8_L,1_R)$, $(8_L,8_R)$ and as $(27_L,1_R)$. In order to establish contact with the phenomenological analysis we carried out in the previous section, it is also conventional to split these contributions in the isospin basis. Therefore, 
\begin{equation}
Q^{(8)}_{1/2}=\left < \left \{ Q_2-Q_1,Q_3,Q_4,Q_5,Q_6 \right \} \right >
\end{equation}
groups the operators which transform as a left octuplet with $\Delta I=1/2$ transitions. Not all of them are independent, actually they are subject to the constraint
\begin{equation}
Q_2-Q_1=Q_4-Q_3
\end{equation}
On the other hand, the following combination yields the $27$-plet component
\begin{equation}
Q^{(27)}=2Q_2+3Q_1-Q_3\equiv \fr{4}{3}\left (Q^{(27)}_{1/2}+5\,Q^{(27)}_{3/2}\right)
\end{equation}
where, contrary to what happens for the octuplet, both $\Delta I =1/2$ and $\Delta I=3/2$ are allowed. The expressions for them read
\begin{equation}
Q^{(27)}_{1/2}=\bar{s}_L\gamma^{\mu}d_L \bar{u}_L\gamma_{\mu}u_L+\bar{s}_L\gamma^{\mu}u_L \bar{u}_L\gamma_{\mu}d_L+2\,\bar{s}_L\gamma^{\mu}d_L \bar{d}_L\gamma_{\mu}d_L-3\,\bar{s}_L\gamma^{\mu}d_L \bar{s}_L\gamma_{\mu}s_L
\end{equation}
\begin{equation}
Q^{(27)}_{3/2}=\bar{s}_L\gamma^{\mu}d_L \bar{u}_L\gamma_{\mu}u_L+\bar{s}_L\gamma^{\mu}u_L \bar{u}_L\gamma_{\mu}d_L-\bar{s}_L\gamma^{\mu}d_L \bar{d}_L\gamma_{\mu}d_L
\end{equation}
The low energy realization of the previous short distance four-quark operators can be achieved with the use of chiral symmetry requirements. This yields, to lowest order,
\begin{equation}\label{DS1}
\Lcal_2^{\Delta S=1}=-\fr{G_F}{\sqrt{2}}\lambda_u \bigg\{ g_{8}\Lcal_8+g_{27}\Lcal_{27}+e^2g_{ew}\langle U\lambda_{32} U^{\dagger}Q_R\rangle\bigg\}
\end{equation}
where
\begin{equation}
\Lcal_8= F_0^4\left< D_{\mu}U^{\dagger}D^{\mu}U \right>_{23}
\end{equation}
groups the operators that transform under $SU(3)_L\times SU(3)_R$ as $(8_L,1_R)$, while 
\begin{equation}
\Lcal_{27}= F_0^4 \left ( \left<U^{\dagger}D_{\mu}U \right>_{23} \left<U^{\dagger}D^{\mu}U \right>_{11}+\fr{2}{3}\left<U^{\dagger}D_{\mu}U \right>_{21} \left<U^{\dagger}D_{\mu}U \right>_{13} \right)
\end{equation}
gathers the operators that fall into the representation $(27_L,1_R)$. The subscripts in the previous equations mean that the expressions inside the brackets are to be multiplied by the spurious matrices ($\delta_{ij}$ is the Kronecker delta)
\begin{equation}
\lambda_{23}=\delta_{i2}\,\delta_{j3}\, , \qquad \lambda_{11}=\delta_{i1}\,\delta_{j1}\, , \qquad \lambda_{21}=\delta_{i2}\,\delta_{j1}\, , \qquad \lambda_{13}=\delta_{i1}\,\delta_{j3}\, , \qquad 
\end{equation}
which project out the right flavour content.

The last term in (\ref{DS1}) is the low-energy realization of the electromagnetic penguin operators $Q_7-Q_{10}$, where
\begin{equation}
Q_R={\textrm{diag}}\,\left(\fr{2}{3},-\fr{1}{3},-\fr{1}{3}\right)
\end{equation}
and thus it transforms as $(8_L,8_R)$.

Experimental numbers on $K\rightarrow \pi \pi$ decays give
\begin{equation}
|g_{27}|\sim 0.16\quad , \quad \left | g_8+\fr{1}{5}g_{27}\right| \sim 5.1
\end{equation}
whereas na{\"i}ve factorization would predict
\begin{equation}
g_8=\fr{3}{5}\quad , \quad g_{27}=\fr{3}{5}
\end{equation}  
The previous discrepancy between theory and experiment is usually coined the {\it{$\Delta I=1/2$ rule}}, of which no convincing account has yet been given. The bulk of the enhancement needed seems to point at the long distance contributions of the $Q_1-Q_2$ operator. A penguin-based explanation in terms of $Q_6$ is ruled out, unless $1/N_C$ corrections turn out to be really sizeable\footnote{See, {\it{e.g.}}, the discussion in \cite{rafael:kaon}.}.  
\subsection{$\Delta S=2$ transitions}

\begin{figure}
\renewcommand{\captionfont}{\small \it}
\renewcommand{\captionlabelfont}{\small \it}
\centering
\psfrag{A}{$\overline{s}_L$}
\psfrag{B}{$d_L$}
\psfrag{D}{$\overline{s}_L$}
\psfrag{H}{$u,c,t$}
\psfrag{G}{$u,c,t$}
\psfrag{C}{$d_L$}
\psfrag{F}{$W^+$}
\psfrag{E}{$W^+$}
\includegraphics[width=2.5in]{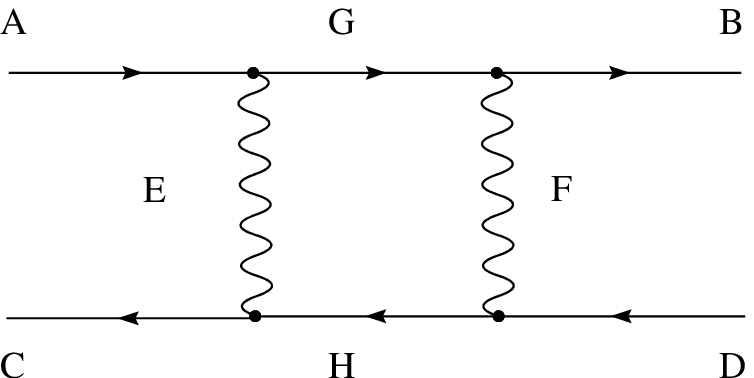}
\hspace{1cm}
\psfrag{A}{$\overline{s}_L$}
\psfrag{B}{$d_L$}
\psfrag{D}{$\overline{s}_L$}
\psfrag{F}{$u,c,t$}
\psfrag{E}{$u,c,t$}
\psfrag{C}{$d_L$}
\psfrag{G}{$W^+$}
\psfrag{H}{$W^+$}
\includegraphics[width=2.5in]{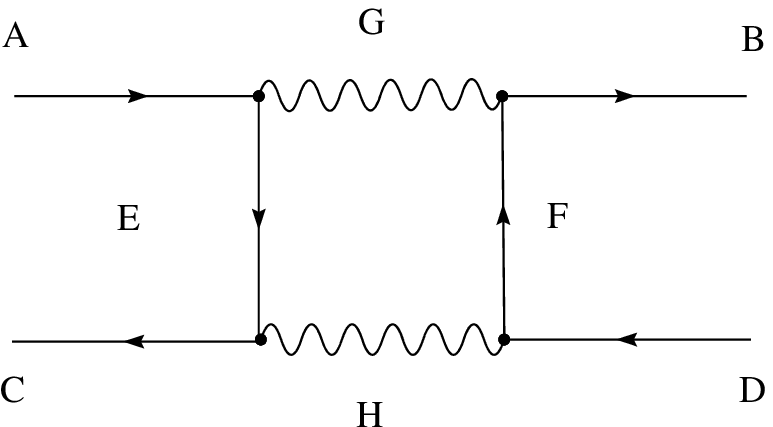}
\caption{Box diagrams for $\Delta S=2$ transitions.}\label{box}
\end{figure}

$\Delta S=2$ transitions arise in the Standard Model from double virtual $W$ exchange, the so-called {\it{box diagrams}}, depicted in figure (\ref{box}).
After integration of heavy particles, the four-quark effective Lagrangian can be written schematically as
\begin{equation}\label{delta2}
\Lcal_2^{\Delta S=2}=\fr{G_F^2}{(4\pi^2)}\bigg[\lambda_c^2\,F_1+\lambda_t^2\,F_2+\lambda_c\lambda_t\,F_3\,\bigg]\,Q\,(\mu)
\end{equation}
Notice that only one operator appears, namely
\begin{equation}
Q\,(\mu)={\bar{s}}_L\gamma_{\mu}d_L{\bar{s}}\gamma^{\mu}d_L
\end{equation}
Its Wilson coefficient has been unfolded in the so-called charm-charm, top-top and charm-top contributions\footnote{This does not mean that the up contribution is missing. To get to (\ref{delta2}) we have set $m_u=0$, and we used whenever necessary the unitarity bound 
\begin{equation}
\lambda_u+\lambda_c+\lambda_t=0
\end{equation}
of the CKM matrix elements to express the result solely in terms of charm and top parameters.}. We have adopted the following conventions
\begin{equation}
\lambda_i= V_{id}V_{is}^*\, , \qquad i\,=\,c,\,t,\,u
\end{equation}
as a short hand notation for the CKM matrix elements. The $F_i$ are functions of the integrated out particles down to the strange quark mass. They are usually cast in the form
\begin{equation}
  F_i=\eta_i\,S_i
\end{equation}
where the $\eta_i$ function collects the strong quantum corrections and the $S_i$ are the Inami and Lim functions, which account for the electroweak leading order corrections \cite{Buras:98}. As already commented on earlier, $\Delta S=2$ transitions are governed by the bag parameter $B_K$, defined as
\begin{equation} 
\langle\, K^0 \,|\,Q(\mu)\,|\,{\bar{K}}^0\,\rangle=\fr{4}{3}\,F_K^2m_K^2B_K(\mu)\end{equation}
Sometimes it is nonetheless more useful to work in terms of the invariant, {\it{i.e.}}, scale-independent, bag parameter ${\widehat{B}}_K$, 
\begin{equation}
\langle\, K^0 \,|\,c_{S=2}(\mu)\,Q(\mu)\,|\,{\bar{K}}^0\,\rangle=\fr{4}{3}\,F_K^2m_K^2{\widehat{B}}_K
\end{equation}
where $c_{S=2} (\mu)$ is the term in square brackets in (\ref{delta2}). In the so-called vacuum saturation hypothesis, using that
\begin{equation}
\langle0|{\bar{s}}_L\,\gamma^{\mu}\,d_L|K^0(p)\rangle=-i\sqrt{2}F_K\,p^{\mu}
\end{equation}
one immediately concludes that
\begin{equation}
\widehat{B}_K\bigg|_{VS}=1
\end{equation}
Different approximations suggest the following bounds,
\begin{equation}
0.3\,\leq\, {\widehat{B}}_K\,\leq\, 0.8
\end{equation}
For instance, the {\it{large-$N_C$ prediction}} yields
\begin{equation}
\widehat{B}_K\bigg|_{N_C}=\fr{3}{4}
\end{equation}
As we will discuss in the next chapter, next-to-leading order corrections in the $1/N_C$ expansion seem to push that value down to ${\widehat{B}}_K\,\sim\, 0.4$, at least in the strict chiral limit. Taking into account that lattice estimates suggest a much bigger value, say ${\widehat{B}}_K\,\sim\,0.8$, one concludes that chiral corrections are clearly not negligible and may account for the bulk of ${\widehat{B}}_K$. Recall that this parameter determines the prediction inside the Standard Model for the kaon mass difference and $\varepsilon_K$, through
\begin{eqnarray}
M_{K_L}-M_{K_S}&\sim & {\textrm{Re}}\,\langle\, K^0\,|{\mathcal{H}}_{eff}^{S=2}(0)\,|{\bar{K}}^0\,\rangle\nonumber\\
\varepsilon_K&\sim&{\textrm{Im}}\,\langle\, K^0\,|{\mathcal{H}}_{eff}^{S=2}(0)\,|{\bar{K}}^0\,\rangle
\end{eqnarray}   
At the same time, at least in the chiral limit, there is a relation connecting $\Delta S=2$ and $\Delta S=1$ transitions. Indeed, $\bigg[(\Delta S=1)\,\times\,(\Delta S=1)\bigg]$ comes as a $\Delta S=2$ long distance effect, as already pointed out. Since the $\Delta S=2$ operator transforms as a $(27_L,1_R)$ under the chiral group, it follows that it is closely related to $g_{27}$, namely
\cite{PichdeR,Donoghue:82}
\begin{equation}\label{pichandco}
    g_{27}= \frac{4}{5}{\hat{B}}_K
\end{equation}
Next chapter is devoted to the determination of the $\Delta S=2$ parameters mentioned above in the chiral limit including $1/N_C$ corrections.


\chapter[Determination of kaon mixing parameters]{Determination of kaon mixing parameters}

\section{The Problem of Matching in Nonleptonic Weak Interactions}
In preceding chapters we have seen that there is a dual description of Standard Model physics. On the one hand, one has the quark-gluon description, accepted as fundamental but of very little use beyond the perturbative regime. For most practical purposes, we are interested in computing matrix elements of hadronic states, and one is faced with an obvious problem of language mismatch. Consider, for instance,
\begin{equation}\label{examplekaons}
\langle K^0|\,{\bar{s}}_L\gamma_{\mu}d_L{\bar{s}}_L\gamma^{\mu}d_L|{\bar{K}}^0\rangle
\end{equation}
Computation of the previous matrix element is far from being straightforward, for we do not know how quarks and gluons assemble to conform $K$ states. Possible way-outs are numerical simulations (lattice QCD), quark models (Constituent Quark Model, ...) or effective field theories. We will adopt the latter strategy, since we want to be analytic until the very end and approximations have the advantage over models that they are systematically improvable.

In previous chapters we have already shown the systematics of EFT's and showed that indeed there exists an alternative formulation of the strong interactions in terms of mesons under the form of chiral lagrangians. It is not known how to do the transition (sometimes people speak of change of variables) between the quark-gluon description and the meson description. However, we saw that symmetry alone was able to relate the quark-gluon operators to their dual counterparts in terms of mesons. This change of language is often referred to as {\it{bosonization}} of operators, and it is precisely what we need to compute matrix elements such as (\ref{examplekaons}). Our remaining ignorance as to how these two worlds are connected is parametrised in terms of a set of low-energy couplings, one for each operator. Therefore, after bosonization, computation of hadronic matrix elements is tantamount to determining the low-energy constants, something that unavoidably requires a matching between the quark-gluon and meson descriptions. 

In full generality, low-energy couplings are related to QCD correlators. We can distinguish between two different behaviours, depending on whether they are strong low energy couplings or electroweak low energy couplings. The parameters of the strong chiral lagrangian modulate the dynamics of strong processes in the presence of external currents. As already stated, they contain the information of the integrated degrees of freedom up to $\sim 1$ GeV. Generically, they are expressible as the coefficients of a Taylor expansion of QCD correlators at low energy. For instance, consider the paradigmatic two-point functions
\begin{equation} 
\Pi^{V,A}_{\mu\nu}(q)=\ i\,\int d^4x\,e^{iqx}\langle
J^{V,A}_\mu(x)\ (J^{V,A}_{\nu})^{\dagger}(0)\rangle
\end{equation}
with the QCD currents given by
\begin{equation}
J^{\mu}_V\ =\ {\bar{d}}(x)\ \gamma^{\mu}\ u(x)\, , \qquad J^{\mu}_A\ =\ {\bar{d}}(x)\ \gamma^{\mu}\gamma^5\ u(x) 
\end{equation}
We define
\begin{equation}
\Pi^{LR}_{\mu\nu}(q)=\fr{1}{2}\bigg(\Pi^V_{\mu\nu}(q)-\Pi^A_{\mu\nu}(q)\bigg)=(q^2g_{\mu\nu}-q_{\mu}q_{\nu})\,\Pi^{LR}(q^2)
\end{equation}
The last line follows from Lorentz invariance and $\Pi^{LR}(q^2)$ has a well-known low energy expansion
\begin{equation}
Q^2\Pi_{LR}(Q^2)=-F_0^2+4L_{10}Q^2+{\cal{O}}\,(Q^4)
\end{equation}
which yields
\begin{equation}
F_0^2=\bigg[Q^2\Pi_{LR}(Q^2)\bigg]\bigg|_{Q^2\rightarrow 0}
\end{equation}
and
\begin{equation}
L_{10}=-\fr{1}{4}\left [ \fr{d}{dQ^2}\bigg(Q^2\Pi_{LR}(Q^2)\bigg)\right]\bigg|_{Q^2\rightarrow 0}
\end{equation}
Consequently, derivatives of Green functions which are order parameters of spontaneous chiral symmetry breaking lead to low energy couplings of the strong chiral Lagrangian.

However, unlike what happens in the strong sector, the low energy couplings arising in the electroweak sector are much more involved. As we saw in chapter 2, they come from the virtual exchange of $W$ and $Z$ particles. Before reaching the non-perturbative regime were bosonisation is required, they have long been integrated out. As a result, the low energy couplings turn out to be integrals of QCD correlators. Following the paradigmatic $\Pi_{LR}$ used above, one can see that the coupling ${\widehat{c}}_2$ introduced in chapter 2 as the lowest-order electromagnetic low energy coupling is expressible as
\begin{equation}\label{ex}
{\widehat{c}}_2=-\fr{3}{32\pi^2}\int_0^{\infty}\, dQ^2\left(1-\fr{Q^2}{Q^2+M_Z^2}\right)\bigg[Q^2\Pi_{LR}(Q^2)\bigg]
\end{equation}
Contrary to the strong case, where the operators were truly products of conserved currents, in the electroweak sector, even though they look like products of currents, they are convoluted currents (in the example above, due to the $Z$ and photon exchange).
Therefore, whereas in the strong sector the determination of the low energy couplings demands knowledge of the low energy properties of certain Green functions, in the electroweak sector the full Euclidean range is needed.
In view of the fact that an exact solution of the problem is out of reach, for it requires full-fledged QCD, we shall have to resort to some approximations. However, whatever approximations we use to the computation of matrix elements, they  have to make sure that the scale dependence of the couplings and the operators entering the expression
\begin{equation}
e^2\,F_0^4\,{\widehat{c}}_2\,(\mu)\langle {\cal{Q}}_L\,{\cal{Q}}_R \rangle (\mu)
\end{equation}
must cancel in order to provide a sound output for any physical observable. Thus, the challenge is not only to find a reliable approximation to the integral (\ref{ex}) above, but one that also guarantees the right matching between short and long distances. In the following sections we will show how to get the right matching. The method basically consists in approximating the integrals of Green functions such as that in (\ref{ex}) by integrals over interpolators based on the MHA we introduced in chapter 3. Recall that such interpolators were determined through matching to the known QCD high and low energies of the QCD Green function we are approximating. It is precisely the right OPE behaviour of our approximant which, eventually, {\it{automatically}} ensures the right matching between long and short distances in nonleptonic weak interactions\footnote{For different applications of the method, we refer to \cite{s1}-\cite{s8}.}.

\section{Determination of kaon mixing parameters}
\subsection{Lowest order Lagrangian}
\begin{figure}
\renewcommand{\captionfont}{\small \it}
\renewcommand{\captionlabelfont}{\small \it}
\centering
\psfrag{A}{$\overline{s}_L$}
\psfrag{B}{$d_L$}
\psfrag{D}{$\overline{s}_L$}
\psfrag{F}{$u,c,t\quad\,\,\,\,\, \longrightarrow$}
\psfrag{E}{$u,c,t$}
\psfrag{C}{$d_L$}
\psfrag{G}{$W^+$}
\psfrag{H}{$W^+$}
\includegraphics[width=1.8in]{box1.eps}\quad \quad \quad
\psfrag{G}{$u,c$}\psfrag{H}{$u,c\quad\,\,\,\,\, +$}\psfrag{M}{${\bar{s}}_L$}\psfrag{N}{$d_L$}\psfrag{O}{$d_L$}\psfrag{L}{${\bar{s}}_L$}
\includegraphics[width=1.3in]{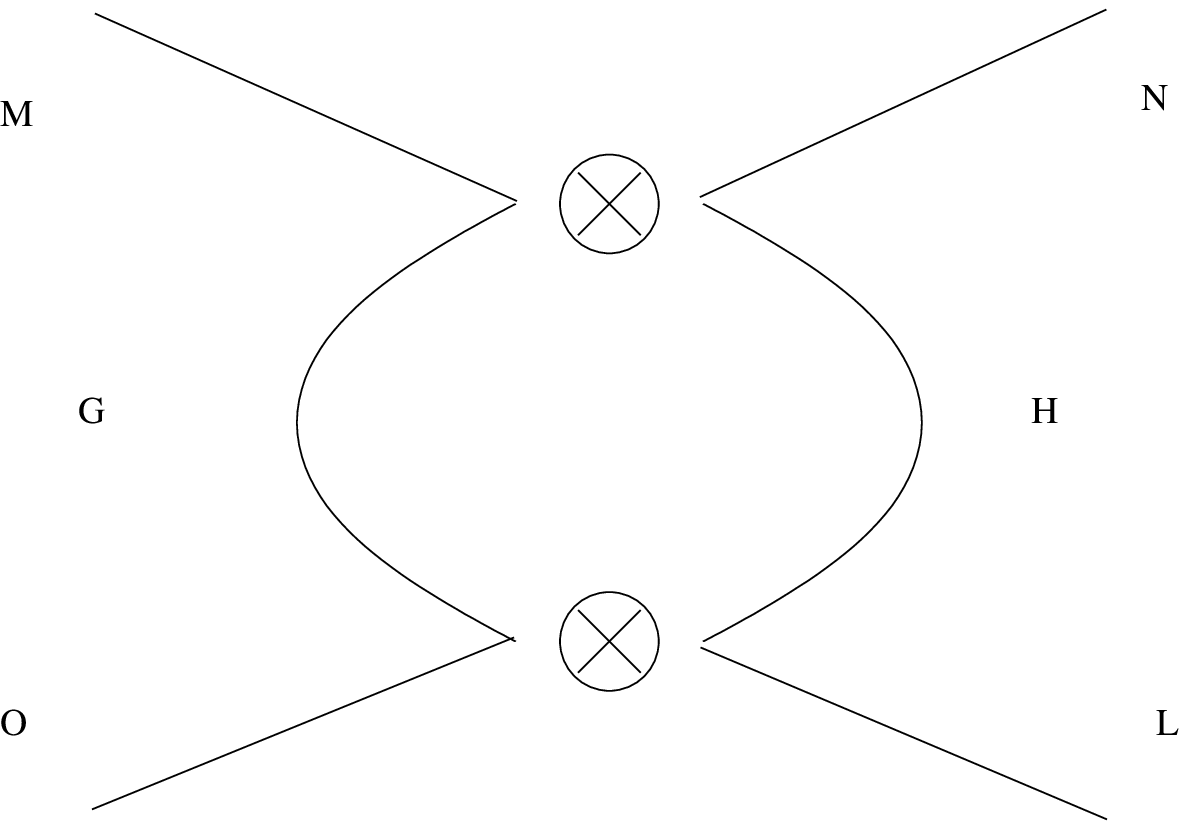}
\qquad \quad
\psfrag{A}{${\bar{s}}_L$}\psfrag{B}{$d_L$}\psfrag{C}{$d_L$}\psfrag{D}{${\bar{s}}_L$}
\includegraphics[width=1.0in]{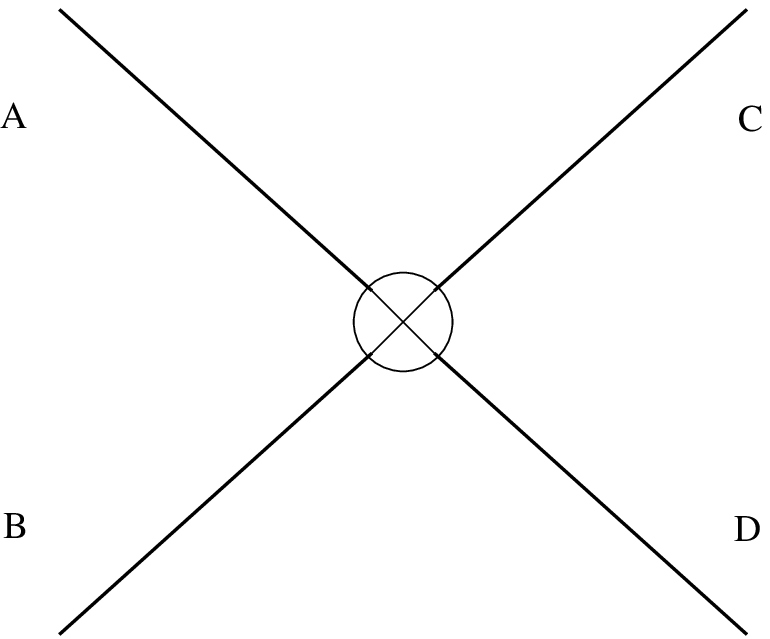}
\caption{Box diagram before (left-hand side) and after (right-hand side) $W$ and top quark integration.}\label{effbox}
\end{figure}

Our purpose is to find the Effective Hamiltonian which describes $K^0-{\bar{K}}^0$ mixing. Our starting point will be the box diagrams of figure (\ref{box}). Following the steps outlined in previous chapters, we have to integrate out the heavy degrees of freedom sequentially, beginning with the top quark and going down in energy. For the sake of simplicity we here reproduce the Gilman-Wise calculation \cite{Gilman-Wise} and consider the $W$ gauge boson heavier than the top quark, since this will not bring any qualitative difference. At the end we will modify our result accordingly to account for the top quark being actually heavier. This means that in our box diagrams every $W$ boson propagator will be shrunk to a point, {\it{i.e.}},
\begin{equation}
\fr{1}{q^2-m_W^2}\sim -\fr{1}{m_W^2}\qquad (q^2\ll m_W^2)
\end{equation}
since the momentum scale is indeed much lower than the particle mass. Box diagrams, after $W$ and top integration, which again can be accomplished by shrinking its propagator, turn into the diagrams depicted in figure (\ref{effbox}).
and we are left with only one four-quark operator
\begin{equation}\label{four}
    \mathcal{L}_{eff}=  \frac{G_F^2}{16\pi^2}
    \  c(\mu)\ \bigg\{\overline{s}(x)\gamma^{\mu}(1-\gamma_5) d(x)\bigg\}^{2}
\end{equation}
whose Wilson coefficient at the top mass threshold $\mu=m_t$ is given by conventional perturbation theory techniques, to yield
\begin{equation}
c(m_t)=-m_t^2\,\lambda_t^2
\end{equation}
This is precisely the matching condition that guarantees that the Effective Lagrangian (\ref{four}) is equivalent to the full-fledged theory below the top quark mass threshold. From the top quark threshold until the charm quark threshold, the Wilson coefficient has to run as dictated by its renormalization group equation,
\begin{equation}\label{five}
    \mu^{2}\frac{d}{d\mu^{2}}\ c(\mu)=-2 m_c^2\, \lambda_c\left(\lambda_c+\lambda_u\right)
     -2 m_u^2\, \lambda_u \left( \lambda_c +  \lambda_u\right)
\end{equation}
which can be easily integrated to yield
\begin{equation}\label{six}
    c(\mu)=-m_t^2\, \lambda_t^2 + 2 m_c^2\, \lambda_c \left(\lambda_c+\lambda_u\right)
    \log\frac{m_t^2}{\mu^2}
\end{equation}
Recall that in the previous expression we have set $m_u=0$, which is a reasonably good approximation. The next step is the integration of the charm quark. Disappearance of the charm from the effective theory requires that the Wilson coefficient picks an extra piece,

\begin{equation}\label{seven}
    c(\mu\lesssim m_c)=-m_t^2\, \lambda_t^2 +2 m_c^2\,\lambda_c
    \left(\lambda_c+\lambda_u\right)\log\frac{m_t^2}{m_c^2}-m_c^2\, \lambda_c^2
\end{equation}
Again, the matching condition has been easily determined by using perturbation theory. We are still above $\Lambda_{\chi}$ and we can run down the Wilson coefficient. However, its evolution below the charm mass threshold is proportional to $m_u$. Thus, to a very good approximation, the Wilson coefficient gets frozen at its $\mu\sim m_c$ value. The full answer for the dimension-six quark operators is then given by the well-known Gilman-Wise Lagrangian
\begin{equation}\label{GilWise}
\Lcal_{eff}=-\fr{G_F^2}{4\pi^2}\left(-m_t^2\, \lambda_t^2 +2 m_c^2\,\lambda_c\lambda_t
    \log\frac{m_t^2}{m_c^2}-m_c^2\, \lambda_c^2\right)\bigg[{\bar{s}}_L\gamma^{\mu}d_L(x)\,{\bar{s}}_L\gamma_{\mu}d_L(x)\bigg]
\end{equation}
In the Effective Field Theory approach the non-logarithmic pieces above are to be interpreted as the matching conditions at the top quark and charm quark mass thresholds, whereas the logarithmic contribution stems from the running between the two scales. 

As mentioned before, the step we made of integrating out almost simultaneously the $W$ gauge boson and the top quark, {\it{i.e.}}, the assumption of considering the W slightly heavier than the top quark can be mended without changing any of the conclusions we reached with the Gilman-Wise Lagrangian (\ref{GilWise}). The full-fledged result for the (leading) $\Delta S=2$ Effective Hamiltonian, once the particle hierarchy is respected and electroweak and strong corrections are included, reads
\begin{eqnarray}\label{oneham}
    &&\mathcal{H}^{S=2}_{eff}=\frac{G_F^2}{4\pi^2}
    \ \overline{s}_L\gamma^{\mu}d_L(x)\
    \overline{s}_L\gamma_{\mu}d_L(x)\,\times \nonumber\\
 && \qquad \times \quad  \ c(\mu)\ \Bigg\{\eta_1\,m^2_c\, \lambda_c^2 + \eta_2\,\left(m^2_{t}\right)_{eff}\,\lambda_t^2 + 2 \eta_3\, m^2_c\, \lambda_c \lambda_t\,\log\frac{m_t^2}{m_c^2}\Bigg\}\ +\ \mathrm{h.c.}
\end{eqnarray}
The appearance of an effective top quark mass, defined as \cite{Buras2}
\begin{equation}\label{two}
\left(m^2_t\right)_{eff}= 2.39 \ M_W^2 \left[\frac{m_t}{167\ GeV}\right]^{1.52}
\end{equation}
is the ammendment to account for the top quark being heavier than the $W$ gauge boson. The remaining $\eta_i$ and $c(\mu)$ functions appear after perturbative corrections are taken into account. Their values are \cite{Buras2}
\begin{eqnarray}\label{three}
&&\eta_1=(1.32\pm 0.32)\left[\frac{1.3\, {\textrm{GeV}}}{m_c(m_c)}\right]^{1.1}\quad ,\quad
\eta_2=0.57\pm 0.01\quad,\quad \eta_3=0.47\pm 0.05\  \nonumber\\
&&c(\mu)= \left(\alpha_s(\mu)\right)^{-\frac{9}{11N_c}}\ \left[1+
\frac{\alpha_s(\mu)}{\pi}\left( \frac{1433}{1936}+ \frac{1}{8}\ \kappa \right)  \right]
\end{eqnarray}
where $\kappa$ is scheme dependent, its value being $0$ or $-4$ depending on whether one is working in the naive dimensional or 't Hooft-Veltman regularization schemes.
\subsection{${\widehat{B}}_K$ with leading short distance OPE constraints}
 
The evaluation of the $K^0-{\bar{K}}^0$ matrix element, as already pointed out, is tantamount to a determination of the bag parameter
\begin{equation}
\langle K^0|\,c(\mu)\,Q(0)\,|{\bar{K}}^0\rangle\equiv \fr{4}{3}\,F_K^2\,m_K^2\,{\widehat{B}}_K
\end{equation}
The computation of the invariant ${\hat{B}}_K$ is however not as straightforward, since there is a mismatch in the above equation, as stressed at the beginning of the chapter. One should be able to express the four-quark operator in terms of meson operators, or, in other words, to know how the four-quark operator changes when strong interactions bind quarks and gluons together so close that they appear as hadrons.

As already emphasised, symmetry alone is capable of constraining this set of meson operators. As with every Effective Field Theory, there are a number of low-energy coupling constants that have to be determined through a matching procedure. Fortunately, there is only one $\Delta S=2$ operator in the leading chiral Lagrangian, to wit
\begin{equation}\label{chiralS=2}
\Lcal_{\chi}^{S=2}= \fr{G_F^2}{16\pi^2}\,F_0^4\, \Lambda_{S=2}^2\, {\mathrm{Tr}}\,\bigg[\lambda_{32}(D^{\mu}U^{\dagger})U\lambda_{32}(D_{\mu}U^{\dagger})U\bigg]+{\mathrm{h.c.}}
\end{equation}
where $\Lambda_{S=2}^2$ is the {\it{a priori}} unknown low-energy coupling constant. For convenience let us factor out the short distance content of $\Lambda_{S=2}^2$,
\begin{equation}
\Lambda_{S=2}^2={\widehat{g}}_{S=2}\left[\eta_1\, m^2_c\, \lambda_c^2+ \eta_2\,
    \left(m^2_{t}\right)_{eff}\, \lambda_t^2+ 2 \eta_3\, m^2_c\, \lambda_c \lambda_t\,
    \log\frac{m_t^2}{m_c^2}\right]
\end{equation} 
Plugging in the previous equation in the definition of the invariant ${\widehat{B}}_K$, one gets
\begin{equation}
{\widehat{B}}_K=\fr{3}{4}\,{\widehat{g}}_{S=2}
\end{equation}
So far, we have changed the problem such that our determination of the bag parameter is now tantamount to knowledge of the low-energy coupling constant ${\widehat{g}}_{S=2}$ (or equivalently $\Lambda_{S=2}$) governing the $\Delta S=2$ transitions. The change in vocabulary has been so abrupt in the bosonization process that the matching procedure is, unlike the ones we have encountered when deriving the Gilman-Wise Hamiltonian, highly non-trivial.

In chapter 2, when dealing with chiral Lagrangians, we emphasised the convenience of introducing external fields in the formalism in order to be able to compute QCD Green functions. Here we will make use of its useful consequences. A close inspection of (\ref{chiralS=2}) reveals that it contains a quadratic term in the right-handed external field. Indeed, recalling that
\begin{equation}
(D^{\mu}U^{\dagger})U=(\partial^{\mu} U^{\dagger})U+iU^{\dagger}r^{\mu}U-i\,l^{\mu}
\end{equation}
it is not difficult to get
\begin{equation}
\lambda_{32}(D^{\mu}U^{\dagger})U\lambda_{32}(D_{\mu}U^{\dagger})U=\dots -r^{\mu}_{{\bar{d}}s}r_{\mu}^{{\bar{d}}s}+\dots
\end{equation}     
The right-handed current $r^{\mu}$ is by construction the same one showing up in the QCD Lagrangian. Furthermore, since this quadratic term in the right-handed external fields is a genuinely $\Delta S=2$ operator, it can solely come from the chiral Lagrangian (\ref{chiralS=2}). The matching condition is therefore imposed by comparing a QCD Green function with the Green function coming from the effective chiral Lagrangian, namely
\begin{equation}\label{match1}
\left(\fr{-i\delta}{\delta r^{\mu}_{{\bar{d}}s}}\right)\left(\fr{-i\delta}{\delta r_{\mu}^{{\bar{d}}s}}\right)\bigg[\Lcal_{QCD}+\Lcal_{eff}^{S=2}\bigg]\doteq\left(\fr{-i\delta}{\delta r^{\mu}_{{\bar{d}}s}}\right)\left(\fr{-i\delta}{\delta r_{\mu}^{{\bar{d}}s}}\right)\Lcal_{\chi}^{S=2}
\end{equation}
Equating the right-hand side of (\ref{match1}), coming from the electroweak chiral Lagrangian (\ref{chiralS=2}), with the left-hand side, computed with the effective Hamiltonian (\ref{oneham}), one obtains \cite{s5}
\begin{equation}\label{match}
{\widehat{g}}_{S=2}=c(\mu)\left[1-\frac{\mu_{had}^2}{32\pi^2F_0^2}
\left(\frac{4\pi\mu^2}{\mu_{had}^2}\right)^{\frac{\epsilon}{2}}
\frac{1}{\Gamma(2-\frac{\epsilon}{2})}\int_0^{\infty}dz\,z^{-\frac{\epsilon}{2}}W(z)\right]
\end{equation}
where $\mu_{had}^2$ is a scale introduced to rewrite the integral in terms of a dimensionless variable $z\equiv Q^2/\mu_{had}^2$. $W(z)$ is a short-hand notation for 
\begin{equation}
W(z)=z\,\fr{\mu_{had}^2}{F_0^2}\,W_{LRLR}^{(1)}(z\mu_{had}^2)
\end{equation}
and $W_{LRLR}^{(1)}(z\mu_{had}^2)$ is the solid angle integral of a four-point Green function with two left-handed currents and two right-handed (soft) currents
\begin{equation}\label{ward}
\int\,d\Omega_q\ g_{\mu\nu}\
{W}^{\mu\alpha\nu\beta}_{LRLR}(q,l)\bigg|_{unfactorized}=
  \left(\frac{l^{\alpha}l^{\beta}}{l^2}-g^{\alpha\beta}\right)W_{LRLR}^{(1)}(Q^2)\quad
  ,\quad Q^2\equiv -q^2 
\end{equation}
where the four-point Green function in the left-hand side is defined as
\begin{equation} \label{Green}
W_{\mu\alpha\nu\beta}^{LRLR}(q,l)=\lim_{l\rightarrow 0}\,i^3
\int\,d^4x\ d^4y\ d^4z\ e^{iq \cdot x}e^{il \cdot (y-z)}
\langle0|T\{L^{\bar{s}d}_{\mu}(x)\,R^{\bar{d}s}_{\alpha}(y)\,L^{\bar{s}d}_{\nu}(0)\,
R^{\bar{d}s}_{\beta}(z)\}|0\rangle
\end{equation}
the currents being
\begin{equation}
L_{{\bar{s}}d}^{\mu}(x)={\bar{s}}(x)\,\gamma^{\mu}\,\fr{1-\gamma_5}{2}\,d(x)\, ,\qquad
R_{{\bar{d}}s}^{\mu}(x)={\bar{d}}(x)\,\gamma^{\mu}\,\fr{1+\gamma_5}{2}\,s(x)
\end{equation}
This four-point Green function with two additional hard left-handed insertions is one of the QCD Green functions underlying $\Delta S=2$ processes, but certainly not the only one. However, it is relatively simple and has the nice feature that it is an order parameter of spontaneous chiral symmetry breaking (S$\chi$SB), which is more than welcome\footnote{Recall that order parameters are preferred if we eventually want to use the MHA (see discussion in section 3.3).}. Furthermore, its physical content is rather transparent: from its very definition, our four-point function is a two-point correlator of left-handed currents, which guarantee the $K^0-{\bar{K}}^0$ transition, together with two soft insertions of right-handed currents required by the matching condition (\ref{match1}).

Equation (\ref{match}) is an example of what we anticipated at the beginning of the chapter, {\it{i.e.}}, low energy couplings of the electroweak sector can be expressed as integrals over the Euclidean of certain Green functions. As already outlined, the procedure hereafter will be the following: we collect as much information about the Green function as possible, on the long distance regime by means of chiral perturbation theory and on the short distance regime by computing the OPE coefficients. We then bridge the energy gap in between by means of the large-$N_C$ version of the Green function. This constrains the analytical form of the Green function to be a meromorphic function. The MHA is then used to truncate the infinite number of mesons to a finite number of resonances where the undetermined parameters (poles and residues) can then be fixed by matching onto the low and high energy behaviour of the original Green function.
 
\begin{figure}
\renewcommand{\captionfont}{\small \it}
\renewcommand{\captionlabelfont}{\small \it}
\centering
\includegraphics[width=1.0in]{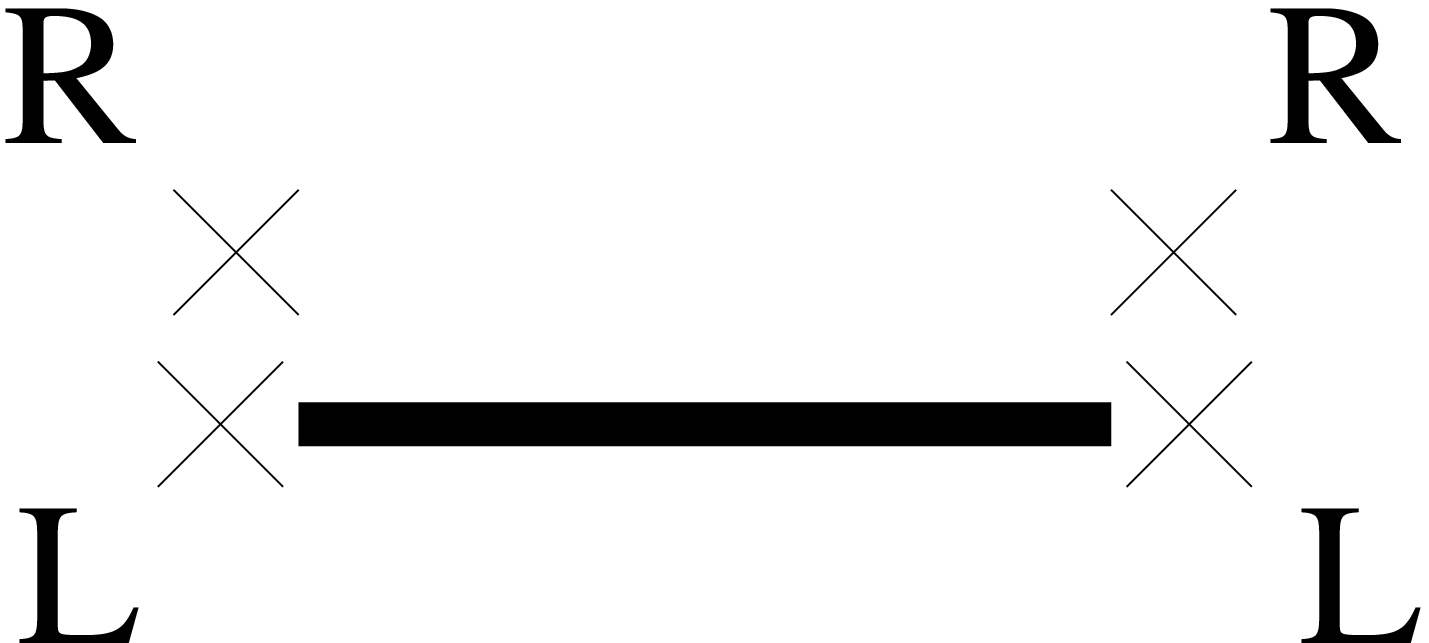}
\hspace{1.in}
\includegraphics[width=1.0in]{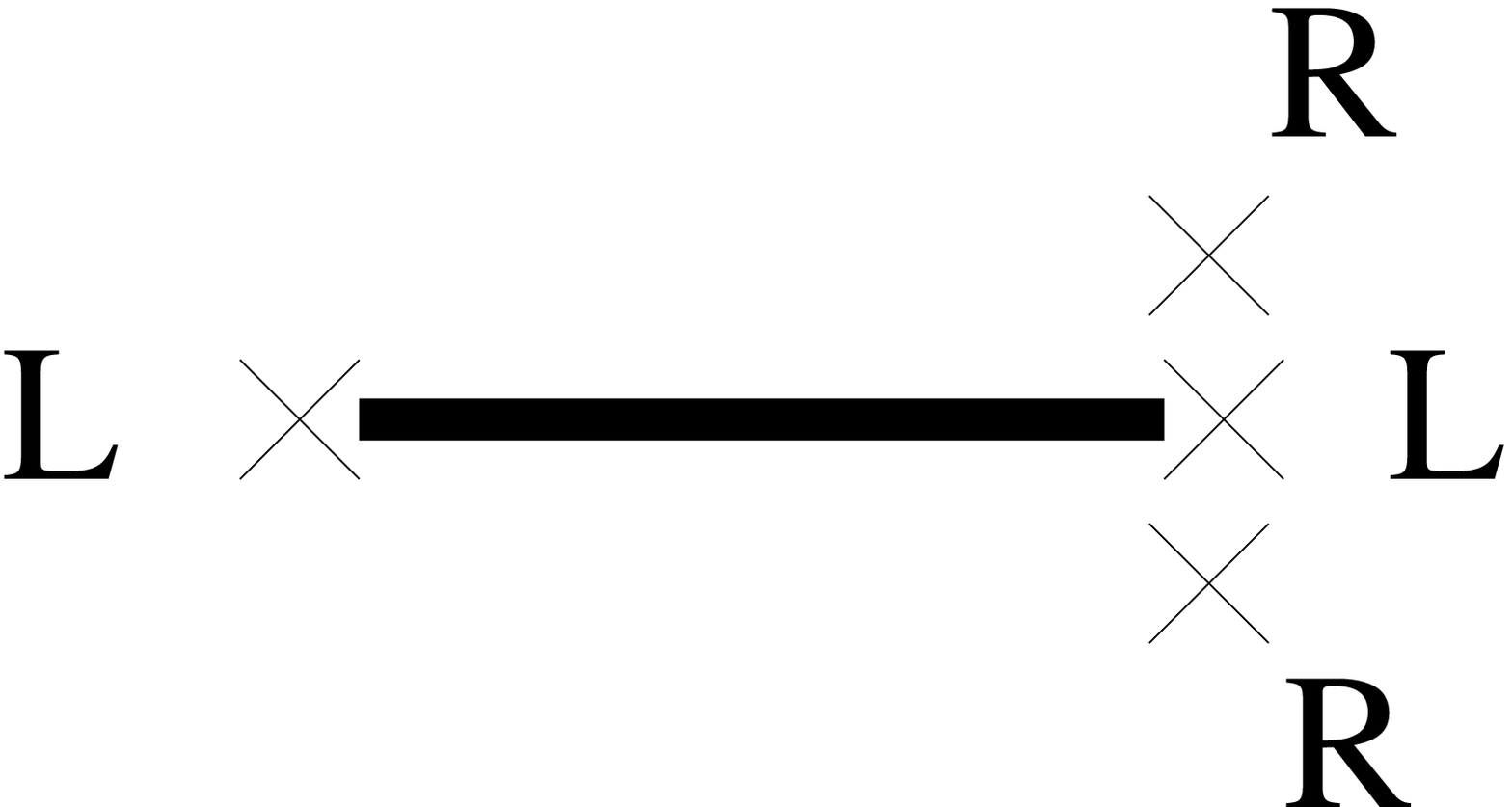}
\hspace{1.in}
\includegraphics[width=1.0in]{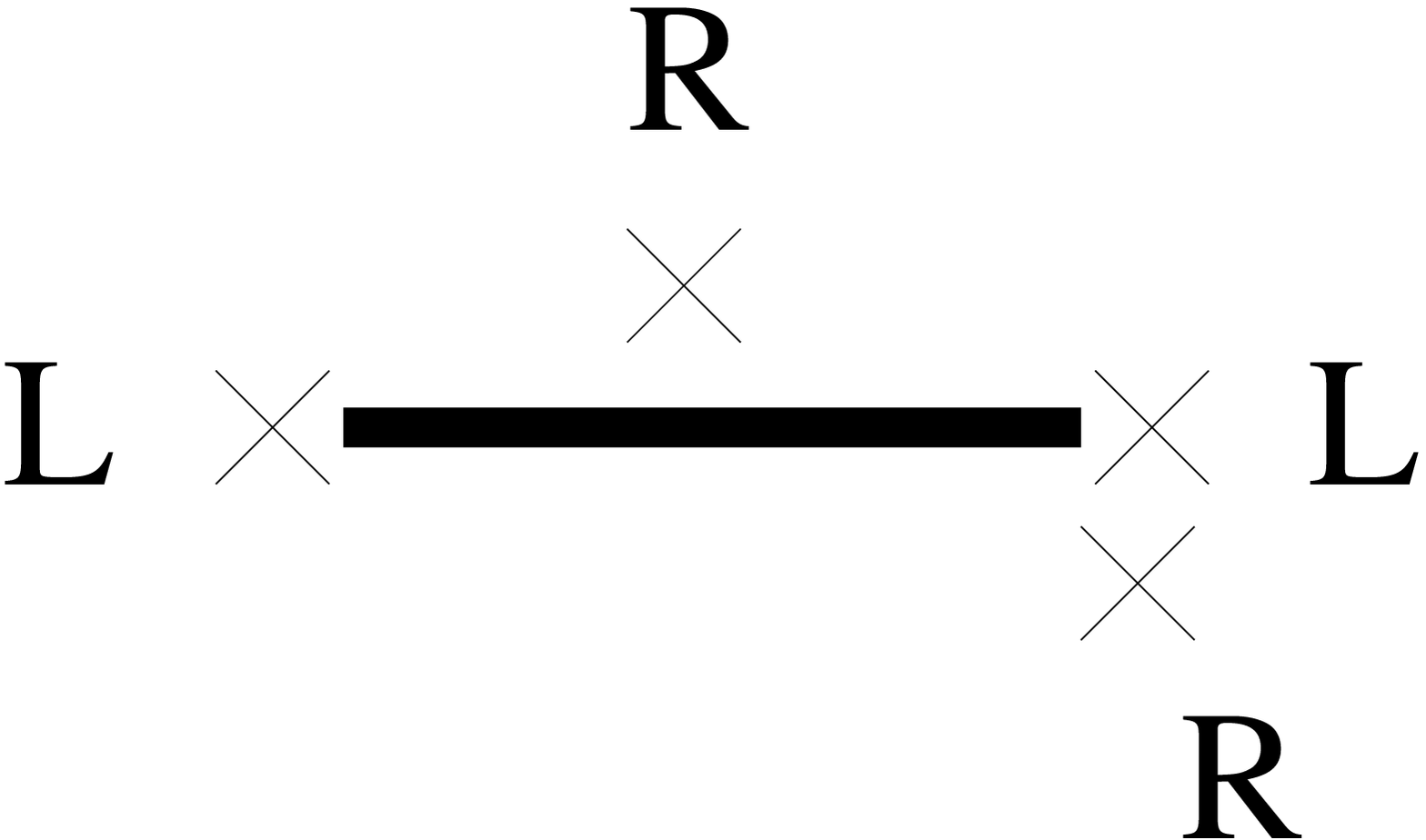}
\hspace{0.5in}
\includegraphics[width=1.0in]{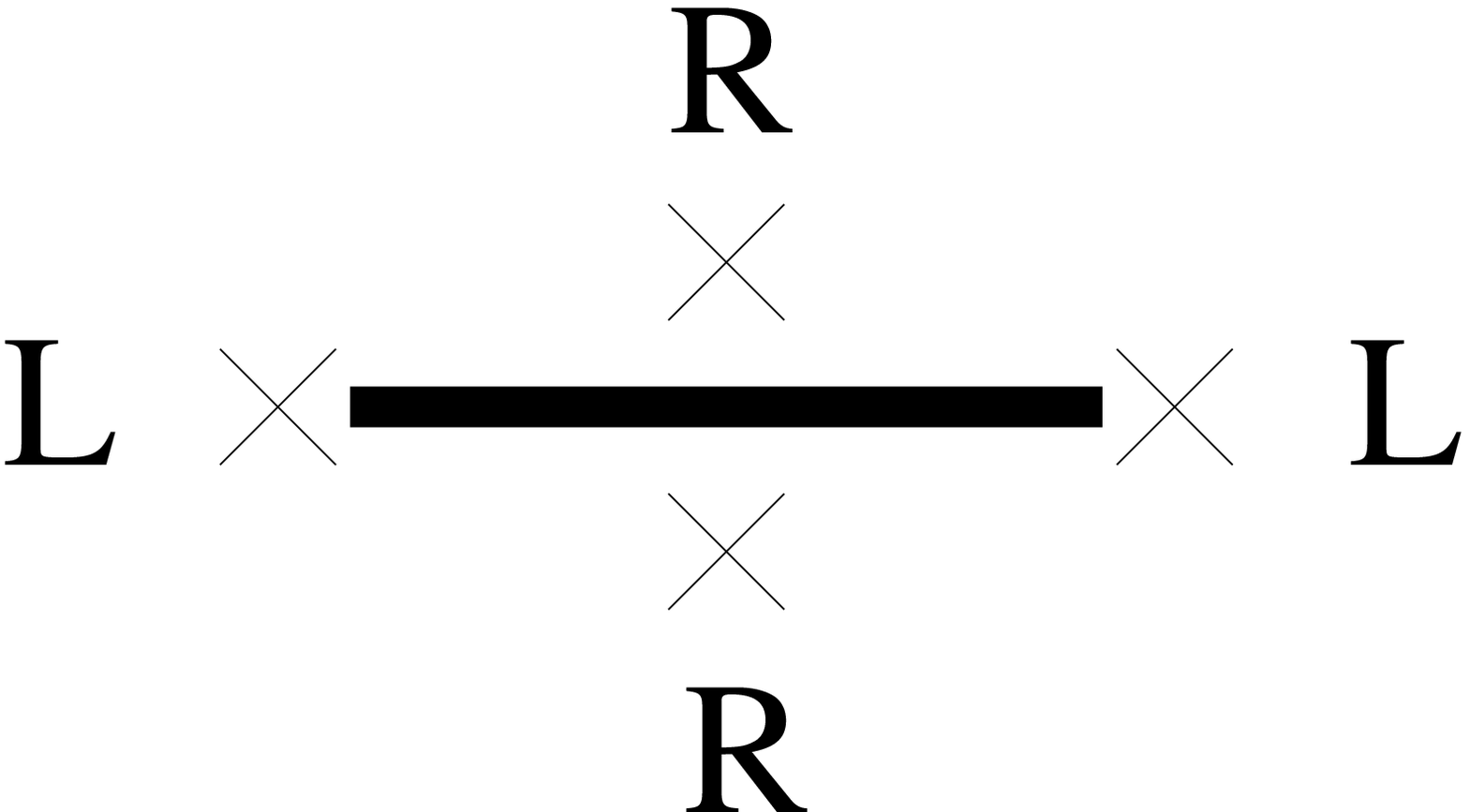}
\hspace{0.5in}
\includegraphics[width=1.0in]{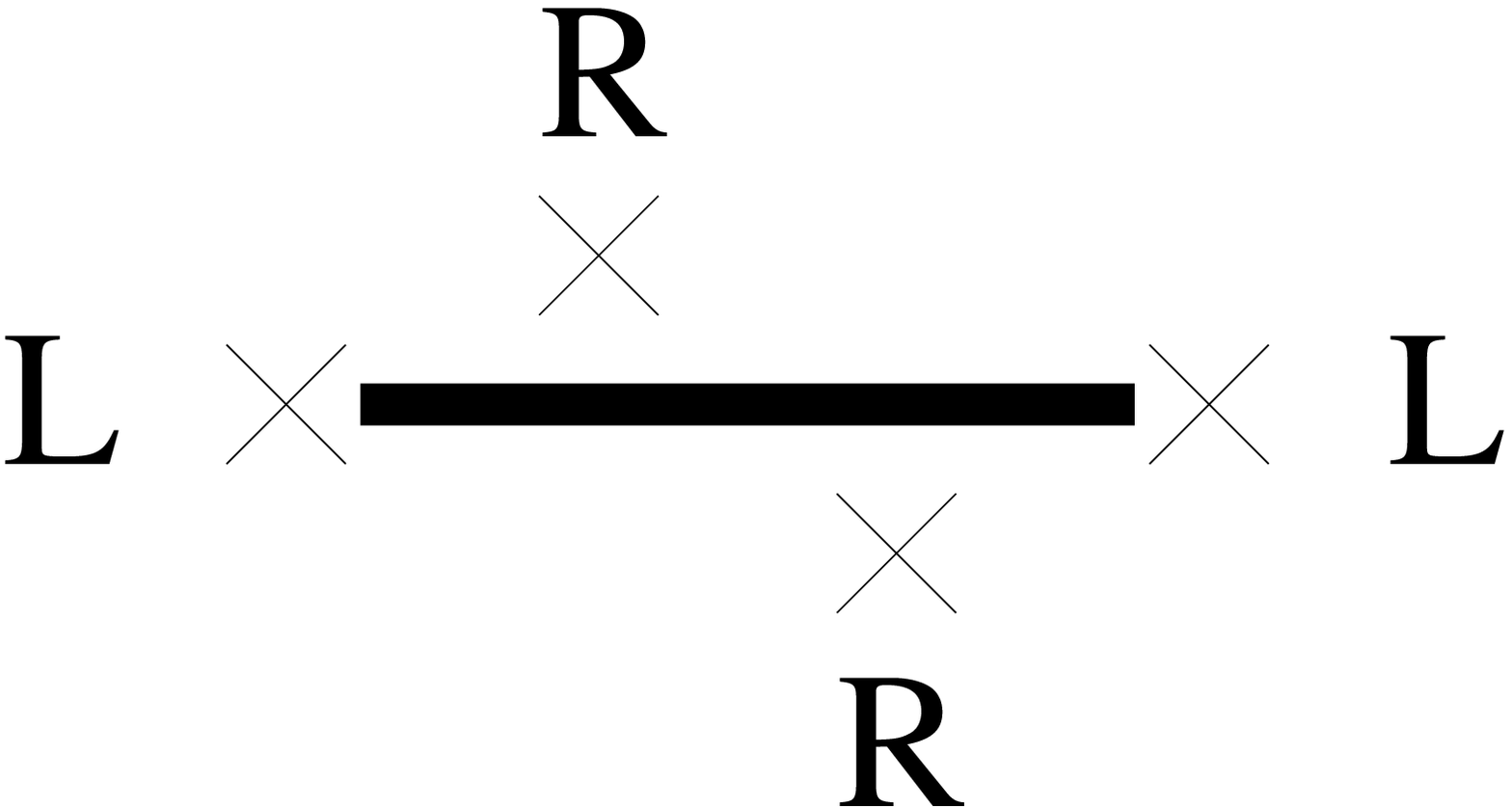}
\caption{{Resonance diagrams contributing to $W(z)$.}}\label{resonance}
\end{figure}

The analytical large-$N_C$ behaviour of the four-point function $W_{LRLR}^{\mu\nu\alpha\beta}$ is depicted in figure (\ref{resonance}), where all possible resonance exchanges are taken into account. This constrains our Green function to be a sum over an infinite number of resonances of at most triple poles if the resonance is a vector or an axial vector and simple poles if they are scalars or pseudoscalars. Therefore, in the MHA to $N_C\rightarrow \infty$ QCD,
\begin{equation} \label{Largeinch5}
W(z)= \sum_{i=1}^{\mathcal{N}_{V,A}}\left(\frac{a_i}{(z+\rho_i)}+
\frac{b_i}{(z+\rho_i)^2}+\frac{c_i}{(z+\rho_i)^3}\right)+\sum_{j=1}^{\mathcal{N}_{S,P}}\frac{a_j}{(z+\rho_j)}
\end{equation}
where $\rho_i= m_i^2/\mu_{had}^2$ is a normalized mass, $m_i$ being the mass of the resonance $i$; $a_{i}$, $b_{i}$, $c_{i}$ are the constants to be determined by matching onto the high and low energy regimes of the $N_C=3$ QCD Green function.
\vskip 1cm
\centerline{\bf{Long Distance Constraints}}
\vskip 1cm
Contributions to be taken into account come from the diagrams coming from the Chiral Lagrangian connecting two left-handed and two right-handed external sources. We quote the result from \cite{s5} 
\begin{equation}\label{lowz}
 W^{\chi PT}(z)= 6 - 24\
\frac{\mu_{had}^2}{F_0^2}\ \bigg(2 L_1+5 L_2+L_3+L_9\bigg)\, z + {\cal{O}} \left(z^2\right)
\end{equation}

The more involved part will be the calculation of the OPE coefficients.
\vskip 1cm
\centerline{{\bf{Short Distance Constraints}}}
\vskip 1cm
Since our four-point Green function has the right-handed currents as soft momentum insertions, one needs to consider just the operator product expansion of the left-handed currents. Thus, this separation of scales allows to factorize the problem to the more manageable OPE two-point computation.

\begin{figure}
\renewcommand{\captionfont}{\small \it}
\renewcommand{\captionlabelfont}{\small \it}
\centering
\psfrag{A}{$\overline{s}_L$}
\psfrag{B}{$d_L$}
\psfrag{D}{$\overline{s}_L$}
\psfrag{C}{$d_L$}
\psfrag{G}{$k$}
\psfrag{F}{$q^2$}
\psfrag{H}{$q+k$}
\psfrag{J}{$q-k$}
\includegraphics[width=2.0in]{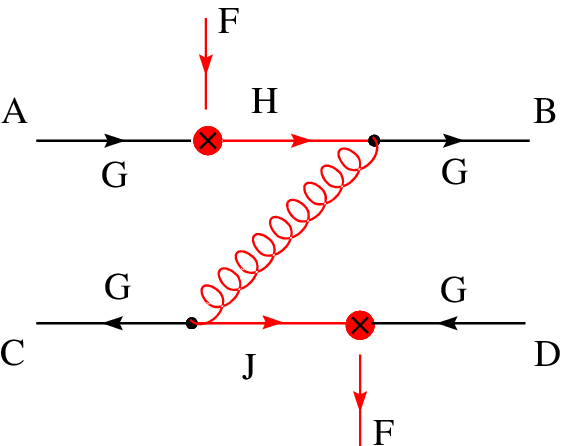}
\hspace{1cm}
\includegraphics[width=2.0in]{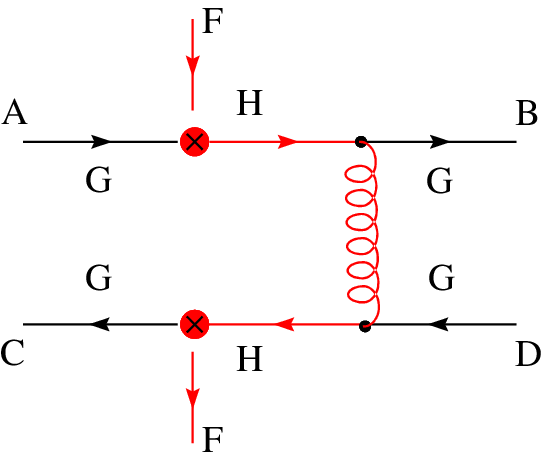}
\newline
\newline
\vspace{0.7cm}
\includegraphics[width=2.0in]{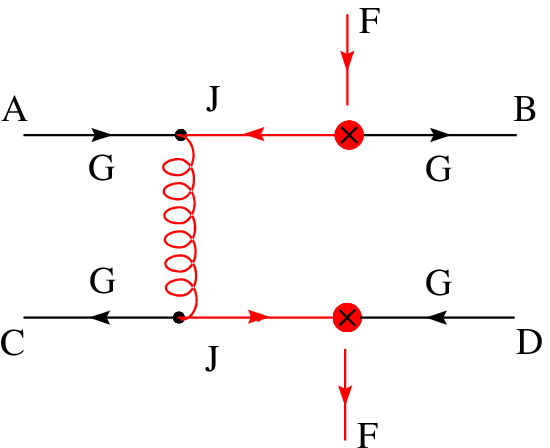}
\hspace{1cm}
\includegraphics[width=2.0in]{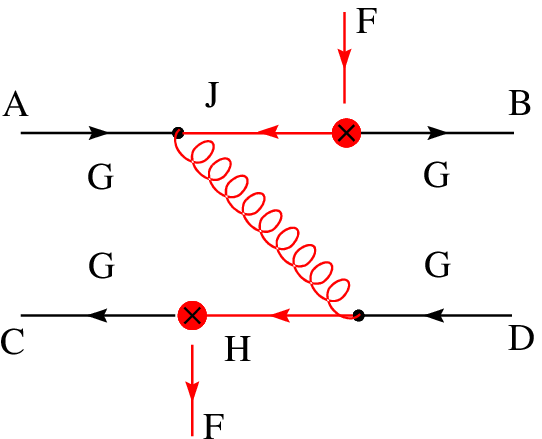}
\caption{Diagrams determining the dimension-six operators in the OPE.}
\label{ope6}
\end{figure}

The diagrams one has to consider are the ones shown in figure (\ref{ope6}), where $q^2$ is a high momentum insertion which flows through the diagram as a gluon exchange in all possible ways. This way we are computing the leading corrections to $K^0-{\bar{K}}^0$ mixing in $\alpha_s$. Obviously there are also electroweak and electromagnetic corrections which enter the Operator Product Expansion, but they are clearly suppressed when confronted with the strong interactions. 

A convenient tool to deal with OPE calculations is the use of the so-called {\it{Schwinger Operator Formalism}} \cite{Shuryak,Shuryak:1,Nov}. Essentially, it is a background field method in coordinate space with the ingredient that covariance is preserved at each step of the computation. A rather detailed introduction into the subject is given in an appendix. 

We set off from the $\Delta S=2$ four quark operator. The four diagrams of figure (\ref{ope6}) come out as an expansion of the product of currents
\begin{equation}\label{currcurr}
{\bar{s}}_L(x)\gamma^{\mu}d_L(x)\,{\bar{s}}_L(0)\gamma^{\mu}d_L(0)
\end{equation}
Following the steps outlined in Appendix B, the previous equation can be cast in the form
\begin{eqnarray}
\qquad \qquad{\bar{s}}^{(0)}_L(x)\gamma^{\mu}\,d^{(0)}_L(x)&&\!\!\!\!\!\!\!\!\!\!\!\!\!\!\!{\bar{s}}^{(0)}_L(0)\gamma^{\mu}\,d^{(0)}_L(0)+\nonumber\\
+{\bar{s}}^{(0)}_L(x)\,\gamma^{\mu}\bigg[\int\,d^4{\tilde{x}}\,S(x-{\tilde{x}})\,ig_sG_{\mu}({\tilde{x}})\,\gamma^{\mu}\,d^{(0)}_L({\tilde{x}})\bigg]&&\!\!\!\!\!\!\!\!\!\!\!\!\!\!\!{\bar{s}}^{(0)}_L(0)\,\gamma^{\mu}\bigg[\int\,d^4{\tilde{y}}\,S(0-{\tilde{y}})\,ig_sG_{\mu}({\tilde{y}})\,\gamma^{\mu}\,d^{(0)}_L({\tilde{y}})\bigg]+\nonumber\\
+{\bar{s}}^{(0)}_L(x)\,\gamma^{\mu}\bigg[\int\,d^4{\tilde{x}}\,S(x-{\tilde{x}})\,ig_sG_{\mu}({\tilde{x}})\,\gamma^{\mu}\,d^{(0)}_L({\tilde{x}})\bigg]&&\!\!\!\!\!\!\!\!\!\!\!\!\!\!\!\bigg[\int\,d^4{\tilde{y}}\,{\bar{s}}^{(0)}_L({\tilde{y}})\,ig_sG_{\mu}({\tilde{y}})\,\gamma^{\mu}\,S(0-{\tilde{y}})\bigg]\gamma_{\mu}\,d^{(0)}_L(0)+\nonumber\\
+\bigg[\int\,d^4{\tilde{x}}\,{\bar{s}}^{(0)}_L({\tilde{x}})\,ig_sG_{\mu}({\tilde{x}})\,\gamma^{\mu}\,S(x-{\tilde{x}})\bigg]\gamma_{\mu}\,\,d^{(0)}_L(x)&&\!\!\!\!\!\!\!\!\!\!\!\!\!\!\!\bigg[\int\,d^4{\tilde{y}}\,{\bar{s}}^{(0)}_L({\tilde{y}})\,ig_sG_{\mu}({\tilde{y}})\,\gamma^{\mu}\,S(0-{\tilde{y}})\bigg]\gamma_{\mu}\,d^{(0)}_L(0)+\nonumber\\
+\bigg[\int\,d^4{\tilde{x}}\,{\bar{s}}^{(0)}_L({\tilde{x}})\,ig_sG_{\mu}({\tilde{x}})\,\gamma^{\mu}\,S(x-{\tilde{x}})\bigg]\gamma_{\mu}\,d^{(0)}_L(x)&&\!\!\!\!\!\!\!\!\!\!\!\!\!\!\!{\bar{s}}^{(0)}_L(0)\,\gamma^{\mu}\bigg[\int\,d^4{\tilde{y}}\,S(0-{\tilde{y}})\,ig_sG_{\mu}({\tilde{y}})\,\gamma^{\mu}\,d^{(0)}_L({\tilde{y}})\bigg]+\nonumber\\
&+&{\mathcal{O}}(\alpha_s^2)
\end{eqnarray}
where the first piece is the unconnected one. We are interested in terms proportional to $\alpha_s$, {\it{i.e.}}, the one gluon exchange. Using the expansion for the gluon and quark propagators given in appendix B, we can systematically extract the different pieces of the operator product expansion order by order in powers of the soft momentum $P$ and inverse powers of the hard momentum $q$. In particular, the $P^0$ terms will give rise to the dimension six pieces of the OPE, while the $P^2$ terms will yield the dimension eight pieces, and so on. Thus, formally, we can determine the OPE to whatever order. However, we eventually would like to compute matrix elements of the resulting operators and, on the phenomenological side, regretfully, little is known about the matrix elements of such operators of increasing complexity. Nonetheless, one expects their impact to be progressively smaller, due to the suppressing $1/q^2$ powers.

In the MHA framework, consistency demands a minimum of terms coming from short distances. In the present case, the leading dimension six contribution to the OPE would suffice to constraint an interpolator with one resonance state, either vector or axial vector. This is precisely what was done in \cite{s5}. However, shortly afterwards there were claims that the neglected dimension eight operators could yield a potentially sizeable effect \cite{Cirigliano:2000ev}, thus affecting the previous determination of ${\widehat{B}}_K$. This motivated a reassessment of ${\widehat{B}}_K$ with the inclusion of next to leading terms in the operator product expansion.   

The full result for dimension six contributions is the following,
\begin{equation}\label{ope6dim}
\int d^{4}x\  e^{iqx}\
T\left\{L^{\overline{s}d}_{\mu}(x)\,L_{\overline{s}d}^{\mu}(0)\right\} = \sum_{i}\ c_i^{(6)}(Q^2)\ {\cal{O}}_i^{(6)}
\end{equation}
where there is only one operator, namely 
\begin{equation}\label{dimsix}
    {\cal{O}}^{(6)}={\bar{s}}_L(0)\,\gamma^{\mu}\,d_L(0)\ {\bar{s}}_L(0)\,\gamma_{\mu}\,d_L(0)
\end{equation}
with the Wilson coefficient given, to lowest order in $\alpha_{s}=g_s^2/4\pi$, by\footnote{See Appendix B.}
\begin{equation}\label{coeffsix}
c^{(6)}(Q^2)= i\,\frac{12\pi\alpha_s}{Q^4}
\end{equation}
The presence of just one operator is precisely what one expects: clearly it is the only possibility for $\Delta S=2$ dimension six operators. We will briefly see that this simplicity no longer holds and combinatorics are larger in dimension eight operators.
Plugging in the previously found OPE for the left-handed currents into the full four-point Green function, one gets
\begin{equation}
W_{\alpha\beta}^{LRLR}(q,l)=\lim_{l\rightarrow 0}\,i^3
\int\,\ d^4y\ d^4z\,e^{il \cdot (y-z)}
\langle\,0\,|\,T\left\{\sum_{i}
c_i^{(6)}(Q^2)\ {\cal{O}}_i^{(6)}\,R^{\bar{d}s}_{\alpha}(y)\,
R^{\bar{d}s}_{\beta}(z)\right\}\,|\,0\,\rangle
\end{equation}
which can be evaluated first using vacuum saturation and then inserting a sum over complete states in all ways allowed by quantum numbers. In the large-$N_C$ limit single particle states are dominant over multiparticle states, and, since by definition the momentum $l$ is soft, this singles out the kaon. The result one gets is,
\begin{eqnarray}
W_{\alpha\beta}^{LRLR}(q,l)&=&2\,c^{6}\lim_{l\rightarrow 0}\,i^3
\langle 0|{\bar{s}}_L\gamma^{\mu}d_L|K(l)\rangle\,\fr{1}{l^2}\,\langle K(l)|\,{\bar{d}}_R\gamma_{\alpha}s_R\,|\,0\rangle\nonumber\\
&&\langle\, 0\,|{\bar{s}}_L\gamma_{\mu}d_L|K(l)\rangle\,\fr{1}{l^2}\,\langle K(l)|{\bar{d}}_R\gamma_{\beta}s_R |0\rangle
\end{eqnarray}
which, together with the definition 
\begin{equation}\label{fk}
\langle 0|{\bar{s}}_L \gamma_{\mu}d_L|K(l)\rangle\doteq -i\sqrt{2}\,F_0\,l_{\mu}
\end{equation}
immediately leads to \cite{s5}
\begin{equation}\label{newope}
W^{OPE}(z)=\frac{24\pi\alpha_sF_0^2}{\mu_{had}^2}\ \frac{1}{z}
    \left[1+\frac{\epsilon}{12}(5+\kappa)\right] +
    \mathcal{O}\left(\frac{1}{z}\right)
\end{equation}
We are provided with two conditions coming from the low-energy side and one condition coming from the high-energy side of our Green function. Since we have already determined the analytical form of our four-point Green function, we can apply the Hadronic Approximation consistently if we include {\it{at least}} one particle, {\it{i.e.}},
\begin{equation} \label{Large}
W(z)_{V}=\frac{a_V}{(z+\rho_V)}+
\frac{b_V}{(z+\rho_V)^2}+\frac{c_V}{(z+\rho_V)^3}
\end{equation}
The authors of \cite{s5} choose the particle to be close to the real $\rho(770)$ state. After fixing the residues $a_V$, $b_V$ and $c_V$ they found a value for ${\widehat{B}}_k$
\begin{equation}\label{primer}
{\widehat{B}}_k=0.33\pm 0.11
\end{equation}
which is in agreement with other estimates \cite{Bijnens:ext,Hans:Ximo} and with chiral extrapolations of lattice simulations \cite{Lee:2004,Aoki:2004}. 
\subsection{Inclusion of next-to-leading short distance constraints on $W_{LRLR}^{\mu\nu\alpha\beta}$} 
Inclusion of next to leading terms in the OPE modifies the expression (\ref{ope6dim}) in the following fashion
\begin{equation}\label{ope}
\int d^{4}x\  e^{iqx}\
T\left\{L^{\overline{s}d}_{\mu}(x)\,L_{\overline{s}d}^{\mu}(0)\right\} = \sum_{i}\ c_i^{(6)}(Q^2)\ {\cal{O}}_i^{(6)}- \sum_{i}\ c_i^{(8)}(Q^2)\ {\cal{O}}_i^{(8)}+ ...
\end{equation}
where the new set of dimension eight operators can be straightforwardly determined (see Appendix B) and read \cite{Cata:2}
\begin{eqnarray}\label{operators}
{\cal{O}}_1^{(8)}&=&\bar{s}\,\overleftarrow{{\cal{D}}_{\mu}}
\overleftarrow{{\cal{D}}^{\mu}}\,\Gamma^{a}_{\nu}\,d\ \bar{s}\,\Gamma^{\nu}_a\, d +
\bar{s}\,\Gamma^{a}_{\nu}\,{\cal{D}}_{\mu}{\cal{D}}^{\mu}\,d\ \bar{s}\,\Gamma^{\nu}_a\, d +
\nonumber  \\
&& \bar{s}\,\Gamma^{\nu}_a\, d\ \bar{s}\,\overleftarrow{{\cal{D}}_{\mu}}
\overleftarrow{{\cal{D}}_{\mu}}\,\Gamma^{a}_{\nu}\,d + \bar{s}\,\Gamma^{\nu}_a\,
d\ \bar{s}\,\Gamma^{a}_{\nu}\,{\cal{D}}_{\mu}{\cal{D}}^{\mu}\,d\nonumber \ ,\\
{\cal{O}}_2^{(8)}&=&\bar{s}\,\Gamma^{\nu}_a\,{\cal{D}}_{\mu} d\
\bar{s}\,\Gamma^{a}_{\nu}\,{\cal{D}}^{\mu}d + \bar{s}
\overleftarrow{{\cal{D}}_{\mu}}\,\Gamma^{a}_{\nu}\,d \
\bar{s}\overleftarrow{{\cal{D}}^{\mu}}\,\Gamma^{\nu}_a\, d
 \nonumber \ ,\\
{\cal{O}}_3^{(8)}&=&\bar{s}\overleftarrow{{\cal{D}}_{\mu}}\,\Gamma^{\nu}_a\, d\
\bar{s}\,\Gamma^{a}_{\nu}{\cal{D}}^{\mu}\,d + \bar{s}\,\Gamma^{a}_{\nu}\,{\cal{D}}_{\mu}d\
{\bar{s}}\overleftarrow{{\cal{D}}^{\mu}}\,\Gamma_{a}^{\nu}\,d\nonumber \ ,\\
{\cal{O}}_4^{(8)}&=&\bar{s}\overleftarrow{{\cal{D}}_{\mu}}\,\Gamma^{\nu}_a\, d\
\bar{s}\overleftarrow{{\cal{D}}_{\nu}}\,\Gamma_{a}^{\mu}\,d +
\bar{s}\,\Gamma^{a}_{\nu}\,{\cal{D}}_{\mu}d\ {\bar{s}}\,\Gamma_{a}^{\mu}\,{\cal{D}}^{\nu}d\nonumber \ ,\\
{\cal{O}}_5^{(8)}&=&\bar{s}\overleftarrow{{\cal{D}}_{\mu}}\,\Gamma^{\nu}_a\, d\
\bar{s}\,\Gamma_{a}^{\mu}\,{\cal{D}}_{\nu}d + \bar{s}\,\Gamma^{a}_{\nu}\,{\cal{D}}_{\mu}d\
{\bar{s}}\overleftarrow{{\cal{D}}^{\nu}}\,\Gamma_{a}^{\mu}\,d\nonumber \ ,\\
{\cal{O}}_6^{(8)}&=&g_s\ \tilde{G}^{a}_{\mu\nu} \ \Big\{\bar{s}\,\Gamma^{\mu}_a\, d \
\bar{s}\,\Gamma^{\nu}\, d - \bar{s}\,\Gamma^{\mu}\, d\ \bar{s}\,\Gamma^{\nu}_a\, d\Big\}
\end{eqnarray}
$\Gamma_a^{\mu}$ is defined as
\begin{equation}
\Gamma^{\mu}_a=\frac{\lambda_a}{2}\ \gamma^{\mu}\ \frac{1-\gamma_5}{2}
\end{equation}
and the following conventions were adopted
\begin{eqnarray}
{\cal{D}}_{\mu} & = &\partial_{\mu}-ig_sA_{\mu}\qquad , \qquad
A_{\mu} = \frac{A_{\mu}^{a} \lambda_{a}}{2} \nonumber\\
G_{\mu\nu}& = &\partial_{\mu}A_{\nu}-\partial_{\nu}A_{\mu}-ig_s[A_{\mu},A_{\nu}]\quad ,
\quad {\tilde{G}}_{\mu\nu}=\frac{1}{2}\,\epsilon_{\rho\sigma\mu\nu}G^{\rho\sigma}\ ,\
\mathrm{with} \quad \epsilon^{0123}= +1
\end{eqnarray}
The corresponding Wilson coefficients are
\begin{equation}\label{coeffs}
    c_i^{(8)}= i\, \frac{4\pi\alpha_s}{Q^6}\ \eta_i^{(8)}
\end{equation}
where the $\eta_i$ are c-numbers, whose values are
\begin{eqnarray}\label{coeffeight}
\eta_1^{(8)}=\frac{5}{3}\qquad , \qquad \eta_2^{(8)}&=&\frac{22}{3}\qquad , \qquad
\eta_3^{(8)}=\frac{8}{3} \nonumber\\
\eta_4^{(8)}=\frac{18}{3}\qquad ,\qquad \eta_5^{(8)}&=&\frac{16}{3}\qquad ,\qquad
\eta_6^{(8)}=\frac{1}{N_c}
\end{eqnarray}
The previous basis of operators satisfy a CPS symmetry inherited from the box diagrams we started from. Indeed, consider the diagrams in position space. Their expressions are
\begin{eqnarray}
T_1= \int d^4x\,d^4x^\prime\,d^4z\,d^4z^\prime\ \bigg[{\bar{s}}_L(z^\prime)\gamma^{\mu}q^{(1)}_L(z^\prime)\ {\bar{q}}^{(1)}_L(x^\prime)\gamma^{\nu}d_L(x^\prime)&&\!\!\!\!\!\!\!\!\!\! {\bar{s}}_L(x)\gamma_{\nu}q^{(2)}_L(x)\ {\bar{q}}^{(2)}_L(z)\gamma_{\nu}d_L(z)\bigg]\cdot\nonumber\\
&&\cdot D(x-x^\prime)\ D(z-z^\prime)\nonumber\\
T_2= \int d^4x\,d^4x^\prime\,d^4z\,d^4z^\prime\ \bigg[{\bar{s}}_L(z^\prime)\gamma^{\mu}q^{(2)}_L(z^\prime)\ {\bar{q}}^{(2)}_L(x^\prime)\gamma^{\nu}d_L(x^\prime)&&\!\!\!\!\!\!\!\!\!\! {\bar{s}}_L(x)\gamma_{\nu}q^{(1)}_L(x)\ {\bar{q}}^{(1)}_L(z)\gamma_{\nu}d_L(z)\bigg]\cdot\nonumber\\
&& \cdot D(x-x^\prime)\ D(z-z^\prime)
\end{eqnarray}
where $D(x-x^\prime)$ is the $W$ propagator between $x$ and $x^\prime$.
A CPS symmetry is a CP transformation followed by an $(s\ \leftrightarrow\ d\,)$ interchange. Using the well-known charge conjugation transformation algebra
\begin{equation}
{\mathbf{C}}^{-1}\ \gamma^{\mu}\ {\mathbf{C}}\ = -(\gamma^{\mu})^T\, , \qquad {\mathbf{C}}^{-1}\ \gamma^{\mu}\gamma_5\ {\mathbf{C}}\ = (\gamma^{\mu}\gamma_5)^T 
\end{equation}
and the one for the parity transformation
\begin{equation}
{\mathbf{P}}^{-1}\ \gamma^{\mu}\ {\mathbf{P}}\ = (\gamma^{\mu})^{\dagger}\, , \qquad {\mathbf{P}}^{-1}\ \gamma^{\mu}\gamma_5\ {\mathbf{P}}\ = (\gamma^{\mu}\gamma_5)^{\dagger}
\end{equation}
one can easily arrive at the CPS transform of the Dirac bilinear
\begin{equation}\label{cpsbilinear}
{\bar{s}}_L(x)\ \gamma^{\mu}\ q_L(x)\ \stackrel{CPS}{\longrightarrow}\ -{\bar{q}}_L({\widetilde{x}})\ \gamma^{\mu}\ d_L({\widetilde{x}})
\end{equation}
where ${\widetilde{x}}$ is the parity-transform of $x$. Integration measures should also be transformed accordingly. The conclusion one reaches is
\begin{equation}
T_1\ \leftrightarrow\ T_2
\end{equation}
Therefore, effective operators at each order have to preserve this original symmetry. The dimension-six operator trivially does. A bit more involved calculation shows that the dimension-eight operator basis (\ref{operators}) is indeed CPS invariant. This is a non-trivial check which, in particular, the basis found in \cite{Pivovarov} does not respect. 
 
The steps to follow hereafter are the same as with the dimension six contribution. However, as we already pointed out, the situation now is much more involved, since we have more operators and their matrix elements are {\it{a priori}} unknown. Use of the large-$N_C$ limit here is extremely advantageous. First, one notices that the contribution coming from ${\mathcal{O}}_6^{(8)}$ is suppressed due to the Wilson coefficient having a $1/N_C$ out front. Second, after colour-fierzing the operators appearing in ${\mathcal{O}}_2^{(8)}$ to ${\mathcal{O}}_5^{(8)}$, one ends up with matrix elements of the form
\begin{equation}
\langle 0|{\bar{s}}_L\gamma_{\mu}D_{\nu}d_L|K(l)\rangle=F_1(l^2)g_{\mu\nu}+F_2(l^2)l_{\mu}l_{\nu}
\end{equation}
whose tensorial behaviour can be inferred from Lorentz invariance. Contracting with the metric tensor $g^{\mu\nu}$ and using the equations of motion in the chiral limit, {\it{i.e.}},
\begin{equation}
D\slash d_L\sim {\mathcal{O}}(m_d)\simeq 0
\end{equation}
one concludes that
\begin{equation}
F_1(l^2)\sim l^2+{\mathcal{O}}(l^4)
\end{equation}
which means that contributions of ${\mathcal{O}}_2^{(8)},\,{\mathcal{O}}_3^{(8)},\,{\mathcal{O}}_4^{(8)},\,{\mathcal{O}}_5^{(8)}$ are suppressed by $l^2$ with respect to the leading terms. The remaining ${\mathcal{O}}_1^{(8)}$ operator can be evaluated by inserting the identity 
\begin{equation}
D^2=D\slash^2 +\fr{1}{2}g_s\sigma_{\mu\nu}G^{\mu\nu}\, ,\qquad \sigma_{\mu\nu}=\fr{i}{2}[\gamma_{\mu},\gamma_{\nu}]
\end{equation}
which after some algebra (see Appendix B) leads to
\begin{equation}
\langle 0|{\bar{s}}_L\gamma_{\mu}D^2d_L|K(l)\rangle=\langle 0|g_s{\bar{s}}_L{\tilde{G}}_{\mu\nu}^a\lambda_a\gamma^{\mu}d_L|K(l)\rangle\doteq (-i\sqrt{2}F_0\,l_{\nu})\,\delta_K^2
\end{equation} 
It is worth emphasising that the previous equation {\it{defines}} the dimensionless $\delta_K^2$ parameter. The notation is due to \cite{NSVZ1}, where the square was chosen {\it{a posteriori}} to stress that $\delta_K^2\geq 0$. By comparing with (\ref{fk}), we see that $\delta_K^2$ measures the relative strength of the dimension-eight quark gluon operator with the dimension-six operator. It turns out, as we will show in the next section, that $\delta_K^2\sim 0.12$ GeV$^2$.

The last equation, in particular, means that ${\mathcal{O}}_1^{(8)}$ is dominant in the chiral expansion, {\it{i.e.}}, it scales in the soft momentum $l$ as the dimension six term does. Therefore, of the six operators that conform the basis of the dimension eight operators, only the first one survives in the large-$N_C$ and chiral limits. We have thus reduced the complexity of the problem by reducing the number of operators to a single one, whose matrix element is governed by the parameter $\delta_K^2$.

\subsection{Determination of $\delta_K^2$}

We will derive a sum rule which will allow us to extract the value of the parameter $\delta_K^2$. It will be a good strategy to consider the whole combination $\delta_K^2 F_0$ as a single parameter. We have already mentioned that $F_0$ is an order parameter which appears in the well-known two-point correlator $\Pi^{\mu\nu}_{LR}$. It is a good idea to follow this analogy and define the two-point Green function\footnote{This came as a suggestion of Eduardo de Rafael in a private communication.}
\begin{equation} \label{GreenLR}
\widetilde{\Pi}_{\mu\nu}^{LR}(Q^2)=i\int\ d^4x\ e^{iq \cdot x}
\langle 0\,|T\left\{\frac{g_s}{2}\bar{s}_L\tilde{G}_{\alpha\mu}\gamma^{\alpha}d_L(x)\
\bar{d}_R\gamma_{\nu}s_R(0)\right\}|0\rangle
\end{equation}
in which, much the same as with $\Pi_{\mu\nu}^{LR}$, Lorentz invariance guarantees that, in the chiral limit, it has the following tensorial
structure
\begin{equation}\label{pitilde}
\widetilde{\Pi}_{\mu\nu}^{LR}(q)=(q_{\mu}q_{\nu}-g_{\mu\nu}q^2)\
\widetilde{\Pi}^{LR}(q^2)
\end{equation}
\begin{figure}\label{opedelta1}
\renewcommand{\captionfont}{\small \it}
\renewcommand{\captionlabelfont}{\small \it}
\centering
\psfrag{A}{${\bar{s}}_L$}
\psfrag{B}{$d_L$}
\psfrag{C}{$k$}
\psfrag{D}{${\bar{s}}_L$}
\psfrag{F}{${\bar{s}}_L$}
\psfrag{G}{$d_L$}
\psfrag{H}{$q^2$}
\psfrag{J}{$q+k$}
\psfrag{M}{$q-k$}
\includegraphics[width=2.5 in]{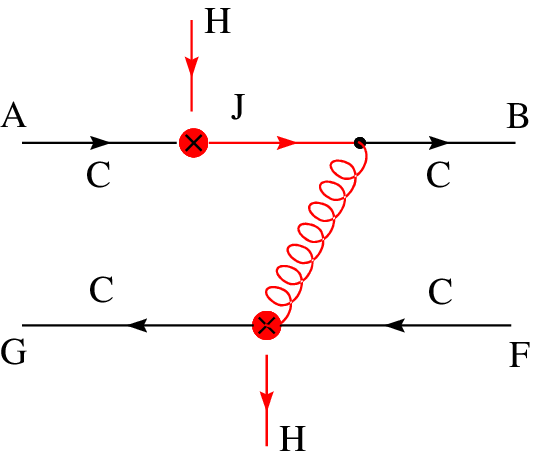}
\hspace{1cm}
\psfrag{F}{$q^2$}\psfrag{G}{$k$}\psfrag{C}{$d_L$}
\includegraphics[width=2.5 in]{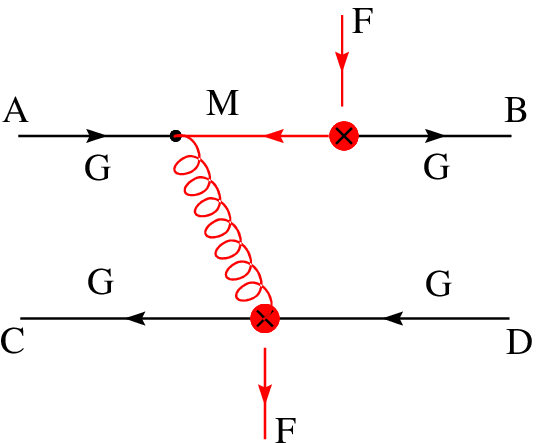}
\caption{Diagrams determining the lowest order contribution to the OPE of (\ref{GreenLR}).}\label{OPEtilde}
\end{figure}
The large-$q^2$ fall-off of the function (\ref{pitilde}) can be straightforwardly
computed to be
\begin{equation} \label{LROPE}
\widetilde{\Pi}^{LR}(q^2)= - \frac{2\pi}{9}\frac{\alpha_{s}<\bar{\psi}\psi>^2}{Q^4}\quad
, \quad Q^2\equiv -q^2
\end{equation}
where factorization of the four-quark condensate in the large-$N_c$ limit, {\it.{i.e.}},\cite{NSVZ1}
\begin{equation}
\langle 0|{\bar{\psi}}\Gamma_i\psi\,{\bar{\psi}}\Gamma_j\psi|0 \rangle=\fr{1}{N^2}\bigg[{\mathrm{Tr}}\,(\Gamma_i)\,{\mathrm{Tr}}\,(\Gamma_j)-{\mathrm{Tr}}\,(\Gamma_i\Gamma_j)\bigg]\,\langle0|{\bar{\psi}}\psi|0\rangle^2
\end{equation}
has been used. The normalization factor $N$ is defined as
\begin{equation}
\langle 0 |{\bar{\psi}}_i\psi_j|0\rangle=\fr{\delta_{ij}}{N}\langle0|{\bar{\psi}}\psi |0\rangle
\end{equation}
such that the above expression is an average over colour and spin degrees of freedom. Diagrams leading to the previous result are depicted in figure (5.4). Using an order parameter as a Green function is a good choice in the sense that it considerably simplifies the calculations, since there is no contribution from the continuum and perturbation theory cancels to all orders. The usual strategy in Finite Energy Sum Rules (FESR) is to make an hadronic ansatz for the spectral function and relate it through a dispersion relation to the values of the Operator Product Expansion\footnote{See \cite{rafael:sr} for a detailed review.}. Our hadronic ansatz will be
\begin{equation} \label{spectralinch5}
\frac{1}{\pi}{\textrm{Im}}\,\widetilde{\Pi}^{LR}(t)=-\frac{F_0^2\delta_{K}^2}{4}\delta(t)+
\frac{f_V^2\delta_V^2}{8}\delta(t-m_V^2)
\end{equation}
and the dispersion relation that ${\widetilde{\Pi}}_{LR}$ obeys is an unsubstracted one,
 \begin{equation} \label{dispersioninch5}
\widetilde{\Pi}^{LR}(Q^2)=\int_0^{\infty}dt\ \frac{1}{t+Q^2-i\,\epsilon}\, \frac{1}{\pi}\, {\textrm{Im}}\,\widetilde{\Pi}^{LR}(t)
\end{equation}
much the same as it happens with $\Pi_{LR}$.
Plugging (\ref{spectralinch5}) in (\ref{dispersioninch5}) and comparing with (\ref{LROPE}), we end up with the following Weinberg-like sum rules
\begin{eqnarray}
\frac{F_0^2\delta_{K}^2}{4}&=&\frac{f_V^2\delta_V^2}{8}\nonumber\\
\frac{f_V^2\delta_V^2}{8}m_V^2&=&\frac{2\pi}{9}\ \alpha_s <\bar{\psi}\psi>^2
\end{eqnarray}
from which the unknown $\delta_V^2$ and $\delta_K^2$ parameters are readily determined to
be
\begin{eqnarray}\label{deltas}
\delta_V^2&=&\frac{16\pi}{9}\ \frac{\alpha_s<\bar{\psi}\psi>^2}{f_V^2m_V^2}\nonumber\\
\delta_{K}^2&=&\frac{8\pi}{9}\ \frac{\alpha_s <\bar{\psi}\psi>^2}{F_0^2 m_V^2}
\end{eqnarray}
Using as input parameters
\begin{equation}
F_0 \simeq 0.087\ {\mathrm{GeV}}\, , \qquad f_V \simeq 0.15\, , \qquad  m_V \simeq 0.77\ {\mathrm{GeV}}\, ,\nonumber
\end{equation}
\begin{equation}\label{param}
\alpha_s(2\
\mathrm{GeV})\simeq 1/3\, , \qquad \langle\bar{\psi}\psi\rangle(2\ \mathrm{GeV})\simeq - (280 \pm
30\ \mathrm{MeV})^3
\end{equation} 
one obtains\footnote{We have added generous error bars in the quark
condensate to include the present spread of values in this quantity.
This error is the dominant one.}
\begin{equation}\label{numbers}
\delta_K^2=(0.12 \pm 0.07)\ \mathrm{GeV}^2\qquad ,\qquad \delta_V^2= (0.06 \pm 0.04)\
\mathrm{GeV}^4
\end{equation}
As a cross-check of stability of (\ref{numbers}), we could construct a Laplace sum rule\footnote{We refer to the review on spectral sum rules of \cite{rafael:sr}.}. Laplace Sum Rules have the advantage over Finite Energy Sum Rules that they exponentially suppress the contributions from high energy resonance states, such that in principle our truncated hadronic ansatz would be more accurate. Taking
\begin{equation} \label{spectr}
\frac{1}{\pi}{\textrm{Im}}\,\widetilde{\Pi}^{LR}(t)= - \frac{F_0^2\delta_{K}^2}{4}\ \delta(t) +
\frac{f_V^2\delta_V^2}{8} \delta(t-m_V^2) + \frac{f_A^2\delta_A^2}{8} \delta(t-m_A^2)
\end{equation}
as our spectral function and applying the Laplace transform to (\ref{spectr}) and to the OPE of (\ref{LROPE}) we find the following Laplace sum rule
\begin{equation}  \label{BSR}
-\frac{F_0^2\delta_{K}^2}{4}+\frac{f_A^2\delta_A^2}{8} e^{-m_A^2\tau}+
\frac{f_V^2\delta_V^2}{8} e^{-m_V^2\tau}=-\frac{2\pi}{9} \alpha_{s}<\bar{\psi}\psi>^2\tau
\end{equation}
Notice that in this Laplace sum rule analysis we have included an axial resonance in our spectral ansatz (\ref{spectr}). Contrary to what happens in Finite Energy Sum Rules, where one has a finite number of relations, Laplace sum rules appear as functions of the Borel parameter $\tau$. Since the parameters one wants to extract should not depend on the Borel parameter $\tau$, one searches for stability in the Borel parameter space.   
Taking as input (\ref{param}), together with 
\begin{equation}
  f_A\sim0.08\, ,\qquad  m_A\sim 1.2\ {\textrm{GeV}}
\end{equation}
We find the Laplace sum rule to hold in the rather wide window $0\lesssim \tau \lesssim 1$ GeV$^{-2}$, as shown in figure (5.5), for values of $\delta_K^2$ and $\delta_V^2$ compatible with the ones in (\ref{numbers}), together with $\delta_A^2\sim 0.05$ GeV$^4$. 

The result we have obtained for $\delta_K^2$ is consistent with two previous analysis \cite{NSVZ1,Nar}. They both determined this $\delta_K^2$ parameter governing dimension-eight effects in kaon mixing with a slightly different Green function, which is not an order parameter. This makes the result depend on the onset of the continuum $s_0$, and some discrepancy arose between both analysis as to the right value for $s_0$. In our opinion, the use of $\widetilde{\Pi}_{LR}$ as the Green function to perform the analysis has the advantage of not having to introduce this extra parameter and thus making the extraction of parameters much cleaner.
\begin{figure}\label{duality}
\renewcommand{\captionfont}{\small \it}
\renewcommand{\captionlabelfont}{\small \it}
\centering
\psfrag{t}{\Large {$\mathbf{\tau}$}}
\includegraphics[width=3 in]{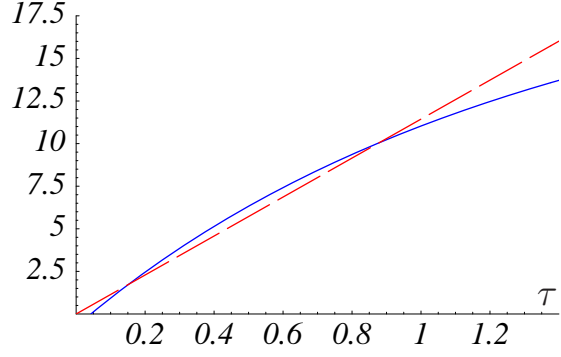}
\caption{Plot of the right and left-hand side of Eq. (\ref{BSR}) (in
arbitrary units), corresponding to the dashed and solid curves,
respectively.}\label{boreltau}
\end{figure}

\subsection{Numerical results for ${\widehat{B}}_K$}

Armed with a precise knowledge of the matrix element relevant for dimension eight contributions, we can parallel the analysis we performed for the dimension six operators and determine the next to leading OPE contributions
\begin{equation}
W_{\alpha\beta}^{LRLR}(q,l)|_{OPE}^{(8)}=\lim_{l\rightarrow 0}\,i^3
\int\,\ d^4y\ d^4z\,e^{il \cdot (y-z)}
\langle\,0|T\bigg\{\sum_{i}
c_i^{(8)}(Q^2) {\cal{O}}_i^{(8)}R^{\bar{d}s}_{\alpha}(y)
R^{\bar{d}s}_{\beta}(z)\bigg\}|0\,\rangle
\end{equation}
Repeating the steps we followed on page 68, we get
\begin{eqnarray}
W_{\alpha\beta}^{LRLR}(q,l)&=&\lim_{l\rightarrow 0}\,i^3
2\,c_1^{(8)}\langle 0|g_s{\bar{s}}_L{\widetilde{G}}^{\nu\rho}\gamma_{\rho}d_L|K(l)\rangle\fr{1}{l^2}\langle K(l)|{\bar{d}}_R\gamma_{\alpha}s_R|0\rangle\nonumber\\&&\langle 0|{\bar{s}}_L\gamma_{\mu}d_L|K(l)\rangle\,\fr{1}{l^2}\,\langle K(l)|{\bar{d}}_R\gamma_{\beta}s_R |0\rangle
\end{eqnarray}
Putting all things together, we obtain
\begin{equation}\label{newope8}
W^{OPE}(z)=\frac{24\pi\alpha_sF_0^2}{\mu_{had}^2}\ \frac{1}{z}
    \left[1+\frac{\epsilon}{12}(5+\kappa)+
    \frac{10}{9} \frac{\delta_{K}^2}{\mu_{had}^2}\ \frac{1}{z}\right] +
    \mathcal{O}\left(\frac{1}{z^3}\right)
\end{equation}
for the OPE up to dimension eight contributions. As expected, dimension eight effects are modulated by the $\delta_K^2$ parameter (compare this expression with (\ref{newope})). Inclusion of these higher dimensionality effects results in an additional condition onto our interpolator (\ref{Largeinch5}). The first possibility is to try to fix the pole of the vectorial resonance we had before. This leads to an imaginary value for the mass, which clearly calls for the introduction of a new resonance in the interpolator to accomodate the new constraint. The simplest possibility is to introduce a scalar particle close to the $f_0(980)-a_0(980)$ doublet.{\footnote{A similar analysis introducing an axial vector particle was also done as a consistency check. Since an axial particle introduces three new parameters to determine, we decided to constrain the interpolator by requiring it to be equal to the OPE at two arbitrary points sufficiently high in energy for the OPE to be trusted. Results as compared with the ones obtained using a scalar particle differ by less than a $5\%$.}} 

Therefore, the interpolator we choose has the form

\begin{equation} \label{Large1}
W(z)_{(V;S)}= \frac{a_V}{(z+\rho_V)}+
\frac{b_V}{(z+\rho_V)^2}+\frac{c_V}{(z+\rho_V)^3}+\fr{a_S}{(z+\rho_S)}
\end{equation}
where the free parameters $a_V$, $b_V$, $c_V$, $a_S$ can be determined by imposing matching onto the two constraints coming from long distances and the two more constraints coming from short distances. This results in the following 4 constraints:
\begin{eqnarray}\label{finalmatching}
a_V+ a_S &=&\frac{24\pi\alpha_s
  F_0^2}{\mu_{had}^2}\left[1+\frac{\epsilon}{12}(5+\kappa)\right]\nonumber\\
b_V- a_V\rho_V- a_S\rho_S&=&\frac{24\pi\alpha_s
F_0^2}{\mu_{had}^4}
\left(\frac{10}{9}\delta_K^2\right)\nonumber\\
\frac{a_V}{\rho_V}+\frac{b_V}{\rho_V^2}+\frac{c_V}{\rho_V^3}+
\frac{a_S}{\rho_S}&=&6\nonumber\\
\frac{a_V}{\rho_V^2}+2\frac{b_V}{\rho_V^3}+
3\frac{c_V}{\rho_V^4}
+\frac{a_S}{\rho_S^2}&=&24\frac{\mu_{had}^2}{F_0^2}\bigg(2L_1+5L_2+L_3+L_9\bigg)
\end{eqnarray}
where the second one is the one coming from next to leading order OPE operators.
Before giving the numbers for ${\widehat{B}}_K$, let us note that the correction we are computing can be understood qualitatively by the following estimation
\begin{eqnarray}\label{estimate}
  \delta {\widehat{g}}_{S=2} &\sim& - \ \frac{\mu_{had}^2}{32 \pi^2 F_0^2}\ \
  \ \frac{24 \pi \alpha_{s}(\mu_{had})F_0^2}{\mu_{had}^2}\ \
  \ \frac{10}{9}\ \frac{\delta_{K}^2}{\mu_{had}^2}\
  \int_{\sim \frac{\mu_{ope}^2}{\mu_{had}^2}}^{\infty}\ \frac{dz}{z^2} \nonumber \\
   & \sim & - \ \frac{15}{18}\frac{\alpha_{s}(\mu_{had})}{\pi}
   \ \frac{\delta_{K}^2}{\mu_{ope}^2}
\end{eqnarray}
where $\mu_{ope}$ is the scale above which the large-$z$ expansion (i.e. the OPE) starts
making sense. The previous expression emphasizes the fact that corrections due to the inclusion of dimension-eight operators in the OPE are governed by the ratio between $\delta_K^2$ and a typical scale $\mu_{OPE}$. Numerically, one obtains $|\delta {\widehat{g}}_{S=2}| \lesssim 0.03$ when $\mu_{ope}\sim 1$
GeV and the value for $\delta_{K}^2$ in Eq. (\ref{numbers}) are used. Plugging in our interpolator in (\ref{match}) and performing the integral one gets the determination for ${\widehat{g}}_{S=2}$, which can be shown to be scale and scheme independent. ${\widehat{g}}_{S=2}$ is trivially related to ${\widehat{B}}_K$, and the final result is the following \cite{Cata:2}
\begin{eqnarray}\label{result}
    \widehat{B}_K&=&\left(\frac{1}{\alpha_s(\mu_{had})}\right)^{\frac{3}{11}}\frac{3}{4}
\Biggl[1-\frac{\alpha_s(\mu_{had})}{\pi}\frac{1229}{1936}
+{\cal{O}}\left(\frac{N_c\alpha_s^2(\mu_{had})}{\pi^2}\right)- \nonumber\\&-&
\frac{\mu_{had}^2}{32\pi^2F_0^2}\left(- a_V \log \rho_V - a_S \log
\rho_S +
\frac{b_V}{\rho_V}+\frac{1}{2}\frac{c_V}{\rho_V^2}\right)\Biggr]
\end{eqnarray}
which is indeed scale and scheme independent.
\begin{figure}
\renewcommand{\captionfont}{\small \it}
\renewcommand{\captionlabelfont}{\small \it}
\centering
\psfrag{A}{\Huge $\frac{W(z)}{W(0)}$}
\includegraphics[width=4.5 in]{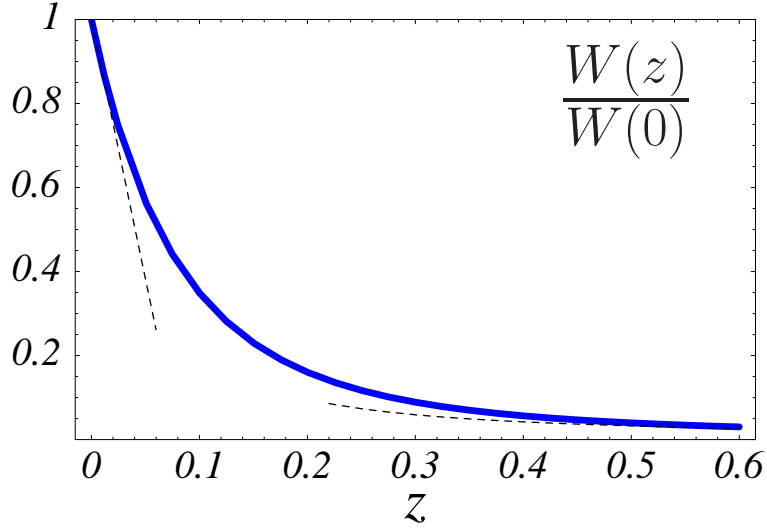}
\caption{Plot of the large- and small-$z$ expansions of
the function $W(z)$ as given by the OPE and Chiral Theory, respectively
(dashed curves). The solid curve corresponds to the interpolating function $W(z)_{(V;S)}$
obtained in Eq. (\ref{Large1}). }\label{Bkplot}
\end{figure}

This is our analytical result for ${\widehat{B}}_K$ in terms of the residues $a_V$, $a_S$, $b_V$ and $c_V$, to be determined using the matching equations of (\ref{finalmatching}). Putting numbers,
\begin{equation}
\alpha_s(2\,{\textrm{GeV}})=0.33\, , \qquad F_0=0.087\,{\textrm{MeV}}\, , \qquad
m_V=0.77\pm 0.03\,{\textrm{GeV}}\nonumber
\end{equation}
\begin{equation}
m_S=0.9\pm 0.4\,{\textrm{GeV}}\, , \qquad 1\,{\textrm{GeV}}\leq \mu_{had}\leq 2\,{\textrm{GeV}}
\end{equation}
we get
\begin{equation}\label{final}
    {\hat{B}}_K= 0.36 \pm 0.15
\end{equation}
for our determination of the invariant $B_K$. Comparison with (\ref{primer}) shows that including next to leading order terms in the OPE does not affect much the determination of ${\widehat{B}}_K$. This is what one should expect, taking into account that $B_K$ comes as the integral under the curve shown in figure (5.6). Changes in the high energy tip, for which the OPE is responsible, should therefore be dramatic to make the area below the curve change notably. Furthermore, we have explicitly shown that the parameter controlling the next to leading OPE contributions ($\delta_K^2$) is rather small.

As already discussed at the end of the previous chapter, in the chiral limit $B_K$ is related to one of the electroweak $\Delta S=1$ couplings, namely $g_{27}$. Therefore, as a by-product, we get a prediction for $g_{27}$, 
\begin{equation}\label{pich}
    g_{27}= \frac{4}{5}{\hat{B}}_K = 0.29 \pm 0.12
\end{equation}
which is in excellent agreement with a recent determination from $K_{e3}$ decays \cite{bij}. 

\subsection{Corrections to the $\Delta S=2$ Lagrangian in inverse powers of the charm mass}

As pointed out at the beginning of the chapter, there are two main contributions to kaon mixing: one coming from direct $\Delta S=2$ transitions, which is conventionally named under {\it{short distance contribution}} and a {\it{long distance contribution}} arising from non-local $\bigg[(\Delta S=1)\times (\Delta S=1)\bigg]$ terms. We already commented on the fact that the latter is estimated to be small. Short distances are meant to be responsible for the bulk of the kaon mass difference, especially through the charm mass contribution, and indeed it is so. However, there are long and short distance contributions arising from the next to leading order operators in the $1/m_c^2$ expansion, where convergence can be dubious and so the impact of these next to leading operators sizeable. Our objective is to assess their actual impact.
 
We already stressed in chapter 2 that one of the requirements for the existence of an effective field theory is the existence of a separation of scales large enough to be able to separate heavy from light degrees of freedom. We already mentioned the potential hazard of considering $m_s\lesssim \Lambda_{\chi}$ as to the convergence of the chiral expansion.  
When we derived the $\Delta S=2$ Effective Lagrangian at leading order, we were implicitly assuming that $m_c\gg \Lambda_{\chi}$, which is not true in the real world. Thus, it seems quite unjustified to truncate the Effective Lagrangian at leading order, since next to leading terms in the form of corrections in the inverse charm mass might as well be sizeable. By dimensional analysis, $1/m_c^2$ corrections in the effective Lagrangian will come as dimension eight operators. Our purpose will be to check if the $1/m_c^2$ expansion in the $\Delta S=2$ Hamiltonian converges good enough. 

From the discussion of the previous section, we already know the basis of dimension eight $\Delta S=2$ operators, and, since we will be working in the large-$N_C$ limit, we can anticipate that this contribution will be eventually proportional to the mixed quark gluon operator
\begin{equation}
g_s\bar{s}_L{\widetilde{G}}_{\mu\nu}\gamma^{\mu}d_L
\end{equation}    
However, this is only the short distance contribution, which we can anticipate that will be proportional to the parameter $\delta_K^2$. There is also an additional long distance contribution, which comes out from computing the up-up contribution to the box diagrams. This contribution is expected to be much bigger and proportional to a hadronic scale $\Lambda_{\chi}\sim 1$ GeV. Their interplay in the determination of the kaon mass difference $\Delta m_K$ and $\varepsilon_K$ is what we will try to determine.
 
\begin{figure}
\renewcommand{\captionfont}{\small \it}
\renewcommand{\captionlabelfont}{\small \it}
\centering
\psfrag{C}{${\mathcal{O}}$}\psfrag{D}{${\mathcal{O}}$}\psfrag{F}{$u$}\psfrag{A}{$R_{\alpha}^{ds}$}\psfrag{B}{$R_{\beta}^{ds}$}
\psfrag{G}{$u,c$}\psfrag{O}{$d_L$}
\includegraphics[width=3.5in]{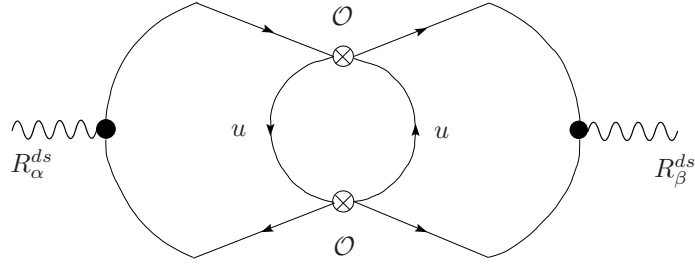}
\caption{Pictorical representation of the up-up exchange contribution.}\label{butterfly}
\end{figure}

\begin{figure}
\renewcommand{\captionfont}{\small \it}
\renewcommand{\captionlabelfont}{\small \it}
\centering
\psfrag{A}{$R_{\alpha}^{ds}$}\psfrag{B}{$R_{\beta}^{ds}$}\psfrag{J}{$q^2$}\psfrag{H}{$u,c$}
\psfrag{G}{$u$}\psfrag{C}{$L_{\mu}^{su}$}
\psfrag{D}{$L_{\mu}^{ud}$}
\psfrag{F}{$L_{\nu}^{ud}$}
\psfrag{H}{$L_{\nu}^{su}$}
\psfrag{K}{$p$}
\includegraphics[width=4.0in]{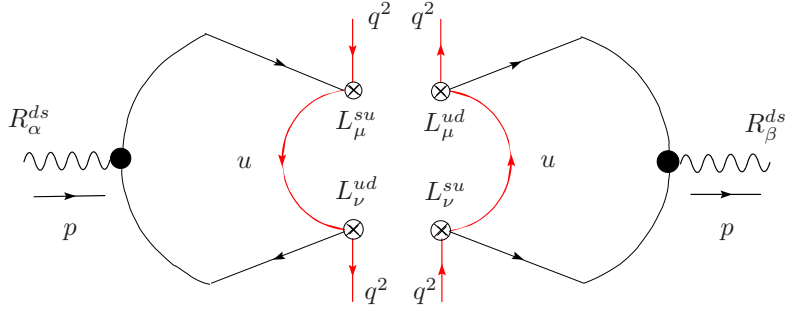}
\caption{Factorization of the up-up contribution leading to $\Gamma_{\mu\nu\alpha}$.}\label{fact}
\end{figure}

As in the derivation of the dimension six effective hamiltonian, we start from the box diagrams after $W$ integration. Dimensional analysis guarantees that no mass terms can appear in the Wilson coefficient. Above the top mass threshold quarks are effectively degenerate and the GIM mechanism is at work, yielding
\begin{equation}
c_8(m_t)=0
\end{equation}
In the energy region $m_c\lesssim \mu \lesssim m_t$ the Wilson coefficient captures the ultraviolet behaviour of top-top, top-charm and top-up diagrams. Due to the absence of masses, this contribution comes in the form
\begin{equation}
\lambda_t^2+2\lambda_t\lambda_c+2\lambda_t\lambda_u=-\lambda_t^2
\end{equation}
where the unitarity relation $\lambda_t+\lambda_c+\lambda_u=0$ has been used. Therefore, there is a GIM-induced suppression in the Wilson coefficients down to the scale of the charm quark. The first non-negligible contribution is then  
\begin{equation}\label{matchcond}
    c_8(m_c) = - \ \frac{7}{6}\ \lambda_u^2 - \frac{13}{6}\ \lambda_c \lambda_u
\end{equation}
at the charm mass threshold. In the equation above we have used $\lambda_c^2+2\lambda_c\lambda_u\simeq -\lambda_u^2$, which is consistent with taking $\lambda_t$ as negligible. Below the charm mass, the Wilson coefficient gets contributions solely from the up-up diagram, with the result \cite{Cata:3}
\begin{equation}\label{twelve}
    c_8(\mu)= -\ \frac{13}{6}\ \lambda_c \lambda_u +
    \lambda_u^2 \left(\log\frac{\mu^2}{m_c^2}- \frac{7}{6}\right)
\end{equation}
Obviously, the above equation ceases to be meaningful below $\Lambda_{\chi}$. Perturbation theory applies no longer and our effective Lagrangian reads 
\begin{equation}\label{twelve-p}
    \mathcal{L}_{eff}=  - \frac{G_F^2}{3 \pi^2}\ c_8(\mu)\
    g_s\overline{s}_L \widetilde{G}^{\mu\nu}\gamma_{\nu}d_L\
    \overline{s}_L\gamma_{\mu}d_L
     - \frac{G_F}{\sqrt{2}} \lambda_u\  \overline{s}_L \gamma^{\mu}u_L\
    \overline{u}_L\gamma_{\mu}d_L + h.c.
\end{equation}
where the second term accounts for the left-over up-up diagram. As it stands, this diagram is a purely non-perturbative object, whose ultraviolet behaviour has to cancel the scale dependence of $c_8(\mu)$, since after all box diagrams are finite. The process of our calculation will eventually show this scale cancellation explicitly. Observe that the evaluation of this long-distance contribution is the only difficulty it remains. The short distance piece can be easily evaluated in the large-$N_C$ limit, where factorization applies, once we know the relevant matrix elements. Not suprisingly, it turns out that the matrix elements we need are precisely the ones involved in the determination of the next to leading terms in the OPE for ${\widehat{B}}_K$, {\it{i.e.}},
\begin{eqnarray}
\langle 0|{\bar{s}}_L \gamma_{\mu}d_L|K(l)\rangle&\doteq& -i\sqrt{2}F_0\,l_{\mu}\nonumber\\
\langle 0|g_s{\bar{s}}_L{\tilde{G}}_{\mu\nu}^a\lambda_a\gamma^{\mu}d_L|K(l)\rangle&\doteq& (-i\sqrt{2}F_0\,l_{\nu})\,\delta_K^2
\end{eqnarray}
Let us therefore focus on the non-perturbative part of our analysis. As it stands in figure (5.7), the up-up diagram is a misrepresentation of the true picture, in which the up-up exchange is a cloud of confined gluons and quarks showing themselves as hadrons. As we did previously, we have to construct an appropriate Green function, schematically shown in figure (\ref{butterfly}), with two left-handed insertions and two soft right-handed ones\footnote{This should come as no surprise. We are eventually computing parameters of $\Delta S=2$ transitions beyond the definition of $B_K$. Recall that $B_K$ is defined as the kaon matrix element of the dimension six operator in the $\Delta S=2$ Lagrangian.}.
In the large-$N_C$ limit, the four-point Green function depicted in figure (\ref{butterfly}) can be split into a convolution of simpler three-point Green functions. This factorization of the Green function is illustrated in figure (\ref{fact}). 
The Green function we have to tackle is therefore
\begin{equation}\label{gamaanom}
\Gamma^{\mathrm{\mathcal{A}}}_{\mu\nu\alpha}(q,p) \equiv \int d^4x\, d^4y\ e^{-i(qx+py)}
    \langle 0|\mathrm{T}\left\{R_{\alpha}^{ds}(y)L_{\mu}^{su}(x)L_{\nu}^{ud}(0)
    \right\}|0\rangle
\end{equation}
where $R_{\alpha}^{ds}(y)$, $L_{\mu}^{su}(x)$ and $L_{\nu}^{ud}$ stand for the QCD left and right-handed currents
\begin{equation}
L_{\mu}^{su}(x)={\bar{s}}(x)\gamma_{\mu}\left(\fr{1-\gamma_5}{2}\right)u(x)\, ,\quad  L_{\nu}^{ud}(0)={\bar{u}}(0)\gamma_{\nu}\left(\fr{1-\gamma_5}{2}\right)d(0)\nonumber
\end{equation}\label{currinch5}
\begin{equation}
R_{\alpha}^{ds}(y)={\bar{d}}(y)\gamma_{\alpha}\left(\fr{1+\gamma_5}{2}\right)s(y)\,
\end{equation}
The subscript ${\mathcal{A}}$ explicitly points at the fact that the Green function is affected by the axial anomaly, a feature which deserves further discussion. 

Anomalies in three-point Green functions were discovered more than thirty years ago in the pioneering papers of Adler, Bell and Jackiw\footnote{See, {\it{e.g.}}, \cite{Bertlmann} for an exhaustive discussion of anomalies in quantum field theory.}. The observation was that certain three-point functions did not obey the classical Ward identities and, even more worrisome, they were ambiguous in the sense that Feynman diagrams were regularization-dependent.
A much closer analysis revealed that this two facts are actually manifestations of the same phenomenon. The triangle diagrams are linearly divergent and depend on the regularization chosen. The simplest way to make this explicit is to compute the triangle with a momentum prescription and then shift it (see, {\it{e.g.}}, \cite{Bertlmann} for details). Therefore, defining $T_{\mu\nu\alpha}$ as the result of the triangle in a given regularization prescription, any ${\widetilde{T}}_{\mu\nu\alpha}$ is also a solution of the triangle if they can be related as
\begin{equation}\label{ambig}
{\widetilde{T}}_{\mu\nu\alpha}=T_{\mu\nu\alpha}+\Delta_{\mu\nu\alpha}
\end{equation}
the last term being a polynomial in momenta. Thus, this is the reason why the classical Ward identities are not fulfilled.
It is conventional to choose the axial Ward identity to be anomalous rather than the vector Ward identity, even though we have just argued that one could as well do the other way round. This means that every triangle with an odd number of axial currents will be anomalous.
Our three-point Green function indeed contains such anomalous pieces. Decomposing the left and right-handed currents (\ref{currinch5}) in terms of vector and axial ones
\begin{eqnarray}
L_{\mu}(x)&=&{\bar{q}}(x)\,\gamma_{\mu}\left(\fr{1-\gamma_5}{2}\right)q(x)\equiv {\mathcal{V}}_{\mu}(x)-{\mathcal{A}}_{\mu}(x)\nonumber\\
R_{\mu}(x)&=&{\bar{q}}(x)\,\gamma_{\mu}\left(\fr{1+\gamma_5}{2}\right)q(x)\equiv {\mathcal{V}}_{\mu}(x)+{\mathcal{A}}_{\mu}(x)
\end{eqnarray}
where
\begin{equation}
{\mathcal{V}}_{\mu}(x)=\fr{1}{2}{\bar{q}}(x)\,\gamma_{\mu}\,q(x)\, , \qquad {\mathcal{A}}_{\mu}(x)=\fr{1}{2}{\bar{q}}(x)\,\gamma_{\mu}\gamma_5\,q(x)
\end{equation}
it is straightforward to check that
\begin{eqnarray}
L_{\mu}L_{\nu}R_{\alpha}&=&{\mathcal{V}}_{\mu}{\mathcal{V}}_{\nu}{\mathcal{V}}_{\alpha}-{\mathcal{A}}_{\mu}{\mathcal{V}}_{\nu}{\mathcal{A}}_{\alpha}+{\mathcal{A}}_{\mu}{\mathcal{A}}_{\nu}{\mathcal{V}}_{\alpha}-{\mathcal{V}}_{\mu}{\mathcal{A}}_{\nu}{\mathcal{A}}_{\alpha}+\nonumber\\
&+&{\mathcal{V}}_{\mu}{\mathcal{V}}_{\nu}{\mathcal{A}}_{\alpha}-{\mathcal{V}}_{\mu}{\mathcal{A}}_{\nu}{\mathcal{V}}_{\alpha}-{\mathcal{A}}_{\mu}{\mathcal{V}}_{\nu}{\mathcal{V}}_{\alpha}-{\mathcal{A}}_{\mu}{\mathcal{A}}_{\nu}{\mathcal{A}}_{\alpha}
\end{eqnarray}
where the first line collects the non-anomalous terms, whereas the second contains the anomaly. It can be readily seen that (\ref{gamaanom}) satisfies de following (anomalous) Ward identities

\begin{eqnarray}\label{cond1}
q_{\mu}\Gamma_{\mathrm{\mathcal{A}}}^{\mu\nu\alpha}(q,p)&=&-i\,\, \Pi^{\nu\alpha}_{LR}(p)
- i \frac{N_c}{48 \pi^2}\
\varepsilon^{\lambda\sigma\nu\alpha} q_{\lambda}p_{\sigma}\nonumber\\
(q+p)_{\nu}\Gamma_{\mathrm{\mathcal{A}}}^{\mu\nu\alpha}(q,p)&=&i\,\,
\Pi^{\mu\alpha}_{LR}(p) + i \frac{N_c}{48 \pi^2}\
\varepsilon^{\lambda\sigma\alpha\mu} q_{\lambda}p_{\sigma} \nonumber\\
p_{\alpha}\Gamma_{\mathrm{\mathcal{A}}}^{\mu\nu\alpha}(q,p)&=& i \frac{N_c}{24 \pi^2}\
\varepsilon^{\mu\nu\lambda\sigma} q_{\lambda}p_{\sigma}
\end{eqnarray}
where $\Pi_{LR}^{\nu\alpha}$ is our well-known two-point function
\begin{equation}
\Pi^{\nu\alpha}_{LR}(p)\equiv \int\,d^4x\,e^{-ip\cdot y}\langle 0 \vert
T\left\{R_{ds}^{\alpha}(y)L_{sd}^{\nu}(0)\right\}\vert 0
\rangle=\left(g^{\nu\alpha}-\frac{p^{\nu}p^{\alpha}}{p^2}\right)\Pi_{LR}(p^2)
\end{equation}
 However, we can now turn the previously discussed ambiguity to our advantage and choose the regularization prescription which makes the na\"ive Ward identities valid. This can be done with a modification of the chronological product $T$ (see, {\it{e.g.}}, \cite{Knecht:2003xy}).
Thus, making use of the inherent ambiguity of the anomaly we can always remove it from our analysis with a proper redefinition of the time-order operator $T$. To this end, we define a non-anomalous Green function
\begin{equation}\label{gama}
\Gamma_{\mu\nu\alpha}(q,p) \equiv \int d^4x d^4y\ e^{-i(qx+py)}
    \langle 0|\mathrm{{\widehat{T}}}\left\{R_{\alpha}^{ds}(y)L_{\mu}^{su}(x)L_{\nu}^{ud}(0)
    \right\}|0\rangle
\end{equation}
where ${\widehat{T}}$ is defined such that $\Gamma_{\mu\nu\alpha}$ satisfy, {\it{de fiat}}, the na\"ive Ward identities 
\begin{eqnarray}\label{cond2}
q_{\mu}\Gamma^{\mu\nu\alpha}(q,p)&=&-i\,\, \Pi^{\nu\alpha}_{LR}(p)\nonumber\\
(q+p)_{\nu}\Gamma^{\mu\nu\alpha}(q,p)&=&i\,\, \Pi^{\mu\alpha}_{LR}(p) \nonumber\\
p_{\alpha}\Gamma^{\mu\nu\alpha}(q,p)&=& 0
\end{eqnarray}
It can be easily shown that our non-anomalous Green functions is related to the anomalous one through
\begin{equation}\label{relat}
    \Gamma^{\mu\nu\alpha}(q,p)= \Gamma_{\mathcal{A}}^{\mu\nu\alpha}(q,p) +
     i \frac{N_c}{24 \pi^2}\
    \varepsilon^{\mu\nu\alpha\lambda} q_{\lambda} +
    i \frac{N_c}{48 \pi^2}\ \varepsilon^{\mu\nu\alpha\lambda} p_{\lambda}
\end{equation}
This is the expression akin to (\ref{ambig}) of our previous discussion.
Having defined our three-point Green function to be anomaly-free, we turn to its characterization. The most general tensorial structure of $\Gamma_{\mu\nu\alpha}$ consists of 22 terms. This number is considerably reduced in the chiral limit, {\it{i.e.}}, if we take $p\rightarrow 0$. The constraints of the above Ward identities (\ref{cond2}) relate them to yield only two independent form factors $I_1(Q^2)$ and $I_2(Q^2)$. The final form for our Green function is
\begin{eqnarray}\label{defthreepoint}
\Gamma_{\mu\nu\alpha}(q,p)&=& \lim_{p\rightarrow 0}\int d^4x d^4y\ e^{-i(qx+py)}\langle 0|\mathrm{\widehat{T}}\left\{R_{\alpha}^{ds}(y)L_{\mu}^{su}(x)L_{\nu}^{ud}(0)\right\}|0\rangle\nonumber\\ 
&=&\Pi_{LR}(0){\cal{T}}_{\mu\nu\alpha}^{S}
   + I_1(Q^2){\cal{T}}_{\mu\nu\alpha}^{ST}
   + I_2(Q^2){\cal{T}}_{\mu\nu\alpha}^A + \mathcal{O}(p)
\end{eqnarray}
where $Q^2=-q^2$, and $\Pi_{LR}(0)$ is the pion pole  
\begin{equation}\label{pifunction}
    \Pi_{LR}(0)= -i\ \frac{F_0^2}{2}
\end{equation}
${\cal{T}}_{\mu\nu\alpha}^{S}$, ${\cal{T}}_{\mu\nu\alpha}^{ST}$ and ${\cal{T}}_{\mu\nu\alpha}^A$ are quite lenghty tensorial structures
\begin{eqnarray}\label{tensors}
{\cal{T}}_{\mu\nu\alpha}^{S}&=&\frac{i}{p^2 q^2}
  \left[p^2 q_{\mu} g_{\alpha\nu}+p^2 q_{\nu}g_{\alpha\mu}-p_{\alpha}p_{\nu}q_{\mu}
  -p_{\alpha} p_{\mu} q_{\nu}+ q_{\mu} q_{\nu}
  \left(\frac{p\cdot q }{q^2}p_{\alpha}-\frac{p^2}{q^2}q_{\alpha}\right)\right]\nonumber\\
{\cal{T}}_{\mu\nu\alpha}^{ST}&=&\left(q^2 g_{\mu\nu} -q_{\mu}q_{\nu}\right)
   \left(\frac{p\cdot q}{p^2}p_{\alpha}-q_{\alpha}\right)\nonumber\\
{\cal{T}}_{\mu\nu\alpha}^A&=&\left[i \varepsilon_{\mu\nu\lambda\sigma}
   q^{\sigma} \frac{p_{\alpha}p^{\lambda}}{p^2}-
   i  \varepsilon_{\mu\nu\alpha\lambda} q^{\lambda}  \right]
\end{eqnarray}
where the superscripts make it explicit the symmetric properties of the tensors with respect to the hard momentum $q$. Thus, ${\cal{T}}_{\mu\nu\alpha}^{S}$ is symmetric under ${\mu} \leftrightarrow {\nu}$ exchange, ${\cal{T}}_{\mu\nu\alpha}^{ST}$ is symmetric and transverse while ${\cal{T}}_{\mu\nu\alpha}^A$ is purely antisymmetric. The expression for $\Gamma_{\mu\nu\alpha}(q,p)$ in its full glory then reads
\begin{eqnarray}\label{nineteen}
  \Gamma_{\mu\nu\alpha}(q,p) &=& \frac{F_0^2}{2 p^2 q^2}
  \left[p^2 q_{\mu} g_{\alpha\nu}+p^2 q_{\nu}g_{\alpha\mu}-p_{\alpha}p_{\beta}q_{\mu}
  -p_{\alpha} p_{\mu} q_{\nu}+ q_{\mu} q_{\nu}
  \left(\frac{p\cdot q }{q^2}p_{\alpha}-\frac{p^2}{q^2}q_{\alpha}\right)\right]\nonumber\\
   &+& I_1(Q^2) \left(q^2 g_{\mu\nu} -q_{\mu}q_{\nu}\right)
   \left(\frac{p\cdot q}{p^2}p_{\alpha}-q_{\alpha}\right)\nonumber\\
   &+& I_2(Q^2) \left[i \varepsilon_{\mu\nu\lambda\sigma}
   q^{\sigma} \frac{p_{\alpha}p^{\lambda}}{p^2}-
   i  \varepsilon_{\mu\nu\alpha\lambda} q^{\lambda}  \right]+ \mathcal{O}(p)
\end{eqnarray}
Let us go back a little to our starting point. The whole dimension eight contributions that we are computing eventually show up in the chiral Lagrangian (\ref{chiralS=2}) as a next to leading order correction to the low energy coupling governing the $\Delta S=2$ transitions $\Lambda_{S=2}^2$. Knowledge of this coupling $\Lambda_{S=2}^2$ is what we eventually need. As already emphasised, a low energy coupling constant appearing in an effective Lagrangian can be related to the parameters of another Lagrangian through a matching condition. This can be accomplished once an appropriate Green function is chosen. When computing the dimension six contributions, {\it{i.e.}} $B_K$, we realized that the $\Delta S=2$ low energy operator
\begin{equation}
\lambda_{32}(D^{\mu}U^{\dagger})U\,\lambda_{32}(D_{\mu}U^{\dagger})U
\end{equation}
contained a kinetic term in the right-handed external sources. We then wrote the matching condition as (\ref{match1})
\begin{equation}
\left(\fr{-i\delta}{\delta r^{\mu}_{{\bar{d}}s}}\right)\left(\fr{-i\delta}{\delta r_{\mu}^{{\bar{d}}s}}\right)\bigg[\Lcal_{QCD}+\Lcal_{ew}\bigg]\doteq\left(\fr{-i\delta}{\delta r^{\mu}_{{\bar{d}}s}}\right)\left(\fr{-i\delta}{\delta r_{\mu}^{{\bar{d}}s}}\right)\Lcal_{eff}^{S=2}
\end{equation}
which is equivalent to compute the contributions from (\ref{chiralS=2}) and (\ref{twelve-p}) to the following Green function
\begin{equation}\label{fourteen}
    \mathcal{G}^{S=2}_{\alpha\beta}(p)=\int d^4x\ e^{ip \cdot x}\
    \langle 0|\mathrm{T}\left\{R_{\alpha}^{ds}(x)R_{\beta}^{ds}(0)\right\} |0\rangle
\end{equation}
where $R_{\alpha}^{ds}(x)= \overline{d}_R \gamma_{\alpha} s_R(x)$. Straightforward calculations with the chiral Effective Lagrangian (\ref{chiralS=2}) yield
\begin{equation}\label{fifteeninch5}
    \mathcal{G}^{S=2}_{\alpha\beta}(p)\bigg|_{\chi PT}= - i \frac{G_F^2}{8 \pi^2}\ F_0^4\ \Lambda_{S=2}^2\
    \left(\frac{p_{\alpha}p_{\beta}}{p^2}-g_{\alpha\beta}  \right) + \mathcal{O}(p^2)
\end{equation}
whereas with the Effective Lagrangian in terms of quarks and gluons (\ref{twelve-p}) one finds
\begin{eqnarray}\label{sixteeninch5}
     \!\!\!\!\!\!\!\!\!\!\!\!\!\!\!\!\!\!\!
     \mathcal{G}^{S=2}_{\alpha\beta}(p)\bigg|_{eff}\!&=\!& - i\ \frac{G_F^2}{3 \pi^2}\ c_8(\mu) \int
     d^4xd^4z\ e^{ipx}\
     \langle0|\mathrm{T}\left\{R_{\alpha}^{ds}(x)
     \mathcal{O}_8(z)  R_{\beta}^{ds}(0)\right\}|0\rangle\nonumber\\
     &&\qquad \quad -i\ \frac{G_F^2}{2 \pi^2} \lambda_u^2 (4\pi \mu^2)^{\epsilon/2}
     \int dQ^2 (Q^2)^{1-\epsilon/2}
     \int d\Omega_q \Gamma_{\mu\nu\alpha}(q,p)\Gamma^{\nu\mu\beta}(q,p)
\end{eqnarray}
where the first line above collects the {\it{short distances}} coming from quark integration and the second line the hadronic contribution coming from the up-up diagram. Equating (\ref{fifteeninch5}) and (\ref{sixteeninch5}) results in the following {\it{matching condition}}
\begin{eqnarray}\label{matchdim8}
 \left(\fr{p_{\alpha}p_{\beta}}{p^2}-g_{\alpha\beta}  \right) \Lambda_{S=2}^2\bigg|_{dim8}&=&\fr{8}{3F_0^4} c_8(\mu) \int
     d^4xd^4z\ e^{ipx}\
     \langle0|\mathrm{T}\left\{R_{\alpha}^{ds}(x)
     {\mathcal{O}}_8(z)  R_{\beta}^{ds}(0)\right\}|0\rangle+\nonumber\\
&&\!\!\!\!\!\!\!\!\!\!\!\!\!\!\!\!\!\!\!\! +\fr{4\lambda_u^2}{F_0^4} (4\pi \mu^2)^{\epsilon/2}
     \int dQ^2 (Q^2)^{1-\epsilon/2}
     \int d\Omega_q \Gamma_{\mu\nu\alpha}(q,p)\Gamma^{\nu\mu\beta}(q,p)
\end{eqnarray}
\begin{figure}
\renewcommand{\captionfont}{\small \it}
\renewcommand{\captionlabelfont}{\small \it}
\centering \psfrag{A}{$R_{\alpha}^{ds}$}\psfrag{B}{$R_{\beta}^{ds}\,\,\,\ \equiv$}
\psfrag{C}{$\Lambda_{S=2}^2$}
\includegraphics[width=1.3in]{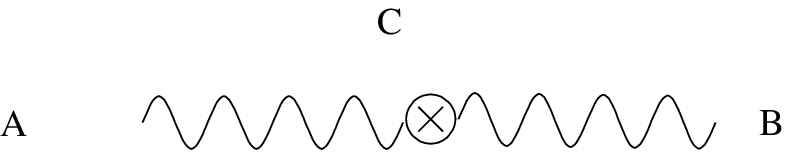}
\hspace{1.5cm}
\psfrag{B}{$R_{\beta}^{ds}\,\,\,\,\,\,\,+$}
\psfrag{O}{${\mathcal{O}}^{(8)}$}
\psfrag{F}{${\widetilde{G}}$}
\includegraphics[width=1.2in]{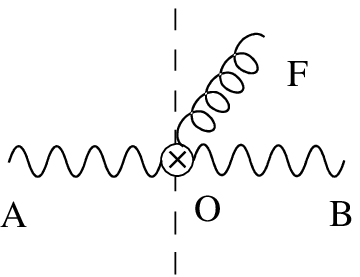}\\
\vskip 0.5cm
\psfrag{A}{$\!\!\!\!\!\!\!\!\!\!\!\!\!\!\!\!\!+\,\quad\ R_{\alpha}^{ds}$}
\psfrag{B}{$R_{\beta}^{ds}$}
\psfrag{K}{$p$}
\psfrag{J}{$q^2$}
\psfrag{G}{$u$}
\psfrag{C}{$L_{\mu}^{su}$}
\psfrag{D}{$L_{\mu}^{ud}$}
\psfrag{F}{$L_{\nu}^{ud}$}
\psfrag{H}{$L_{\nu}^{su}$}
\includegraphics[width=3.5in]{greenfunction.eps}
\caption{Pictorical illustration of the matching condition of (\ref{matchdim8}).}\label{matchingfig}
\end{figure}
which is an expression for the dimension eight contribution to $\Lambda_{S=2}$. Therefore, our problem from now onwards reduces to the determination of this low energy coupling, which, recall, is expressed as a matrix element of the dimension eight operators and an integral over the Euclidean regime of the square of the Green function $\Gamma_{\mu\nu\alpha}$. This is precisely how short distances, {\it{i.e.}}, heavy quarks but also hadronic resonances down to the kaon scale, are encoded in low energy couplings. 
The matching condition (\ref{matchdim8}) is shown schematically in figure (\ref{matchingfig}). The convolution of the three-point Green function $\Gamma_{\mu\nu\alpha}$ with itself can be done without much effort to give
\begin{equation}\label{twenty}
    \int d\Omega_q \Gamma_{\mu\nu\alpha}(q,p) \Gamma^{\nu\mu\beta}(q,p)=
    \left(\frac{p_{\alpha} p_{\beta}}{p^2}-g_{\alpha\beta}\right)\ W(Q^2)+ \mathcal{O}(p^2)
\end{equation}
where the tensorial structure is dictated by Lorentz symmetry. $W(Q^2)$ is given by
\begin{equation}\label{21}
    W(Q^2)= \frac{3}{4}\ I_1^2(Q^2)\ Q^6-
    \frac{3}{2}\ I_2^2(Q^2)\ Q^2
    + \frac{7}{16}\ \frac{F_0^4}{Q^2}
\end{equation}

\begin{figure}
\renewcommand{\captionfont}{\small \it}
\renewcommand{\captionlabelfont}{\small \it}
\centering \psfrag{A}{$R_{ds}$}\psfrag{B}{$L$}
\psfrag{C}{$L$}\psfrag{D}{$R_{ds}$}\psfrag{F}{$L$}\psfrag{G}{$L$}
\includegraphics[width=1.5in]{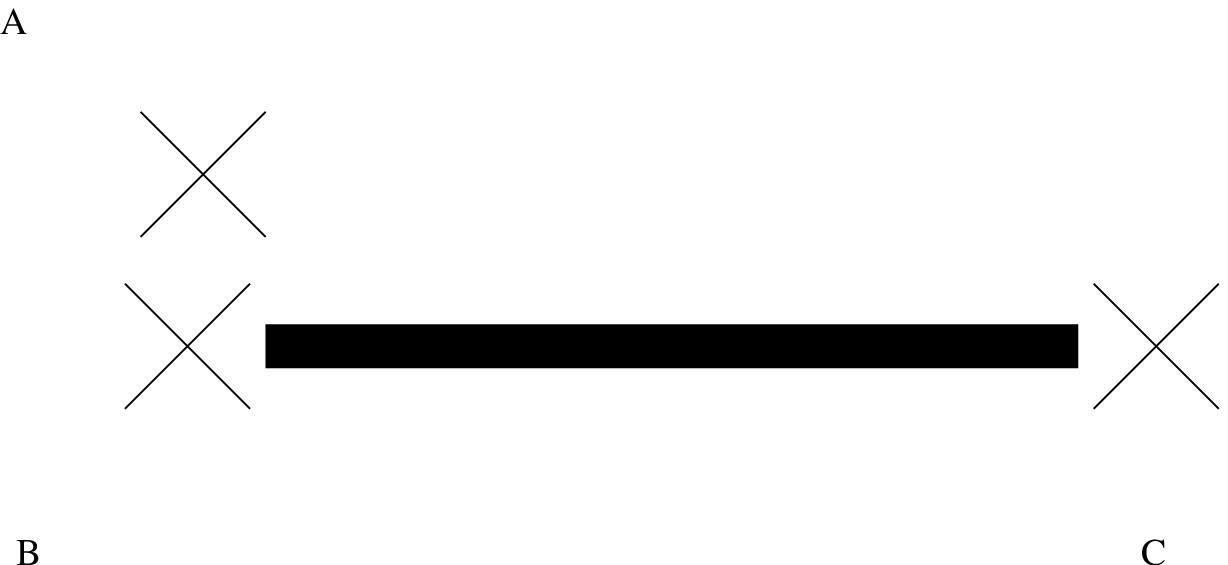}
\hspace{1cm}
\includegraphics[width=1.5in]{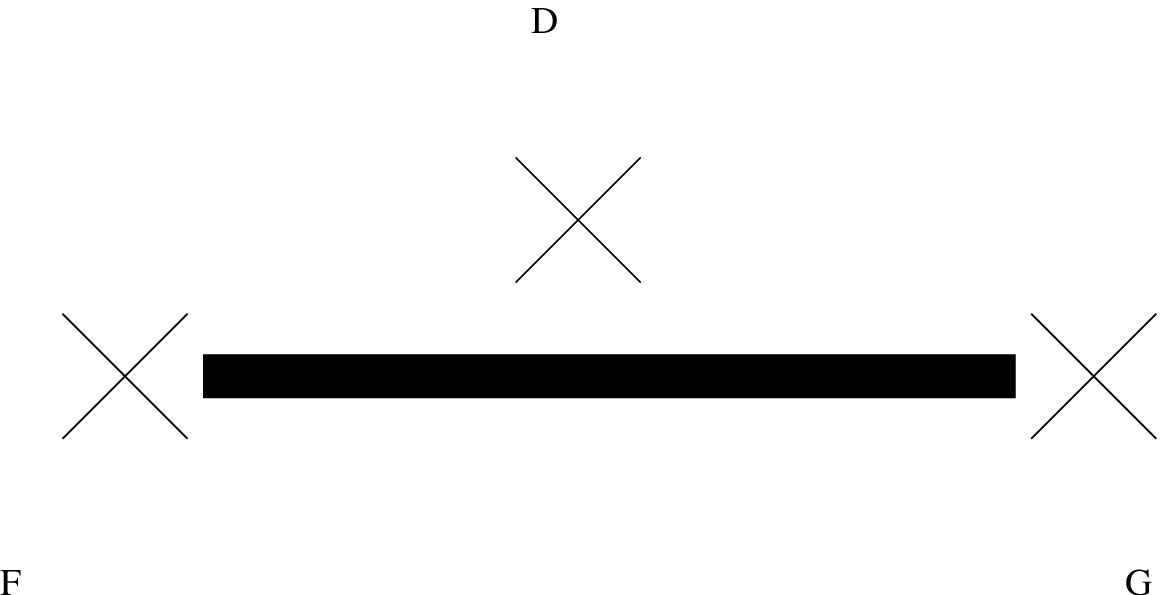}
\caption{Pole structure of the function $\Gamma_{\mu\nu\alpha}\bigg|_{N_C}$ in Eq. (\ref{defthreepointNC}).
``$L$'' stands for $L_{ud}$ or $L_{su}$.}\label{polestructure}
\end{figure}

\begin{figure}
\renewcommand{\captionfont}{\small \it}
\renewcommand{\captionlabelfont}{\small \it}
\centering
\psfrag{F}{$q$}\psfrag{D}{$q+p$}\psfrag{A}{$r_{\alpha}^{{\bar{d}}s}$}\psfrag{B}{$l_{\nu}^{{\bar{u}}d}$}\psfrag{H}{$p$}\psfrag{C}{$l_{\mu}^{{\bar{s}}u}$}\psfrag{K}{$\pi$}
\psfrag{G}{$p$}\psfrag{O}{$d_L$}
\includegraphics[width=1.2in]{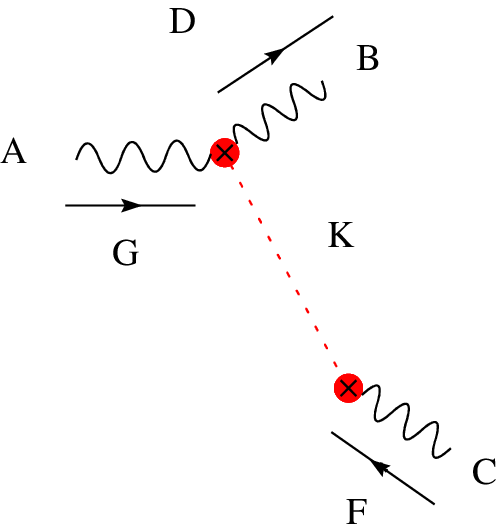}
\hspace{1.1cm}
\psfrag{F}{$p$}\psfrag{G}{$q$}\psfrag{H}{$\pi$}
\includegraphics[width=1.2in]{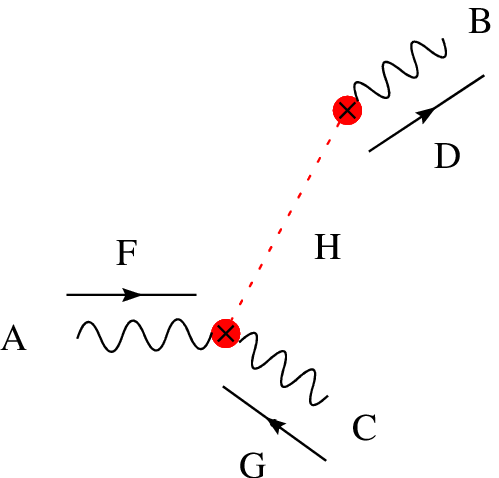}
\hspace{1.1cm}
\psfrag{F}{$q+p$}\psfrag{D}{$q$}\psfrag{H}{$\pi$}\psfrag{G}{$p$}\psfrag{J}{$\pi$}
\includegraphics[width=1.5in]{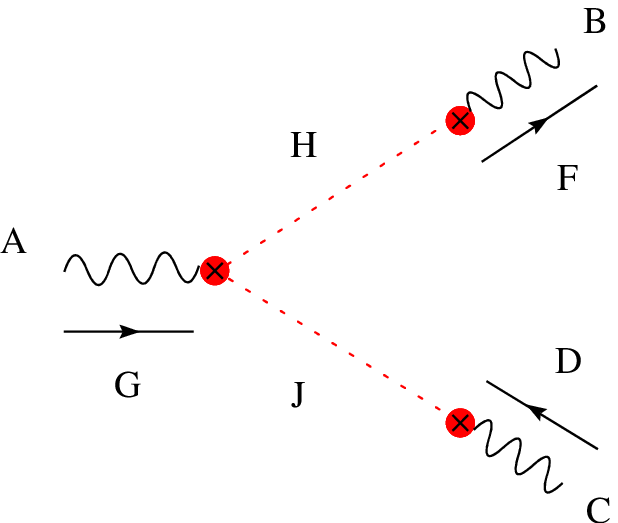}
\newline
\newline
\newline
\newline
\vspace{1.2cm}
\psfrag{F}{$q$}\psfrag{G}{$p$}\psfrag{H}{$\pi$}\psfrag{D}{$q+p$}
\includegraphics[width=1.5in]{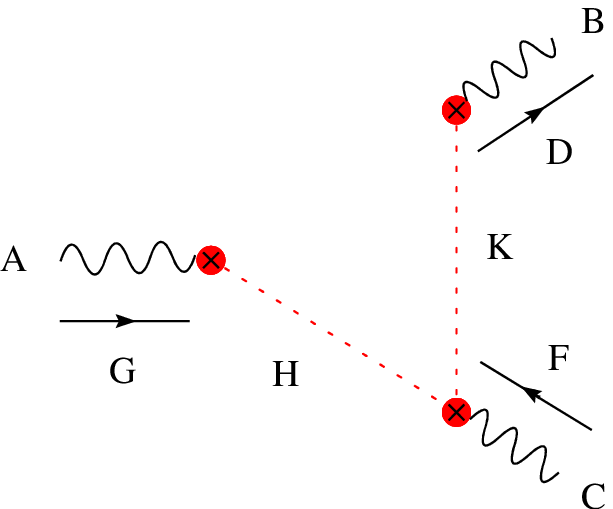}
\hspace{1.1cm}
\psfrag{F}{$q$}\psfrag{G}{$q+p$}\psfrag{H}{$p$}
\includegraphics[width=1.5in]{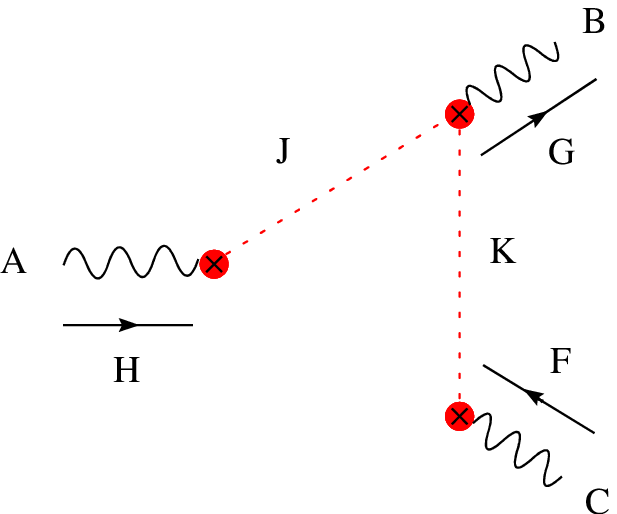}
\caption{Diagrams contributing to the leading order in the chiral expansion of $\Gamma_{\mu\nu\alpha}$.}\label{chip2}
\end{figure}

\begin{figure}
\renewcommand{\captionfont}{\small \it}
\renewcommand{\captionlabelfont}{\small \it}
\centering
\psfrag{F}{$q$}\psfrag{D}{$q+p$}\psfrag{A}{$r_{\alpha}^{{\bar{d}}s}$}\psfrag{B}{$l_{\nu}^{{\bar{u}}d}$}\psfrag{H}{$p$}\psfrag{C}{$l_{\mu}^{{\bar{s}}u}$}
\psfrag{G}{$p$}\psfrag{O}{$d_L$}\psfrag{K}{$\pi$}
\vspace{-0.6cm}
\includegraphics[width=1.2in]{chi4.eps}
\hspace{1.1cm}
\psfrag{F}{$p$}\psfrag{G}{$q$}\psfrag{H}{$q+p$}\psfrag{K}{$\pi$}
\includegraphics[width=1.5in]{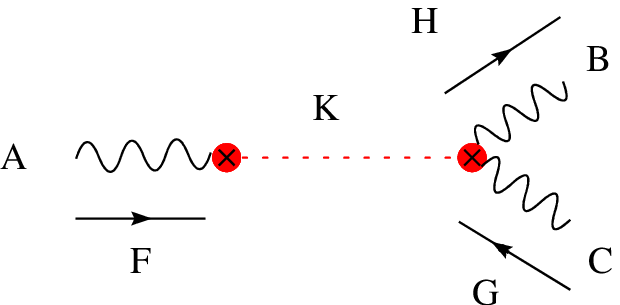}
\hspace{1.1cm}
\psfrag{F}{$p$}\psfrag{G}{$q$}\psfrag{H}{$\pi$}
\includegraphics[width=1.2in]{chi7.eps}
\newline
\newline
\vspace{0.8cm}
\psfrag{F}{$p$}\psfrag{D}{$q$}\psfrag{H}{$q+p$}\psfrag{J}{$\pi$}
\includegraphics[width=1.0in]{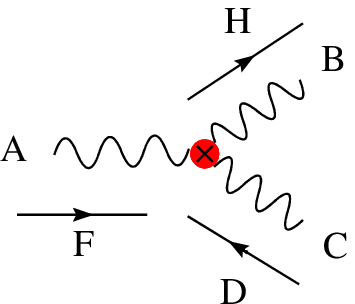}
\hspace{1.0cm}
\psfrag{F}{$q+p$}\psfrag{D}{$q$}\psfrag{H}{$\pi$}\psfrag{G}{$p$}
\includegraphics[width=1.5in]{chi1.eps}
\hspace{1.0cm}
\psfrag{F}{$q$}\psfrag{G}{$p$}\psfrag{H}{$\pi$}\psfrag{D}{$q+p$}
\includegraphics[width=1.5in]{chi2.eps}
\hspace{1.0cm}
\psfrag{F}{$q$}\psfrag{G}{$q+p$}\psfrag{H}{$p$}
\includegraphics[width=1.5in]{chi3.eps}
\caption{Diagrams contributing to the next to leading order even parity processes in the chiral expansion of $\Gamma_{\mu\nu\alpha}$.}\label{chip4}
\end{figure}

\begin{figure}
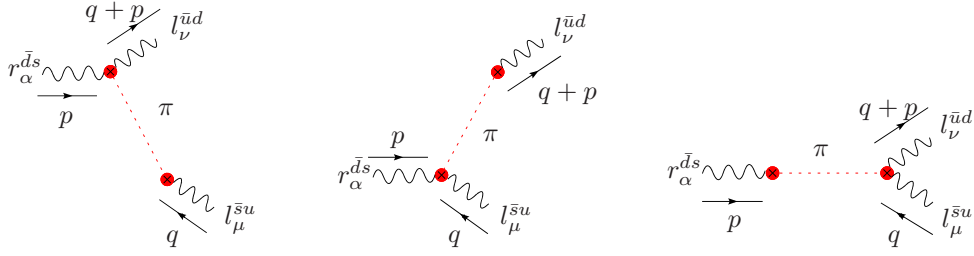

\renewcommand{\captionfont}{\small \it}
\renewcommand{\captionlabelfont}{\small \it}
\centering
\psfrag{F}{$q$}\psfrag{D}{$q+p$}\psfrag{A}{$r_{\alpha}^{{\bar{d}}s}$}\psfrag{B}{$l_{\nu}^{{\bar{u}}d}$}\psfrag{H}{$p$}\psfrag{C}{$l_{\mu}^{{\bar{s}}u}$}
\psfrag{G}{$p$}\psfrag{O}{$d_L$}\psfrag{K}{$\pi$}
\includegraphics[width=1.2in]{chi4.eps}
\hspace{1.1cm}
\psfrag{F}{$p$}\psfrag{G}{$q$}\psfrag{H}{$\pi$}
\includegraphics[width=1.2in]{chi7.eps}
\hspace{1.1cm}
\psfrag{F}{$p$}\psfrag{G}{$q$}\psfrag{H}{$q+p$}\psfrag{K}{$\pi$}
\includegraphics[width=1.5in]{chi5.eps}
\caption{Diagrams contributing to $\Gamma_{\mu\nu\alpha}$ coming from the WZW term.}\label{WZWinch5}
\end{figure}

Recall that, as a consequence of the symmetry properties of the tensors $\mathcal{T}_{\mu\nu\alpha}$, there appear no crossed terms of the form factors in (\ref{21}). Once we have our function $W(Q^2)$, the usual procedure is to find out its low energy and high energy expressions by means of Chiral Perturbation Theory and the Operator Product Expansion, respectively. Afterwards we shall connect them through an interpolator of the analytical form dictated by large-$N_C$ QCD. A relatively simple analysis of the three-point Green function $\Gamma_{\mu\nu\alpha}\bigg|_{N_C}$ reveals that it can only consist of simple and double poles,
\begin{equation}\label{defthreepointNC}
\Gamma_{\mu\nu\alpha}\bigg|_{N_C}=\sum_i\bigg\{\fr{a_i}{Q^2+M_i^2}+\fr{b_i}{(Q^2+M_i^2)^2}\bigg\}
\end{equation}
 as shown in figure (\ref{polestructure}). Therefore, its square $W(Q^2)$ has as much as quadruple poles. Its expression in the large-$N_C$ limit looks like
\begin{equation}\label{25inch5}
    Q^2 W(Q^2)_{N_C}= a + \sum_i\left\{\frac{A_i}{Q^2+M_i^2}+
    \frac{B_i}{(Q^2+M_i^2)^2}+\frac{C_i}{(Q^2+M_i^2)^3}+
    \frac{D_i}{(Q^2+M_i^2)^4}\right\}
\end{equation}
The above expression, though analytically simple, still requires an infinite amount of information. The strategy to truncate the above series is provided by the Minimal Hadronic Approximation. The number of resonances to be included in our analysis depends strongly on our knowledge of the high energy QCD behaviour of $W(Q^2)$. At large momentum, for the form factors, one has 
\begin{equation}\label{high}
    I_1(Q^2)= \frac{F_0^2}{2Q^4}+ \mathcal{O}\left(\frac{1}{Q^6}\right)\quad ,
    \quad I_2(Q^2)= - \frac{F_0^2}{2 Q^2}+ \mathcal{O}\left(\frac{1}{Q^4}\right)
\end{equation}
It is rather difficult to get further in the OPE expansion. Knowledge of matrix elements such as 
\begin{equation}
\langle 0|{\bar{s}}_L\gamma^{\mu}\gamma^{\nu}\gamma^{\rho}{\widetilde{G}}_{\nu\rho}d_L|0\rangle
\end{equation}
is needed and, to the best of our knowledge, such matrix elements have not been determined. However, what we eventually need for our calculation is the function $W(Q^2)$, for which the OPE gets simplified. The final result is
\begin{equation}\label{23}
    Q^2 W(Q^2)_{OPE} \approx  \frac{F_0^4}{4}+ \frac{F_0^4}{3}\frac{\delta_K^2}{Q^2}
    \left(1+ \frac{\epsilon}{6} \right)+
    \mathcal{O}\left(\frac{1}{Q^4}\right)
\end{equation}
Recalling that\footnote{see the discussion in chapter 3.}
\begin{equation}
{\mathcal{N}}-{\mathcal{P}}=-p_{OPE}
\end{equation}
we conclude that one can do with just one resonance in the interpolator (\ref{25inch5}).
At low energies, the contribution comes from the diagrams listed in figures (\ref{chip2}), (\ref{chip4}) and (\ref{WZWinch5}). 
At low energies both momenta $q$ and $p$ are soft. However, taking into account the definition of our Green function $\Gamma_{\mu\nu\alpha}$, $p$ has been sent to zero accordingly. At leading order (figure (\ref{chip2})), the surviving diagrams are the first three ones. At ${\mathcal{O}}(p^4)$, only $L_9$ and $L_{10}$ yield some contribution. Nonetheless, all terms containing $L_{10}$ vanish when $p\rightarrow0$. Actually, terms containing $L_{10}$ can only appear in $\Pi_{LR}(p)$ at ${\mathcal{O}}(p^2)$ and therefore consistently cancel as $p\rightarrow 0$. Even-parity contributions feed $\Pi_{LR}(0)$ and $I_1(Q^2)$. As for the Wess-Zumino-Witten term, its contribution is absorbed entirely in the odd-parity form factor $I_2(Q^2)$. This contribution is non-anomalous, since we already subtracted the anomaly. The presence of non-zero WZW contributions in non-anomalous processes is sometimes called the {\it{Cheshire cat smile}} of the WZW Lagrangian. At low energies, therefore,
\begin{equation}\label{low}
    I_1(Q^2)= 2 \frac{L_9}{Q^2}+ \mathcal{O}(Q^0)\quad ,
    \quad I_2(Q^2)= - \frac{N_c}{24 \pi^2}+ \mathcal{O}(Q^2)
\end{equation}
which, after squaring, yields
\begin{equation}\label{22}
    Q^2 W(Q^2)_{\chi} \approx  \frac{7}{16}F_0^4 -
    \left[\frac{3}{2} \left(\frac{N_c}{24\pi^2}\right)^2 -3
    L_9^2\right]Q^4+ \mathcal{O}(Q^6)
\end{equation}
Thus, equations (\ref{23}) and (\ref{22}) are the ones onto which we will match our interpolating function, to wit 
\begin{equation}\label{25}
    Q^2 W(Q^2)_{HA}= a + \frac{A}{Q^2+M_V^2}+
    \frac{B}{(Q^2+M_V^2)^2}+\frac{C}{(Q^2+M_V^2)^3}+
    \frac{D}{(Q^2+M_V^2)^4}
\end{equation}
This will result in the following set of matching equations, 
\begin{eqnarray}\label{conditions}
  \frac{A}{M_V^2}+ \frac{B}{M_V^4}+\frac{C}{M_V^6}+\frac{D}{M_V^8} &=&
  \frac{7}{16} F_0^4 -a \nonumber \\
  \frac{A}{M_V^4}+ \frac{2B}{M_V^6}+\frac{3C}{M_V^8}+\frac{4D}{M_V^{10}} &=& 0\nonumber \\
  \frac{A}{M_V^6}+ \frac{3B}{M_V^8}+\frac{6C}{M_V^{10}}+\frac{10D}{M_V^{12}} &=& -\frac{3}{2}
  \ \frac{N_c^2}{(24 \pi^2)^2}+  3\ L_9^2 \nonumber\\
  a &=& \frac{F_0^4}{4} \nonumber\\
  A &=& \frac{F_0^4 \delta_K^2}{3} \left(1+ \frac{\epsilon}{6} \right)
\end{eqnarray}
out of which we can extract the unknown residues $A$, $B$, $C$ and $D$ in terms of $L_9$, $F_0$, $M_V$ and $\delta_K^2$. Figure (5.14) shows the interpolating function (\ref{25}) for a given choice of input parameters together with its chiral and OPE extrapolations. Notice that the parameter $a$ above, which accounts for the pion contribution, does not contribute to the integral of (\ref{matchdim8}) (in dimensional regularisation). This is why we have subtracted it away in figure (5.14).
\subsection{Numerical analysis for $\Delta M_K$ and $\varepsilon_K$.}
Using our expression for the Green function (\ref{25}) and plugging it into (\ref{matchdim8}) we get
\begin{equation}\label{26}
    \Lambda^2_{S=2}\bigg|_{dim8}= \frac{4}{3}\delta_K^2 \left\{\lambda_u^2 \left( \log\frac{\mu^2}{M_V^2}  -
    \frac{1}{3}\right)- c_8(\mu)\right\}+ 4 \frac{\lambda_u^2}{F_0^4}
    \left[\frac{B}{M_V^2}+ \frac{C}{2 M_V^4} +
    \frac{D}{3 M_V^6} \right]
\end{equation}
as our analytical result for the low energy $\Delta S=2$ coupling, once corrections coming from the next to leading order terms in the charm mass expansion are included. Recall that our result is scale and scheme independent, as it should. Indeed, the scale dependence coming from the integral of the $A$ parameter in (\ref{conditions}) is exactly the one coming from the Wilson coefficient $c_8(\mu)$ (see (\ref{twelve})). Taking as input parameters
\begin{equation}
L_9=7\times 10^{-3}\, , \qquad F_0=0.087\,{\textrm{MeV}}\, , \qquad M_V=0.77\,{\textrm{GeV}}\nonumber
\end{equation}
\begin{equation}
\delta_K^2=0.12\pm 0.07\,{\textrm{GeV}}^2\, , \qquad m_c(m_c)=1.3\,{\textrm{GeV}}
\end{equation}
 we get
\begin{equation}\label{resultinch5}
\Lambda^2_{S=2}\bigg|_{dim8}= \left\{\frac{26}{9}\ \lambda_c \lambda_u\  \delta_K^2
+ \lambda_u^2\ (0.95\ \mathrm{GeV})^2\right\}\left( 1 \pm 0.3\right)
\end{equation}
which shows the contribution from the two scales involved in the problem: $\delta_K\sim 0.35$ GeV, weighting the short distances and an hadronic scale $\Lambda_{\chi}\sim 1$ GeV, responsible for the long distances coming from the up-up exchange box diagrams. The previous equation already shows that, since only $\lambda_c$ has an imaginary part, contrary to $\lambda_u$, $\varepsilon_K$ will be unaffected by the up-up diagram and will only pick a small correction coming from the short distance scale $\delta_K^2$. However, the kaon mass difference picks the real part of $\Lambda_{S=2}^2$ and so gets corrected mainly by the hadronic scale $\Lambda_{\chi}$. The equations relating our previously-determined  $\Lambda_{S=2}$ low energy coupling to $\varepsilon_K$ and $\Delta m_K$ are as follows
\begin{eqnarray}\label{epsilo}
    \epsilon_K &=& e^{i\pi/4}\
    \frac{G_F^2 F_0^2 M_K}{6\sqrt{2}\pi^2 \left(M_{K_{L}}-M_{K_{S}}\right)}\ \times\\
    && \mathrm{Im}\ \left\{\frac{3}{4}\ {\widehat{g}}_{S=2} \Big[\eta_1 \lambda^{*2}_{c} m^2_c+ \eta_2
    \lambda^{*2}_t \left(m^2_{t}\right)_{eff}+ 2 \eta_3 \lambda^{*}_c \lambda^{*}_t m^2_c
    \log\frac{m_t^2}{m_c^2}\Big]+ \frac{3}{4}\ \Lambda^{*2}_{S=2}|_{dim8}
    \right\}\nonumber
\end{eqnarray}
for $\epsilon_K$, whereas for the kaon mass difference, one obtains
\begin{eqnarray}\label{mass}
    M_{K_{L}}-M_{K_{S}}&=&\frac{1}{M_K}\ \mathrm{Re}
    \langle K^0|\mathcal{H}_{eff}^{S=2}(0)|\overline{K}^0\rangle =
    \frac{G_F^2}{3\pi^2}F_0^2
    M_K \times \\
    &&\!\!\!\!\!\!\!\!\!\!\!\!\!\!\!\!\!\!\!\!\!\!\!\!\!\!\!\!\!\!\!\!\!\!\!\!
    \ \mathrm{Re}\ \left\{ \frac{3}{4}\ {\widehat{g}}_{S=2} \Big[\eta_1 \lambda_c^{*2} m^2_c
    + \eta_2
    \lambda_t^{*2} \left(m^2_{t}\right)_{eff}+ 2 \eta_3 \lambda_c^* \lambda_t^* m^2_c
    \log\frac{m_t^2}{m_c^2}\Big]+ \frac{3}{4}\ \Lambda^{*2}_{S=2}|_{dim8}\right\}
    \nonumber
\end{eqnarray}
The first term in the curly brackets accounts for the dimension six contribution to $\Lambda_{S=2}^2$, whereas the second term is the result found in (\ref{resultinch5}).
We stress that the previous results are in the chiral and large-$N_C$ limits. Some chiral and $1/N_C$ corrections are known for the dimension-six part, which amounts to the replacement
\begin{equation}
{\widehat{g}}_{S=2}\rightarrow \fr{4}{3}{\widehat{B}}_K\fr{F_K^2}{F_0^2}
\end{equation}
with ${\widehat{B}}_K$ given in (\ref{final}) and $F_K=0.114$ GeV. Lattice calculations of ${\widehat{B}}_K$ with inclusion of chiral corrections push its value to ${\widehat{B}}_{K}=0.86\pm 0.15$ \cite{Lellouch}. Making the previous replacements, together with the input
\begin{equation}
m_t=175\,{\textrm{GeV}}\, , \qquad M_K=0.498\,{\textrm{GeV}}
\end{equation}
and the CKM matrix elements as given in \cite{Buras2}, results in a $0.5\%$ correction to $\varepsilon_K$. As for the kaon mass difference, our determination yields
\begin{equation}
M_{K_L}-M_{K_S}=(3.1\pm 0.5)\times 10^{-15}\,{\textrm{GeV}}
\end{equation}
to be compared with the experimental value \cite{hagi}
\begin{equation}
M_{K_L}-M_{K_S}\bigg|_{exp}=(3.490\pm 0.006)\times 10^{-15}\,{\textrm{GeV}}
\end{equation}
\begin{figure}
\renewcommand{\captionfont}{\small \it}
\renewcommand{\captionlabelfont}{\small \it}
\centering \psfrag{T}{\Large $Q^2$} \psfrag{Expressio}{\Large \!\!\!\!\!\!\!\!\!\!\!\!$\
\frac{16}{3}\frac{Q^2}{F_0^4}\ W_{HA}(Q^2)$} \vspace{1cm}
\includegraphics[width=3.5in]{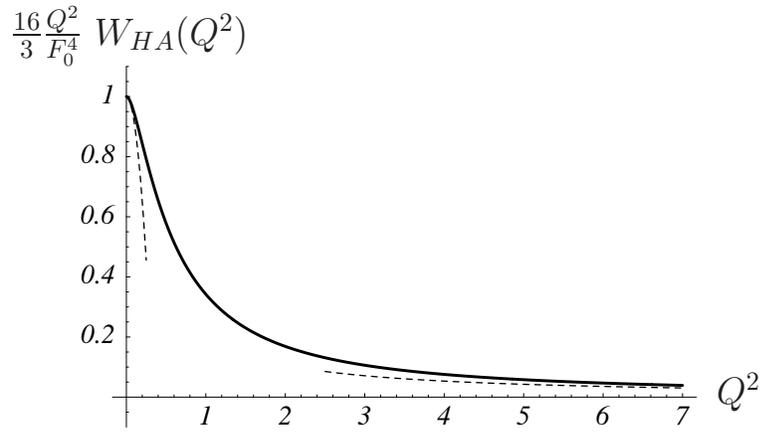}
\caption{Profile of the function $Q^2\ W(Q^2)_{HA}$ together with the chiral and OPE
behaviour at low and high $Q^2$ (in $D=4$), respectively. The pion contribution has been
subtracted away (the $a$ parameter in Eq. (\ref{25inch5})), according to the discussion in the main
text.}\label{fig:plotmatching}
\end{figure}
We therefore find that contributions arising from direct $\Delta S=2$ transitions account for the $(90\pm15)\%$ of the observed kaon mass difference. The remaining $\bigg[\Delta S=1\times \Delta S=1\bigg]$ long distance contributions might account for the rest. We already know that they cancel at tree level due to the Gell-Mann-Okubo mass relation. However, calculations at one loop are stopped by the fact that there is no determination of the low energy couplings involved (See, for instance, the discussion in \cite{John}).


\chapter[Duality Violations]{Duality Violations}
\section{Motivation}

The OPE \cite{SVZ} is one of the most important tools with which one can extract information about the QCD parameters and, obviously, the condensates. These parameters are however not unrelated, and, for instance, the condensates of the OPE of the two-point correlator $\Pi_{LR} (q^2)$ are related to kaon physics parameters\footnote{See, for instance, \cite{relationship}-\cite{relation}.} The OPE is known to be a valid expansion in the deep Euclidean, but physical information is only available on the Minkowski plane. Both regions can nonetheless be related with the use of Cauchy's integral theorem 
\begin{equation}\label{Cauchy}
\int_0^{s_0}\, dt\, t^n\,\fr{1}{\pi}\, {\mathrm{Im}}\, \Pi_{LR}(t)=-\fr{1}{2\pi i} \oint_{|q^2|=s_0}\,dq^2 q^{2n}\,\Pi_{LR}(q^2)
\end{equation}
over the contour depicted in figure (6.1). Since knowledge of the full Green function is not available, the common practice is to replace the correlator by its OPE. Therefore, strictly speaking, we can rewrite Cauchy's theorem in the following fashion
\begin{equation}\label{CauchyOPE}
\int_0^{s_0}\, dt\, t^n\,\fr{1}{\pi}\, {\mathrm{Im}}\, \Pi_{LR}(t)=-\fr{1}{2\pi i} \oint_{|q^2|=s_0}\,dq^2 q^{2n}\,\Pi_{LR}^{OPE}(q^2)+{\mathcal{D}}^{[n]}(s_0)
\end{equation}
The substitution of the full Green function by its OPE is what goes under the name of {\it{duality}}. Differences arising from the above substitution, collected under ${\mathcal{D}}^{[n]}(s_0)$, are therefore coined {\it{duality violations}}\footnote{For a review, see \cite{Shifman:00}.}
Since the OPE is believed to be a good approximation at large $s_0$, there is hope that duality violations are small enough to be neglected. However, extraction of the condensates for $\Pi_{LR}$ has recently showed discrepancies between the different approaches \cite{c1}-\cite{c9}, pointing at the importance of gaining insight into these duality violations.

Unfortunately, there is no theory behind duality violations and one has to resort to models. Our aim will be therefore to study duality violations in the two-point correlator function $\Pi_{LR} (q^2)$ with a model based on large-$N_C$ QCD. We will start with a model in the strict large-$N_C$ limit to later on move to a more realistic scenario by adding widths through leading $1/N_C$ corrections. Our hope is to be able to single out the features which are model-independent and therefore can be extrapolated to the real QCD case. Our discussion essentially follows \cite{Cata:4}.    

\begin{figure}
\renewcommand{\captionfont}{\small \it}
\renewcommand{\captionlabelfont}{\small \it}
\centering
\psfrag{A}{${\textrm{Re}}\,q^2$}
\includegraphics[width=2.1in]{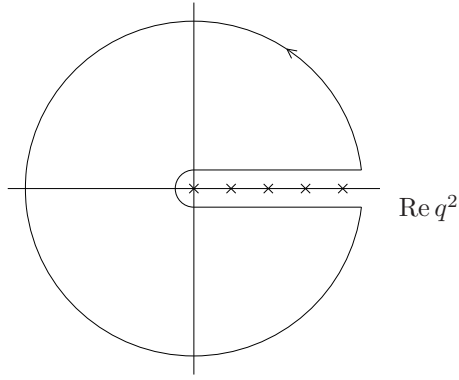}
\caption{Contour integral leading to (\ref{Cauchy}). The crosses on the real axis stand for the resonance poles in large-$N_C$ or the physical cut when resonance widths are taken into account.}\label{circle}
\end{figure}

\section{A toy model for duality violations}
We define the axial and vector vacuum polarization Green functions as
\begin{equation}\label{correlator} 
\Pi^{V,A}_{\mu\nu}(q)=\ i\,\int d^4x\,e^{iqx}\langle
J^{V,A}_\mu(x)\ (J^{V,A}_{\nu})^{\dagger}(0)\rangle
\end{equation}
with the QCD currents
\begin{equation}
J^{\mu}_V\ =\ {\bar{d}}(x)\ \gamma^{\mu}\ u(x)\, , \qquad J^{\mu}_A\ =\ {\bar{d}}(x)\ \gamma^{\mu}\gamma^5\ u(x) 
\end{equation}
We will hereafter work in the strict chiral limit. Lorentz invariance then fixes the tensorial structure of (\ref{correlator}) to be
\begin{equation}\label{Lorentz}
\Pi_{\mu\nu}^{V,A}(q) = \left(q_{\mu} q_{\nu} - g_{\mu\nu} q^2
\right)\Pi^{V,A}(q^2) 
\end{equation}
Moreover, both $\Pi_{V}$ and $\Pi_{A}$ can be shown to obey a dispersion relation (up to one subtraction) of the form
\begin{equation}\label{dispersion}
\Pi_{V,A}(q^2)= \int_0^{\infty} \frac{dt}{t-q^2-i\epsilon}\ \frac{1}{\pi} {\rm
Im}\,\Pi_{V,A}(t)
\end{equation}
We choose as our spectral functions the following {\it{ans\"atze}} \cite{s2}
\begin{eqnarray}\label{spectrum} 
\frac{1}{\pi} {\rm Im}\,\Pi_V(t)&=& 2 F_{\rho}^2
\delta(t-M_{\rho}^2) + 2 \sum_{n=0}^{\infty} F_V^2\delta(t-M^2_V(n))\nonumber\\\frac{1}{\pi} {\rm Im}\,\Pi_A(t)&=& 2 F_{0}^2 \delta(t) + 2 \sum_{n=0}^{\infty}
F_A^2\delta(t-M^2_A(n))
\end{eqnarray}
which describe axial and vectorial towers of narrow resonance states, whose decay constants $F_{V,A}\ (n)$ and masses $M_{V,A}\ (n)$ show an n-dependence to be specified. We take the $\rho$ and $\pi$ analogs out of the tower to make our model more realistic. $F_{\rho}$ and $m_{\rho}$ are therefore the electromagnetic decay constant and mass pole of the $\rho$ and $F_0$ is the pion decay constant in the chiral limit. Notice that (\ref{spectrum}) can be interpreted as an {\it{ansatz}} for the spectrum of large-$N_c$ QCD. The n-dependence of masses and couplings is then constrained and a possible choice is
\begin{equation}\label{twoinch6}
    F^2_{V,A}(n)= \mathrm{const.}\quad ,
    \quad M_{V,A}^2(n) = m_{V,A}^2 + n \ \Lambda_{V,A}^2
\end{equation}
which means that our spectrum now displays a Regge-like behaviour of equally-spaced resonance states, much the same as what happens in QCD$_2\ (N_c\rightarrow \infty)$\footnote{See the discussion in chapter 3.}. Furthermore, we choose $F_V^2=F_A^2=F^2$, which will prove to be a convenient choice when we later on want to provide the resonances with a non-zero width. Matching our model to QCD short distances (see next page) will additionally lead to $\Lambda_V=\Lambda_A=\Lambda$. We will somewhat anticipate things and use hereafter a single decay constant and a single resonance spacing for both channels. 

We consider the following two-point Green function
\begin{equation}
\Pi_{LR}\ (q^2)=\frac{1}{2}\ \bigg[\ \Pi_V(q^2)-\Pi_A(q^2)\ \bigg]
\end{equation}
whose explicit expression, obtained from (\ref{dispersion}) and (\ref{spectrum}), reads
\begin{equation}\label{one}
    \Pi_{LR}(q^2)=\frac{F^2_0}{q^2}+\frac{F_{\rho}^2}{-q^2+M_{\rho}^2}-
    \sum_{n=0}^{\infty} \frac{F^2}{-q^2+M^2_A(n)}+ \sum_{n=0}^{\infty}
    \frac{F^2}{-q^2+M^2_V(n)}
\end{equation}
The infinite sums in (\ref{one}) are by itself divergent and a regulator has to be defined to render them finite\footnote{See discussion in \cite{GP02}.}. We introduce a cutoff ${\widehat{\Lambda}}$ as our regulator. In order for chiral invariance to be preserved, the cutoff has to satisfy
\begin{equation}
{\widehat{\Lambda}}^2\ \sim N_V\ \Lambda_V^2\ \sim N_A\ \Lambda_A^2\ \qquad {\textrm{as}}\ \qquad N_{V,A}\rightarrow \infty
\end{equation} 
As it was already emphasised in \cite{GP02}, this choice yields a well-defined answer, with infinities in each channel cancelling each other. Using the identity
\begin{equation}\label{sum}
\lim_{N\rightarrow
\infty}\left\{\sum_{n=1}^{N}\frac{1}{z+n}-\sum_{n=1}^{N}\frac{1}{n}\right\}=-\psi(z)-\frac{1}{z}-
\gamma_E
\end{equation}
where $\gamma_E$ is the Euler-Mascheroni constant and $\psi(z)$ the Digamma function, {\it{i.e.}},
\begin{equation}\label{definitioninch6}
\psi(z)=\int_0^{\infty} \left(\frac{e^{-t}}{t}-\frac{e^{-zt}}{1-e^{-t}} \right)\,
dt=\frac{d}{dz}\log{\Gamma(z)}
\end{equation}
we can perform the sums in (\ref{one}) explicitly to yield
\begin{equation}\label{onecompact}
    \Pi_{LR}(q^2)=\frac{F^2_{0}}{q^2}+\frac{F_{\rho}^2}{-q^2+M_{\rho}^2}+
    \frac{F^2}{\Lambda^2}\left[\psi\left(\frac{-q^2+m_A^2}{\Lambda^2}\right)-
    \psi\left(\frac{-q^2+m_V^2}{\Lambda^2}\right)\right]
\end{equation}
In order to ensure the right short distance behaviour of our model, we match (\ref{onecompact}) to the first OPE terms of QCD. This induces the following constraints:
\begin{equation}
\fr{F^2}{\Lambda^2}=\fr{N_C}{24\pi^2}
\end{equation}
to guarantee that the parton model logarithm is recovered;
\begin{equation}
F_{\rho}^2=F^2\left(\fr{m_V^2}{\Lambda^2}-\fr{1}{2}\right)\qquad , \qquad F_{0}^2=F^2\left(\fr{m_A^2}{\Lambda^2}-\fr{1}{2}\right) 
\end{equation}
and
\begin{equation}
-2F_{\rho}^2\ M_{\rho}^2+\ F^2\Lambda^2\left( \fr{m_V^4}{\Lambda^4}-\fr{m_V^2}{\Lambda^2}+\fr{1}{6}\right)=F^2\Lambda^2\left(\fr{m_A^2}{\Lambda^4}-\fr{m_V^2}{\Lambda^2}+\fr{1}{6}\right)
\end{equation}
which are the constraints arising from the Weinberg-like sum rules, {\it{i.e.}}, the absence of $Q^{-2}$ and $Q^{-4}$ terms in the OPE.

We are now in a position to evaluate the OPE as a $1/Q^2$ expansion for our model, defined as ($Q^2=-q^2$)
\begin{equation}
\Pi_{LR}^{OPE}\ (-Q^2)\ \sim \sum_{k=1,2,3,\cdots}\ \fr{C_{2k}}{Q^{2k}}
\end{equation}
Using the derivative of (\ref{definitioninch6}) and the well-known expansion
\begin{equation}
 e^{xz}\frac{z}{e^z -1}=\sum_{n=0}^{\infty}B_{n}(x)\frac{1}{n!}z^n
\end{equation}
where $B_n(x)$ are the Bernoulli polynomials defined in Appendix C, the analytical expressions for the OPE coefficients then read,
\begin{eqnarray}\label{opecoef}
     C_{2k} &=& -F_{0}^2 \ \delta_{k,1}+\nonumber \ ,\\
     &&\!\!\!\!\!\!\!\!\!\!\!\!
     (-1)^{k+1} \left[ F_{\rho}^2M_{\rho}^{2k-2}  -\frac{1}{k}F^2 \Lambda^{2k-2}
  \left\{B_{k}\left(\frac{m_V^2}{\Lambda^2}\right) -
  B_{k}\left(\frac{m_A^2}{\Lambda^2}\right)\right\}\right]
\end{eqnarray}
the first ones of which are, explicitly 
\begin{eqnarray}\label{examples}
 C_2 &=& + F_{\rho}^2 - \,\, F_{0}^2 -\,\,
  F^2\left\{B_1\left(\frac{m_V^2}{\Lambda^2}\right) -
  B_1\left(\frac{m_A^2}{\Lambda^2}\right)\right\}\nonumber \\
  C_4 &=&  - F_{\rho}^2 M_{\rho}^2+
  \frac{1}{2}F^2 \Lambda^2\left\{B_2\left(\frac{m_V^2}{\Lambda^2}\right) -
  B_2\left(\frac{m_A^2}{\Lambda^2}\right)\right\}\nonumber \\
  C_6 &=&  + F_{\rho}^2 M_{\rho}^4-
  \frac{1}{3}F^2 \Lambda^4\left\{B_3\left(\frac{m_V^2}{\Lambda^2}\right) -
  B_3\left(\frac{m_A^2}{\Lambda^2}\right)\right\} \nonumber \\
  C_8 &=&  - F_{\rho}^2 M_{\rho}^6+
  \frac{1}{4}F^2 \Lambda^6\left\{B_4\left(\frac{m_V^2}{\Lambda^2}\right) -
  B_4\left(\frac{m_A^2}{\Lambda^2}\right)\right\} 
  \end{eqnarray}
A na\"ive $1/Q^2$ expansion of the spectral function in the dispersion relation
\begin{equation}\label{spectral}
\Pi_{LR}\ (-Q^2)\ =\ \int_0^{\infty}\ \fr{dt}{t+Q^2-i\epsilon}\ \rho(t)
\end{equation}
where $\rho(t)$ is a common short-hand for
\begin{equation}
\rho(t)\ =\ \fr{1}{\pi}\ {\textrm{Im}}\, \Pi_{LR}(t)
\end{equation}
would yield to the so-called spectral moments
\begin{equation}\label{moments}
 M_{n}(s_0)=\int_{0}^{s_0}\ dt \ t^{n} \rho(t)
\end{equation}
which are {\it{not}} the coefficients of the OPE. To establish the actual expression for the OPE coefficients in terms of the spectral function, consider the Laplace transform of (\ref{spectral}),
\begin{equation}\label{sixinch6}
    \Pi(-Q^2)= \int_{0}^{\infty} d\tau\ e^{- \tau Q^2}\ \widehat{\rho}(\tau)\quad
    \mathrm{with}\quad
\widehat{\rho}(\tau)= \int_{0}^{\infty} dt\ e^{-t\tau}\ \rho(t)
\end{equation}
Plugging in the OPE expansion one gets
\begin{equation}\label{eight}
    \sum_{k=1}^{\infty} \frac{C_{2k}}{(k-1)!}\ \tau^{k-1}=
    \int_{0}^{\infty} dt\ e^{-t\tau} \rho(t)
\end{equation}
and finally,
\begin{equation}\label{nine}
    C_{2k}= \lim_{\tau\rightarrow 0}\left\{(-1)^{k-1}\int_{0}^{\infty} dt\ t^{k-1}\ e^{-t\tau}
    \rho(t)\right\}
\end{equation}
showing that coincidence between spectral moments and OPE coefficients is accomplished if and only if the spectral function fall-off is of an exponential type. This is neither the case in QCD nor in our model. Actual computation yields
\begin{eqnarray}\label{eleven}
  M_0(s_0) &=& C_2-F^2 \Big[B_1(x_V)-B_1(x_A)\Big]\nonumber \\
  M_1(s_0) &=&- C_4 -F^2 \Big[B_1(x_V)-B_1(x_A)\Big]s_0 + \frac{1}{2}F^2\Lambda^2
  \Big[B_2(x_V)-B_2(x_A)\Big]\nonumber \\
  M_2(s_0) &=& C_6 - F^2 \Big[B_1(x_V)-B_1(x_A)\Big]s_0^2+
  F^2\Lambda^2\Big[B_2(x_V)-B_2(x_A)\Big] s_0 \nonumber\\
  &&- \frac{1}{3} F^2 \Lambda^4 \Big[B_3(x_V)-B_3(x_A)\Big] \nonumber \\
  M_3(s_0) &=& -C_8 -F^2  \Big[B_1(x_V)-B_1(x_A)\Big] s_0^3+
  \frac{3}{2} F^2\Lambda^2  \Big[B_2(x_V)-B_2(x_A)\Big] s_0^2 \nonumber\\
&&- F^2\Lambda^4 \Big[B_3(x_V)-B_3(x_A)\Big] s_0 +\frac{1}{4}F^2 \Lambda^6
\Big[B_4(x_V)-B_4(x_A)\Big]
\end{eqnarray}
where $0 \leq x_{V,A} \leq 1$ is the fractional part of $(s_0-m_{V,A}^2)/\Lambda^2$. Spectral moments show therefore an oscillatory behaviour around the OPE coefficient. This oscillatory behaviour is provided by the Bernoulli polynomial terms in (\ref{eleven}), in the form of a multistep function, which pushes upwards at vectorial mass thresholds ($s_0=m_V^2+n\Lambda^2$) and downwards at axial mass thresholds  ($s_0=m_A^2+n\Lambda^2$). The expressions above are valid for $s_0\geq m_V^2$, {\it{i.e.}}, above the threshold for axial and vectorial towers. The result is depicted in figure (6.2), with our set of free parameters chosen to be
\begin{eqnarray}\label{nature}
\hspace{-2. cm} F_{0}= 85.8 \,\,{\mathrm{MeV}}\ , \quad F_{\rho}= 133.884
\,\,{\mathrm{MeV}}\ ,
\quad F= 143.758 \,\,{\mathrm{MeV}}\, ,\qquad \qquad \nonumber\\
M_{\rho}= 0.767 \,\,{\mathrm{GeV}}, \quad m_A= 1.182 \,\,{\mathrm{GeV}}, \quad m_V=
1.49371 \,\,{\mathrm{GeV}}\, , \quad \Lambda= 1.2774 \,\, {\mathrm{GeV}}
\end{eqnarray}
\begin{figure}
\renewcommand{\captionfont}{\small \it}
\renewcommand{\captionlabelfont}{\small \it}
\centering
\includegraphics[width=2.5in]{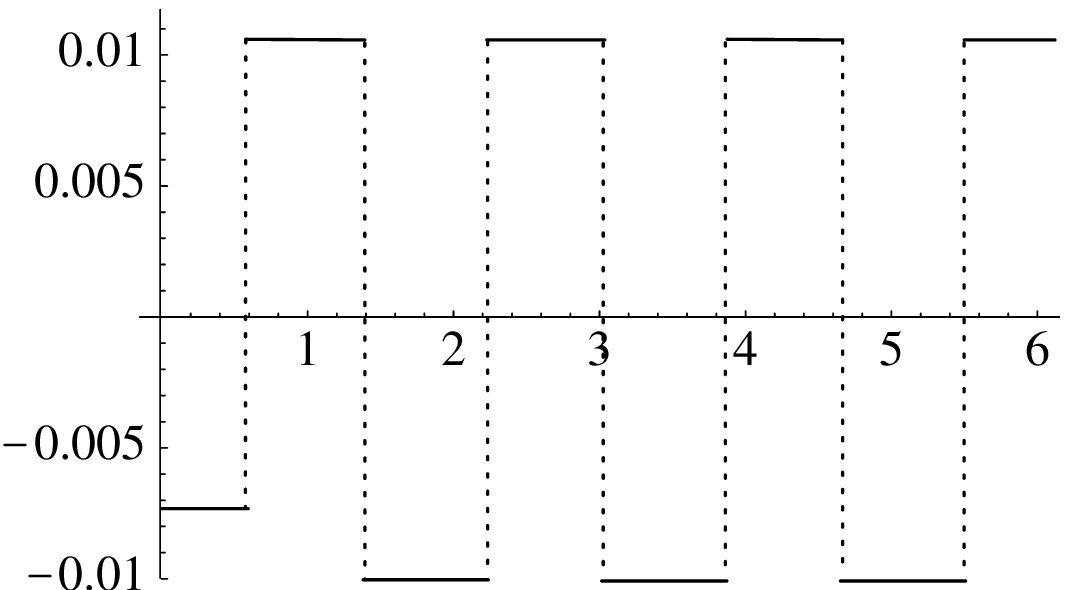}
\hspace {1cm}
\includegraphics[width=2.5in]{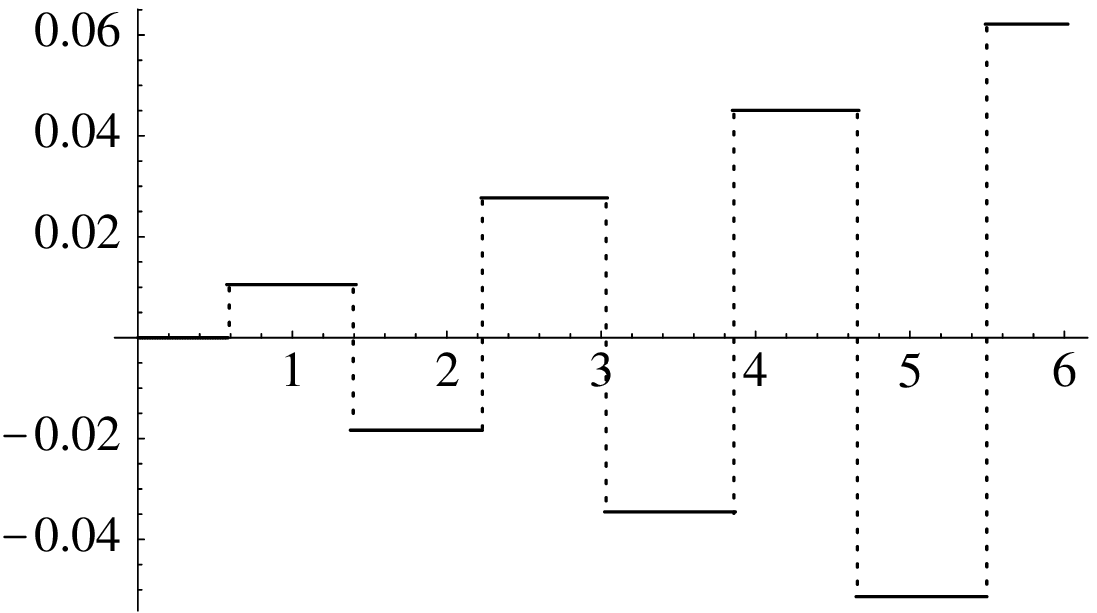}\\
\includegraphics[width=2.5in]{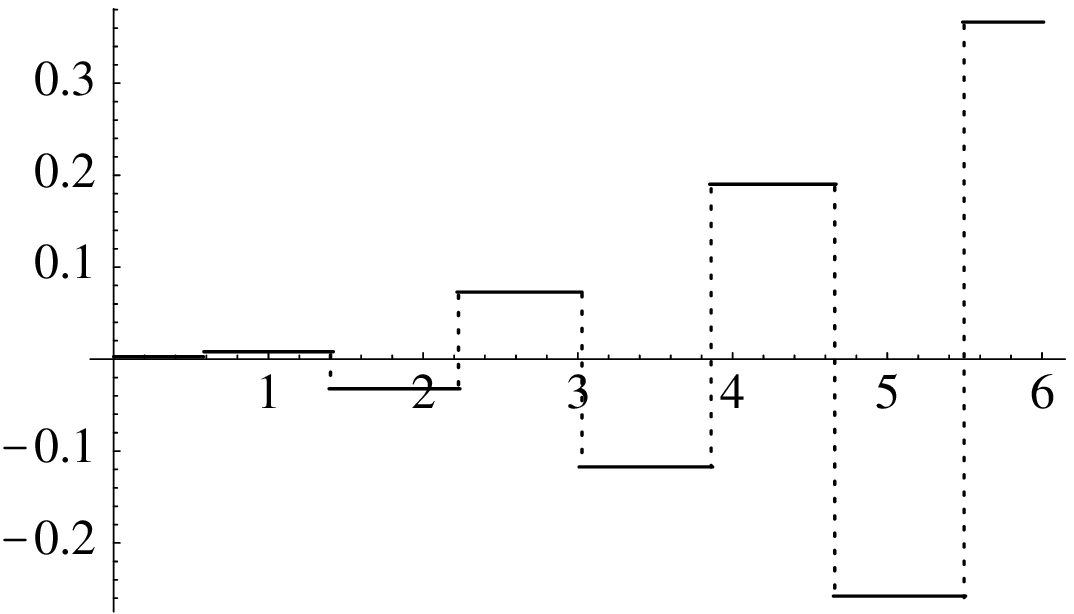}
\hspace {1cm}
\includegraphics[width=2.5in]{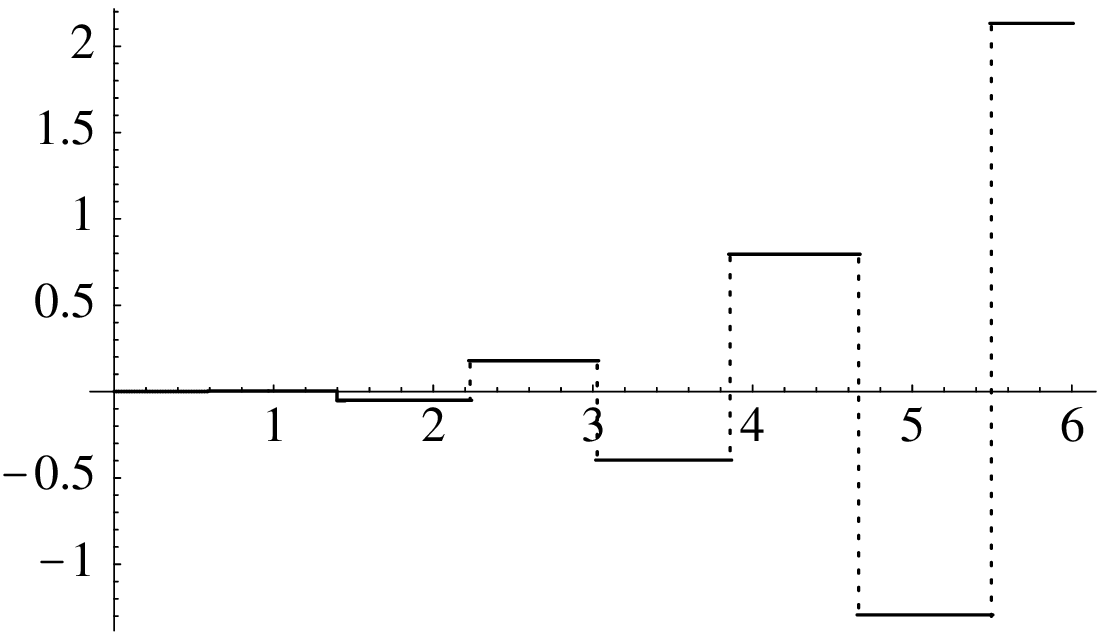}
\caption{From left to right: $M_0$, $M_1$ (first row), $M_2$ and $M_3$ (second row)
as defined in the main text.}\label{momentsNC}
\end{figure}
As it can be readily seen, duality points $s_0^*$, which would correspond to cuts in the $s_0$ axis of figure (6.2), are absent\footnote{Strictly speaking, duality points happen when $M_0(s_0^*)=C_2$, $M_1(s_0^*)=C_4$ and so on. However, it happens that the OPE coefficients are much smaller than the typical size of the oscillations in figure (6.2), so that the duality points actually lie very close to the $s_0$ axis.}. Step-like discontinuities happen at mass pole thresholds for all the moments, but without any predictability: the OPE value is bounded by the maximum and minimum of the oscillations, but otherwise completely undetermined. Moreover, as the moment increases the oscillation gets amplified, because moments scale as $M_n\ \sim s_0^n$. Even if no prediction is available, the qualitative information we get from this zero-width model hints at some general features to bear in mind. First, it provides a rationale for FESR, in the sense that duality points are common to all moments and exactly at mass poles. We will later on see that with the introduction of widths, this picture gets blurred: slopes become milder and therefore predictions are possible, but duality points get dissaligned. Second, the amplification of oscillations seems to select low duality points as the best candidates to make predictions for the OPE values. Again, the passage to adding widths will show that this picture gets distorted.

The previous analysis concentrated on the comparison of moments and OPE coefficients. We already mentioned that duality violations are much bigger than the actual OPE values. A sound strategy would then be to reduce duality violations to a minimum. An approach in this direction is based on a slightly more sophisticated use of the Cauchy theorem 
\begin{equation}
\int_0^{s_0}\, dt\, w(t)\,\fr{1}{\pi}\, Im \Pi_{LR}(t)=-\fr{1}{2\pi i} \oint_{|q^2|=s_0}\,dq^2 w(q^2)\,\Pi_{LR}(q^2)
\end{equation}
where $w(t)$ is a polynomial chosen so as to suppress the duality violating oscillations where they are more harmful, {\it{i.e.}}, on the Euclidean real axis. Such a strategy is usually coined pinched-weight sum rules (pFESR) \cite{c8,Cirig:2}. Due to the recursiveness we observe in the oscillations in (\ref{eleven}), appropriate combinations of moments would be, by inspection,
\begin{eqnarray}\label{cancel}
M_0-\fr{1}{s_0}M_1&\sim& F^2\,\Lambda^2\,[\, B_2\, (x_V)-B_2\,(x_A)\, ]\nonumber\\
M_0-\fr{2}{s_0}M_1+\fr{1}{s_0^2}M_2&\sim& F^2\,\Lambda^4\,[\,B_3\, (x_V)- B_3\, (x_A)\, ]\nonumber\\
M_0-\fr{3}{s_0}M_1+\fr{3}{s_0^2}M_2-\fr{1}{s_0^3}M_3&\sim& F^2\,\Lambda^6\,[\, B_4\, (x_V) - B_4\, (x_A)\, ]
\end{eqnarray}
which would correspond to pinched-weights of the form
\begin{eqnarray}
w(t)&=&\left( 1-\fr{t}{s_0}\right)\nonumber\\
w(t)&=&\left( 1-\fr{t}{s_0}\right)^2\nonumber\\
w(t)&=&\left( 1-\fr{t}{s_0}\right)^3
\end{eqnarray}
As one can see, such combinations indeed kill the dominant terms of the duality oscillations, leaving a remnants shown in (\ref{cancel}). Quite generally, a pinched-weight of the form
\begin{equation}
w(t)\ =\ \left( 1-\fr{t}{s_0}\right)^n 
\end{equation} 
would involve the first $M_{n+1}$ moments, killing duality violations up to ${\mathcal{O}}(F^2\, \Lambda^n\, B_{n+1})$. Other pinched weights proposed in the literature are \cite{Cirig:2}
\begin{eqnarray}
w_1(t)&=&\left( 1-3\fr{t}{s_0}\right)\ \left(1-\fr{t}{s_0}\right)^2\nonumber\\
w_2(t)&=&\fr{t}{s_0}\ \left(1-\fr{t}{s_0}\right)^2
\end{eqnarray}
which are not optimal in our model, as figure (6.3) shows, but still will be useful for later numerical comparison.  The usual procedure in the pFESR approach is to make a fit over an $s_0$ window between the pinched-weight, which is a linear combination of spectral moments, and the same combination of OPE coefficients to determine them all. The latter combination is obviously unable to reproduce any oscillatory behaviour (see figure (6.3)). Therefore, even though the oscillations have been atenuated, one expects the fit to be unreliable and very sensitive to the window chosen. However, notice that in figure (6.3) the OPE curve seems to interpolate between the pinched-weight oscillation. This is a consequence of the fact that $B_3(x)$ and $B_4(x)$ display a sinusoidal-like behaviour. This suggests the possibility of extracting the OPE coefficients by fitting over a {\it{full}} oscillation. Later on, with the introduction of widths, we will check whether this strategy is still effective.
\begin{figure}
\renewcommand{\captionfont}{\small \it}
\renewcommand{\captionlabelfont}{\small \it}
\centering
\includegraphics[width=2.5in]{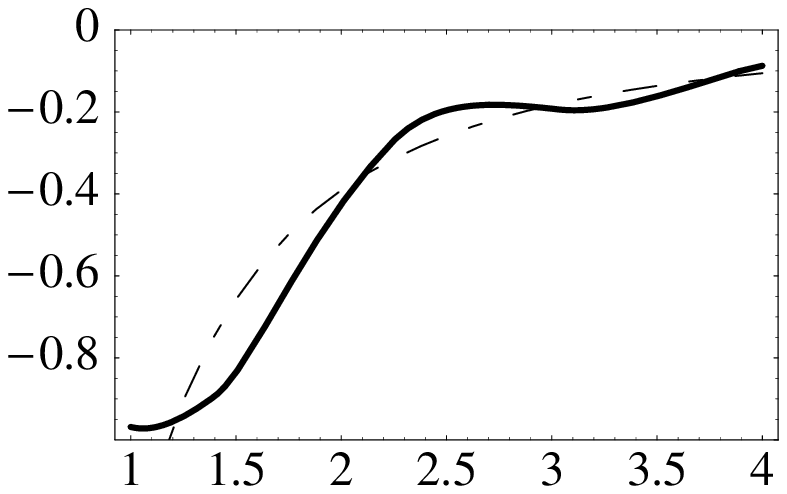}
\includegraphics[width=2.5in]{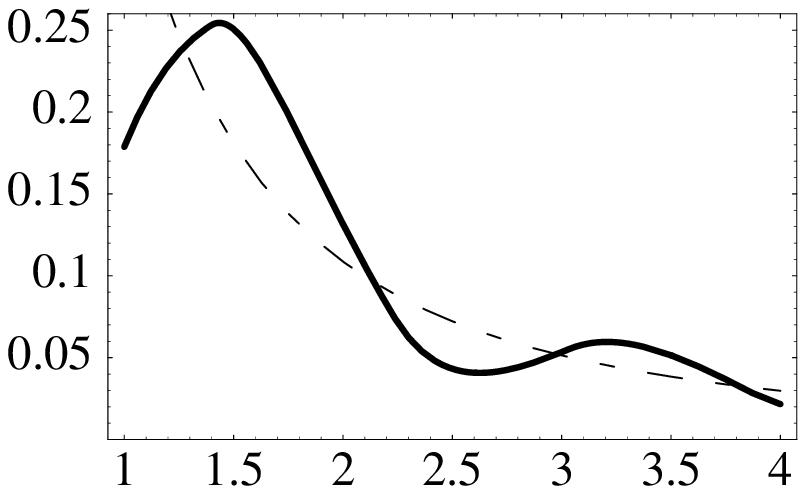}
\includegraphics[width=2.5in]{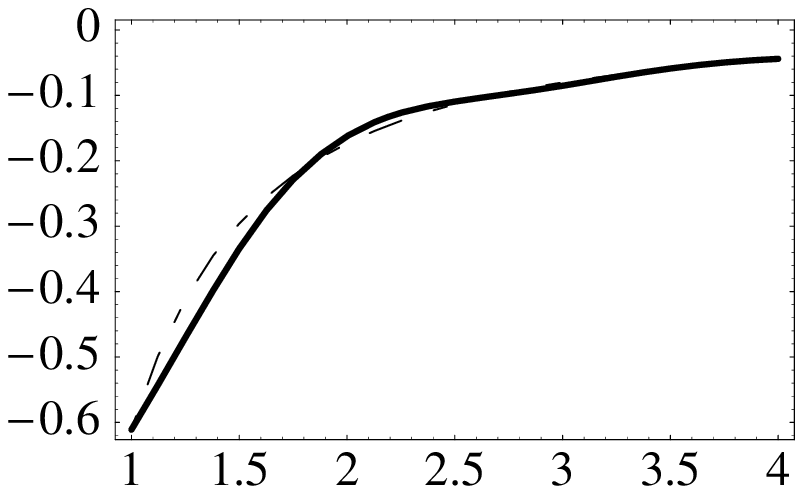}
\caption{Integrals of the pinched weights, $w_{1,2,3}(t)$ defined in
the text (solid lines from left to right and top to bottom)  and the corresponding
OPE prediction (dashed lines). As it can be seen, $w_3(t)$ shows the smallest
oscillations, i.e., the smallest duality violations.}\label{pwzero}
\end{figure}

So far we have discussed two of the most employed methods in the literature. However, having a model of $\Pi_{LR}\ (q^2)$ we can determine the exact form of duality violations in our model. Recalling (\ref{CauchyOPE})
\begin{equation}\label{ddelta}
    \int_{0}^{s_0}dt\ t^{n}\ \rho(t)\ =
    -\frac{1}{2\pi i}\oint_{|q^2|=s_0}dq^2 \ q^{2n}\
    \Pi_{LR}^{OPE}(q^2)+ \mathcal{D}^{[n]}(s_0)
\end{equation}
where $\mathcal{D}^{[n]}(s_0)$ measures the duality violations. Remember that the OPE expansion is not convergent, at least on the Minkowski axis. Determination of the analytical form for the duality violations will be done with the help of the reflection property of the Digamma function
\begin{equation}\label{reflexion}
    \psi(z)=\psi(-z)-\pi\ \cot(\pi z)-\frac{1}{z}
\end{equation} 
\begin{figure}
\renewcommand{\captionfont}{\small \it}
\renewcommand{\captionlabelfont}{\small \it}
\centering
\psfrag{A}{$q^2$}
\psfrag{B}{$+i\,s_0$}
\psfrag{C}{$-i\,s_0$}
\psfrag{D}{$-i\,\epsilon$}
\psfrag{F}{$+i\,\epsilon$}
\includegraphics[width=2.1in]{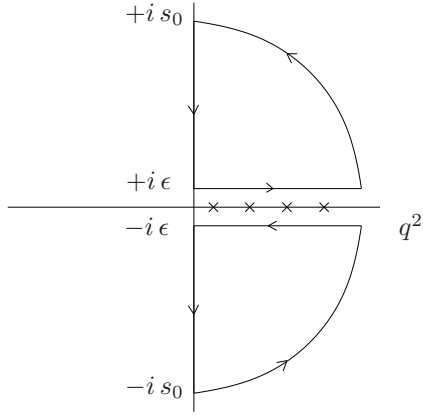}
\caption{Contour over the complex $q^2$ plane used to split the duality violating term in the two pieces of (\ref{oscill}) and (\ref{const}).}\label{dualdual}
\end{figure}
On the negative half plane of figure (6.1) we can use the asymptotic expansion of the Digamma function at large $|z|$, to wit
\begin{equation}\label{asymp}
\psi(z)\sim \log{z}-\frac{1}{2z}-\sum_{n=1}^{\infty}\frac{B_{2n}}{2n\ z^{2n}}\quad
,\qquad |\arg(z)|<\pi 
\end{equation}
On the positive axis we can use (\ref{reflexion}) such that we find an expression for $\Pi_{LR}\ (q^2)$ in terms of the OPE, to wit
\begin{equation}\label{DVinf}
\Pi_{LR}(q^2)\approx \Bigg\{
\begin{array}{ll}
    \!\!\!\Pi_{LR}^{OPE}(q^2) +\mathcal{O}
    \left( e^{-2 \pi|q^2|/\Lambda^2}\right)& ,
    \  \hbox{$Re(q^2)\leq0$} \\
    \!\!\!\Pi_{LR}^{OPE}(q^2)+\Delta_\infty(q^2)+
    \mathcal{O}\left( e^{-2 \pi|q^2|/\Lambda^2}\right)&,\  \hbox{$Re(q^2)\geq0$} \\
\end{array}
\end{equation}
where the exponential terms appear as a consequence of the OPE being an asymptotic expansion. The duality violating function $\Delta_\infty(q^2)$ appears only in the Minkowski region and takes the form
\begin{equation}\label{delta}
    \Delta_\infty(q^2)=\frac{\pi F^2}{\Lambda^2}\ \Bigg\{\cot \left[\pi\frac{-q^2+m_V^2}{\Lambda^2}
    \right]- \cot \left[\pi\frac{-q^2+m_A^2}{\Lambda^2}\right]\!\Bigg\}
\end{equation}
This contribution does not vanish nor die out as $q^2$ increases, as it can be readily seen in figure (6.2).

In view of (\ref{DVinf}) and (\ref{ddelta}), the family of functions $\mathcal{D}^{[n]}(s_0)$ can be expressed as a counter-clockwise integral over the semicircle $|q^2|=s_0\ ({\textrm{Re}}\ q^2\geq\ 0)$ as follows
\begin{equation}\label{dualviol}
    \mathcal{D}^{[n]}(s_0)=-\frac{1}{2\pi i}\
    \int_{_{_{_{\!\!\!\!\!\!\!\!\!\!\!\!\!\!\!\!\!\!\small{\begin{array}{c}
      |q^2|=s_0 \\ Re(q^2)\geq 0
    \end{array}}}}}} \!\!\!\!\!\!\!\!\!\!\!\!\!\!\! dq^2\ q^{2n}\
    \Delta_\infty(q^2)
    + \mathcal{O}\left( e^{-2\pi s_0/\Lambda^2}\right)
\end{equation}
It is specially convenient to split the previous integral into a spectral integral and a counterterm, as shown pictorically in figure (6.4). Applying Cauchy's residue theorem, one sees that the duality violation can be expressed as two integrals, one over ${\textrm{Re}}\ (q^2)$ and the other over ${\textrm{Im}}\ (q^2)$, 
\begin{equation}\label{dualspectral}
    \mathcal{D}^{[n]}(s_0)=\mathcal{D}^{[n]}_{oscill.}(s_0)+\mathcal{D}^{[n]}_{const.}(s_0)
\end{equation}
with
\begin{eqnarray}
\mathcal{D}^{[n]}_{oscill.}(s_0)&=&\int_0^{s_0}  dt\ t^n
\ \frac{1}{\pi}\ \mathrm{Im} \Delta_\infty(t+i\varepsilon)\label{oscill}\\
   \mathcal{D}^{[n]}_{const.}(s_0)&=& - \frac{1}{2\pi i}\left\{\int_{-is_0}^{-i\epsilon}
   +\int_{i\epsilon}^{is_0}\right\}
   dq^2\ q^{2n}\ \Delta_\infty(q^2)\label{const}
\end{eqnarray}
where in (\ref{oscill}) use has been made of Schwartz reflection theorem. $\mathcal{D}^{[n]}_{oscill.}(s_0)$ is the responsible for the oscillations appearing in the spectral function $\rho(t)$ which the OPE is unable to capture. Mathematically, the singularities of the Digamma function are absorbed into the cotangent of (\ref{reflexion}), which has the form
\begin{equation}\label{cot}
    \pi \cot\left(\pi z\right)=\frac{1}{z}+ 2z\sum_{n=1}^{\infty}\frac{1}{z^2-n^2}
\end{equation}
On the contrary, $\mathcal{D}^{[n]}_{const.}(s_0)$ gets contributions mainly from the low $q^2$ regime, decaying exponentially to a constant
\begin{equation}\label{counterterms}
    \mathcal{D}^{[n]}_{const.}(s_0)\ = \ \mathcal{C}^{[n]} +
    \ \mathcal{O}\left(e^{-2 \pi s_0/\Lambda^2}\right)
\end{equation}
whose explicit expression is
\begin{equation}\label{constants}
\mathcal{C}^{[n]}=\frac{F^2\Lambda^{2n}}{n+1}\left\{B_{n+1}\left(\frac{m_V^2}{\Lambda^2}\right)-
  B_{n+1}\left(\frac{m_A^2}{\Lambda^2}\right)\right\}
  -  F^2\left(m_V^2-\Lambda^2\right)^n
\end{equation}
As it should, plugging in the results of (\ref{dualspectral})-(\ref{constants}) in (\ref{ddelta}), one recovers the results for the moments $M_i$ of (\ref{eleven}).

\section{$1/N_C$ corrections}
In this section we will move to a more realistic model departing from the zero-width model and allowing the resonances to have some width. In the light of large-$N_C$ QCD, this would correspond to including the $1/N_C$ corrections in a consistent way. Lacking a solution as to how to implement this corrections, we have to incorporate them in an {\it{ad hoc}} way. Obviously, this has to be done in a somehow sophisticated way if we want $\Pi^{LR}\ (q^2)$ to preserve its analytic properties. For instance, the cut in the complex $q^2$ plane for ${\textrm{Re}}\ q^2$ is not achieved through a na\"ive Breit-Wigner function. A convenient choice was found in \cite{Shifman:00,Blok}, which, translated to our model, would yield
\begin{equation}\label{fifteen}
    \Pi_{LR}(q^2)=\left(1- \frac{a}{\pi N_c}\right)^{-1}
    \left\{- \frac{F^2_{0}}{z}+\frac{F_{\rho}^2}{z+M_{\rho}^2}- \sum_{n=0}^{\infty}
    \frac{F^2}{z+M_A^2(n)}+\sum_{n=0}^{\infty} \frac{F^2}{z+M_V^2(n)}\right\}
\end{equation}
where the variable $z$ reads
\begin{equation}\label{sixteen}
    z=\Lambda^2 \left(\frac{-q^2-i\epsilon}{\Lambda^2}\right)^{\zeta}\quad , \quad
    \zeta=1-\frac{a}{\pi N_c}
\end{equation}
As in the previous section, the sums can be performed and expressed in terms of Digamma functions,
\begin{equation}\label{fifteencompact}
\Pi_{LR}(q^2)=\frac{1}{\zeta}
    \left\{-\frac{F^2_{0}}{z}+\frac{F_{\rho}^2}{z+M_{\rho}^2}+
    \frac{F^2}{\Lambda^2}\left[\psi\left(\frac{z+m_A^2}{\Lambda^2}\right)-
    \psi\left(\frac{z+m_V^2}{\Lambda^2}\right)\right]\right\}
\end{equation}
Notice that this expression provides a smooth transition to the zero-width case through letting the parameter ${\it{a}}$ go to zero (or equivalently, $\zeta$ to 1). This parameter ${\it{a}}$ is proportional to the width of {\it{all}} the resonances. Indeed, resonance propagators aquire an imaginary part of the form
\begin{eqnarray}\label{seventeen}
    \Gamma_{V,A}(n)&=&\frac{a}{N_c} M_{V,A}(n)\ \left(1+ \mathcal{O}\left(\frac{a}{N_c}\right)\right)\nonumber\\
\Gamma_{\rho}(n)&=&\frac{a}{N_c} M_{\rho}(n)\ \left(1+ \mathcal{O}\left(\frac{a}{N_c}\right)\right)
\end{eqnarray}
meaning that resonance widths increase with the excitation number $n$. Therefore, the width of every resonance is modulated by the parameter ${\it{a}}$ and by the way that masses grow with the excitation number (cf. (\ref{twoinch6})). The resulting behaviour is precisely the one found in the {\it{'t Hooft}} model. Comparison with experimental data suggests a value for $a$ 
\begin{equation}\label{aa}
    a=0.72
\end{equation}
while the other free parameters are the ones inherited from the zero-width scenario and are given in (\ref{nature}). The shape of the spectral function of $\Pi_{LR}\ (q^2)$  as compared with data is shown in figure (6.5).
\begin{figure}
\renewcommand{\captionfont}{\small \it}
\renewcommand{\captionlabelfont}{\small \it}
\centering
\includegraphics[width=2.5in]{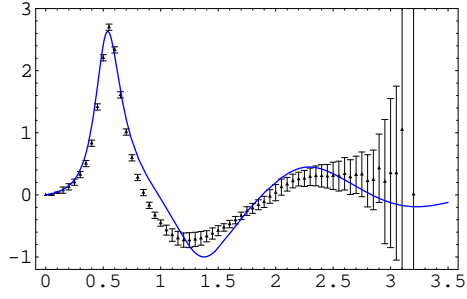}
\caption{Comparison of the data from ALEPH \cite{aleph} and OPAL \cite{opal} (overlaid on a
single curve) with the model, $\fr{1}{\pi}\,\mathrm{Im}\,\Pi(q^2)$. As with the experimental data, the kinematical factors for tau decays are included and the pion
contribution has been subtracted.}\label{fig3}
\end{figure}

 The analytical structure of $\Pi_{LR}(q^2)$ has not changed except for the replacement $q^2\ \longrightarrow\ z$. This means that $\Pi_{LR}(q^2)$ has an expression very close to the one given in (\ref{DVinf}), namely
\begin{equation}\label{DV}
\Pi_{LR}(q^2)\approx \Bigg\{
\begin{array}{ll}
    \!\!\!\Pi_{LR}^{OPE}(z) +\mathcal{O}\left(e^{-2\pi|q^2/\Lambda^2|^{\zeta}}\right)& ,
    \  \hbox{$Re(q^2)\leq0$} \\
    \!\!\!\Pi_{LR}^{OPE}(z)+\Delta(q^2)+\mathcal{O}\left(e^{-2\pi|q^2/\Lambda^2|^{\zeta}}\right)&,\  \hbox{$Re(q^2)\geq0$} \\
\end{array}
\end{equation}
where the duality violating term $\Delta (q^2)$ now reads
\begin{equation}\label{deltaa}
    \Delta(q^2)=\frac{\pi F^2}{\Lambda^2}\ \frac{1}{\zeta}\ \Bigg\{\cot
    \left[\pi\left(\frac{-q^2}{\Lambda^2}\right)^{\zeta}+\pi \frac{m_V^2}{\Lambda^2}
    \right]- \cot \left[\pi\left(\frac{-q^2}{\Lambda^2}\right)^{\zeta}+
    \pi \frac{m_A^2}{\Lambda^2}\right]\!\Bigg\}
\end{equation}
with $\zeta$ as given in (\ref{sixteen}). The OPE has an akin expansion to the one in (\ref{opecoef}), but the variable $z$, once expanded in terms of $q^2$, gives rise to logarithmic corrections in the OPE coefficients, to wit
\begin{equation}\label{eighteen}
    C_{2k}(Q^2)^{N_c=3} = C_{2k}^{N_c=\infty}
    \left(1+\frac{a}{\pi N_c}+ \frac{ka}{\pi N_c}\log\frac{Q^2}{\Lambda^2}
    +\mathcal{O}\left(\frac{a^2}{N_c^2}\right)\right)
\end{equation}
where $C_{2k}^{N_c=\infty}$ are the zero-width OPE coefficients as given in (\ref{opecoef}). These logarithmic corrections mimic those found in real QCD. Therefore, it will prove convenient to split the OPE coefficients as follows
\begin{equation}\label{ab}
     C_{2k}(Q^2)^{N_c=3}= a_{2k}+b_{2k}\
     \log\frac{Q^2}{\Lambda^2}+ \mathcal{O}\left(\frac{a^2}{N_c^2}\right)
\end{equation}
We must note, however, that $b_{2k}$ coefficients, which yield the anomalous dimensions in QCD, do not arise there as a $1/N_C$ effect, unlike in our model. With this caveat in mind, we can plug in our set of parameters (\ref{nature},\ref{aa}) to obtain the numerical values for the OPE coefficients
\begin{eqnarray}\label{opewidths}
a_2=b_2=&0&=a_4=b_4\nonumber\\
a_6= -2.8 \times 10^{-3} \mathrm{GeV}^6\quad &,&
\quad b_6=-5.9 \times 10^{-4} \mathrm{GeV}^6\nonumber\\
a_8= +1.8 \times 10^{-3} \mathrm{GeV}^8 \quad &,&\quad b_8= +5.1 \times
10^{-4}\mathrm{GeV}^8
\end{eqnarray}
\begin{figure}
\renewcommand{\captionfont}{\small \it}
\renewcommand{\captionlabelfont}{\small \it}
\centering
\includegraphics[width=2.5in]{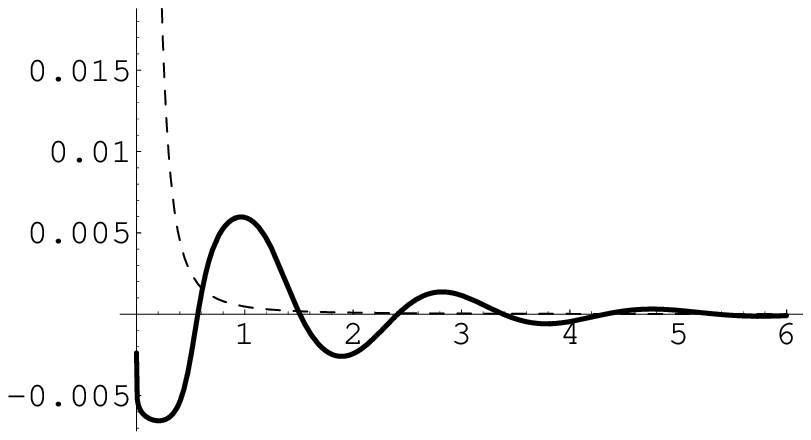}
\hspace {1cm}
\includegraphics[width=2.5in]{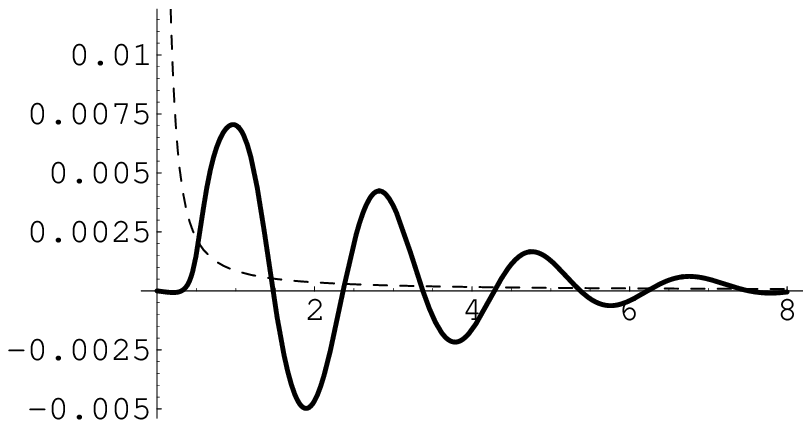}\\
\includegraphics[width=2.5in]{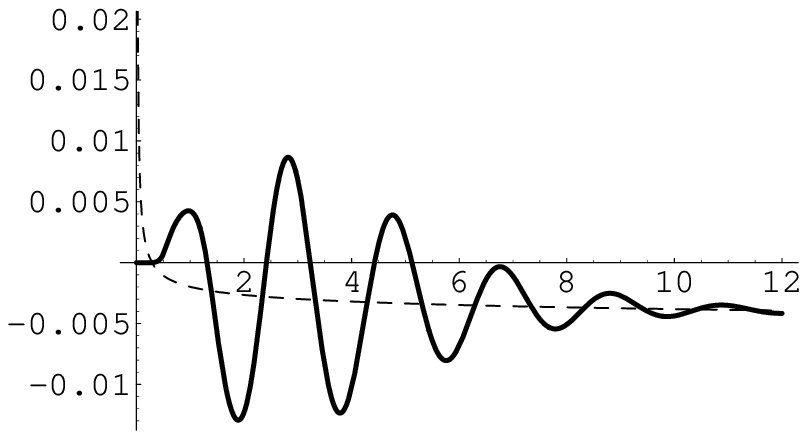}
\hspace {1cm}
\includegraphics[width=2.5in]{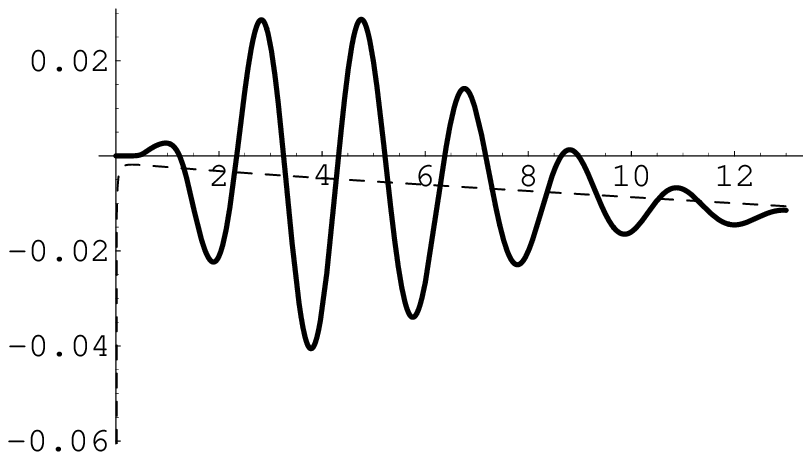}
\caption{From left to right: $M_0$, $M_1$ (solid curves, first row), $M_2$ and $M_3$ (solid curves, second row) as a function of $s_0$ together with the OPE prediction (dashed curves) as it stands in the square brackets of (\ref{works0}) to (\ref{works3}).}\label{widthaway}
\end{figure}
Contrary to what happens with the OPE coefficients, {\it{i.e.}}, in the Euclidean, where we have seen that $1/N_C$ corrections are rather mild, the spectral function in the Minkowski changes dramatically with the inclusion of widths. This follows from the fact that the $N_C\rightarrow \infty$ and $q^2\rightarrow \infty$ limits do not commute, as already pointed out in \cite{Shifman:00}. Indeed, in the Minkowski region, ($q^2=t\geq 0$), the variable $z$ takes the form
\begin{equation}\label{imaginary}
z = -t\left(1-\frac{a}{\pi
N_c}\log{\left(\frac{t}{\Lambda^2}\right)}+i\frac{a}{N_c}+
\mathcal{O}\left(\frac{a^2}{N_c^2}\right)\right)
\end{equation}
Taking the imaginary part of $\Pi(q^2)$ in the Minkowski, one finds
\begin{equation}\label{oscillation}
    \mathrm{Im}\Pi(t)= \ \mathrm{Im}\Pi_{OPE}(t)\  +\
    \mathrm{Im}\Delta(t)
\end{equation}
where the term coming from the OPE is precisely the analytical continuation to the Minkowski of the OPE in the Euclidean,  
\begin{equation}\label{imope}
    \mathrm{Im}\Pi_{OPE}(t)= \frac{3 a }{N_C}\ \frac{b_6}{t^3}
    \left(1+ \mathcal{O}\left(\frac{a}{N_c}\right)\right)+
    \mathcal{O}\left(\frac{1}{t^4}\right)
\end{equation}
Thus, the absence of $t^{-1}$ and $t^{-2}$ terms above is closely related to the absence of dimension-two and dimension-four operators in the OPE, a condition we have enforced with the two Weinberg sum rules. The last term in (\ref{oscillation}) reads
\begin{eqnarray}\label{imdv}
&&\mathrm{Im}\Delta(t)=\nonumber\\
&&\!\!\!\!\!\!\!\!\!\!\frac{4\pi F^2}{\Lambda^2}\ \  e^{-\frac{2\pi a}{N_c}\frac{t}{
\Lambda^2 }}\ \ \sin\left(\pi \frac{2t-m_A^2-m_V^2}{\Lambda^2}\right) \
\sin\left(\pi \frac{m_V^2-m_A^2}{\Lambda^2}\right)\left(1+\mathcal{O}\left(\frac{a}{N_c}\right)
\right) +\nonumber
\\&& +\  \mathcal{O}\left(e^{-\frac{4\pi a}{N_c}\frac{t}{ \Lambda^2 }}\right)
\end{eqnarray}
and stems from the duality violating piece $\Delta (q^2)$. It is illustrative at this point to see how the picture has changed for the moments as one has gone from the $N_C\rightarrow \infty$ case of figure (6.2) to the finite $N_C$ case depicted in figure (6.6). Expression (\ref{imdv}) illustrates our previous statement of non-commutativity of the $N_C\rightarrow \infty$ and $t\rightarrow \infty$ limits. Indeed, taking $N_C\rightarrow \infty$ first leads to the delta function spectrum of the large-$N_C$ limit, whereas reverting the order yields an exponential damping which eventually dies out (see figure (6.6)).

This dramatic change in the imaginary part of $\Delta (q^2)$ also translates into its real part. Actually, the full function $\Delta (q^2)$ behaves as follows
\begin{equation}\label{falloff}
    \Delta(q^2)\ \sim_{_{_{\!\!\!\!\!\!\!\!\!\! \!\!\!|q^2|\ \mathrm{large}}}}
    e^{-2\pi\left(\frac{|q^2|}{\Lambda^2}\right)^{\zeta}\  \big|\sin\big\{\zeta
    (\varphi-\pi)\big\}\big|}\ ,\quad q^2=|q^2|\ e^{i\varphi}\ ,\quad
\left\{%
\begin{array}{ll}
     0\leq \varphi\leq
    \frac{\pi}{2}\\
    \frac{3 \pi}{2}\leq \varphi\leq 2\pi \\
\end{array}%
\right.
\end{equation}
Recall that the $N_C\rightarrow \infty$ limit ({\it{i.e.}}, $\zeta\rightarrow 1$) in the previous equation kills the exponential fall-off for $\phi=0,\ 2\pi$, but it is exponentially suppressed otherwise. This is in accordance with what we observed previously in figure (6.2): on the Minkowski axis oscillations are undamped.
Therefore, in the finite $N_C$ case, contrary to what happens in the $N_C\rightarrow \infty$ case, the duality function $ \mathcal{D}^{[n]}(s_0)$ displays an exponential suppression everywhere
\begin{equation}\label{falloff2}
    \mathcal{D}^{[n]}(s_0) \sim e^{-\frac{2 \pi a}{N_c} \frac{s_0}{\Lambda^2}}
\end{equation}
which allows one to write
\begin{eqnarray}\label{a}
    \mathcal{D}^{[n]}(s_0)&=&  \mathcal{D}_{oscill.}^{[n]}(s_0)+
    \mathcal{D}_{const.}^{[n]}(s_0)\qquad  \sim \quad \mathcal{O}
    \left(e^{-\frac{2 \pi a}{N_c}\frac{ s_0}{\Lambda^2}}\right)\\
 \mathcal{D}_{oscill.}^{[n]}(s_0)&=&\int_{0}^{s_0} dt\ t^n\ \frac{1}{\pi}
    \mathrm{Im}\Delta(t+i\varepsilon)\qquad \sim \quad -{\mathcal{C}}^{[n]}+
    \mathcal{O}\left(e^{-\frac{2 \pi a}{N_c}\frac{s_0}{\Lambda^2}}\right)\label{b}\\
   \mathcal{D}_{const.}^{[n]}(s_0) &=&\ \mathcal{C}^{[n]}\ +\
      \mathcal{O}\left(e^{-2 \pi \frac{s_0}{\Lambda^2}}\right)\label{c}
\end{eqnarray}    
Therefore, combining the previous equations and taking the large $s_0$ limit results in the following
\begin{equation}\label{d}
    \mathcal{C}^{[n]}=- \int_{0}^{\infty} dt\ t^n\ \frac{1}{\pi}
    \mathrm{Im}\Delta(t+i\varepsilon)
\end{equation}
which also means that $\mathcal{D}^{[n]}(s_0)$ can finally be cast in the simple form\footnote{All along we have used that 
\begin{equation}\label{hierarchy}
    e^{-\frac{2 \pi a }{N_c}\frac{s_0}{\Lambda^2}}\gg e^{-2 \pi
    \frac{s_0}{\Lambda^2}}
\end{equation}
which can be shown to be satisfied from a relatively low $s_0$ upwards.}
\begin{equation}\label{finally}
    \mathcal{D}^{[n]}(s_0)=- \int_{s_0}^{\infty} dt\ t^n\ \frac{1}{\pi}
    \mathrm{Im}\Delta(t+i\varepsilon)\ +\
    \mathcal{O}\left(e^{-\frac{2 \pi s_0}{\Lambda^2}}\right)
\end{equation}

Armed with this expression for the duality violating term, we can write the analog of (\ref{ddelta}) for the finite $N_C$ version of our model in the following fashion
\begin{eqnarray}\label{works0}
 \int_{0}^{s_0}\!\!\!\!\!dt\ \rho(t)+\left[\frac{b_6}{2 s_0^2}-\frac{b_8}{3s_0^3}+...  \right]
  \!\!\!\!&=&\!\!\!  \mathcal{D}^{[0]}(s_0) \\\label{works1}
  \int_{0}^{s_0}\!\!\!\!\!dt\ t\ \rho(t)+\left[\frac{b_6}{s_0}-\frac{b_8}{2 s_0^2}+...  \right]
  \!\!\!&=&\!\!\!  \mathcal{D}^{[1]}(s_0)  \\\label{works2}
  \int_{0}^{s_0}\!\!\!\!\!dt\ t^2\rho(t)-\left[a_6+ b_6 \log\frac{s_0}{\Lambda^2}+
  \frac{b_8}{s_0}+...\right] \!\!\!&=& \!\!\!\mathcal{D}^{[2]}(s_0)
  \\\label{works3}
  \int_{0}^{s_0}\!\!\!\!\!dt\ t^3\rho(t)+\left[a_8+b_8 \log\frac{s_0}{\Lambda^2}-
   b_6 s_0 +  \frac{b_{10}}{s_0}+...  \right]\!\!\!&=& \!\!\!
   \mathcal{D}^{[3]}(s_0) \\
&\vdots & \nonumber \\
 \!\!\!\!\!\!\!\!\!\!\!\!\! \int_{0}^{s_0}\!\!\!\!\!dt\ t^{7}\rho(t)+\left[a_{16}+b_{16}
 \log\frac{s_0}{\Lambda^2}-
\frac{b_6 s_0^5}{5}+ \frac{b_8 s_0^4}{4}+\ldots+ \frac{b_{18}}{s_0}+...
\right]\!\!\! &=&\!\!\!
    \mathcal{D}^{[7]}(s_0) \label{works5}
\end{eqnarray}
where the OPE integrals (the terms in brackets above) get contributions from the $b_{2k}$ parameters, according to the following formula
\begin{equation}
-\fr{1}{2\pi i}\ \oint_{|q^2|=s_0}\ dq^2\ (q^2)^n\ \Pi_{LR}^{OPE}(q^2)=(-1)^n\ \left[a_{2n+2}+b_{2n+2}\ \log{\fr{s_0}{\Lambda^2}}+\sum_{k\neq n}^{\infty}\ (-1)^k\ \fr{b_{2k+2}}{|n-k|}s_0^{(n-k)}\right]
\end{equation} 
The extraction of the $a_{2k}$ OPE coefficients therefore gets polluted not only from the duality violation term but also from {\it{all}} $b_{2k}$ terms. The existing analyses for the extraction of OPE coefficients in QCD do not include the influence of these $b_{2k}$ terms. Certainly this seems reasonable for the $b_{2k}$ which come with inverse powers of $s_0$, even though one should be aware of their growth with increasing dimension (eventually, they get bigger than the $a_{2k}$ parameters)
\begin{equation}
b_{2k}\ \sim \fr{ka}{N_C}\ C_{2k}^{N_C\rightarrow \infty}  
\end{equation}
What certainly deserves more attention are the $b_{2k}$ terms coming with positive powers of $s_0$ out front. Despite their small absolute value, at $s_0=m_{\tau}^2\, \sim 3.15$ GeV$^2$ the $s_0$ term alone can bring a huge number.
In the particular case of the dimension-six condensate, (\ref{works2}) shows that there are no harmful $b_{2k}$ terms coming with a positive power of $s_0$. Moreover, in real QCD (only) $b_6$ has been determined so far and shown to be negligible. This might very well explain the overall agreement in $a_6$ of all the existing determinations\footnote{See, for instance, \cite{Davier:1,s4}, \cite{c1}-\cite{c9},\cite{Narison:05}.}. Concerning the dimension-eight condensate, the discrepancies found in the different analyses on the market may signal at sizeable $b_{2k}$ terms not taken into account.

As already pointed out, the $b_{2k}$ terms in our model do not have the right $1/N_C$ scaling. This results in a value for the $b_6$ parameter which is far bigger than the observed one. Probably the same happens with the remaining $b_{2k}$ terms, so that our model would be overexaggerating the weight of the $b_{2k}$ terms. Should this be true, they would have a milder impact in QCD than they have in the model. Anyway, at least for consistency, one should try to assess their actual impact.

We now move to the analysis of the duality violating terms, which in our opinion bear the main source of difficulty in the determination of the OPE condensates. To show that point clear, we will compare a FESR and pFESR analysis without taking into account duality violations (which is what present determinations do) to later on show how to improve on that.
\subsection{Finite Energy Sum Rules}
The strategy to follow will be to identify the duality points $s_0^*$ in the moments $M_0$ and $M_1$, {\it{i.e.}}, those points satisfying $M_0\ (s_0^*)=0$ and $M_1\ (s_0^*)=0$. Present experimental data cover the first two duality points, which we will refer as $s_0^{(1)}$ and $s_0^{(2)}$. Contrary to what happens in the large-$N_C$ version of our model, duality points need not be the same for the different moments, and actually they are not. However, and this is the main motivation of the {\it{finite energy sum rule}} approach\footnote{For a review, see \cite{rafael:sr}.}, they should lie close enough to allow for a reliable extraction of the OPE coefficients. In the light of (\ref{works0}), finite energy sum rules take the following scenario
\begin{equation}
M_0(s_0^*)\ =\ 0\ =\ M_1(s_0^*)\, , \qquad M_2(s_0^*)\ \sim\ a_6\, , \qquad M_3(s_0^*)\sim\ a_8
\end{equation}
as a good enough approximation to the physical picture. The first duality point in our model sits at $s_0^{(1)}=1.472$ GeV$^2$, and one obtains the following predictions
\begin{equation}\label{fesr1}
A_6^{FESR}=-4.9 \cdot 10^{-3} \,\,{\mathrm{GeV}}^6 \quad , \quad A_8^{FESR}=+9.3
\cdot 10^{-3} \,\, {\mathrm{GeV}}^8
\end{equation}
which are to be contrasted with the OPE expressions, which have to be truncated at some point. We choose   
\begin{eqnarray}\label{maarten}
    A_6(s_0^*)&\equiv& a_6+ b_6 \log \frac{s_0^*}{\Lambda^2}+ \frac{b_8}{s_0^*} \ , \nonumber \\
    A_8(s_0^*)&\equiv& a_8+ b_8 \log \frac{s_0^*}{\Lambda^2}- b_6 s_0^* +
    \frac{b_{10}}{s_0^*}
\end{eqnarray}
to yield
\begin{equation}
A_6(1.47 \ {\mathrm{GeV}}^2)= -2.4\cdot
10^{-3} \mathrm{GeV}^6\, , \qquad
A_8(1.47\ {\mathrm{GeV}}^2)= +2.6\cdot 10^{-3}
\mathrm{GeV}^8
\end{equation}
Contrary to what happens to $s_0^{(1)}$, the second duality point does not simultaneously satisfy $M_0(s_0^*)=0=M_1(s_0^*)$. This is quite similar to what happens in real QCD. Following \cite{c5}, we choose $M_1(s_0^{(2)})=0$ to define the duality point with which to make our predictions. This corresponds to $s_0^{(2)}=2.363$ GeV$^2$. This strategy seems to be well justified, since one expects duality points to shift minimally between neighbouring moments. The results we find are
\begin{equation}\label{fesr2}
A_6^{FESR}=-2.0 \cdot 10^{-3} \,\,{\mathrm{GeV}}^6 \quad , \quad A_8^{FESR}=-1.6
\cdot 10^{-3} \,\, {\mathrm{GeV}}^8
\end{equation} 
to be compared to the true OPE values
\begin{equation}
A_6(2.36 \ {\mathrm{GeV}}^2)= -2.8\cdot 10^{-3} \mathrm{GeV}^6\, , \qquad
A_8(2.36\ {\mathrm{GeV}}^2)= +3.4\cdot 10^{-3} \mathrm{GeV}^8
\end{equation}
In view of the preceding results, one can hardly argue in favour of a preferred duality point. In principle, one should expect that a larger $s_0$ would reduce the amount of duality violation, thus improving the predictions. Indeed, the prediction for $A_6$ is slightly better in the second duality point, but clearly this is not the case for $A_8$. Most probably, the amount of duality violation is still big enough to ruin this picture. Remarkably, the scenario we find in going to the first duality point to the second one, {\it{i.e.}}, $A_6$ reducing to half its value and $A_8$ changing sign mimics the one observed in real QCD. However, regardless of the duality point, uncertainties in the predictions amount to, at least, a $30\%$.
\begin{figure}
\renewcommand{\captionfont}{\small \it}
\renewcommand{\captionlabelfont}{\small \it}
\centering
\includegraphics[width=2.5in]{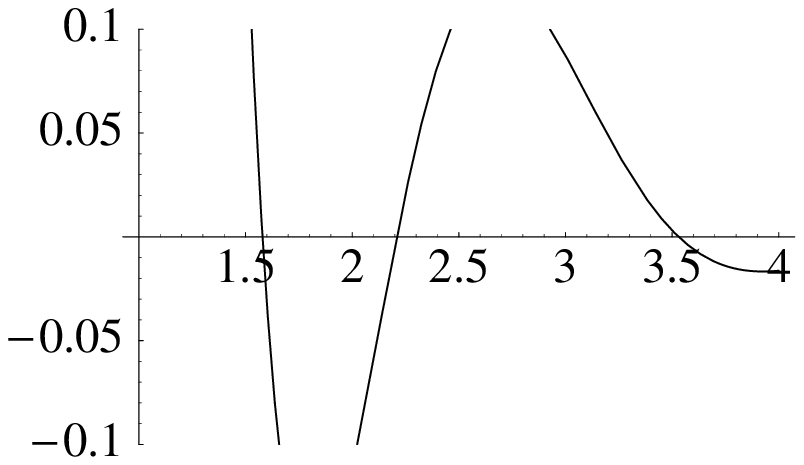}
\includegraphics[width=2.5in]{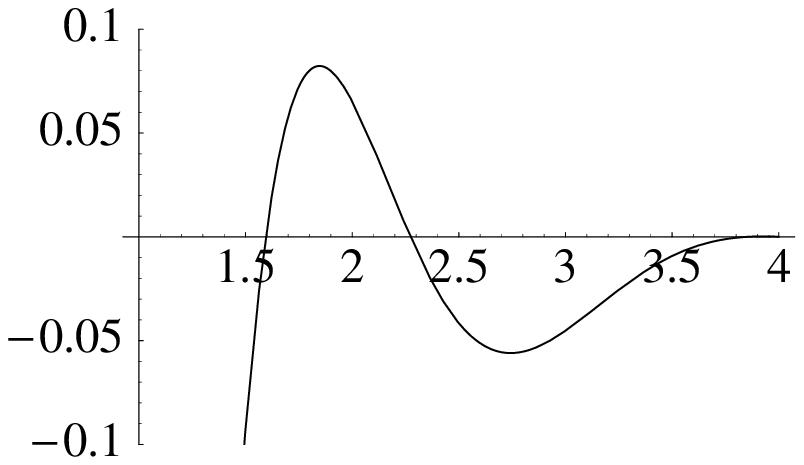}
\includegraphics[width=2.5in]{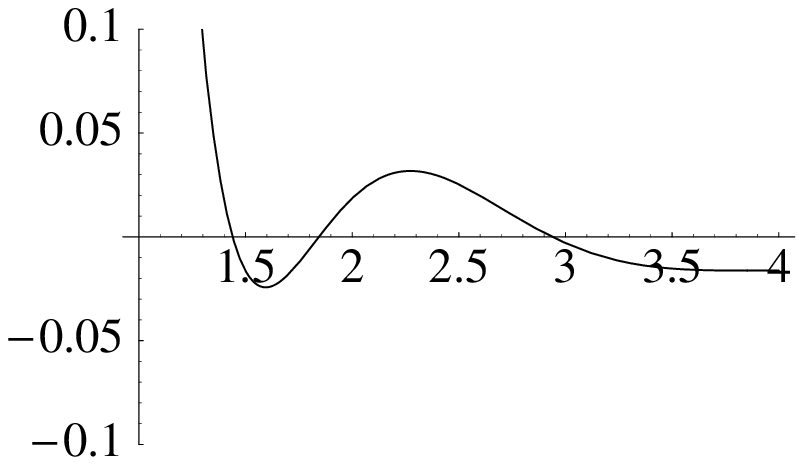}
\caption{Comparison of the pinched weights $w_1(t)$, $w_2(t)$ and $w_3(t)$ as defined in the main text with their corresponding OPE contributions subtracted. Vertical axis in arbitrary units.}\label{YY}
\end{figure}
In the large-$N_C$ version of our model we saw that pinched weights indeed suppress the duality violating terms, but do not get rid of them entirely. We here reproduce an analysis in the same spirit as that of \cite{Cirig:2}. We fit the OPE contributions to the pinched-weights $w_1(t)$ and $w_2(t)$, choosing a window of equally-spaced 20 points ranging between $1.5\ {\textrm{GeV}}^2\ \leq s_0\ \leq 3.5\ {\textrm{GeV}}^2$. The upper end is chosen so as to coincide with the would-be $m_{\tau}$ mass threshold\footnote{Obviously we have not included the tau lepton in our model. In real QCD it falls around the $\rho(1700)$ resonance, which in our model would correspond to $m_{\rho\prime\prime}^2\simeq 3.8\ {\textrm{GeV}}^2$.}. The values obtained this way are
\begin{equation}\label{pw}
    A_6^{pinch}= -3.8\times 10^{-3}\ \mathrm{GeV}^6 \quad ,
    \quad A_8^{pinch}= 6.5 \times 10^{-3}\ \mathrm{GeV}^8
\end{equation} 
which also fail to reproduce the true values better than a 30$\%$.
\subsection{Other approaches}
We can also use our model to test the {\it{minimal hadronic approximation}} we have been using in previous chapters as an interpolating function for different Green functions. As a starting point, we will restrict our spectrum to consist of a pion, a $\rho$ resonance and an axial $a_1$ resonance, all of them infinitely narrow, namely
\begin{equation}\label{spectrMHA}
\fr{1}{\pi}{\textrm{Im}}\,\Pi_{LR}^{MHA}(t)=-F_0^2\,\delta(t)+{\widehat{F}}_{V}^2\,\delta(t-{\widehat{M}}_V^2)-{\widehat{F}}_A^2\,\delta(t-{\widehat{M}}_A^2)
\end{equation}
It then follows that  
\begin{equation}\label{piMHA}
    \Pi_{LR}^{MHA}(q^2)=\frac{F_{0}^2}{q^2}+
\frac{{\widehat{F}}_V^2}{{\widehat{M}}_{V}^2-q^2}-\frac{{\widehat{F}}_{A}^2}{{\widehat{M}}_{A}^2-q^2}
\end{equation}
with free parameters $F_0$, ${\widehat{F}}_V$, ${\widehat{F}}_A$, ${\widehat{M}}_V$ and ${\widehat{M}}_A$. We fix $F_0$ to take its value from our model of previous sections, {\it{i.e.}},
\begin{equation}\label{fpi}
F_0=85.8\ {\mathrm{MeV}}
\end{equation}
and determine the remaining four parameters by demanding the two Weinberg sum rules to hold
\begin{equation}\label{weisr}
{\widehat{F}}_V^2-{\widehat{F}}_A^2=F_0^2\, , \qquad {\widehat{F}}_V^2\ {\widehat{M}}_V^2-{\widehat{F}}_A^2\ {\widehat{M}}_A^2=0
\end{equation}
Additionally, we impose constraints on the slope at the origin and area below $\Pi_{LR}(q^2)$ in the Euclidean. These are given by $L_{10}$ and the electromagnetic pion mass difference, defined as
\begin{eqnarray}
L_{10}&=&-\fr{1}{4}\bigg[\fr{d}{dq^2}\bigg(q^2\Pi^{LR}(q^2)\bigg)\bigg]\bigg|_{q^2\rightarrow 0}\nonumber\\
m_{\pi}^+-m_{\pi}^0&=&-\fr{3\alpha}{8\pi^2 m_{\pi}F_0^2}\int_0^{\infty}\,dQ^2\,\bigg[Q^2\Pi^{LR}(Q^2)\bigg]\, , \qquad (Q^2=-q^2)
\end{eqnarray}
The resulting matching equations for the previous two parameters as given by the model of (\ref{fifteen}) and the Minimal Hadronic Approximation read
\begin{eqnarray}\label{constraints}
    L_{10}&=& -\frac{1}{4}F_{0}^2\frac{\widehat{M}_{A}^2+
\widehat{M}_{V}^2}{\widehat{M}_{V}^2 \widehat{M}_{A}^2}=-(5.2\pm 0.5) \cdot 10^{-3}\nonumber\\
\!\!\!\!m_{\pi}^{+}-m_{\pi}^{0}\!\!\!\!&=&\!\!\!\!\! \left(\frac{3\alpha}{8\pi^2
m_{\pi}F_{0}^2}\right)F_{0}^2 \frac{\widehat{M}_{A}^2
\widehat{M}_{V}^2}{\widehat{M}_{A}^2-
\widehat{M}_{V}^2}\log\left(\frac{\widehat{M}_{A}^2}{\widehat{M}_{V}^2}\right)= (4.2\pm
0.4)\cdot 10^{-3}\,\,{\mathrm{GeV}}
\end{eqnarray}
where we have taken $\alpha=1/137$ and $m_{\pi}=135$ MeV. Combining (\ref{weisr}) and (\ref{constraints}) yields
\begin{eqnarray}\label{solMHA}
\widehat{M}_{V}=0.70\pm 0.01\,\,{\mathrm{MeV}}\quad &,& \quad
\widehat{M}_{A}=1.00\pm0.03\,\,{\mathrm{GeV}}\nonumber\\
\widehat{F}_{V}=122\pm 6\,\,{\mathrm{MeV}}\quad &,& \quad \widehat{F}_{A}=84\pm 7
\,\,{\mathrm{MeV}}
\end{eqnarray}
\begin{figure}
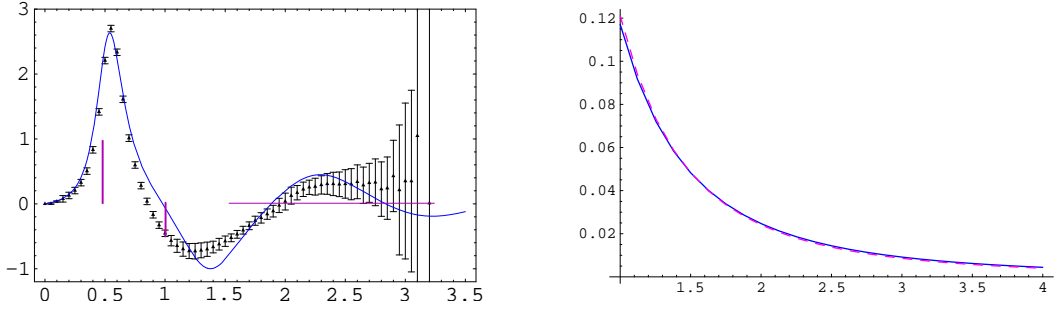
\label{MHAplot}
\renewcommand{\captionfont}{\small \it}
\renewcommand{\captionlabelfont}{\small \it}
\centering
\includegraphics[width=2.5in]{data2MHA.eps}
\hspace{1cm}
\includegraphics[width=2.5in]{euclideanMHA.eps}
\caption{Comparison of $\Pi_{LR}(q^2)\bigg|_{MHA}$ as it stands in (\ref{spectrMHA}) with the experimental data of ALEPH \cite{aleph} and OPAL \cite{opal}. The kinematical factors for $\tau$ decay are included and the pion pole subtracted away. As already emphasised, being a bad approximation to the Minkowski, the MHA is however a very good approximation to the model in the Euclidean region.}
\end{figure}
Plugging (\ref{solMHA}) and (\ref{fpi}) into (\ref{piMHA}), we obtain the following predictions for the OPE coefficients
\begin{equation}\label{numbersMHA}
a_6^{MHA}=- (3.6\pm 0.3) \cdot 10^{-3}\,\,{\mathrm{GeV}}^6\quad , \quad
a_8^{MHA}=(5.4 \pm0.7) \cdot 10^{-3}\,\,{\mathrm{GeV}}^8
\end{equation}
which again are comparable to the previous estimates using FESR and pFESR. 

Just for illustrative purposes, figure (6.8) shows $\Pi_{LR}(q^2)$ of our model as compared to $\Pi_{LR}^{MHA}(q^2)$. We here reproduce the well-known fact that the MHA to Green functions, despite being extremely different from the experimentally known QCD Green function in the Minkowski region, is a very good approximation in the Euclidean. Remember that the Euclidean region is precisely where we have been using the MHA as an interpolator to compute nonleptonic electroweak couplings in the previous chapter. Recall that electroweak couplings have the geometrical interpretation of being the area under the curve, and indeed the agreement is remarkable. 

Another commonly used method to extract the OPE coefficients relies on the Laplace transform of the Cauchy theorem. They are usually coined {\it{Laplace sum rules}} or sometimes {\it{Borel sum rules}}. Following \cite{Narison:05}, we can write  
\begin{eqnarray}\label{laplace}
a_6^L - a_{10}^L\ \frac{\tau^2}{12} +...&=& \frac{6}{\tau^2} \int_0^{s_0}dt\ e^{-t
\tau}\ \rho(t)+
\frac{2}{\tau} \int_0^{s_0}dt\ t\ e^{-t \tau}\ \rho(t)\nonumber \\
a_8^L+ a_{10}^L\ \frac{\tau}{2} +...&=& -\frac{12}{\tau^3} \int_0^{s_0}dt\ e^{-t
\tau}\ \rho(t)- \frac{6}{\tau^2} \int_0^{s_0}dt\ t\ e^{-t \tau}\ \rho(t)
\end{eqnarray}
which yield the dimension six $a_6^L$ and dimension eight condensates $a_8^L$ in terms of the (exponentially weighted) moments. The strategy in Laplace sum rules is to look for a stability region of the parameters $a_6^L$ and $a_8^L$ in the Borel parameter ($\tau$) space. We follow \cite{Narison:05} and fix $s_0$ to be close to a duality point. As it can be seen in figure (6.9), no such stability region is found. We have also moved the $s_0$ and nowhere stability was found. Laplace sum rules therefore do not work in our model, even though in real QCD they seem to do.  
\begin{figure}
\renewcommand{\captionfont}{\small \it}
\renewcommand{\captionlabelfont}{\small \it}
\centering
\includegraphics[width=2.5in]{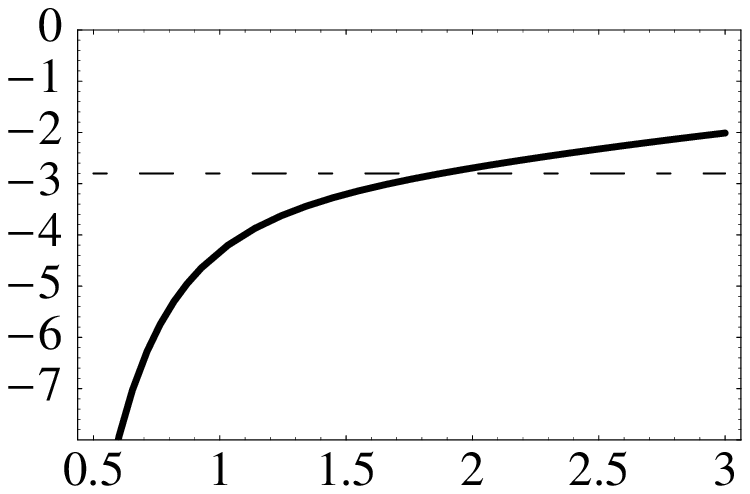}
\hspace {1cm}
\includegraphics[width=2.5in]{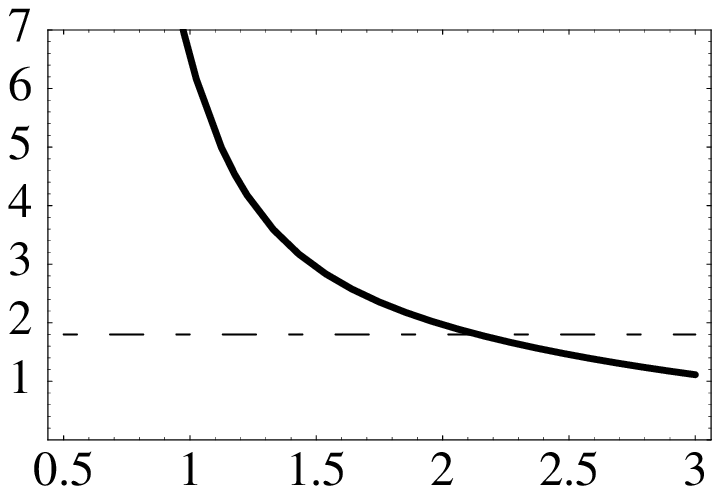}\\
\caption{Plot of the coefficients $a_6^L$ (left) and $a_8^L$ (right) for a fixed $s_0$ close to a duality point, together with
the true values $a_{6,8}$ from Eqs. (\ref{opewidths}), as a function of the Laplace
variable, $\tau$.}\label{fig7}
\end{figure}

\subsection{Modelling duality violations}
The previous analyses suggest that indeed the inclusion of duality violations is necessary in order to improve our determinations beyond $30\%$. The challenge is then how to model duality violations in a generic way. Recalling (\ref{imdv}), we take as our {\it{ansatz}}
\begin{equation}\label{fit}
    \frac{1}{\pi} \mathrm{Im}\Delta(t)= \kappa\  e^{-\gamma t}\  \sin\left(\alpha +
    \beta t\right)
\end{equation}
One possible strategy is to use (\ref{fit}) to fit the right-hand and left-hand sides of (\ref{works0}), (\ref{works1}), where we set the $b_{2k}$ coefficients to zero. This is obviously an approximation, but it seems reasonable as all those coefficients come with inverse powers of $s_0$. The fit results in the following values for the parameters
\begin{equation}\label{fitsinch6}
\kappa=0.026\, , \qquad \gamma=0.591\ {\textrm{GeV}}^{-2}\, , \qquad \alpha=3.323\, , \qquad \beta=3.112\ {\textrm{GeV}}^{-2} 
\end{equation} 
\begin{figure}
\renewcommand{\captionfont}{\small \it}
\renewcommand{\captionlabelfont}{\small \it}
\centering
\includegraphics[width=2.3in]{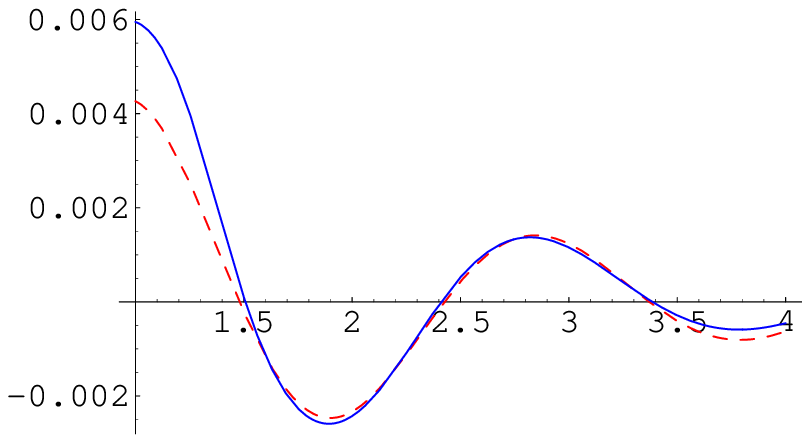}
\hspace{1cm}
\includegraphics[width=2.3in]{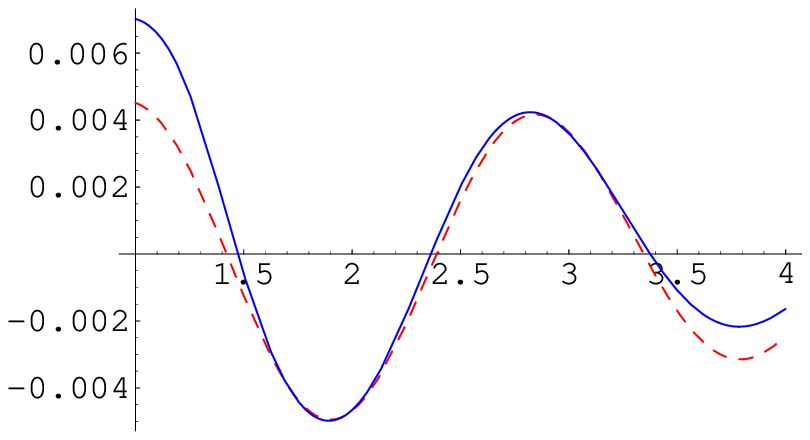}
\includegraphics[width=2.3in]{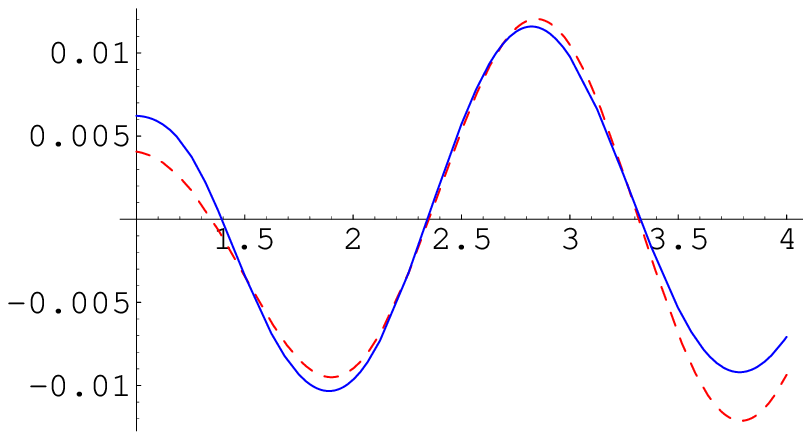}
\hspace{1cm}
\includegraphics[width=2.3in]{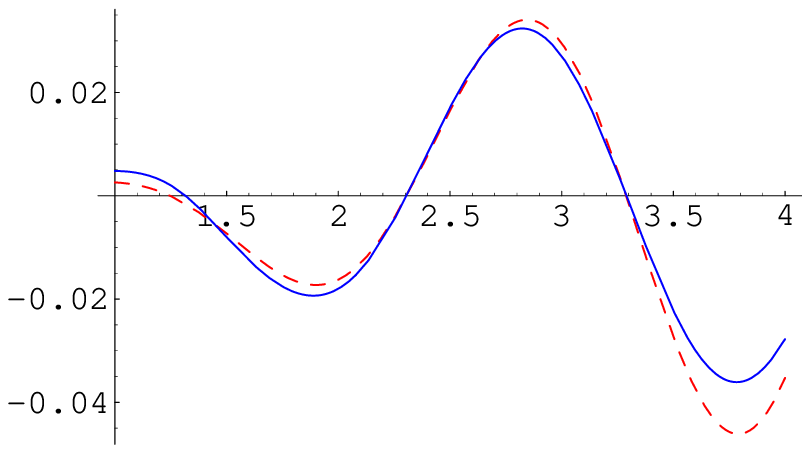}
\caption{Upper row: Plot of the left-hand side (solid line) and
right-hand side (dashed line) of Eqs.~(\ref{works0},\ref{works1}), neglecting the $b$
coefficients, as a function of $s_0$ (in $\mathrm{GeV^2}$). Lower row: Plot of the
left-hand side (solid line) and right-hand side (dashed line)
of Eqs.~(\ref{works2},\ref{works3}), as a function of $s_0$ (in
$\mathrm{GeV^2}$). The right-hand sides are obtained with the fitted parameters of (\ref{fitsinch6}).}\label{fits}
\end{figure}
which produce the first two plots of figure (\ref{fits}). The prediction for $M_2$ and $M_3$ is given in the last two plots, where one can see that the agreement between the true and predicted curves is remarkable, improving in the intermediate regions where $M_2$ and $M_3$ cancel. Once we have modelled the duality violations, we can use the predicted curves to find the duality points for each moment. For the sake of illustration, we show how this strategy works with $M_2$, $M_3$ and $M_7$. Predicted duality points sit at
\begin{eqnarray}
{\mathcal{D}}^{[2]}(s_0^*)=0\, &,& s_0^*=2.350\ {\textrm{GeV}}^2\nonumber\\
{\mathcal{D}}^{[3]}(s_0^*)=0\, &,& s_0^*=2.307\ {\textrm{GeV}}^2\nonumber\\
{\mathcal{D}}^{[7]}(s_0^*)=0\, &,& s_0^*=2.130\ {\textrm{GeV}}^2
\end{eqnarray}
Therefore, (\ref{works2}), (\ref{works3}) and (\ref{works5}) give
\begin{equation}\label{result1}
    a_6 + b_6 \log \frac{s_0^*}{\Lambda^2}+ \frac{b_8}{s_0^*}- \frac{b_{10}}{2 s_0^*{^2}}=\int_0^{s_0^*} dt\
    t^2 \rho(t)\ =\  -0.00251 \ \mathrm{GeV}^6
\end{equation}
\begin{equation}\label{result2}
  \!\!\!\!\!\!\!\!\!\!\!\!\!\! a_8 +
   b_8 \log \frac{s_0^*}{\Lambda^2}- b_6\, s_0^*+ \frac{b_{10}}{s_0^*}=
    - \int_0^{s_0^*}\!\!\!\! dt\ t^3 \rho(t) = 0.00329\ \mathrm{GeV}^8
\end{equation}
\begin{equation}
 a_{16} + b_{16} \log \frac{s_0^*}{\Lambda^2}- \frac{b_6 s_0^*{^5}}{5}+ \frac{b_{8}s_0^*{^4}}{4}-\dots-b_{14}s_0^*+ \frac{b_{18}}{s_0^*}=- \int_0^{s_0^*}\!\!\!\!\! dt\ t^7 \rho(t) =  0.0179 \
 \mathrm{GeV}^{16}
\end{equation}
whereas the true value on the left hand sides should be $-0.00281\ {\textrm{GeV}}^6$, $0.00344\ {\textrm{GeV}}^8$ and $0.0161\ {\textrm{GeV}}^{16}$, respectively, meaning a $10\%$, $4\%$ and $11\%$ error, clearly much precise determinations than those obtained by any method neglecting duality violations.

\chapter{Conclusions}

In this work we have exploited some of the features of an approximation to QCD, called large-$N_C$ QCD, in order to provide physical insight into the non-perturbative aspects of the hadronic world. In chapter 3 we introduced a Lagrangian designed to go one step further than chiral Lagrangians in the sense of being capable of describing the interactions between the pseudo-Goldstone bosons (the pion octet fields) and the lowest lying resonance states. The Lagrangian admitted an interpretation in the framework of large-$N_C$ QCD. This turned out to be a powerful tool which allowed a well-defined computation of quantum corrections, in powers of $1/N_C$. We therefore tested the validity of the Lagrangian with a simple example, the prediction for $L_{10}$ at the quantum level. The tree level determination of that parameter with the same Lagrangian was long ago shown to be in good agreement with experimental data. We however found that this agreement no longer persists at the quantum level and the Lagrangian is unable to yield the right evolution for $L_{10}(\mu)$, dictated by chiral perturbation theory. We suggest this as one of the easiest non-trivial exercises to put to the test every Lagrangian purported to go beyond chiral perturbation theory.

In subsequent chapters we set the stage for the study of kaon physics parameters. In particular, we stressed the fact that in nonleptonic weak interactions scale-independence in the final physical results is not easy to achieve. This has to do with the fact that matching long and short distances is highly non-trivial: Wilson coefficients encode the ultraviolet aspects of the calculation, {\it{i.e.}}, the short distances, and can be computed with perturbative techniques; as for the long distances, one has first to find the chiral realization of the operators involved. It can be shown that the complicated part of the calculation thereafter amounts to a determination of the (associated) low energy chiral couplings, which can always be expressed as integrals over certain Green functions. Non-perturbative tools are then needed to further proceed. Yet they have to make sure that these two pieces, Wilson coefficients and hadronic matrix elements, are smoothly connected in order to make any spurious scale dependence disappear from the final physical answer. We showed how a well-defined approximation to large-$N_C$ QCD, termed the {\it{Minimal Hadronic Approximation}}, solves in a very natural way the problem.

The advantage of the Minimal Hadronic Approximation is that it provides an analytic expression for those Green functions lying under the low energy chiral couplings, expressions which are valid at {\it{all}} energies. Those expressions can be thought as interpolating functions (rational approximants) between low energies and high energies, where calculational methods exist ($\chi$PT and the OPE). In particular, we have shown explicitly how the correct implementation of the OPE constraints is the feature that ensures the right matching and renders scale-independent results.

We have applied this method to the determination of the parameters governing kaon phenomenology: we have provided a new value for ${\widehat{B}}_K$ in the chiral limit and to next to leading order in the $1/N_c$ expansion, taking into account the effect of including dimension eight operators in the OPE of the underlying Green function. This was motivated by a recent analysis concluding that these dimension eight operators were potentially sizeable, therefore suggesting its inclusion. Our result shows that ${\widehat{B}}_K$ is essentially unaffected by such terms.

Another issue was the evaluation of the corrections in the charm mass expansion to the kaon mass difference, $\Delta m_K$, and $\epsilon_K$. The charm mass expansion is known to have slower convergence due to its proximity to $\Lambda_{\chi}$ and therefore it is a source of potentially sizeable contributions which deserve attention. This calculation in inverse powers of the charm mass is tantamount to computing the leading short and long distance corrections to the $\Delta S=2$ Lagrangian. Such corrections give rise to two different scales: a long distance hadronic scale $\Lambda_{\chi}\sim 1$ GeV due to the (non-perturbative) up-up exchange box diagram and a short distance parameter $\delta_K\sim 0.35$ GeV. It turns out that $\Lambda_{\chi}$ only has impact on $\Delta m_K$, while $\varepsilon_K$ receives a small $0.5\%$ correction proportional to $\delta_K$. The net result is that, whereas $\Delta m_K$ has a sizeable correction of a roughly $10\%$, which pushes its value closer to the experimental one, $\varepsilon_K$ remains basically unaffected. It is worth stressing here that our results were performed in the chiral limit, and thus there is room for potential improvement with the addition of chiral corrections. Furthermore, recall that we have only computed the short distance contribution to the kaon mass difference. There still remains an evaluation of the so-called long distance $\bigg[\Delta S=1 \times \Delta S=1\bigg]$ contributions, which lies beyond the scope of the present work.  

In the last chapter we have addressed the issue of quark-hadron duality violations, which arise as a consequence of the lack of validity of the OPE over the whole complex $q^2$ plane, most blatantly in the Minkowski axis. We have focussed our attention on the two-point Green function $\Pi_{LR} (q^2)$, whose OPE coefficients are related to kaon physics parameters. Armed with a model, we have tested the most relevant approaches to the extraction of the OPE coefficients and have shown how in principle one can improve on them. Our conclusion is that they all lie in the same ballpark of values and to really make an improvement, at least in our model, one has to take duality violations into account. We have put forward a proposal to model the duality violating pieces, which in our opinion goes beyond our particular toy model and can be considered as generic. As such, we suggest that it be incorporated in QCD analyses of $\Pi_{LR}(q^2)$ in order to reassess the existing approaches used to determine the OPE coefficients.
\appendix

\renewcommand{\theequation}{A.\arabic{equation}} \setcounter{equation}{0}

\chapter[$SU(3)$ algebra]{$SU(3)$ algebra }
\section{SU(3) algebra}
The eight generators of the algebra are usually written down in terms of the following matrices, known as the {\it{Gell-Mann matrices}},
\begin{equation}
\lambda_1=\left(\begin{array}{ccc}
0 & 1 & 0\nonumber\\
1 & 0 & 0\nonumber\\
0 & 0 & 0\nonumber
\end{array}\right) \quad , \quad
\lambda_2=\left(\begin{array}{ccc}
0 & -i & 0\nonumber\\
i & 0 & 0\nonumber\\
0 & 0 & 0\nonumber
\end{array}\right) \quad , \quad
\lambda_3=\left(\begin{array}{ccc}
1 & 0 & 0\nonumber\\
0 & -1 & 0\nonumber\\
0 & 0 & 0
\end{array}\right)
\end{equation}
\begin{equation}
\lambda_4=\left(\begin{array}{ccc}
0 & 0 & 1\nonumber\\
0 & 0 & 0\nonumber\\
1 & 0 & 0\nonumber 
\end{array}\right) \quad , \quad
\lambda_5=\left(\begin{array}{ccc}
0 & 0 & -i\nonumber\\
0 & 0 & 0\nonumber\\
i & 0 & 0\nonumber
\end{array}\right) \quad , \quad 
\lambda_6=\left(\begin{array}{ccc}
0 & 0 & 0\nonumber\\
0 & 0 & 1\nonumber\\
0 & 1 & 0
\end{array}\right)
\end{equation}
\begin{equation}
\lambda_7=\left(\begin{array}{ccc}
0 & 0 & 0\\
0 & 0 & -i\\
0 & i & 0
\end{array}\right) \quad , \quad
\lambda_8=\fr{1}{\sqrt{3}}\left(\begin{array}{ccc}
1 & 0 & 0\\
0 & 1 & 0\\
0 & 0 & -2
\end{array}\right)
\end{equation}
It is relatively easy to check that they satisfy the following algebra
\begin{equation}\label{deff1}
[\lambda_a,\lambda_b]=2i\,f_{abc}\,\lambda_c
\end{equation}
\begin{equation}\label{deff2}
\{\lambda_a,\lambda_c\}=\fr{4}{N_C}\delta_{ab}{\mathbf{1}}+2\,d_{abc}\,\lambda_c
\end{equation}
which leads to
\begin{equation}\label{deff3}
\lambda_a\lambda_b=\fr{2}{N_C}\,\delta_{ab}{\mathbf{1}}+d_{abc}\,\lambda_c+if_{abc}\,\lambda_c
\end{equation}
from where it is immediate to verify that
\begin{equation}
\textrm{Tr}\, (\,\lambda_a\lambda_b\,)=2\, \delta_{ab}
\end{equation}
Another useful relation is the following
\begin{equation}
\left(\fr{\lambda_a}{2}\right)_{ij}\left(\fr{\lambda_a}{2}\right)_{kl}= \fr{1}{2}\left(\delta_{il}\delta_{jk}-\fr{1}{N_C}\delta_{ij}\delta_{kl}\right)
\end{equation}
The values of the (completely antisymmetric) structure constants $f_{abc}$ appearing in (\ref{deff1},\ref{deff3}) can be inferred from their expressions in terms of the Gell-Mann matrices
\begin{equation}
f_{abc}=\fr{1}{4i}\,{\textrm{tr}}\,[\lambda_a,\lambda_b]\,\lambda_c
\end{equation}
to yield
\begin{eqnarray}
f_{123}\!\!\!\!&=&\!\!\!\!1\nonumber\\
f_{147}\!\!\!\!&=&\!\!\!\!-f_{156}=f_{246}=f_{257}=f_{345}=-f_{367}=\fr{1}{2}\nonumber\\ 
f_{458}\!\!\!\!&=&\!\!\!\!f_{678}=\fr{\sqrt{3}}{2}
\end{eqnarray}
while the (completely symmetric) coefficients $d_{abc}$ are defined as
\begin{equation}
d_{abc}=\fr{1}{4}\,{\textrm{tr}}\,\{\lambda_a,\lambda_b\}\,\lambda_c
\end{equation}
whose values are listed below
\begin{eqnarray}
d_{118}\!\!\!\!&=&\!\!\!\!d_{228}=d_{338}=-d_{888}=\fr{1}{\sqrt{3}}\nonumber\\
d_{146}\!\!\!\!&=&\!\!\!\!d_{157}=-d_{247}=d_{256}=d_{344}=d_{355}=-d_{366}=-d_{377}=\fr{1}{2}\nonumber\\
d_{448}\!\!\!\!&=&\!\!\!\!d_{558}=d_{668}=d_{778}=-\fr{1}{2\sqrt{3}}
\end{eqnarray}
As for the $SU(2)$ group, their generators are usually written in terms of the Pauli matrices, to wit
\begin{equation}
\tau_1=\left(\begin{array}{cc}
0 & 1\nonumber\\
1 & 0\nonumber
\end{array}\right) \quad , \quad
\tau_2=\left(\begin{array}{cc}
0 & -i \nonumber\\
i & 0 \nonumber
\end{array}\right) \quad , \quad
\tau_3=\left(\begin{array}{cc}
1 & 0\nonumber\\
0 & -1
\end{array}\right)
\end{equation}
which satisfy the $SU(2)$ algebra
\begin{equation}
[\tau_i,\tau_j]=2i\,\varepsilon_{ijk}\,\tau_k
\end{equation}
\section{Fierz transformations}
Consider a product of Dirac bilinears of the form
\begin{equation}\label{fierzone}
\bar{\psi}_4\,{\Gamma}\,\psi_1\,\bar{\psi}_3\,{\Gamma}\,\psi_2 
\end{equation}
where $\Gamma$ is to be understood as the most general matrix in Dirac space,
\begin{equation}
\Gamma=c_S\, \Gamma_S+c_V\, \Gamma_{V}+c_T\, \Gamma_{T}+c_A\, \Gamma_{A}+ c_P\, \Gamma_P
\end{equation}
where
\begin{equation}
\Gamma_S={\mathrm{1}}\, , \qquad \Gamma_V=\gamma_{\mu}\, , \qquad \Gamma_T=\sigma_{\mu\nu}\doteq\fr{i}{2}[\gamma_{\mu},\gamma_{\nu}]\nonumber
\end{equation}
\begin{equation}
\Gamma_A=\gamma_5\gamma_{\mu}\, , \qquad \Gamma_P=\gamma_5
\end{equation}
Sometimes one may want to express (\ref{fierzone}) as
\begin{equation}
\bar{\psi}_3\,{\Gamma}\,\psi_1\,\bar{\psi}_4\,{\Gamma}\,\psi_2 
\end{equation}
This amounts to a change of basis. Defining the Dirac bilinears as
\begin{eqnarray}
{\cal{O}}_S(i,j;k,l)&\doteq& {\bar{\psi}}_i\psi_j\,{\bar{\psi}}_k\psi_l\nonumber\\
{\cal{O}}_V(i,j;k,l)&\doteq& {\bar{\psi}}_i\gamma_{\mu}\psi_j\,{\bar{\psi}}_k\gamma^{\mu}\psi_l\nonumber\\
{\cal{O}}_T(i,j;k,l)&\doteq& {\bar{\psi}}_i\sigma_{\mu\nu}\psi_j\,{\bar{\psi}}_k\sigma^{\mu\nu}\psi_l\nonumber\\
{\cal{O}}_A(i,j;k,l)&\doteq& {\bar{\psi}}_i\gamma_5\gamma_{\mu}\psi_j\,{\bar{\psi}}_k\gamma_5\gamma^{\mu}\psi_l\nonumber\\
{\cal{O}}_P(i,j;k,l)&\doteq& {\bar{\psi}}_i\gamma_5\psi_j\,{\bar{\psi}}_k\gamma_5\psi_l
\end{eqnarray}
one can write
\begin{equation}
\sum_{M=S,V,T,A,P} {\widehat{c}}_M\, (\Gamma_M)_{ij}\, (\Gamma_M)_{kl}\, {\cal{O}}_M(i,j;k,l)=\sum_{N=S,V,T,A,P} c_N\, (\Gamma_N)_{il}\, (\Gamma_N)_{kj}\, {\cal{O}}_N(i,l;k,j)
\end{equation}
with the coefficients ${\widehat{c}}_M$ given by
\begin{equation}
{\widehat{c}}_M=\Lambda_{MN}\, c_N
\end{equation}
where
\begin{equation}
\Lambda_{MN}=\fr{1}{8}\left (\begin{array}{ccccc}
2 & 8 & 24 & 8 & 2\\
2 & -4 & 0 & -4 & -2\\
1 & 0 & -4 & 0 & 1\\
-2 & -4 & 0 & -4 & 2\\
2 & -8 & 24 & 8 & 2
\end{array}\right)
\end{equation}
is the matrix associated to the change of basis.
\renewcommand{\theequation}{B.\arabic{equation}}
\setcounter{equation}{0}

\chapter[Schwinger operator formalism]{Schwinger operator formalism}
This
formalism \cite{Shuryak} is a very convenient method to do calculations involving covariant derivatives
because it preserves gauge invariance at all stages of the calculation. This is unlike
ordinary diagrammatic perturbation theory where the $\partial_{\mu}$ and the $g_s
G_{\mu}$ sitting in a covariant derivative appear at different orders in the expansion.
 The advantages provided by this technique are difficult to emphasize, since working in coordinate space in a covariant way is especially suited for extracting without much effort the operators and coefficients that build up the OPE. Unlike a Feynman diagram approach, there is no need to make any surgery: the formalism singles out the Wilson coefficient and its associated operator in a most natural way.

\section{The background field method}
We start from the QCD Lagrangian
\begin{equation}
\Lcal_{QCD}=-\fr{1}{4}\,G_{\mu\nu}^{(a)}\,G^{\mu\nu}_{(a)}+{\bar{q}}\,(i\,D\slash-m)\,q
\end{equation}
where
\begin{eqnarray}
D_{\mu}&=&\partial_{\mu}-ig_s\,A_{\mu}^a\,\fr{\lambda^a}{2}\nonumber\\
G_{\mu\nu}^{(a)}&=&\partial_{\mu}G_{\nu}^{(a)}-\partial_{\nu}G_{\mu}^{(a)}+g_s\,f^{abc}G_{\mu}^{(b)}\,G_{\nu}^{(c)}
\end{eqnarray}
Gauge fields are split in the usual way in any background field method
\begin{equation}\label{split}
G_{\mu}(x)={\widehat{G}}_{\mu}(x)+g_{\mu}(x)\, ,\qquad g_{\mu}(x)=\fr{\lambda_a}{2}\,g_{\mu}^{a}(x)
\end{equation}
where ${\widehat{G}}_{\mu} (x)$ is the classical field, which will play the r{\^ o}le of a static background upon which the quantum $g_{\mu}(x)$ will propagate.
The field strength tensor can then be rewritten as 
\begin{equation}
G_{\mu\nu}^{(a)}={\widehat{G}}_{\mu\nu}^{(a)}+D_{\mu}g_{\nu}^a-D_{\nu}g_{\mu}^a+g_sf^{abc}g_{\mu}^b g_{\nu}^c
\end{equation}
where ${\widehat{G}}_{\mu\nu}^{(a)}$ is the classical field strength
\begin{equation}
{\widehat{G}}_{\mu\nu}^{(a)}=D_{\mu}{\widehat{G}}_{\mu}^{(a)}-D_{\nu}{\widehat{G}}_{\nu}^{(a)}+g_sf^{abc}{\widehat{G}}_{\mu}^{(b)}{\widehat{G}}_{\nu}^{(c)}
\end{equation}
and the covariant derivative is redefined to contain only the (strong) background field
\begin{equation}
D_{\mu}=\partial_{\mu}-ig_s{\widehat{G}}_{\mu}
\end{equation}
It is worth emphasising that the background field in the equation above is of gluonic nature because we are interested in the strong corrections. Other interactions can in principle be accounted for once the proper background fields are provided. 
The QCD Lagrangian with the splitting (\ref{split}) can be written
\begin{equation}\label{inicial}
\Lcal_{QCD}=-\fr{1}{4}\bigg({\widehat{G}}_{\mu\nu}^{(a)}+D_{\mu}g_{\nu}^{(a)}-D_{\nu}g_{\mu}^{(a)}+g_s\,f^{abc}\,g_{\mu}^{(b)}\,g_{\nu}^{(c)}\bigg)^2+{\bar{q}}(iD\slash+g_s\,\gamma^{\mu}g_{\mu})q
\end{equation}
\section{The Schwinger formalism}
One starts by defining quantum mechanical operators $\widehat{X}_{\mu},
\widehat{P}_{\nu}$ satisfying the eigenvalue equation
\begin{equation}\label{eigen}
    \widehat{X}_{\mu}|x\rangle=x_{\mu}|x\rangle\ , \qquad {\widehat{P}}_{\mu}|x\rangle=iD_{\mu}|x\rangle
\end{equation}
from where it is immediate to compute the usual commutation relations
\begin{eqnarray}\label{qm}
    \left[\widehat{X}_{\mu},\widehat{P}_{\nu}\right]&=&-i g_{\mu\nu}\nonumber\\
\left[\widehat{P}_{\mu},\widehat{P}_{\nu}\right]&=&ig_s\frac{\lambda^{a}}{2}G_{\mu\nu}^{a}
\end{eqnarray}
One then has that
\begin{equation}\label{definition}
\langle x \vert {\widehat{P}\slash}\vert y\rangle=i{D}\slash_{\!\!\!x}\ \delta{(x-y)}
\end{equation}
where the covariant derivative is understood as acting upon the delta function. With this
definition the quark propagator can be expressed as
\begin{equation}\label{quarkprop}
S_i(k)=\int\,d^4x e^{ik\cdot x}\langle x\vert \frac{1}{\widehat{P}\slash-m_i}\vert
0\rangle=\int\, d^4x \langle x\vert \frac{1}{\widehat{P}\slash+k\slash-m_i}\vert 0\rangle
\end{equation}
where the second equality follows from the relation
\begin{equation}\label{relations}
    e^{ik\cdot\widehat{X}}\widehat{P}_{\nu}=(\widehat{P}_{\nu}+k_{\nu})
    e^{ik\cdot\widehat{X}}
\end{equation}
Analogously, the gluon propagator can be computed from (\ref{inicial}) by conventional means to yield
\begin{equation}\label{gluonprop}
g_{\mu}^{a}(x)g_{\nu}^{b}(y)=\langle x|\left [\fr{-i}{P^2g_{\mu\nu}-2g_sf^{abc}{\widehat{G}}_{\mu\nu}^b}\right]^{ab}|y\rangle
\end{equation}
Following (\ref{eigen}), any quark field $q(x)$ is to be understood as $q(\widehat{X})$, so that one has
\begin{equation}\label{quarkdef}
q(\widehat{X})|y\rangle=q(y)|y\rangle
\end{equation}
\section{Dimension six and dimension eight operators in the weak OPE}
 With these definitions we can apply the formalism to $\Delta S=2$ transitions. As an example of how the method works, we will derive the leading and next to leading order operators in the OPE of the function $W_{LRLR}(q^2)$ as defined in section 5.2. All along we shall regulate divergent integrals in dimensional regularization using
dimensional reduction \cite{Altarelli:1980}. In this regularization scheme one keeps the algebra of Dirac
matrices in four dimensions while the integrals over momenta are performed in $d=4-\varepsilon$ dimensions. Consider the four quark operator
\begin{equation}\label{ssdd}
{\bar{s}}_L(x)\gamma^{\mu}d_L(x)\,{\bar{s}}_L(0)\gamma^{\mu}d_L(0)
\end{equation}
where the quark fields ${\bar{s}}(x)$ and $d(x)$ are to be understood as the full ones, in contrast with the perturbative ones ${\bar{s}}^{(0)}(x)$, $d^{(0)}(x)$. Their relation is the following
 \begin{eqnarray}\label{fullfields}
{\bar{s}}_L(x)&=&{\bar{s}}_L^{(0)}(x)+ig_s\int\,d^4{\tilde{x}}\ {\bar{s}}_L^{(0)}({\tilde{x}})\,G_{\mu}({\tilde{x}})\gamma^{\mu}\,S(x-{\tilde{x}})\nonumber\\
d_L(x)&=&d_L^{(0)}(x)+ig_s\int\,d^4{\tilde{x}}\ S(x-{\tilde{x}})\,G_{\mu}({\tilde{x}})\gamma^{\mu}\,d_L^{(0)}({\tilde{x}})
\end{eqnarray}
Plugging the previous expressions in (\ref{ssdd}) leads to the following expansion
\begin{eqnarray}\label{expan}
\qquad \qquad{\bar{s}}^{(0)}_L(x)\gamma^{\mu}\,d^{(0)}_L(x)&&\!\!\!\!\!\!\!\!\!\!\!\!\!\!\!{\bar{s}}^{(0)}_L(0)\gamma^{\mu}\,d^{(0)}_L(0)+\nonumber\\
+{\bar{s}}^{(0)}_L(x)\,\gamma^{\mu}\bigg[\int\,d^4{\tilde{x}}\,S(x-{\tilde{x}})\,ig_sG_{\mu}({\tilde{x}})\,\gamma^{\mu}\,d^{(0)}_L({\tilde{x}})\bigg]&&\!\!\!\!\!\!\!\!\!\!\!\!\!\!\!{\bar{s}}^{(0)}_L(0)\,\gamma^{\mu}\bigg[\int\,d^4{\tilde{y}}\,S(0-{\tilde{y}})\,ig_sG_{\mu}({\tilde{y}})\,\gamma^{\mu}\,d^{(0)}_L({\tilde{y}})\bigg]+\nonumber\\
+{\bar{s}}^{(0)}_L(x)\,\gamma^{\mu}\bigg[\int\,d^4{\tilde{x}}\,S(x-{\tilde{x}})\,ig_sG_{\mu}({\tilde{x}})\,\gamma^{\mu}\,d^{(0)}_L({\tilde{x}})\bigg]&&\!\!\!\!\!\!\!\!\!\!\!\!\!\!\!\bigg[\int\,d^4{\tilde{y}}\,{\bar{s}}^{(0)}_L({\tilde{y}})\,ig_sG_{\mu}({\tilde{y}})\,\gamma^{\mu}\,S(0-{\tilde{y}})\bigg]\gamma_{\mu}\,d^{(0)}_L(0)+\nonumber\\
+\bigg[\int\,d^4{\tilde{x}}\,{\bar{s}}^{(0)}_L({\tilde{x}})\,ig_sG_{\mu}({\tilde{x}})\,\gamma^{\mu}\,S(x-{\tilde{x}})\bigg]\gamma_{\mu}\,\,d^{(0)}_L(x)&&\!\!\!\!\!\!\!\!\!\!\!\!\!\!\!\bigg[\int\,d^4{\tilde{y}}\,{\bar{s}}^{(0)}_L({\tilde{y}})\,ig_sG_{\mu}({\tilde{y}})\,\gamma^{\mu}\,S(0-{\tilde{y}})\bigg]\gamma_{\mu}\,d^{(0)}_L(0)+\nonumber\\
+\bigg[\int\,d^4{\tilde{x}}\,{\bar{s}}^{(0)}_L({\tilde{x}})\,ig_sG_{\mu}({\tilde{x}})\,\gamma^{\mu}\,S(x-{\tilde{x}})\bigg]\gamma_{\mu}\,d^{(0)}_L(x)&&\!\!\!\!\!\!\!\!\!\!\!\!\!\!\!{\bar{s}}^{(0)}_L(0)\,\gamma^{\mu}\bigg[\int\,d^4{\tilde{y}}\,S(0-{\tilde{y}})\,ig_sG_{\mu}({\tilde{y}})\,\gamma^{\mu}\,d^{(0)}_L({\tilde{y}})\bigg]+\nonumber\\
&+&{\mathcal{O}}(\alpha_s^2)
\end{eqnarray}
The first term is the non-interacting one, whereas the following four lines account for the one-gluon exchange contribution, depicted in figure (5.3). Remember that $G_{\mu}$ can be factorized as a background added to a dynamical gluon field. Plugging in the expressions for the quark (\ref{quarkprop}) and gluon (\ref{gluonprop}) propagators, and using (\ref{quarkdef}) for the quark fields, the second line of (\ref{expan}), {\it{i.e.}}, the first diagram of figure (5.3), reads
\begin{eqnarray}
\int\,d^4x\,e^{iq \cdot x}\int\,d^4{\tilde{x}}&&\!\!\!\!\!\!\!\!\!\!\int\,d^4{\tilde{y}}\,\langle x|\,{\bar{s}}_L^{(0)}\,\gamma_{\mu}\,\fr{i}{P\slash}\,ig_s\gamma^{\alpha}\fr{\lambda^a}{2}\,d_L^{(0)}|{\tilde{x}}\rangle\cdot\nonumber\\
&& \cdot\langle{\tilde{x}}| \left[\fr{-i}{P^2g_{\alpha\beta}-2g_sf_{abc}{\widehat{G}}_{\alpha\beta}^b}\right]^{ac}|{\tilde{y}}\rangle\langle 0|\,{\bar{s}}_L^{(0)}\gamma^{\mu}\,\fr{i}{P\slash}\,ig_s\gamma^{\beta}\fr{\lambda^b}{2}\,d_L^{(0)}|{\tilde{y}}\rangle=\nonumber\\
=\int\,d^4{\tilde{x}}\int\,d^4{\tilde{y}}&&\!\!\!\!\!\!\!\!\!\!\langle x|\,{\bar{s}}_L^{(0)}\,\gamma_{\mu}\,\fr{i}{P\slash+q\slashs}\,ig_s\gamma^{\alpha}\fr{\lambda^a}{2}\,d_L^{(0)}|{\tilde{x}}\rangle\,\langle{\tilde{x}}|\left [\fr{-i}{(P+q)^2g_{\alpha\beta}-2g_sf^{abc}{\widehat{G}}_{\alpha\beta}^b}\right]^{ac}|{\tilde{y}}\rangle\nonumber\\
&&\,\,\,\,\,\,\,\,\,\,\,\, \langle 0|\,{\bar{s}}_L^{(0)}\,\gamma^{\mu}\,\fr{i}{P\slash-q\slashs}\,ig_s\gamma^{\beta}\fr{\lambda^b}{2}\,d_L^{(0)}|{\tilde{y}}\rangle
\end{eqnarray}
where in the second equality (\ref{relations}) has been used. Now, expanding the propagators at large external momentum
\begin{eqnarray}\label{proppp}
\fr{1}{{P\slash}+q\slashs}&=&\fr{1}{q\slashs}\sum_n\left[{P\slash}\fr{1}{q\slashs}\right]^n=\fr{1}{q\slashs}-\fr{1}{q\slashs}{P\slash}\fr{1}{q\slashs}+\fr{1}{q\slashs}{P\slash}\fr{1}{q\slashs}{P\slash}\fr{1}{q\slashs}-\cdots\nonumber\\
\fr{1}{(P+q)^2g_{\alpha\beta}-2g_sf^{abc}{\widehat{G}}_{\alpha\beta}^b}&=&\fr{g_{\alpha\beta}}{q^2}-2\fr{P\cdot q}{q^4}g_{\alpha\beta}+\nonumber\\
&\quad+&\bigg[4\fr{(P\cdot q)^2}{q^6}g_{\alpha\beta}-\fr{P^2}{q^4}g_{\alpha\beta}+\fr{2g_sf^{abc}{\widetilde{G}}_{\alpha\beta}}{q^4}\bigg]+\cdots
\end{eqnarray}
and keeping the ${\mathcal{O}}(P^0)$ terms we end up with the dimension six contribution. Colour-fierzing can be done with the aid of 
\begin{equation}
\left(\fr{\lambda^a}{2}\right)_{ij}\,\left(\fr{\lambda^a}{2}\right)_{kl}=\fr{1}{2}\left(\delta_{il}\,\delta_{jk}-\fr{1}{N_C}\delta_{ij}\delta_{kl}\right)
\end{equation}
while Dirac algebra can be simplified with the following identity (in 4 dimensions)
\begin{eqnarray}
\gamma^{\mu}\gamma^{\nu}\gamma^{\lambda}=g^{\mu\nu}\gamma^{\lambda}-g^{\mu\lambda}\gamma^{\nu}+g^{\nu\lambda}\gamma^{\nu}+i\varepsilon^{\mu\nu\lambda\rho}\gamma_{\rho}\gamma_5\, , \qquad \varepsilon^{0123}=+1
\end{eqnarray}
From the equation above it is straightforward to derive the useful relations
\begin{eqnarray}
\gamma^{\mu}\gamma^{\nu}\gamma^{\lambda}\otimes \gamma_{\mu}\gamma_{\sigma}\gamma_{\lambda}=2\delta^{\nu}_{\ \sigma}\gamma^{\alpha}\otimes\gamma_{\alpha}+2\gamma^{\sigma}\otimes\gamma_{\nu}+2\delta^{\nu}_{\ \sigma}\gamma^{\alpha}\gamma_5\otimes\gamma_{\alpha}\gamma_5-2\gamma^{\sigma}\gamma_5\otimes\gamma_{\nu}\gamma_5\nonumber\\
\gamma^{\mu}\gamma^{\nu}\gamma^{\lambda}\otimes \gamma_{\lambda}\gamma_{\sigma}\gamma_{\mu}=2\delta^{\nu}_{\ \sigma}\gamma^{\alpha}\otimes\gamma_{\alpha}+2\gamma^{\sigma}\otimes\gamma_{\nu}-2\delta^{\nu}_{\ \sigma}\gamma^{\alpha}\gamma_5\otimes\gamma_{\alpha}\gamma_5+2\gamma^{\sigma}\gamma_5\otimes\gamma_{\nu}\gamma_5
\end{eqnarray}
where use has been made of the well-known expression
\begin{equation}
\varepsilon^{\mu\nu\lambda\rho}\varepsilon_{\mu\nu\alpha\beta}=-2(\delta^{\lambda}_{\ \alpha}\delta^{\rho}_{\ \beta}-\delta^{\lambda}_{\ \beta}\delta^{\rho}_{\ \alpha})
\end{equation}
The procedure can be straightforwardly extended to the remaining diagrams of figure (5.3). The extraction of dimension eight operators (${\mathcal{O}}(P^2)$ in (\ref{expan})) is formally identical to the one for dimension six, except for the fact that now one has to deal with covariant derivatives and a bit more complicated Dirac structures. Both points are illustrated in the next section.  
\section{Determination of dimension eight operators in the $1/m_c^2$ expansion}

We concentrate on the box diagram of figure (5.1) with, {\it{e.g.}}, charm-charm exchange. The procedure can be readily exported to the remaining diagrams. The
contribution from this diagram can be obtained in coordinate space from the following
expression:
\begin{eqnarray}
&&\frac{G_F^{2}}{2}M_W^4 \lambda_u^2 \int d^4x\, d^4y\, d^4z\, d^4w\ \langle x\vert
\bar{s} (1+\gamma_5)\gamma^{\mu}\frac{1}{\widehat{P}\slash +q\slashs -
  m_i}\gamma^{\nu}(1-\gamma_5)d \vert y\rangle \nonumber \\
  &&\langle x\vert \frac{1}{q^{2} - M_W^{2}}\vert z\rangle
\langle w\vert\bar{s}(1+\gamma_5)\gamma_{\nu}\frac{1}{\widehat{P}\slash
+q\slashs-m_i}\gamma_{\mu}(1-\gamma_5)d\vert z\rangle\,\langle y\vert \frac{1}{q^{2} -
M_W^{2}}\vert w\rangle
\end{eqnarray}
Integration of the W particle amounts to the expansion :
\begin{equation}
\langle x\vert \frac{1}{q^{2} - M_W^{2}}\vert z\rangle=\frac{-1}{M_W^{2}}\ \delta{(x-z)}+
...
\end{equation}
This has to be done twice. One of the Dirac deltas can be used to reduce the number of
integrals, whereas for the second one we use its representation in momentum space:
\begin{equation}
\delta{(x-y)}=\int \frac{d^4 q}{(2\pi)^4}e^{iq\cdot(x-y)}
\end{equation}
which will eventually give rise to the loop integral.

Now we can expand the quark propagators to any order in the soft momentum
$\widehat{P}\slash$ in the following way:
\begin{eqnarray}
\frac{1}{\widehat{P}\slash + q\slashs -m_i}=\fr{1}{q\slashs-m_i}\sum_n\left[P\slash \fr{1}{q\slashs-m_i}\right]^n&=&\frac{1}{q\slashs
  -m_i}-\frac{1}{q\slashs -m_i}\widehat{P}\slash \frac{1}{q\slashs
  -m_i}+\nonumber\\
&&\!\!\!\!\!\!\!\!\!\!\!\!\!+\,\frac{1}{q\slashs -m_i}\widehat{P}\slash \frac{1}{q\slashs -m_i}\widehat{P}\slash
\frac{1}{q\slashs -m_i}+ \dots
\end{eqnarray}
Recall that this is the massive counterpart of the first line of (\ref{proppp}).
Keeping terms of ${\cal{O}}(P^2)$, since we are interested in dimension-eight operators,
 and after some diracology, we end up with:
\begin{eqnarray}
&&\langle x \vert P_L \left[\frac{1}{\widehat{P}\slash +q\slashs - m_i}\right]P_L\vert 0\rangle
=\langle x \vert P_L \left[\frac{q\slashs}{q^{2} -m_i^{2}}\right]P_L\vert
0\rangle +\nonumber\\
&&\qquad +\ \langle x \vert P_L \left[\frac{\widehat{P}\slash}{q^{2}
-m_i^{2}}-\frac{2(\widehat{P}\cdot q)q\slashs}{(q^{2}
  -m_i^{2})^{2}}\right]P_L\vert 0\rangle+ \\
&&\!\!\!\!\!\!\!\!\!\!\!\!\!\!+\ \langle x \vert P_L \left[\frac{4 q\slashs
(\widehat{P}\cdot
  q)^{2}}{(q^{2} -m_i^{2})^{3}}-\frac{\widehat{P}^{2} q\slashs}{(q^{2} -m_i^{2})^{2}}
  -\frac{(\widehat{P}\cdot q)
  \widehat{P}\slashs}{(q^{2} -m_i^{2})^{2}}-
  \frac{\widehat{P}\slash (\widehat{P}\cdot q)}{(q^{2} -m_i^{2})^{2}}-
  i\frac{\epsilon^{\alpha\beta\gamma\delta}
  \widehat{P}_{\alpha}\widehat{P}_{\beta}q_{\gamma}
  \gamma_{\delta}\gamma_5}{(q^{2} -m_i^{2})^{2}}
  \right]P_L \vert 0\rangle
  \nonumber
\end{eqnarray}
which is the expression to be employed for the quark propagator hereafter.

Now we use (\ref{definition}) to convert $\widehat{P}_{\mu}$ operators into covariant
derivatives. Integration by parts shifts the derivatives to the quark fields and delta
functions can then be integrated in a trivial way. Thus, one can write, for instance:
\begin{eqnarray}\label{oscar}
\int \frac{d^D q}{(2\pi)^D}\int d^4x\, d^4y\, \bar{s}_L(x)\gamma^{\mu}\langle x \vert
  \frac{q\slashs}{q^2-m_i^2}\vert 0 \rangle
  \gamma^{\nu}d_L(0)\bar{s}_L(y)\gamma_{\nu}\langle
  y \vert \frac{\widehat{P}^2 q\slashs}{(q^2-m_i^2)^2}\vert 0 \rangle
  \gamma_{\mu}d_L(0)=&&\nonumber\\
=\bar{s}_L(0)\overleftarrow{{D}^2}\gamma_{\nu}d_L(0)\bar{s}_L(0)\gamma_{\mu}d_L(0)\int
\frac{d^D
  q}{(2\pi)^D}\frac{q^{\mu}q^{\nu}}{(q^2-m_i^2)^3} &&\nonumber\\
=g_s\bar{s}_L(0){\tilde{G}}_{\nu\sigma}\gamma^{\sigma}d_L(0)
\bar{s}_L(0)\gamma_{\mu}d_L(0)\int \frac{d^D
  q}{(2\pi)^D}\frac{q^{\mu}q^{\nu}}{(q^2-m_i^2)^3} \qquad \qquad  &&
\end{eqnarray}
where the Dirac structure has already been simplified and the operator in the second line
has been rewritten in terms of the dual gluon field strength tensor
\begin{equation}
{\tilde{G}}^{\mu\nu}=\frac{1}{2}\varepsilon^{\mu\nu\alpha\beta}G_{\alpha\beta}, \qquad
G_{\mu\nu}=  \frac{\lambda^a}{2}G_{\mu\nu}^a, \qquad \varepsilon^{0123}=+1
\end{equation}
In order to get to the result in Eq. (\ref{oscar}) one must employ the identity
$\overleftarrow{D}^2 = \frac{g_s}{2}\ \sigma_{\mu\nu} G^{\mu\nu}+
{\overleftarrow{D}\slash}^2$, and the equations of motion for the quark fields in the
chiral limit. This will leave us with two operators, one with the gluon field strength and one with its dual. Out of them, the former is chiral-suppressed. Indeed, using
\begin{equation}\label{ident}
    \bar{s}_L{G}^{\alpha\nu}\gamma_{\nu}d_L=\frac{-i}{g}
    \left\{\bar{s}_L\left(D_{\alpha}D\slash- \overleftarrow{D}\slash D_{\alpha}\right)d_L -
    \partial_{\mu}\left( \bar{s}_L \gamma_{\nu} D_{\alpha}d_L\right) \right\}
\end{equation}
one finds that
\begin{equation}\label{ident2}
    \langle \overline{K}^0(p)|\overline{s}_L\gamma^\nu G_{\alpha\nu} d_L|0\rangle =
    \frac{i}{g}\ \partial_{\mu}\langle \overline{K}^0(p)|\overline{s}_L\gamma^\mu
    D_{\alpha}d_L|0\rangle \sim \mathcal{O}(p^2p_{\alpha})
\end{equation}
and, thus, it can be neglected.

Upon performing the integral over $q$ in Eq. (\ref{oscar}) one can easily extract the
$\mu$ dependence. After adding up all the contributions, one obtains the result
(\ref{twelve})-(\ref{twelve-p}) in the main text.



\renewcommand{\theequation}{C.\arabic{equation}}
\setcounter{equation}{0}

\chapter[Bernoulli Numbers and Bernoulli Polynomials]{Bernoulli Numbers and Bernoulli Polynomials}
\section{Bernoulli Polynomials}
A sequence of polynomials $A_n(x)$, where $n$ is a positive integer, is called an {\it{Appell sequence}} if they satisfy the following properties
\begin{equation}
{\textrm{deg}}(A_n)=n
\end{equation}
\begin{equation}
A_n^\prime(x)=n\,A_{n-1}(x)
\end{equation}
The last equation provides a method for recursively determine the full Appell sequence, given an initial $A_0$. Therefore, in general
\begin{equation}
A_n(x)=\sum_{r=0}^n
\left (\begin{array}{c}
n\\
r
\end{array} \right)
a_r\, x^{n-r}
\end{equation}
where the combinatorial numbers are defined as usual
\begin{equation}
\left (
\begin{array}{c}
n\\
r
\end{array} \right)=\fr{\Gamma(n+1)}{\Gamma(r+1)\,\Gamma(n-r+1)}
\end{equation}
As with most sequences, one can define the whole sequence through its generating function, to wit
\begin{equation}
G(x,z)=\sum_{n=0}^{\infty}A_n(x)\,\fr{z^n}{n!}
\end{equation}
Given the above definitions, the Bernoulli polynomials $B_n(x)$ can be defined as the Appell sequence with
\begin{equation}
B_0(x)=1
\end{equation}
which satisfies the additional constraint
\begin{equation}
\int_0^1\,B_n(x)\,dx=0\,\, ,\qquad \qquad  n=1,2,...
\end{equation}
It is relatively easy to show that their generating function is
\begin{equation}
\fr{z}{e^z-1}\,e^{xz}=\sum_{n=0}^{\infty}B_n(x)\,\fr{z^n}{n!}
\end{equation}
Quite notably, Bernoulli polynomials have a reflection symmetry relation 
\begin{equation}\label{refl}
B_n(1-x)=(-1)^n\,B_n(x)
\end{equation}
and a related one
\begin{equation}
B_n(x+1)-B_n(x)=n\,x^{n-1}
\end{equation}
In particular, using the above equation for certain values of $x$ it is easy to conclude that odd Bernoulli polynomials satisfy
\begin{equation}\label{zeros}
B_{2n+1}(x)=0\quad , \quad (n=1,2,...) \quad\qquad  {\mathrm{when}} \quad x=0,\,\fr{1}{2},\,1\end{equation}
Sometimes it is also useful to express Bernoulli polynomials in terms of their trigonometric expansion
\begin{equation}
B_{2n}(x)=(-1)^{n+1}\,2\,(2n)!\,\fr{1}{(2\pi)^{2n}}\sum_{k=1}^{\infty}\fr{\cos{(2\pi kx)}}{k^{2n}}
\end{equation}
\begin{equation}
B_{2n+1}(x)=(-1)^{n+1}\,2\,(2n+1)!\,\fr{1}{(2\pi)^{2n+1}}\sum_{k=1}^{\infty}\fr{\sin{(2\pi kx)}}{k^{2n+1}}
\end{equation}
which follow from the expansion in Fourier series
\begin{equation}
B_n(x)=-\fr{n!}{(2\pi i)^n}\sum_{k=-\infty}^{\infty}k^{-n}\,e^{2\pi ikx}
\end{equation}
The first Bernoulli polynomials read
\begin{eqnarray}
B_0(x)&=&1\nonumber\\
B_1(x)&=&x-\fr{1}{2}\nonumber\\
B_2(x)&=&x^2-x+\fr{1}{6}\nonumber\\
B_3(x)&=&x^3-\fr{3}{2}x^2+\fr{1}{2}x\nonumber\\
B_4(x)&=&x^4-2x^3+x^2-\fr{1}{30}\nonumber\\
B_5(x)&=&x^5-\fr{5}{2}x^4+\fr{5}{3}x^3-\fr{1}{6}x\nonumber\\
B_6(x)&=&x^6-3x^5+\fr{5}{2}x^4-\fr{1}{2}x^2+\fr{1}{42}
\end{eqnarray} 
\section{Bernoulli numbers}
Bernoulli numbers $B_r$ are defined as the coefficients $a_r$ of the Appell sequence for the Bernoulli polynomials. Therefore 
\begin{equation}
B_n(x)=\sum_{r=0}^n
\left (\begin{array}{c}
n\\
r
\end{array} \right)
B_r\, x^{n-r}
\end{equation}
It is easy to show that indeed Bernoulli numbers can be determined from Bernoulli polynomials in a straightforward way, namely
\begin{equation}
B_n=B_n(0)\quad {\mathrm{or\ analogously}} \quad B_n=(-1)^nB_n(1)
\end{equation}
by using the reflection property (\ref{refl}). This justifies the use of the same symbol for both Bernoulli polynomials and Bernoulli numbers. However, sometimes it is also useful to determine the Bernoulli numbers without having to resort to Bernoulli polynomials. The following formulae are then quite useful. Bernoulli numbers can be expressed as an integral of the generating function,
\begin{equation}
B_n=\fr{n!}{2\pi i}\oint\fr{z}{e^z-1}\,\fr{dz}{z^{n+1}}
\end{equation}
as derivatives of the generating function
\begin{equation}
B_n=\lim_{x\rightarrow 0}\,\fr{d^n}{dx^n}\left(\fr{x}{e^x-1}\right)
\end{equation}
or in terms of the Riemann zeta function
\begin{equation}
B_n=(-1)^{n+1}\,n\,\zeta(1-n)
\end{equation}
Recalling (\ref{zeros}), we can immediately conclude that only even Bernoulli numbers are non-trivial, {\it{i.e.}}
\begin{equation}
B_{2n+1}=0\quad , \quad (n=1,2,...)
\end{equation}
For even Bernoulli numbers there exist the following representation
\begin{equation}
B_{2n}=\fr{(-1)^{n-1}\,2\,(2n)!}{(2\pi)^{2n}}\,\zeta(2n)
\end{equation}
whose asymptotic behaviour can be determined readily if one uses the Stirling asymptotic formula
\begin{equation}
B_{2n}\sim (-1)^{n-1}4\sqrt{\pi n}\left(\fr{n}{\pi e}\right)^{2n}
\end{equation}
The first Bernoulli numbers read
\begin{eqnarray}
B_0&=&1\nonumber\\
B_1&=&-\fr{1}{2}\nonumber\\
B_2&=&\fr{1}{6}\nonumber\\
B_4&=&-\fr{1}{30}\nonumber\\
B_6&=&\fr{1}{42}\nonumber\\
B_8&=&-\fr{1}{30}\nonumber\\
B_{10}&=&\fr{5}{66}
\end{eqnarray}

\newpage
\listoffigures

\newpage
\listoftables

\end{document}